\documentclass[11pt]{article}

\input{epsf}
\usepackage{epsfig}
\usepackage{amssymb}
\usepackage{amsfonts}
\usepackage{amsbsy}
\usepackage[all]{xy}
\usepackage{amsmath}
\usepackage{esint}
\usepackage{xcolor}
\definecolor{burgundy}{rgb}{0.5, 0.0, 0.13}
\usepackage[linktocpage=true,colorlinks=true,linkcolor=burgundy,citecolor=black!20!blue,urlcolor=violet]{hyperref}

\usepackage{amssymb,amscd}
\usepackage{mathrsfs}
\usepackage{amsmath,amsthm}
\usepackage{dsfont}
\usepackage{soul}
\usepackage{comment}
\usepackage{cite} 

\def\be{ \begin{equation} }
\def\ee{ \end{equation}}

\def\Co0{{\rm Co}_0}

\def\exp{{\rm exp}}

\def\I{{\rm i}}

\def\log{{\rm log}}

\def\Tr{{\rm Tr}}

\def\p{\partial}

\def\bpsi{\bar{\psi}}

\def\CA{{\cal A}}
\def\CB{{\cal B}}
\def\CC {{\cal C}}
\def\CD {{\cal D}}
\def\CE {{\cal E}}
\def\CF {{\cal F}}
\def\CG {{\cal G}}
\def\CH {{\cal H}}
\def\CI {{\cal I}}
\def\CJ {{\cal J}}
\def\CK {{\cal K}}
\def\CL {{\cal L}}
\def\CM {{\cal M}}
\def\CN {{\cal N}}
\def\CO {{\cal O}}
\def\CP {{\cal P}}
\def\CR {{\cal R}}
\def\CV {{\cal V}}
\def\CW {{\cal W}}

\def\CO {{\cal O}}
\def\CZ {{\cal Z}}
\def\CE {{\cal E}}
\def\CG {{\cal G}}
\def\CH {{\cal H}}
\def\CI {{{\cal I}}}
\def\CB {{\cal B}}
\def\CQ {{\cal Q}}
\def\CS {{\cal S}}
\def\CT {{\cal T}}
\def\CU {{\cal U}}

\def\CY {{\cal Y}}
\def\CZ{{\cal Z}}

\def\IC{\mathbb{C}}

\def\II{\mathbb{I}}

\def\IN{\mathbb{N}}

\def\IP{\mathbb{P}}
\def\IR{{\mathbb{R}}}
\def\IS{{\mathbb{S}}}

\def\IZ{{\mathbb{Z}}}

\def\fa{\mathfrak{a}}
\def\fb{\mathfrak{b}}

\def\fD{\mathfrak{D}}

\def\fg{\mathfrak{g}}

\def\fl{\mathfrak{l}}
\def\fm{\mathfrak{m}}

\def\fq{\mathfrak{q}}
\def\fr{\mathfrak{r}}
\def\fs{\mathfrak{s}}

\def\fq{\mathfrak{q}}
\def\fr{\mathfrak{r}}
\def\fs{\mathfrak{s}}

\def\fB{\mathfrak{B}}

\def\fH{\mathfrak{H}}
\def\fG{\mathfrak{G}}

\def\fJ{\mathfrak{J}}

\def\fV{\mathfrak{V}}
\def\fQ{\mathfrak{Q}}
\def\fX{\mathfrak{X}}

\def\bPhi{{\boldsymbol{\Phi}}}

\def\bpsi{{\boldsymbol{\psi}}}

\def\Dslash{\,{\raise.15ex\hbox{/}\mkern-12mu \CD}}

\def\rmk#1{\bigskip\noindent{\bf Remarks} }

\usepackage{tikz}
\usepackage{tikz-cd}
\usetikzlibrary{arrows}
\usetikzlibrary{arrows.meta}
\usetikzlibrary{positioning}
\usetikzlibrary{shapes,snakes}
\usetikzlibrary{fit}
\usetikzlibrary{decorations.pathmorphing,decorations.pathreplacing,decorations.markings}

\def\lm{\limits}
\def\nn{\nonumber}

\hoffset= - 0.9in

\numberwithin{equation}{section}
\numberwithin{theorem}{section}

\begin{document}

\pagenumbering{Alph} 
\begin{titlepage}
	
	\begin{center}
		\hfill\break
		
		\vspace{2.5cm}
		
		{\bf\Large{On Supersymmetric Interface Defects,\\ \vskip 0.2in Brane Parallel Transport, Order-Disorder Transition \\ \vskip 0.2in and Homological Mirror Symmetry}}
		\vskip 1cm 
		\renewcommand{\thefootnote}{}
		{Dmitry Galakhov\footnote[2]{e-mail: d.galakhov.pion@gmail.com; galakhov@itep.ru}} 
		\vskip 0.2in 
		
		\renewcommand{\thefootnote}{\roman{footnote}}
		
		{\small{ 
				\textit{Kavli Institute for the Physics and Mathematics of the Universe (WPI),\\ }
				\textit{University of Tokyo, Kashiwa, Chiba 277-8583, Japan}\\
				\vskip 0.4 cm 
				\textit{Institute for Information Transmission Problems,\\}
				\textit{ Moscow, 127994, Russia}
		}}
	
	\end{center}
	
	\vskip 0.5in
	\baselineskip 16pt
	
	\begin{abstract}
		We concentrate on a treatment of a Higgs-Coulomb duality as an absence of manifest phase transition between ordered and disordered phases of 2d $\CN=(2,2)$ theories. 
		We consider these examples of QFTs in the Schr\"odinger picture and identify Hilbert spaces of BPS states with morphisms in triangulated categories of D-brane boundary conditions.
		As a result of Higgs-Coulomb duality D-brane categories on IR vacuum moduli spaces are equivalent, this resembles an analog of homological mirror symmetry.
		Following construction ideas behind the Gaiotto-Moore-Witten algebra of the infrared one is able to introduce interface defects in these theories and associate them to D-brane parallel transport functors.
		We concentrate on surveying simple examples, analytic when possible calculations, numerical estimates and simple physical picture behind curtains of geometric objects.
		Categorification of hypergeometric series analytic continuation is derived as an Atiyah flop of the conifold. 
		Finally we arrive to an interpretation of the braid group action on the derived category of coherent sheaves on cotangent bundles to  flag varieties as a categorification of Berry connection on the Fayet-Illiopolous parameter space of a sigma-model with a quiver variety target space. 
		
	\end{abstract}
	
	\date{\today}
\end{titlepage}
\pagenumbering{arabic} 

\newpage
\setcounter{tocdepth}{2}
\tableofcontents
\newpage

\section{Introduction}
\subsection{Supersymmetric interfaces}
Since being discovered in the late 80's mirror symmetry has acquired a lot of attention of both physicist and mathematician communities. 
We are not aiming to give a detailed profound review of this huge topic in the modern string theory referring the reader to canonical literature sources on this subject \cite{D-book_1,D_book_2,Strominger:1996it}, as well as modern reviews of mirror symmetry and Langlands correspondence physical applications \cite{Bullimore:2016nji,Dimofte:2019zzj,Rimanyi:2019zyi,Aganagic:2016jmx,Braverman:2016pwk,Gu:2018fpm,Fan:2007ba}, and references therein.
Instead we would like to narrow our current scope to a relation between homological mirror symmetry as it is understood in algebraic geometry and its physical avatar -- duality of D-brane boundary conditions in 2d $\CN=(2,2)$ supersymmetric theories. 
In practice, homological mirror symmetry \cite{Kontsevich:1994dn} relates certain triangulated categories on a mirror pair of Calabi-Yau manifolds. 
A physical interpretation of this symmetry refers to a duality between observables of different phases of the same theory where the manifolds of the mirror pair represent vacuum moduli spaces.

Our aim is to look at this problem in a perspective of the algebra of the infrared discovered in \cite{GMW} and developed its mathematical counterpart in \cite{Kapranov:2014uwa} (see also \cite{Kapranov:2020zoa,Khan:2020hir,2018arXiv181008776S}). 
A common proposal for physical categorification of various quantities indicates that such geometric objects as cohomology appear in consideration of topologically protected Hilbert subspaces of quantum theories with supersymmetry \cite{Witten:1982im}. 
An approach of \cite{GMW} refers to a canonical consideration of a 2d $\CN=(2,2)$ theory as a quantum system with a Hilbert space of physical states where the evolution is driven by a Hamiltonian evolution operator in the Schr\"odinger picture. 
An alternative approach to $d$-dimensional quantum field theory is the standard path integral approach where the resulting partition function is calculated as a continual integral over maps from a $d$-dimensional world-volume $\CV_d$ to the target space of fields $\CT$. 
To pass to the mechanics of quantum systems -- quantum mechanics -- one chooses a Killing vector in $\CV_d$ as a temporal direction and splits the world-volume $\CV_d=\IR_t\times \CV_{d-1}$, then the quantum mechanics configuration space is given by field maps 
\be\label{maps}
{\rm Map}\left(\CV_{d-1}\to \CT\right).
\ee

In general, $\CV_{d-1}$ could be both non-compact or have boundaries, in either case constraints on the field asymptotic behavior or boundary conditions are in order. 
Naively, admissible boundary conditions form an abstract set, however relations in QFT could produce certain structures on these sets. 
In particular, in 2d $\CN=(2,2)$ theories we will be interested in the boundary conditions will have a structure of a \emph{category}. 
In more general situations the boundary conditions of $d$-dimensional topological field theory are expected to form a $d-1$-category \cite{Carqueville:2016kdq,Leinster,Gukov:2017kmk}.

In the concrete case of 2d $\CN=(2,2)$ theories we are going to consider $\CV_d$ given by a 2d strip, and $\CV_{d-1}$ is a segment of length, say, $L$:
\be\nn
\begin{array}{c}
	\begin{tikzpicture}
		\draw[dashed,fill=blue, opacity=0.4] (0,0) -- (4,0) -- (4,2) -- (0,2) -- cycle;
		\node at (2,1) {interface $p(x^1)$};
		\draw[ultra thick] (0,0) -- (0,2) (4,0) -- (4,2);
		\node[left] at (0,1) {$\CA$};
		\node[right] at (4,1) {$\CB$};
		\draw[<->] (-1,2) -- (-1,-0.5) -- (4.5,-0.5);
		\node[right] at (4.5,-0.5) {$x^1$};
		\node[above] at (-1,2) {$x^0$};
		\draw[ultra thick] (0,-0.4) -- (0,-0.6) (4,-0.4) -- (4,-0.6);
		\node[below] at (0,-0.6) {$0$};
		\node[below] at (4,-0.6) {$L$};
	\end{tikzpicture}
\end{array}
\ee
Maps from 1d spatial segment $[0,L]$ to $\CT$ may be called ``strings'' in $\CT$. 
The standard reasoning \cite{D-book_1} leads to a conclusion that admissible boundary conditions can be reinterpreted as a permission for string ends to move only inside special loci in $\CT$ called D(irichlet)-branes. 
In addition, D-branes will be allowed to carry some complementary information  -- Chan-Paton factors -- that are complexes of vector bundles in this case. 
We will denote the corresponding category of D-branes as $\fD$.

In general, we would like to consider families of theories fibered over parameter space $\CP$ spanned by fugacities and couplings.
In the literature $\CP$ is often called a moduli space when $\CT$ is Calabi-Yau, and fugacity parameters are its K\"ahler moduli. 
To a generic path $\wp$ in the parameter space $\CP$ one is able to associate  parallel transport induced by a Berry connection on a fibration of Hilbert spaces over $\CP$ \cite{Moore:2017byz}, in the case of 2d $\CN=(2,2)$  Landau-Ginzburg models this connection is also known as $tt^*$-connection \cite{Cecotti:1992rm}, in the case of Gromov-Witten theory this connection is also known as a Casimir \cite{tarasov2002duality} or quantum connection \cite{MaulikOkounkov,Okounkov:2016sya}.
A physical counterpart of categorification for the Berry connection  is
an interface defect $\fJ_{\wp}$ preserving a half of the initial supersymmetries. 
Given a family of 2d $\CN=(2,2)$ theories and two supercharges $\CQ$ and $\CQ^{\dagger}$ of the initial $\CN=(2,2)$ supersymmetry that needs to be preserved by an interface, such an observable could be defined for a generic $\wp$ following guidelines of \cite{GMW} (see also Section \ref{sec:loc_GLSM} for details).
The basic idea is to consider the supecharge operator as an integral of a local charge density $\fq$:
\begin{equation}\label{charge_density}
	\CQ=\int d^{d-1}x\,\fq\left(\phi(x),\p\phi(x),p\right)\,,
\end{equation}
depending on fields $\phi$, their derivatives $\p\phi$ and parameter values $p$.
Simply promoting $p$ to a function $p(x)$ of a spacial coordinate defines a new supercharge corresponding to a system with an interface defect inserted.
One could go further and reconstruct first the Hamiltonian from the supersymmetry algebra and then a Lagrangian for a theory with an interface by an inverse Legendre transform.

The category of D-branes $\fD_p$ depends on a choice of point $p\in \CP$. 
If a path $\wp$ interpolates between points $p_1$ and $p_2$ in $\CP$ then we choose D-brane boundary conditions for the left and the right strip edges as objects in corresponding categories:
\be
\CA\in \fD_{p_1},\quad \CB\in\fD_{p_2}.
\ee

Certain topological properties of theories in consideration follow form the superalgebra properties.
We will concentrate on topologically protected states.
These states saturate a Bogomolny-Prasad-Sommerfeld (BPS)  lower bound on the Hamiltonian eigenvalues and are annihilated by two supercharges $\CQ$ and $\CQ^{\dag}$:
\be
\CQ|\Psi_{\rm BPS}\rangle=\CQ^{\dagger}|\Psi_{\rm BPS}\rangle=0.
\ee

For the Hamiltonian $\CH$ eigenvalue we have:
\be
\CH|\Psi_{\rm BPS}\rangle=|\CZ|\cdot|\Psi_{\rm BPS}\rangle,
\ee
where $\CZ$ is a central element of the superalgebra.

The BPS states saturating the BPS bound form a subspace of the Hilbert space of all states we call a BPS Hilbert space $\fG_{\rm BPS}$. 
It depends on all incoming data including $\fJ_{\wp}$ and $\CA$, $\CB$ and, in general, has a grading by various flavor charges preserved by the interface supersymmetry. 
The main observation \cite{GMW,Aganagic:2020olg,aganagic2021knot} we are going to explore and justify throughout this paper is the following relation between interface BPS Hilbert spaces and morphisms in D-brane categories:
\be\label{main}
\boxed{
	\fG_{\rm BPS}^{(*,*,\ldots)}\left(\fJ_{\wp}|\CA,\CB\right)\cong {\rm Hom}_{\fD_{p_2}}^{*,*,\ldots}\left(\beta_{\wp}(\CA),\CB\right)},
\ee
where $\beta_{\wp}$ is a {\bf parallel transport} functor acting in D-brane categories:
$$
\beta_{\wp}:\quad \fD_{p_1}\longrightarrow \fD_{p_2},
$$
satisfying the parallel transport relation:
$$
\beta_{\wp_2}\circ\beta_{\wp_1}=\beta_{\wp_1\circ\wp_2}.
$$
Eigenvalues of physical charges grading the BPS Hilbert space in the l.h.s. of \eqref{main} are identified with various grading degrees of the r.h.s.

Another property we expect from functor $\beta_{\wp}$ is ``flatness'' -- a reflection of the fact that the BPS Hilbert space on the supersymmetry side is protected from small deformations of $\wp$, so that for two homotopic paths $\wp_1$ and $\wp_2$:
$$
\beta_{\wp_1}= \beta_{\wp_2}.
$$

\subsection{Higgs-Coulomb duality, order-disorder transition and mirror symmetry}\label{sec:Intro_mirror}

The quantum mechanics approach to the quantum field theory allows one to treat supercharge operators $\CQ$ and $\CQ^{\dagger}$ following \cite{Witten:1982im} as a differential and its Hodge dual on the cotangent bundle to the space of maps \eqref{maps}.
This observation, in turn, incorporates the QFT framework into the Morse and equivariant localization techniques \cite{Cordes:1994fc} reducing a generic problem of constructing wave-functions to a simplified counting of BPS classical field configurations and only few loop quantum corrections to them. 
The localization techniques allow one to construct physical quantities that are invariant under the action of the renormalization group, in other words those quantities are  independent of the Plank constant $\hbar$, or, alternatively, of the Yang-Mills coupling constant $g_{\rm YM}^2$. 
Therefore one can calculate them in the limit:
$$
\hbar\longrightarrow 0,
$$
where classical field configuration give the major contribution.  See \cite{Pestun:2016qko} for a review of the localization paradigm.

A basic model we will consider in this paper is a 2d $\CN=(2,2)$ supersymmetric gauged linear sigma-model (GLSM) with a matter content encoded in a quiver diagram $\fQ$ \cite{Douglas:1996sw}.  
A quiver is an oriented graph, we denote sets of quiver nodes and quiver arrows as $\fQ_0$ and $\fQ_1$ respectively. 
Nodes of the quiver label gauge multiplets consisting of a gauge connection, a complex scalar, fermion partners and an auxiliary field:
$$
A_v,\quad \sigma_v,\quad \lambda_{\alpha,v},\quad {\bf D}_v,\quad v\in\fQ_0,
$$
arrows label bi-fundamentally charged chiral multiplets consisting of a complex scalar, fermion partners and an auxiliary field:
$$
\phi_a,\quad \psi_{\alpha,a},\quad {\bf F}_a,\quad a\in\fQ_1.
$$
To fix the theory we have to choose parameters $\CP$ that can be thought of as additional elements entering data associated to the quiver. 
So to the quiver nodes $\fQ_0$ one associates the rank of the gauge group $U(n)$, a theta-angle and a Fayet-Illiopolous parameter:
$$
n_v\in \IN,\quad \theta_v\in \IR,\quad r_v\in\IR,\quad v\in \fQ_0
$$ 
To assign to chiral field multiplets non-trivial masses one could use a ``framing'' procedure \cite{nakajima1996varieties}. 
A quiver node with flavor group $U(n_f)$ is claimed to be a framing node, to construct flavor symmetry it suffices to ``freeze'' gauge degrees of freedom to fixed expectation values:
$$
\sigma={\rm diag}(\mu_1,\ldots,\mu_{n_f}),
$$
here $\mu_i$ are complex masses assigned to the chiral multiplet, and the gauge group $U(n_f)$ becomes a flavor symmetry group of the chiral multiplet.

For example, a theory with $n_f$ chiral fields charged fundamentally with respect to the gauge group $U(k)$ can be schematically depicted by the following quiver:
$$
\begin{array}{c}
	\begin{tikzpicture}
		\node at (0,0) {$k$};
		\node at (2,0) {$n_f$};
		\draw (0,0) circle (0.3);
		\begin{scope}[shift={(2,0)}]
			\draw (-0.3,-0.3) -- (-0.3,0.3) -- (0.3,0.3) -- (0.3,-0.3) -- cycle;
		\end{scope}
		\draw[->] (1.7,0) -- (0.3,0);
	\end{tikzpicture}
\end{array}
$$

Models described by quivers with loops allow introducing a gauge-invariant superpotential function $W(\phi_a)$ that is a holomorphic function of complex fields $\phi_a$.

The localization procedure forces quantum field expectation values to approach classical vacua.
Classical vacua form a moduli space -- zero locus of D-term and F-term constraints:
\be\label{D-term}
\begin{split}
	&{\bf D}_v=r_v-\sum\lm_{\substack{w\in \fQ_0\\ (a:v\to w)\in\fQ_1}}\phi_a\phi_a^{\dagger}+\sum\lm_{\substack{w\in \fQ_0\\ (a:w\to v)\in\fQ_1}}\phi_a\phi_a^{\dagger}=0,\quad \forall v\in \fQ_0\\
	&{\bf F}_a=-\p_{\phi_a}W=0,\quad a\in \fQ_1
\end{split}
\ee
modulo the action of the gauge group. 
Due to mixing of the gauge fields $A_v$ and the scalar fields $\sigma_v$ the initial unitary gauge group is complexified:
$$
\prod\lm_{v\in \fQ_0}U(n_v)\longrightarrow \prod\lm_{v\in \fQ_0}GL(n_v,\IC).
$$
An enhancement of a Riemann manifold structure of the moduli space to a complex variety structure of stable quiver representations is usually referred to as Narasimhan-Seshadri-Hitchin-Kobayashi correspondence \cite{donaldson1983new}.

The complex gauge field delivers a complex equivariant action on the moduli space with associated Killing vector fields:
$$
V\sim\sum\lm_{(a:v\to w)\in\fQ_1}\Tr\left[(\sigma_v\phi_a-\phi_a\sigma_w)\frac{\p}{\p q_a}\right].
$$

Equivariant localization demands the points of the moduli spaces contributing to the vacua to be fixed points of $V$.
This constraint in terms of field expectation values has two generic solutions usually referred to as a Higgs branch and a Coulomb branch in the literature \cite{deBoer:1996mp}:
\be\label{branches}
\mbox{Higgs}:\; \langle \phi\rangle\neq0,\; \langle\sigma\rangle=0,\quad \mbox{Coulomb}:\; \langle \phi\rangle=0,\; \langle\sigma\rangle\neq 0,
\ee
we will also identify these branches to physical phases of the theory as in \cite{Hori:2013ika,Witten:1993yc}.
Surely, in practice one may encounter all sorts of mixed branches.

\begin{figure}[h!]
	\begin{center}
		\begin{tikzpicture}
			\draw[<->,thick] (0,3.2) -- (0,0) -- (6.2,0);
			\draw[fill=white!80!blue] (0,0.5) -- (6,0.5) -- (6,3) -- (0,3) -- cycle;
			\begin{scope}[shift={(0,3)}]
				\begin{scope}[xscale=0.8]
					\draw[fill=white!80!green] (0,0) -- ([shift=(270:2.5)]0,0) arc (270:360:2.5) -- cycle;
				\end{scope}
			\end{scope}
			\begin{scope}[shift={(6,3)}]
				\begin{scope}[xscale=-0.8]
					\draw[fill=white!80!green] (0,0) -- ([shift=(270:2.5)]0,0) arc (270:360:2.5) -- cycle;
				\end{scope}
			\end{scope}
			\draw[fill=white!80!gray] (0,0) -- (0,0.8) -- (6,0.8) -- (6,0) -- cycle;
			\draw[thick] (0,0.8) -- (6,0.8);
			\begin{scope}[shift={(0,3)}]
				\begin{scope}[xscale=0.8]
					\draw[thick] ([shift=(270:2.5)]0,0) arc (270:360:2.5);
				\end{scope}
			\end{scope}
			\begin{scope}[shift={(6,3)}]
				\begin{scope}[xscale=-0.8]
					\draw[thick] ([shift=(270:2.5)]0,0) arc (270:360:2.5);
				\end{scope}
			\end{scope}
			\draw[dashed] (3,0) -- (3,3);
			\node at (3,2.5) {Coulomb};
			\node at (1,2.5) {Higgs I};
			\node at (5,2.5) {Higgs II};
			\node at (3,0.4) {???};
			\node[right] at (6.2,0) {$r$};
			\node[above] at (0,3.2) {$\hbar^{-1}$, $g_{\rm YM}^{-2}$};
			\node[left] at (0,0.8) {$\hbar_0^{-1}$};
			\node[left] at (0,0.4) {\color{orange} strong $\downarrow$};
			\node[left] at (0,1.9) {\color{orange} weak $\uparrow$};
			\draw[purple,<->,thick,>=stealth'] (2,1) -- (1,2);
			\draw[purple,<->,thick,>=stealth'] (4,1) -- (5,2);
			\node[below left,purple] at (1.5,1.5) {\tiny cHC duality};
			\node[below right,purple] at (4.5,1.5) {\tiny cHC duality};
			\node[below] at (3,0) {$0$};
		\end{tikzpicture}
		\caption{Simple gauged sigma-model phase diagram.}\label{fig_phase}
	\end{center}
\end{figure}
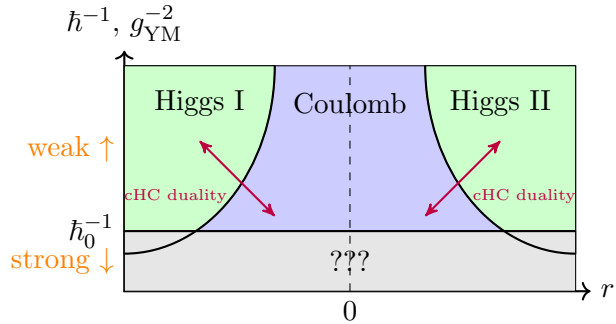

Despite both the Higgs and Coulomb branches solve apparently the fixed point constraint they can not simultaneously solve the D-term constraint \eqref{D-term}. 
If the FI parameters $r$ are non-zero we see the chiral field has to acquire an expectation value of order $\sim\sqrt{r}$, therefore the Higgs branch is realized in the IR. 
On the other hand, if $r=0$ there is no vev for chiral fields, therefore the Coulomb branch is realized in the IR. 
From the practical point of view there is no need to put strictly $r=0$ to get the Coulomb branch, rather it suffices to make it small of the order of quantum fluctuations $\sim \hbar^{\frac{1}{2}}$. 
To mark the weak coupling region contingently let us assume:
$$
\hbar<\hbar_0\ll 1.
$$
In this region depending if $|r|>\sqrt{\hbar}$ or $|r|<\sqrt{\hbar}$ the BPS Hilbert space is described effectively in the IR by either the Higgs branch or the Coulomb branch.
So that one can depict the phase diagram as in Fig.\ref{fig_phase}.

Consider a path in $\hbar$ connecting Higgs branch and Coulomb branch regions as in Fig.\ref{fig_phase}.
According to \cite{Witten:1982im} (see also Section \ref{sec:Heis_loc} for details) the BPS Hilbert spaces for different values of $\hbar$ are isomorphic.
We promote this isomorphism to an isomorphism between Higgs and Coulomb branch descriptions:
\be\label{HC_du}
\fG_{\rm BPS,\; Clmb}\cong\fG_{\rm BPS,\; Higgs}\,.
\ee
This relation is a direct analog of a Coulomb-Higgs duality for theories in other dimensions (see e.g. \cite{Manschot:2013sya,deBoer:1996mp}).

We should stress that our semi-classical description of branches \eqref{branches} is drastically affected by the quantum corrections.
If quiver $\fQ$ is chiral -- implying that for some pair of nodes the amount of arrows flowing in one direction is not equal to the number of arrows flowing in the opposite direction excluding arrows starting and ending on the same node -- the resulting theory might have an anomaly in the axial symmetry $U(1)_A$ (see \cite[Section 4.2]{HHP}).
Simultaneously, in the case of chiral quivers the FI parameters $r_v$ run under the RG flow.
Due to these both effects $r_v$ (or, more generally, complexified parameters $t_v=r_v-\I\theta_v$) are not good parameters for the theory.
In the cases considered in Sections \ref{sec:cat_contin_hyper} and \ref{sec:quiver} we will apply the reasoning presented in this section to non-chiral quivers only.
On the contrary, the toy model of Section \ref{sec:equiv_CP_1} is going to be described by a chiral quiver.
In the latter case we work with an extreme regime ${\rm Re}\,t\gg 0$ where Higgs and Coulomb branch overlap regardless corrections, relation \eqref{HC_du} holds, and we will not consider traveling across the parameter space.

Let us note that the Higgs and Coulomb branch descriptions are quite different. 
The Coulomb branch description reduces to an effective field theory of twisted chiral superfield $\Sigma=\sigma+\ldots$. 
2d $\CN=(2,2)$ twisted chiral multiplet is quite similar \cite{Hori:2000kt} to a chiral multiplet, only some fermion variables of opposite chiralities should be swapped, so this description is equivalent to a Landau-Ginzburg model of a chiral field spanning the IR target space $X_{\rm Clmb}$. 
In the Higgs branch description when FI parameters are non-zero the D-term constraint can be substituted by a stability condition \cite{donaldson1983new}. 
So that the Higgs branch $X_{\rm Higgs}$ is given by a stable quiver representation moduli space \cite{Alim:2011kw}.

Returning back to \eqref{main} we see that the Higgs-Coulomb duality applied to 2d GLSM provides two equivalent descriptions of the categories of the D-brane boundary conditions.

A D-brane category for the Landau-Ginzburg theory is usually identified with a Fukaya-Seidel category of $X_{\rm Coulomb}$, in general, with a superpotential $W$, and boundary conditions in the GLSM are identified with a derived category of coherent sheaves on $X_{\rm Higgs}$, thus we will arrive to the following form of the Higgs-Coulomb duality for BPS Hilbert spaces:
\be\label{mirror}
\mbox{Fukaya-Seidel}(X_{\rm Clmb},W)\cong
D^b{\rm Coh}(X_{\rm Higgs})\,.
\ee

In the literature \cite{Hori:2000kt} authors also use an ``intermediate'' model description of the Landau-Ginzburg theory in addition to twisted chiral field $\Sigma$ considering a twisted chiral field traditionally denoted $Y$. 
We will identify the chiral field $\phi$ with an \emph{order} parameter on the Higgs branch. 
Then the dual \emph{disorder} parameter field representing an insertion of a vortex singularity gives rise in a supersymmetric theory to a twisted chiral field $Y$. 
This duality is direct analog of an order-disorder transition appearing in the 2d Ising model \cite{Polyakov:1987ez}. 

The order-disorder duality is also related to T-duality. 
Applying T-duality \cite{Strominger:1996it,Ooguri:1996ck} one also exchanges twisted chiral fields in a B-twisted model to chiral fields in an A-twisted model \cite{Hori:2000kt}.
\emph{Mirror symmetry} relates B-twisted effective non-linear sigma-model of order parameter $\phi$ to an A-twisted Landau-Ginzburg model of disorder parameter $Y$.
A further elaborated analog of relation \eqref{mirror} for triangulated categories on corresponding varieties is usually referred to as \emph{homological mirror symmetry} \cite{Kontsevich:1994dn,Kontsevich:2008fj,Gukov:2010sw,2008arXiv0811.1228F}.

To return to the original model of fields $\Sigma$ one has to integrate out fields $Y$ to an effective theory in the IR, however this operation comes with its price.
In general, integration produces singularities in the effective action with a possibility to spoil the homotopy properties of interface observables and mirror symmetry.
These obstructions can be resolved either using a model with a UV completed superpotential without singularities -- in other words, a superpotential with both fields $\Sigma$ and $Y$ (see \cite[Section 7]{Galakhov:2016cji}) -- or dualizing the theory back to gauged sigma-model of field $\phi$.
Fortunately, in our examples we will not encounter these obstructions directly, so we leave them beyond the scope of this paper and will return to this issue elsewhere.
Nevertheless, keeping in mind this obstacle and the fact that $X_{\rm Clmb}$ and $X_{\rm Higgs}$  are not ideal mirrors in our construction in full generality we will refer to relation \eqref{mirror} as a \emph{categorified Higgs-Coulomb duality}, or simply a cHC duality for brevity.

\subsection{Decategorification, categorification and parallel transport}\label{sec:dec_cat_par}

A concept simpler than a triangulated category $\CC$ capturing its partial behavior is its 
Grothendieck group $K_0(\CC)$. 
A transition from $\CC$ to $K_0(\CC)$ may be called a \emph{decategorification} process, and the inverse process is usually referred to as a \emph{categorification} process \cite{Chun:2015gda}. 
Physically, a relation between a category and its Grothendieck group is similar to a relation between the graded BPS Hilbert space and its supersymmetric index:
\be
\begin{array}{c}
	\begin{tikzpicture}
		\node(A)[left] at (0,0) {$\fG_{\rm BPS}^{\left(\CF,J_1,J_2,\ldots\right)}$};
		\node(B)[right] at (4,0) {$\Tr_{\fG_{\rm BPS}}(-1)^{\CF}\cdot y_1^{J_1}\cdot y_2^{J_2}\cdot\ldots$};
		\draw[-stealth] (0.2,0.06) -- (3.8,0.06);
		\draw[stealth-] (0.2,-0.06) -- (3.8,-0.06);
		\node[above] at (2,0.06) {\tiny decategorification};
		\node[below] at (2,-0.06) {\tiny categorification};
	\end{tikzpicture}
\end{array},
\ee
where $\CF$ is a fermion number, $J_i$ are charges commuting with the supercharges and $y_i$ are introduced fugacities.
Apparently, the index does not distinguish a contribution of a boson-fermion pair of BPS states with other identical quantum numbers. 

The supersymmetric index can be re-interpreted in the path integral formulation as a partition function.
To do so we substitute the temporal dimension $\IR_t$ with a thermal circle $S^1_{\rm thrm}$, apply the Wick rotation to the theory, the fermionic fields are periodic on the thermal circle $S^1_{\rm thrm}$. 
The resulting world-volume manifold is $\CV^{\rm E}_d=S_{\rm thrm}^1\times \CV_{d-1}$. 
In our case $\CV_2^{\rm E}$ is a cylinder of width $L$.

Let us make  a brief digression and review properties of partition functions of theories in question. 
We start with a partition function on a disk. 
To identify this partition function $Z$ we have to choose some parameter values $p\in\CP$ and a D-brane boundary condition $\CA\in\fD_p$. 
Mathematically $Z$ forms a functor from the category of D-branes to a vector space -- its Grothendieck group:
\be
Z:\quad \fD_p\longrightarrow K_0\left(\fD_p\right)
\ee

The disk partition function can be calculated \cite{Hori:2013ika} for various choices of boundary conditions with the help of localization techniques.
Again, one has two options to localize on either the Higgs or the Coulomb branch, therefore as we mentioned in the previous section there are two dual choices of categories:
$$
\fD_p^{\rm Clmb},\quad \fD_p^{\rm Higgs},
$$
corresponding to the Fukaya-Seidel category and the derived category of coherent sheaves as well as two partition functions. 

The Coulomb branch partition function is a partition function of a Landau-Ginzburg theory with a superpotential $W$. The D-branes are represented  by Lagrangian loci in $X_{\rm Clmb}$. 
The corresponding expression reads:
\be\label{Clmb_br_pf}
Z_{\rm Clmb}[p,\CA_{\rm Clmb}]=\int\lm_{\CA_{\rm Clmb}}\Omega\; e^{2\pi\I\; W},\quad \CA_{\rm Clmb}\in\fD_p^{\rm Clmb}.
\ee
where $\Omega$ is  a holomorphic top form.

On the Higgs branch the effective theory is a sigma-model with the target space given by $X_{\rm Higgs}$, the D-brane boundary conditions correspond to a choice of a complex of holomorphic vector bundles, or, in a more precise mathematical formulation, of a coherent sheaf $\CA_{\rm Higgs}$. 
The corresponding partition function reads (see also \cite{Honda:2013uca}):
\be
Z_{\rm Higgs}[p,\CA_{\rm Higgs}]=\int\lm_{X_{\rm Higgs}} e^{B+\frac{\I}{2\pi}\omega} {\rm ch}\left(\CA_{\rm Higgs}\right)\sqrt{{\rm Td}\left(X_{\rm Higgs}\right)},\quad \CA_{\rm Higgs}\in\fD_p^{\rm Higgs},
\ee
where $B$ and $\omega$ are a B-field and a K\"ahler form correspondingly, $\rm Td$ denotes the Todd class, and ${\rm ch}(*)$ is a Chern class. 

In these terms Higgs-Coulomb duality manifests itself as an equality between Higgs and Coulmb partition functions:
\be
Z_{\rm Clmb}[p,\CA_{\rm Clmb}]\mathop{\scalebox{6}[1]{$=$}}\lm^{\rm HC} Z_{\rm Higgs}[p,\CA_{\rm Higgs}],
\ee
for some choices of branes $\CA_{\rm Clmb}$ and $\CA_{\rm Higgs}$. 
The categories we will consider are triangulated, in certain cases one can choose a ``\emph{basis}'' among exceptional category objects, so that all the objects can be reconstructed by degree shifts and cones \cite{Seidel_book, gorodentsev2004helix}. 
Similarly, a basis can be chosen in the vector space $K_0(\fD)$. 
It is rather simple to observe this phenomenon in the case of the partition function on the Coulomb branch. 
Indeed according to \eqref{Clmb_br_pf} $Z_{\rm Clmb}$ is a holomorphic integral and depends only on the homotopy class of $\CA_{\rm Clmb}$ preserving the asymptotic behavior of the integrand so that the integral converges. 
A good choice of a basis of such integration cycles is Lefschetz thimbles that are in one-to-one correspondence with critical points of $W$ if $W$ is Morse. 
On the other hand, $Z_{\rm Higgs}$ is resembling a matrix integral with an inserted operator defined by $\CA_{\rm Higgs}$. 
Such a set of ``interesting'' insertions is finite in examples we will consider, has a linear structure and, therefore, a basis.
Moreover both are expected to be flat sections of the Berry connection on the parameter space $\CP$ so that $Z_{\rm Clmb}[p,\CA_{\rm Clmb}]$ and $Z_{\rm Higgs}[p,\CA_{\rm Higgs}]$ are solutions of the same set of differential equations.

Over a generic points $p$ of $\CP$ we expect $K_0(\fD_p)$ to be of some fixed dimension. 
Therefore on $\CP$ minus some singular loci $K_0(\fD_p)$ forms a vector bundle. 
We would like to consider a \emph{parallel transport} induced by the Berry connection of a vector $Z$ along a path $\wp$ from some point $p_1$ to a new point $p_2$. 
Geometrically, we employ the topological properties of the theory and bend a world-sheet disk in such a way that it resembles a vial, then extend its neck. 
We assume that parameters are slowly varying along this neck, in other words we insert interface defect $\fJ_\wp$ into this neck. 
One can cut the resulting world-sheet into  a smaller vial and a long neck cylinder. 
Accordingly, we glue the resulting disk partition function out of a smaller disk partition function and the interface partition function:
\be\nn
\begin{split}
	\begin{array}{c}
		\begin{tikzpicture}
			\begin{scope}[xscale=0.5]
				\draw circle (0.3);
			\end{scope}
			\draw (0,0.3) to[out=180,in=90] (-1,0) to[out=270,in=180] (0,-0.3);
			\begin{scope}[shift={(0.5,0)}]
				\begin{scope}[xscale=0.5]
					\draw circle (0.3);
				\end{scope}
			\end{scope}
			\begin{scope}[shift={(2.5,0)}]
				\begin{scope}[xscale=0.5]
					\draw circle (0.3);
				\end{scope}
			\end{scope}
			\draw (0.5,0.3) -- (2.5,0.3) (0.5,-0.3) -- (2.5,-0.3);
			\draw[dashed] (0,0.3) -- (0.5,0.3) (0,-0.3) -- (0.5,-0.3);
			\begin{scope}[shift={(6,0)}]
				\begin{scope}[xscale=0.5]
					\draw circle (0.3);
				\end{scope}
				\draw (0,0.3) to[out=180,in=90] (-1,0) to[out=270,in=180] (0,-0.3);
			\end{scope}
			\draw[->] (3,0) -- (4.7,0);
			\node[above] at (-0.1,0.3) {$\CA$};
			\node[above] at (0.6,0.3) {$\CA$};
			\node[above] at (2.5,0.3) {$\CA'$};
			\node[above] at (6,0.3) {$\CA'$};
			\node[below] at (1.5,-0.3) {$\fJ_\wp$};
			\node[below] at (0.25,-0.3) {$p_1$};
			\node[below] at (2.5,-0.3) {$p_2$};
			\node[below] at (6,-0.3) {$p_2$};
		\end{tikzpicture}
	\end{array}\\
	Z[p_2,\CA']=\sum\lm_{\CA}Z_{\rm intf}[\fJ_{\wp}|\CA,\CA']Z[p_1,\CA].
\end{split}
\ee
The role of the interface partition function -- Witten index -- $Z_{\rm intf}$ is a linear map between fibers of the partition function bundle associated to the parallel transport along $\wp$:\footnote{The careful reader, especially familiar with the phenomenon of wall-crossing \cite{Kontsevich:2008fj,Aganagic:2009kf,Gaiotto:2008cd} may argue that dimensions entering this expression are only piece-wise constant functions of parameters $\CP$, therefore the parallel transport will not be a parallel transport on a differentiable bundle governed by some connection. 
	This is not the case. 
	Indeed we are able to identify the categories and calculate some dimensions of the BPS Hilbert spaces by localization only in the IR limit $\hbar\to 0$, for partition functions this limit corresponds to a semi-classical or WKB limit, or a calculation of their asymptotic behavior. 
	Asymptotic develops natural discontinuities on certain loci of the parameter space even when the very function at all finite values of the parameters is continuous, this phenomenon is known as a Stokes phenomenon.}
\be
\begin{split}
	Z_{\rm intf}[\fJ_{\wp}|\CA,\CB]=\Tr_{\fG_{\rm BPS}}(-1)^{\CF}=&\sum\lm_{\CF}(-1)^{\CF}{\rm dim}\,\fG_{\rm BPS}^{\left(\CF,\ldots\right)}\left(\fJ_{\wp}|\CA,\CB\right)=\\
	&\sum\lm_{\CF}(-1)^{\CF}{\rm dim}\,{\rm Hom}_{\fD_{p_2}}^{\CF,\ldots}\left(\beta_{\wp}(\CA),\CB\right).
\end{split}
\ee
Therefore we call the functor $\beta_{\wp}$ \emph{categorified parallel transport} functor. Obviously, the only partition function does not have enough information to calculate this functor, this is why we are aiming to exploit QFT machinery to calculate explicitly $\fG_{\rm BPS}$ and to apply this information to a restoration of the $\beta_{\wp}$ action on categories of D-branes.

The description of parallel transport in terms of Grothendieck groups we presented so far is universal, however by lifting it to the level of categories we will end up in either a Fukaya-Seidel category or a derived category of coherent sheaves. 
Similarly, in the literature \cite{HHP,GMW,Khan:2020hir,Galakhov:2017pod,Kerr:2017usa,Clingempeel:2018iub,Brunner:2020miu,Brunner:2021cga,Chen:2020iyo,Brunner:2009zt,Brunner:2008fa,Aspinwall:2004jr,Knapp:2016rec,Erkinger:2017aaa,Knapp:2020oba} in the process of calculating the parallel transport preferences are devoted mostly to one side of the duality. 
Correspondingly, the practical approaches are also quite different -- the algebra of the infrared (see Section \ref{sec:Web} for a review) for Landau-Ginzburg theories and the Fourier-Mukai transform (see Appendix \ref{sec:categories} for a review) for GLSMs.
In this note we will make an attempt to synthesize the best options from the two worlds and apply techniques of the algebra of the infrared \cite{GMW} to calculate \emph{Fourier-Mukai kernel} for associated categorified parallel transport functor $\beta_{\wp}$.

\subsection{Atiyah flop and hypergeometric series}\label{sec:Atiyah_intro}
We would like to concentrate on a simple and yet non-trivial model of conifold transition.
Consider a $U(1)$ gauged sigma model with $n_f$ chirals. 
We assume that chirals have charges $Q_i$ under this $U(1)$ and masses $\mu_i$ where the index runs over a set $i=1,\ldots,n_f$. 
For the Higgs branch disk partition function the localization calculation produces the following result:
\be\label{Z_B}
Z[t,\CE]=\int\lm_{-\I\infty}^{\I\infty}d\sigma \; f_{\CE}\left(e^{2\pi\I \sigma}\right)\; e^{t\sigma}\prod\lm_{i=1}^{N_f}\Gamma(Q_i\sigma+\mu_i),
\ee
where
$$
t=r-\I\theta
$$
is a complexified combination of Fayet-Illiopolous parameter $r$ and theta-angle $\theta$, and $f_{\CE}(z)$ is a polynomial in $z$ defining so called D-brane data of a coherent sheaf $\CE\in\fD_{p}$. This expression is a well-known Barnes representation for generalized hypergeometric series (see Appendix \ref{sec:hypergeom}):
$$
{}_pF_q\left[\begin{array}{c}
	a_1,\ldots,a_p\\
	b_1,\ldots,b_q
\end{array}\right](z),\quad \mbox{where}\; p=\sum\lm_{i:\;Q_i>0}|Q_i|,\;q=\sum\lm_{i:\;Q_i<0}|Q_i|-1,
$$
where $z=e^{2\pi t}$ and parameters $a_i$ and $b_i$ are linearly related to masses
$\mu_i$. In other words partition function \eqref{Z_B} is annihilated by a differential operator:
\be
D_{p,q}\left(z,\frac{d}{dz}\right)=z\prod\lm_{i=1}^p\left(z\frac{d}{dz}+a_i\right)-z\frac{d}{dz}\prod\lm_{i=1}^q\left(z\frac{d}{dz}+b_i-1\right)\,.
\ee
This differential operator has a mathematical interpretation of a quantum connection in Gromov-Witten theory \cite{MaulikOkounkov}, and the physical interpretation is an induced Berry connection on the parameter space. 
The parallel transport induced by the supersymmetric interfaces is compatible with this connection.

The simplest non-trivial case is the ordinary hypergeometric series ${}_2F_1$. 
The solutions of the hypergeometric equations are known to develop singularities in three marked points $0$, $1$ and $\infty$ on $\CP$ that is a Riemann sphere parameterized by $z$ in this case. 

Corresponding theory is $U(1)$ gauged sigma-model with $n_f=4$ chiral multiplets with charges:
$$
(+1,+1,-1,-1)\,.
$$
This model describes a conifold resolution. The conifold is a singular hypersurface in $\IC^4$ defined by an algebraic equation as a set of degenerate 2 by 2 $\IC$-valued matrices:
\be\label{S-mat}
S:=\left(\begin{array}{cc}
	M_{11} & M_{12}\\
	M_{21} & M_{22}
\end{array}\right),\quad {\rm det}\;S=M_{11}M_{22}-M_{12}M_{21}=0\,.
\ee
To resolve it we rewrite $\IC^4$ coordinates $M_{ij}$ in terms of chiral fields:
\be
M_{11}=\phi_2\phi_4,\quad M_{12}=-\phi_1\phi_4,\quad M_{21}=-\phi_2\phi_3,\quad M_{22}=\phi_1\phi_3.
\ee
This assignment is invariant under the action of the complexified gauge group $\IC^{\times}$: 
$$
(\phi_1,\phi_2,\phi_3,\phi_4)\to \left(\lambda \phi_1,\lambda \phi_2,\lambda^{-1} \phi_3,\lambda^{-1} \phi_4\right).
$$
Consider an obvious consequence of this identification:
\be\label{res_+}
\left(\begin{array}{cc}
	M_{11} & M_{12}\\
	M_{21} & M_{22}
\end{array}\right)\left(\begin{array}{c}
	\phi_1\\ \phi_2
\end{array}\right)=0.
\ee
If $|\phi_1|^2+|\phi_2|^2>0$ then a pair $(\phi_1:\phi_2)$ modulo the action of $\IC^\times$ can be associated with a projective coordinate of $\IC\IP^1$. 
In this case \eqref{res_+} represents a canonical \emph{blowup} resolution of the conic singularity \cite{Hartshorne}. 
The D-term constraint in this theory reads:
\be
|\phi_1|^2+|\phi_2|^2-|\phi_3|^2-|\phi_4|^2=r.
\ee
In the case $r>0$ the moduli space is exactly confined to a locus $|\phi_1|^2+|\phi_2|^2>0$, we call this blowup resolution $X_+$ so that:
\be
X_{\rm Higgs}(r>0)=:X_+\,.
\ee

Another possible blowup resolution of the conic singularity is as follows:
\be\label{res_-}
\left(\begin{array}{cc}
	\phi_3 & \phi_4
\end{array}\right)\left(\begin{array}{cc}
	M_{11} & M_{12}\\
	M_{21} & M_{22}
\end{array}\right)=0,\quad (\phi_3:\phi_4)\in \IC\IP^1\,.
\ee
This resolution corresponds to the region $r<0$ and we call this variety $X_-$ respectively:
\be
X_{\rm Higgs}(r<0)=:X_-\,.
\ee

The most intriguing situation is $r=0$, in this case the conifold remains unresolved. 
However from the physical point of view as we discussed in the previous section it is not correct to talk about the Higgs branch in $\hbar$-neighborhood of point $r=0$, in this area the theory is in its Coulomb phase.

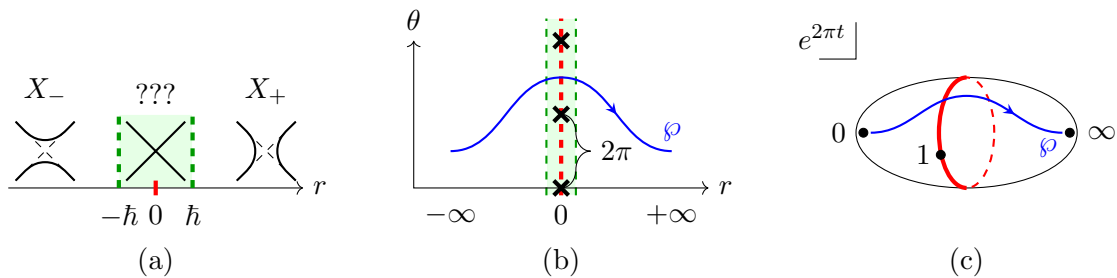
\begin{figure}[h!]
	\begin{center}
		\begin{tikzpicture}[scale=0.98]
			\begin{scope}[shift={(-2,-0.5)}]
				\draw[white, fill=green, opacity=0.1] (1.5,0) -- (1.5,1) --(2.5,1) -- (2.5,0) -- cycle;
				\draw[->] (0,0) -- (4,0);
				\node[right] at (4,0) {$r$};
				\draw[dashed, ultra thick, black!40!green] (1.5,0) -- (1.5,1);
				\draw[dashed, ultra thick, black!40!green] (2.5,0) -- (2.5,1);
				\draw[ultra thick, red] (2,-0.1) -- (2,0.1);
				\node[below] at (2,-0.1) {$0$};
				\node[below] at (1.5,-0.1) {$-\hbar$};
				\node[below] at (2.5,-0.1) {$\hbar$};
				\begin{scope}[shift={(2,0.5)}]
					\draw[thick] (-0.4,-0.4) -- (0.4,0.4) (0.4,-0.4) -- (-0.4,0.4);
				\end{scope}
				\begin{scope}[shift={(0.5,0.5)}]
					\draw[dashed] (-0.4,-0.4) -- (0.4,0.4) (0.4,-0.4) -- (-0.4,0.4);
					\draw[thick] (-0.4,0.4) to[out=315,in=180]  (0,0.15) to[out=0,in=225] (0.4,0.4);
					\draw[thick] (-0.4,-0.4) to[out=45,in=180]  (0,-0.15) to[out=0,in=135] (0.4,-0.4);
				\end{scope}
				\begin{scope}[shift={(3.5,0.5)}]
					\begin{scope}[rotate=90]
						\draw[dashed] (-0.4,-0.4) -- (0.4,0.4) (0.4,-0.4) -- (-0.4,0.4);
						\draw[thick] (-0.4,0.4) to[out=315,in=180]  (0,0.15) to[out=0,in=225] (0.4,0.4);
						\draw[thick] (-0.4,-0.4) to[out=45,in=180]  (0,-0.15) to[out=0,in=135] (0.4,-0.4);
					\end{scope}
				\end{scope}
				\node[above] at (0.5,1) {$X_-$};
				\node[above] at (3.5,1) {$X_+$};
				\node[above] at (2,1) {???};
			\end{scope}
			\begin{scope}[shift={(3.5,-0.5)}]
				\draw[white, fill=green, opacity=0.1] (2.2,-0.1) -- (2.2,2.3) -- (1.8,2.3) -- (1.8,-0.1) -- cycle;
				\draw[black!40!green,dashed,thick] (2.2,-0.1) -- (2.2,2.3) (1.8,-0.1) -- (1.8,2.3);
				\draw[<->] (0,2) -- (0,0) -- (4,0);
				\node[right] at (4,0) {$r$};
				\node[above] at (0,2) {$\theta$};
				\draw[dashed, ultra thick, red] (2,-0.1) -- (2,2.3);
				\begin{scope}[shift={(2,0)}]
					\draw[ultra thick] (-0.1,-0.1) -- (0.1,0.1) (0.1,-0.1) -- (-0.1,0.1);
				\end{scope}
				\begin{scope}[shift={(2,1)}]
					\draw[ultra thick] (-0.1,-0.1) -- (0.1,0.1) (0.1,-0.1) -- (-0.1,0.1);
				\end{scope}
				\begin{scope}[shift={(2,2)}]
					\draw[ultra thick] (-0.1,-0.1) -- (0.1,0.1) (0.1,-0.1) -- (-0.1,0.1);
				\end{scope}
				\node[below] at (0.5,-0.1) {$-\infty$};
				\node[below] at (2,-0.1) {$0$};
				\node[below] at (3.5,-0.1) {$+\infty$};
				\draw[thick,blue, postaction={decorate},decoration={markings, 
					mark= at position 0.75 with {\arrow{stealth}}}] (0.5,0.5) to[out=0,in=180] (2,1.5) to[out=0,in=180] (3.5,0.5);
				\node[above,blue] at (3.5,0.5) {$\wp$};
				\draw (2,1) to[out=0,in=90] (2.2,0.75) to[out=270,in=180] (2.4,0.5) to[out=180,in=90] (2.2,0.25) to[out=270,in=0] (2,0);
				\node[right] at (2.4,0.5) {$2\pi$}; 
			\end{scope}
			\begin{scope}[shift={(11,0.25)}]
				\begin{scope}[yscale=0.5]
					\draw(0,0) circle (1.5);
				\end{scope}
				\begin{scope}[xscale=0.5]
					\draw[thick, red, dashed] (0,0) circle (0.75); 
					\draw[ultra thick, red] ([shift=(90:0.75)]0,0) arc (90:270:0.75);
				\end{scope}
				\draw[fill=black] (-1.4,0) circle (0.06) (1.4,0) circle (0.06) (-0.35,-0.3) circle (0.06);
				\node[left] at (-1.5,0) {$0$};
				\node[right] at (1.5,0) {$\infty$};
				\node[left] at (-0.35,-0.3) {$1$};
				\draw[blue,thick,postaction={decorate},decoration={markings, 
					mark= at position 0.75 with {\arrow{stealth}}}] (-1.3,0) to[out=0,in=180] (0,0.5) to[out=0,in=180] (1.3,0);
				\node[blue] at (1.1,-0.2) {$\wp$};
				\draw (-2,1) -- (-1.5,1) -- (-1.5,1.5);
				\node[above left] at (-1.5,1) {$e^{2\pi t}$};
			\end{scope}
			\node at (0,-1.5) {(a)};
			\node at (5.5,-1.5) {(b)};
			\node at (11,-1.5) {(c)};
		\end{tikzpicture}
		\caption{Different representations of the path $\wp$ on the parameter space $\CP$. In all these three representations the path travels between Higgs branch descriptions through a narrow region around values $r=0$, $|e^{2\pi t}|=1$ where a Coulomb branch description of the IR physics is in order.}\label{fig:wp}
	\end{center}
\end{figure}

The transition between $X_\pm$ is known in the literature as a conifold transition or Atiyah flop \cite{Flop,Segal:2010cz,2012arXiv1206.0219D}. 
Let us consider corresponding path $\wp$ connecting regions $r<0$ and $r>0$ in the parameter space and performing this transition (see Fig.\ref{fig:wp}(a)). 
This path starts in a Higgs branch sector and ends in a Higgs branch sector intersecting a Coulomb branch sector in the middle (see Fig.\ref{fig:wp}(b)).
Therefore in our description of parallel transport functor $\beta_{\wp}$ we have to sandwich these phases through cHC duality transitions.

However the disk partition function given by the hypergeometric series ${}_2F_1$ is a smooth function of $e^{2\pi t}$, therefore we do not expect any problems with parallel transport on the level of Grothendieck groups. 
On the Riemann sphere parameterized by a complex variable $z=e^{2\pi t}$ the locus $r=0$ corresponds to a unit circle (see Fig.\ref{fig:wp}(c)). 
And parallel transport along path $\wp$ is a well-known and well-studied problem in the theory of hypergeometric series known as \emph{analytic continuation} of the hypergeometric series outside the unit circle.
As long as this path does not hit the singularity located at $e^{2\pi t}=1$ the parallel transport is smooth. 

The corresponding parallel transport functor $\beta_{\wp}$ can be calculated by alternative means \cite{HHP} using a brane grade restriction rule. The result coincides with the mathematical description of the conifold transition functor \cite{Kapranov:2014uwa}:
\be
\beta_{\wp}:\quad D^b{\rm Coh}(X_-)\longrightarrow D^b{\rm Coh}(X_+)\,.
\ee
This transition is given by Fourier-Mukai transform $\Phi_{\CK}$ (see Appendix \ref{sec:App_FM} for details on this definition) with the kernel:
\be
\CK=\CO_{\{S_{(-)}=S_{(+)}\}}\,,
\ee
where $S_{(\pm)}$ are values of matrix $S$ \eqref{S-mat} for resolutions $X_{\pm}$, or $r\gtrless 0$, respectively.
We will calculate this kernel and Fourier-Mukai transform explicitly using the algebra of the infrared \cite{GMW}.

\subsection{Braid group categorification, affine Grassmannians, crystal melting, solidifying}

A natural way to generalize the framework of analytic continuation of hypergeometric series and Atiyah flops discussed in the previous section is to consider applications to a categorification of the braid group.
Naturalness is dictated by observations that simple objects related to the braid group action like the Drinfeld associator \cite{DuninBarkowski:2012na} and simple conformal blocks with a degenerate at level 2 vertex operator insertion in 2d conformal field theories \cite{Belavin:1984vu} are described by hypergeometric series.

To physicists the braid group is mostly known due to appearance in problems related to anyon statistics, FQHE, braiding vertex operators in 2d conformal field theory \cite{DiFrancesco:639405} as well as an impact on knotted Wilson line observables in 3d Chern-Simons theory \cite{Witten:1988hf} and its mathematical counterpart -- knot invariants and braided modular categories of representations of quantum groups \cite{Kirillov:191317,drinfeld1989quasi}.

A categorification of Chern-Simons link invariants \cite{Khovanov:1999qla,KR} opened a vast amount of opportunities for applications in both string theories \cite{Ooguri:1999bv,Gukov:2004hz,Gukov:2007tf,Haydys:2010dv,WItten:2011pz,Witten:2011zz,Anokhina:2017iui,Anokhina:2021lbh,Dolotin:2013osa,Arthamonov:2017oxw,2007arXiv0708.2228M} and pure mathematics \cite{2005math......8510R,2016arXiv160807308G,2019arXiv190208281G,2019arXiv190506511O,2007arXiv0710.4300O,Abouzaid:2017clg} as well as their profound synthesis to discover underlying structures in relations between physics and geometry.
Unfortunately, we are unable to cover literature for this popular topic even partially and indicate just few sources the reader could use to find a particular subject interesting for a concrete application.

Some approaches \cite{2008arXiv0803.4121K,2008arXiv0812.5023R,seidel2001braid,gorodentsev2004helix,2011arXiv1104.0352C,2012arXiv1212.6076L,CautisKamnitzerI,seidel2001braid,2013arXiv1306.6242M,2015arXiv150206011M,2014arXiv1409.4461W} consider as an intermediate step a categorification of braid group representations induced by quantum groups.

A geometric way to categorify braid group representations is to consider coherent sheaves on algebraic varieties with permutable elements, then to identify permutations with functor actions.
A physical setup of \cite{Gaiotto:2015zna,GMW} proposes to consider a braid as a defect operator in a 5d supersymmetric theory.
A consideration of the IR dynamics in this theory translates the problem to a language of a categorified Berry connection in a certain Landau-Ginzburg model.
A BPS Hilbert space for the corresponding interface will deliver the desired categorification \cite{Galakhov:2016cji}.
As it is explained in \cite{Aganagic:2020olg} the most suitable dual counterpart incorporating categories of coherent sheaves on algebraic varieties is a field theoretic description of affine Grassmannians related to moduli spaces of monopole-like solutions of \cite{Gaiotto:2015zna}.

We will consider a parallel transport in moduli spaces of cotangent bundles to flag varieties representing specific slices in affine Grassmannians.
Physically this parallel transport is represented by a supersymmetric interface defect $\fJ_{\wp}$ in a 2d $\CN=(2,2)$ gauged linear sigma-model with a target space given by a Nakajima quiver variety \cite{nakajima1994instantons,nakajima1998quiver,nakajima2001quiver,2009arXiv0905.0686G}.
The transport path $\wp$ represents a tangle in the parameter space spanned by Fayet-Illiopolous parameters complexified with the help of topological angles.
Boundary conditions on interval ends are defined by coherent sheaves $\CA$ and $\CB$.
Hilbert space $\fG_{\rm BPS}(\fJ_{\wp}|\CA,\CB)$ is a categorification of the parallel transport along $\wp$ induced by a Berry connection  and categorifies a braid group representation associated to $\wp$.

Our plan is to describe the Berry parallel transport of brane boundary conditions associated to simple braids as a Fourier-Mukai transform on the derived category of coherent sheaves.
The resulting Fourier-Mukai transform is in a complete agreement with mathematical constructions of \cite{CautisKamnitzerI}.

We should stress that this phenomenon finds an intuitively natural description in a language of condensed matter physics.
So classical vacua in certain QFTs with quiver target spaces representing D-brane systems on Calabi-Yau 3-folds have a labeling by crystal lattices \cite{Ooguri:2008yb,Okounkov:2003sp,Yamazaki:2010fz,Li:2020rij,Aganagic:2010qr}.
Higgs branch varieties associated to initial and final points of path $\wp$ are isomorphic, so the Berry parallel transport along $\wp$ is in practice a \emph{Berry holonomy} from the crystal phase to itself.
However the very path $\wp$ overlaps inevitably with a parameter space region where the Higgs branch develops a conic singularity. 
The crystal melts as the path enters this region to a partially ``liquid'' state and solidifies back afterwards as the path exits the critical region.
An impact of this hysteresis on the BPS Hilbert space can be calculated explicitly due to localizing properties of supersymmetry.

\subsection{Outline of the paper}

The paper is organized as follows.

In Section \ref{sec:Preamble} we present a slow pace review of the cHC duality on a cylinder $S^1\times \IR_t$ quantizing this system using the Hamiltonian evolution along temporal direction $\IR_t$.
$S^1$ does not have boundaries, there is no need to consider complicated non-trivial boundary conditions except periodic ones. 
Notions of order and disorder parameters are introduced.

In Section \ref{sec:Localization} we make all the preliminary work discussing localization to the Higgs, Coulomb and mixed phases. 
We describe how the effects of the RG group flow on the BPS Hilbert space may be taken into account exactly and how the notion of interface defects can be taken into consideration.
Eventually, we discuss the low energy dynamics of mentioned phases and give a brief review of a web formalism of \cite{GMW} used for the IR description of solitons and instantons in 2d $\CN=(2,2)$ theories with massive vacua.

In Section \ref{sec:equiv_CP_1} we consider the IR dynamics in a GLSM describing equivariant $\IC\IP^1$. 
This model is an elementary building block for more complicated models.
We compare the equivariant Higgs and Coulomb branch soliton spectra and find their agreement.
Some scattering soliton vertices contributing to instanton transitions are calculated.

In Section \ref{sec:cat_contin_hyper} we construct a categorification of the analytic continuation for hypergeometric series and calculate an associated Fourier-Mukai kernel.

In Section \ref{sec:quiver} we consider a categorified braid group action on Nakajima quiver varieties and cotangent bundles to flag varieties related by Maffei's isomorphism.
Associated Fourier-Mukai kernels are calculated.
An accompanying  physical intuition of melting and solidifying crystals during Berry parallel transport is discussed.

In Section \ref{sec:future} we discuss open problems and possible future directions.

\section{Preamble: a simple version of Higgs-Coulomb duality on a cylinder}\label{sec:Preamble}

\subsection{Order parameter}

We would like to start with a very simple setup of a single uncharged complex field with a complex mass $\mu=\mu_{\IR}+\I\,\mu_{\II}$. 
Rather than considering this theory on a strip world-sheet we put it on a cylinder. 
Fields have periodic boundary conditions:
\be\label{periodic}
\phi(x^1+L)=\phi(x^1).
\ee
There is no need to consider D-brane boundary conditions and D-brane category in this case. 
The BPS Hilbert space is a vector space spanned by wave functions satisfying the BPS constraint:
\be
\CQ_B|\Psi_{\rm BPS}\rangle=\bar \CQ_B|\Psi_{\rm BPS}\rangle=0.
\ee 
The supercharge $\bar \CQ_B$ in this simple model reads (see Appendix \ref{sec:App_SUSY}):
\be\label{sc}
\begin{split}
	\bar \CQ_B=\int\lm_0^L dx^1\Big[\bar\psi_1\left(-\I\delta_{\bar\phi}-\I \mu_{\IR}\phi \right)+\bar\psi_2(-\p_{1}\phi+\mu_{\II}\phi)\Big].
\end{split}
\ee

Using periodicity of the field functions \eqref{periodic} it is natural to decompose bosonic and fermionic fields over normalized Fourier modes:
\be
\phi(x^1)=\sum\lm_{n=-\infty}^{\infty}\phi_n\frac{ e^{\I \kappa_n x^1}}{\sqrt{L}},\quad \psi_a(x^1)=\sum\lm_{n=-\infty}^{\infty}\psi_{a,n}\frac{ e^{\I \kappa_n x^1}}{\sqrt{L}},\quad \kappa_n=\frac{2\pi n}{L}.
\ee
The variation operators are decomposed in a similar way:
\be
\frac{\delta}{\delta \phi(x^1)}=\sum\lm_{n=-\infty}^{\infty}\frac{ e^{-\I \kappa_n x^1}}{\sqrt{L}}\p_{\phi_n},\quad \frac{\delta}{\delta \bar\phi(x^1)}=\sum\lm_{n=-\infty}^{\infty}\frac{ e^{\I \kappa_n x^1}}{\sqrt{L}}\p_{\bar\phi_n}.
\ee
The supercharge expression in terms of normalized Fourier modes has the following form:
\be
\bar \CQ_B=\sum\lm_{n=-\infty}^{\infty}\Big[\bar\psi_{1,n}\left(-\I\p_{\bar\phi_n}-\I\mu_{\IR}\phi_n\right)+\bar\psi_{2,n}\left(-\I \kappa_n+\mu_{\II}\right)\phi_n\Big].
\ee

There are two fundamentally different cases $\mu_{\II}\neq 0$ and $\mu_{\II}=0$.
In the case $\mu_{\II}\neq 0$ there is just a single BPS state, and its normalized wave function reads:\footnote{We assume that the fermion vacuum $|0\rangle$ is annihilated by all fermion annihilation operators $\psi$, and $\bar\psi$ create new states.} 
\be\label{wave1}
|\Psi_{\rm BPS}\rangle=\prod\lm_{n=-\infty}^{\infty}e^{-\sqrt{\kappa_n^2+|\mu|^2}|\phi_n|^2}\frac{\left(\mu_R-\sqrt{\kappa_n^2+|\mu|^2}\right)\bar{\psi}_{1,n}+(\kappa_n+\I\mu_I)\bar{\psi}_{2,n}}{\sqrt{2\pi(\sqrt{\kappa_n^2+|\mu|^2}-\mu_R)}}|0\rangle.
\ee

In the case $\mu_{\II}=0$ the ground state is infinitely degenerate, and the zero mode can create a condensate. 
The corresponding BPS wave-function expression admits a choice of a pair of arbitrary holomorphic functions $g_b$ and $g_f$:
\be\label{wave2}
\begin{split}
	|\Psi_{\rm BPS}\rangle=\left\{\begin{array}{ll}
		g_b(\phi_0)+g_f(\phi_0)\bar\psi_{2,0}, & {\rm if}\;\mu_{\IR}>0 \\
		(g_b(\bar\phi_0)+g_f(\bar\phi_0)\bar\psi_{2,0})\bar\psi_{1,0}, & {\rm if}\;\mu_{\IR}<0 \\
	\end{array} \right\}e^{-|\mu_{\IR}||\phi_0|^2}\times\\
	\times\prod\lm_{n\neq 0}e^{-\sqrt{\kappa_n^2+\mu_{\IR}^2}|\phi_n|^2}\frac{\left(\mu_{\IR}-\sqrt{\kappa_n^2+\mu_{\IR}^2}\right)\bar{\psi}_{1,n}+\kappa_n\bar{\psi}_{2,n}}{\sqrt{2\pi(\sqrt{\kappa_n^2+\mu_{\IR}^2}-\mu_{\IR})}}|0\rangle.
\end{split}
\ee

The theory of a complex scalar has a global $U(1)$ symmetry. 
The single BPS state \eqref{wave1} at $\mu_{\II}\neq 0$ is preserved by $U(1)$ rotations of field modes.
On the other hand $U(1)$-rotations of $\phi_0$ in \eqref{wave2} vary functions $g_b$ and $g_f$ and are not symmetries of this state. 
Therefore in this theory we distinguish two phases: the phase with \emph{unbroken} global $U(1)$ symmetry corresponding to $\mu_{\II}\neq 0$, the phase with \emph{broken} global $U(1)$ symmetry corresponding to $\mu_{\II}= 0$.

Let us consider the expectation value of the scalar field:
\be
\langle\phi\rangle:=\langle\Psi_{\rm BPS}|\phi|\Psi_{\rm BPS}\rangle.
\ee

We easily calculate this expectation value in both phases:
\be
\begin{array}{lll}
	\mbox{unbroken:}& \mu_{\II}\neq 0, & \langle\phi\rangle=0;\\
	\mbox{broken:}& \mu_{\II}= 0, & \langle\phi\rangle=\frac{\int d^2\phi\; \left(|g_b|^2+|g_f|^2\right)\;\phi\; e^{-|\mu_{\IR}||\phi|^2}}{\int d^2\phi\; \left(|g_b|^2+|g_f|^2\right)\; e^{-|\mu_{\IR}||\phi|^2}}.
\end{array}
\ee

As we see in this example operator $\phi$ can be used as a measure of symmetry breaking.
Therefore we identify it with an \emph{order parameter} of this theory.

Let us consider the following two operators -- fermion number and electric charge respectively:
\be
\hat\CF=\int\lm_0^L dx^1\left[\bar\psi_+\psi_++\bar\psi_-\psi_-\right],\quad 
\hat Q=\int\lm_0^L dx^1\left[\left(\bar\phi\delta_{\bar\phi}-\phi\delta_{\phi}\right)+\bar\psi_+\psi_++\bar\psi_-\psi_-\right].\label{el_ch}
\ee
The BPS Hilbert space is stratified under the action of this operators into one-dimensional subspaces:
\be
\fG_{\rm BPS}=\bigoplus\lm_{\CF,Q}\IC\,|\CF,Q\rangle\,,
\ee
where $\CF$ and $Q$ are possible eigenvalues of operators $\hat\CF$ and $\hat Q$ correspondingly.

Using this identification we can summarize the result of this section in describing the BPS Hilbert spaces of the theory of the order parameter in the following way:
\be
\begin{split}
	\fG_{\rm BPS}^{({\rm ord})}\left(\mu_{\II}\neq 0\right)=\IC|0,0\rangle,\\
	\fG_{\rm BPS}^{({\rm ord})}\left(\mu_{\II}= 0,\mu_{\IR}>0\right)=\bigoplus\lm_{n=0}^{\infty}\left(\IC|0,-n\rangle\oplus \IC|-1,-n-1\rangle\right),\\
	\fG_{\rm BPS}^{({\rm ord})}\left(\mu_{\II}= 0,\mu_{\IR}<0\right)=\bigoplus\lm_{n=0}^{\infty}\left(\IC|0,n\rangle\oplus \IC|1,n+1\rangle\right).
\end{split}
\ee

\subsection{Disorder parameter}
Now we would like to turn to a dual description. 
We could call an operator dual to the order operator a \emph{disorder} operator in analogy with the Ising model \cite{Polyakov:1987ez}.
Physical role of the disorder operator is to insert a vortex defect into the theory so that the phase of the order parameter winds around the vortex core, similarly we could have called it a \emph{vortex operator} (compare also to 3d monopole operator definition in \cite{Borokhov:2002ib}).

A duality in this case is a mere Fourier transform exchanging the field phase and the field winding number operator -- T-duality.

Let us extract explicitly the phase contribution into quantum fields:
\be
\phi=e^{\rho+\I\vartheta},\quad \psi_a=e^{\I\vartheta}\chi_a.
\ee
In these new variables the supercharge expression \eqref{sc} can be rewritten as:
\be\label{bar_Q_B}
\bar\CQ_B=\int\lm_0^L dx^1\left[-\I\bar{\chi}_1\left(\frac{e^{-\rho}}{2}\left(\delta_{\rho}+\I\delta_{\vartheta}\right)+\mu_{\IR} e^{\rho}\right)+V_3+\bar{\chi}_2e^{\rho}\left(-\p_{1}\rho-\I\p_{1}\vartheta+\mu_{\II}\right) \right],
\ee
where
\be\label{VV_3}
V_3=\frac{\I}{2} e^{-\rho}\bar{\chi}_1\bar{\chi}_2\chi_2
\ee
is a new cubic fermion interaction term produced by quadratic fermion term in $\delta_{\vartheta}$.\footnote{
	Let us choose the following polarization for the fermions so that $\psi_a$ annihilate the vacuum, then the wave functional has the following form:
	$$
	\Psi[\phi,\bar\phi,\bar\psi_a]|0\rangle\,.
	$$
	In these terms the variations with respect to the phase fields read:
	$$
	\delta_{\rho}=e^{\rho+\I\vartheta}\delta_{\phi}+e^{\rho-\I\vartheta}\delta_{\bar\phi},\quad
	\delta_{\vartheta}=\I e^{\rho+\I\vartheta}\delta_{\phi}-\I e^{\rho-\I\vartheta}\delta_{\bar\phi}-\I e^{-\I\vartheta}\sum\lm_a\bar\chi_a\delta_{\bar\chi_a}\,.
	$$
	Using fermion anti-commutation relations we substitute $\delta_{\bar\chi_a}=\chi_a$. 
	Then solving for $\delta_{\bar\phi}$ one finds:
	$$
	\delta_{\bar\phi}=\frac{1}{2}e^{-\rho+\I\vartheta}(\delta_{\rho}+\I\delta_{\vartheta})+\frac{\I}{2}e^{-\rho}\sum\lm_a\bar\chi_a\chi_a\,.
	$$
	Simply substituting this expression into \eqref{sc} one derives \eqref{bar_Q_B}.
}

To pass to the disorder operator description we dualize phase operator $\vartheta$ in terms of a new field $\IS$.
We will describe properties of field $\IS$ momentarily, first review the properties of phase field $\vartheta$.

First notice a large global phase shift:
\be\label{global}
\vartheta(x^1)\longrightarrow \vartheta(x^1)+2\pi n,\quad n\in\IZ,
\ee
is a shift symmetry of the theory.

Then notice as well that rather having purely periodic boundary conditions the phase field has a shifting twist:
\be\label{twisted}
\vartheta(x^1+L)=\vartheta(x^1)+2\pi k,\quad k\in\IZ.
\ee

Function $\p_1\vartheta(x^1)$ is a periodic function on $[0,L]$, therefore we can decompose it over normalized Fourier modes ($\kappa_n=2\pi n/L$):
\be
\p_1\vartheta(x^1)=\frac{\alpha_0}{\sqrt{L}}+\sum\lm_{n=1}^{\infty}\frac{\alpha_n}{\sqrt{2L}}\cos\kappa_n x^1+\sum\lm_{n=1}^{\infty}\frac{\beta_n}{\sqrt{2L}}\sin\kappa_n x^1.
\ee
Integrating both sides of this equality we find a decomposition for the field $\vartheta(x^1)$ over modes:
\be\label{vartheta}
\vartheta(x^1)=\vartheta_0+\frac{\alpha_0}{\sqrt{L}}x^1+\sum\lm_{n=1}^{\infty}\frac{\alpha_n}{\kappa_n}\frac{1}{\sqrt{2L}}\sin\kappa_n x^1-\sum\lm_{n=1}^{\infty}\frac{\beta_n}{\kappa_n}\frac{1}{\sqrt{2L}}\cos\kappa_n x^1.
\ee
Global symmetry \eqref{global} and twisted periodic boundary conditions \eqref{twisted} impose a periodicity constraint on BPS wave functions of this theory:
\be
\Psi(\vartheta_0+2\pi)=\Psi(\vartheta_0)\,.
\ee
As well an operator in front of the $x^1$-linear term in the expansion \eqref{vartheta} acquires only discrete eigenvalues, so we can consider an eigen basis of this operator:
\be
\frac{\alpha_0}{\sqrt{L}}\Psi_k=\frac{2\pi k}{L}\Psi_k,\quad k\in\IZ.
\ee
Due to this discreteness operator $\alpha_0$ does not contribute to variations of the field $\vartheta$, so for the variation we have the following expansion:
\be
\frac{\delta}{\delta\vartheta(x^1)}=\frac{1}{L}\frac{\p}{\p\vartheta_0}+\sum\lm_{n=1}^{\infty}\kappa_n\frac{1}{\sqrt{2L}}\sin\kappa_nx^1\frac{\p}{\p\alpha_n}-\sum\lm_{n=1}^{\infty}\kappa_n\frac{1}{\sqrt{2L}}\cos\kappa_nx^1\frac{\p}{\p\beta_n}.
\ee
Let us consider a Fourier transform of the wave function:
\be\label{Fourier}
\hat\Phi\left[\Psi\right]_m(\IS_0,\gamma_i,\eta_i):=\int\lm_0^{2\pi}e^{-\I n\vartheta_0}d\vartheta_0\sum\lm_{k=-\infty}^{+\infty}e^{2\pi i k\IS_0}\prod\lm_{a=1}^{\infty}\int\lm_{-\infty}^{+\infty}d\alpha_ad\beta_a\; e^{-\I\frac{\alpha_a\eta_a+\beta_a\gamma_a}{\kappa_a}}\Psi_k(\vartheta_0,\alpha_i,\beta_i).
\ee
For operators we have:
\be\label{rel}
\hat\Phi\left[\delta_{\vartheta}\right]=-\I\,\p_{1}\IS,\quad \hat\Phi\left[\p_{1}\vartheta\right]=\I\,\delta_{\IS},
\ee
where field $\IS$ has the following expansion:
\be
\IS(x^1)=\IS_0+\frac{ m}{L}x^1+\sum\lm_{n=1}^{\infty}\frac{\gamma_n}{\kappa_n}\frac{1}{\sqrt{2L}}\sin\kappa_n x^1-\sum\lm_{n=1}^{\infty}\frac{\eta_n}{\kappa_n}\frac{1}{\sqrt{2L}}\cos\kappa_n x^1.
\ee
Also notice that the resulting wave-function is a periodic function of $\IS_0$ with an integer period. Summarizing these observations we conclude that field $\IS$ has twisted periodic boundary conditions:
\be\label{twisted_S}
\IS(x^1+L)=\IS(x^1)+m,\quad m\in \IZ,
\ee
as well as an overall global shift
\be\label{global_S}
\IS(x^1)\to\IS(x^1)+l,\quad l\in\IZ
\ee
is a symmetry of the theory.

We combine operators $\rho$ and $\IS$ into a complex twisted chiral field: 
\be
Y=Y_{\IR}+\I Y_{\II}:=e^{2\rho}-\I\,\IS,
\ee
we will call a \emph{disorder} operator. It is simple to rewrite the supercharge in new terms:
\be\label{dual_Q_B}
\begin{split}
	\hat\Phi\left[\bar \CQ_B\right]=\int\lm_0^L dx^1&\left[-2\I\sqrt{Y_{\IR}}\bar\chi_+\left(\delta_Y-\frac{\I}{4Y_{\IR}}\p_1\bar Y+\frac{\mu}{2}\right)-\right.\\
	&\left.-2\I\sqrt{Y_{\IR}}\bar\chi_-\left(\delta_{\bar Y}+\frac{\I}{4Y_{\IR}}\p_1Y+\frac{\bar \mu}{2}\right)+V_3\right],
\end{split}
\ee
where the following notations for fermion fields are introduced:
$$
\chi_1=\chi_-+\chi_+,\quad \chi_2=\chi_--\chi_+.
$$

If the triple fermion term $V_3$ is ignored this supercharge is equivalent to a twisted form\footnote{Dynamical descriptions of chiral and twisted chiral field are more or less alike, one suffices to swap fermions $\psi_+$ and $\bar\psi_-$.} of Landau-Ginzburg supercharge \eqref{nlin_sup} for a 1d complex K\"ahler manifold with the following expressions for the K\"ahler potential and an induced superpotential \cite{Hori:2000kt} (for a definition of Landau-Ginzburg model with a K\"ahler target space see Appendix \ref{sec:App_LG}):
\be\label{Kahler}
\begin{split}
	&K(Y,\bar Y)=-\frac{1}{2}(Y+\bar Y)\log(Y+\bar Y),\\
	&W=\mu Y.
\end{split}
\ee

Let us observe that in the proposed BPS vacuum the disorder operator $Y$ acquires a non-zero vacuum expectation value:
\be\label{disorder}
\langle Y\rangle:=\Big\langle\Psi_{\rm BPS} \Big||\phi|^2-\frac{\delta}{\delta(\p_1\vartheta)}\Big|\Psi_{\rm BPS}\Big\rangle=\sum\lm_n \frac{1}{L} \langle\Psi_{n} ||\phi_n|^2-\p_{\kappa_n}|\Psi_{n}\rangle.
\ee

It is easy to calculate expectation value \eqref{disorder} in the limit $L\to \infty$ using \eqref{wave1}.
The summation in the r.h.s. of \eqref{disorder} is substituted by an integration\footnote{A rule for substituting a summation by an integration is simply canonical:
	$$
	\sum\lm_{n=-\infty}^{\infty}\frac{1}{L}\;f\left(\frac{n}{L}\right)\mathop{\longrightarrow}^{L\to \infty}\int\lm_{-\infty}^{\infty}d\nu\;f(\nu)\,.
	$$} and diverges. This is a standard divergence of QFT loop calculations and it is needed to be regularized. For regularization we introduce a cutoff parameter $\Lambda$:
\be\label{Y_vev}
\begin{split}
	\langle Y\rangle =\frac{1}{4\pi}\int\lm_{-\Lambda}^{\Lambda}dk\;\frac{1}{\sqrt{k^2+|\mu|^2}}\left(1+\frac{\I\mu_{\II}}{\sqrt{k^2+|\mu|^2}-\mu_{\IR}}\right)=\\
	=\frac{1}{2\pi}\left[\log\frac{2\Lambda}{|\mu|}+\I\left(\arctan\frac{\Lambda}{\mu_{\II}}+\arctan\frac{\mu_{\IR}}{\mu_{\II}}\right)\right]+O\left(\frac{|\mu|}{\Lambda}\right)=\\
	=\frac{1}{2\pi}\left(\log\frac{2\Lambda}{\mu}+\pi\I\right)+O\left(\frac{|\mu|}{\Lambda}\right).
\end{split}
\ee

The resulting expression is a multivalued function. The logarithm multivaluedness is inherited from physical symmetry \eqref{global_S}. In principle, field $Y$ is not a physical observable, rather $e^{-2\pi Y}$ is. So we can rewrite an equation defining theory vacua in an invariant form:
\be\label{defect_vev}
\mu=-2 \Lambda\, e^{-2\pi \langle Y\rangle}.
\ee
This vacuum equation has obvious $\IZ$ symmetry $Y\to Y+2\pi \I$ shifting the sheet of the logarithm cover that is \emph{broken} by the vacuum solution \eqref{Y_vev}, this broken phase is described by a non-zero expectation value of the disorder operator.

We should note that the operator
$$
e^{-2\pi\, Y(y)}
$$
is a vortex operator inserting a vortex defect singularity in a point $y$ on the world-sheet. It is easy to observe this if one performs a Wick rotation $x^0=-\I\, x^2$ to make the action Euclidean, then relations \eqref{rel} can be treated semi-classically as the following:
$$
\p_i\IS=-\I\epsilon_{ij}\p_j\vartheta.
$$
Let us consider a ``Dirac string'' integration path $\wp(y)$ going from infinity to point $y$. Insertion of $e^{-2\pi \, Y(y)}$ in the Euclidean path integral performs a modification of the action by a new term:
$$
-2\pi \I\, \IS= \int\lm_{\wp(y)}2\pi\epsilon_{ij}\p_i\vartheta\,dx^j.
$$
Such term in the action introduces a boundary condition for phase $\vartheta$ that jumps across the Dirac string $\wp$ by $2\pi$, the vortex defect core located in point $y$:
$$
\begin{array}{c}
	\begin{tikzpicture}
		\draw[fill=black] (0,0) circle (0.08);
		\draw[thick, dashed] (-4,-0.5) to[out=300,in=240] (-3,-1);
		\draw[thick] (-3,-1) to[out=60,in=180] (-2,0.5) to[out=0,in=180] (-1,-0.5) to[out=0,in=180] (0,0);
		\node[left] at (-4,-0.5) {$\wp(y)$};
		\node[right] at (0,0) {$y$};
		\draw[->] (-0.2,0) to[out=270,in=180] (0.1,-0.3) to[out=0,in=270] (0.4,0) to[out=90,in=0] (0.1,0.3) to[out=180,in=90] (-0.2,0);
		\node[right] at (0.4,0) {$\Delta\vartheta=2\pi$};
	\end{tikzpicture}
\end{array}
$$

To complete dualization of the supercharge $\CQ_B$ in \eqref{dual_Q_B} we need to calculate the contribution of the triple fermion operator $V_3$ \eqref{VV_3}. Again we do it up to the first loop order substituting double fermion contributions by their expectation values:
\be\label{V_3}
V_3\to \frac{\I}{2\sqrt{Y_{\IR}}}\bar\chi_-\left(\langle:\bar\psi_2\psi_2:\rangle-\langle\bar\psi_1\psi_2\rangle \right)+\frac{\I}{2\sqrt{Y_{\IR}}}\bar\chi_+\left(\langle:\bar\psi_2\psi_2:\rangle+\langle\bar\psi_1\psi_2\rangle \right).
\ee
For the expectation values we have:
\be
\begin{split}
	\langle:\bar\psi_2\psi_2:\rangle=\frac{\mu_{\IR}}{4\pi}\int\lm_{-\Lambda}^{\Lambda}\frac{dk}{\sqrt{k^2+|\mu|^2}}=\frac{\mu_{\IR}}{2\pi}\log\frac{2\Lambda}{|\mu|}=-Y_{\IR}\;{\rm Re}\left(2 \Lambda\, e^{-2\pi  Y}\right),\\
	\langle\bar\psi_1\psi_2\rangle=-\frac{\I\mu_{\II}}{4\pi}\int\lm_{-\Lambda}^{\Lambda}\frac{dk}{\sqrt{k^2+|\mu|^2}}=-\frac{\I\mu_{\II}}{2\pi}\log\frac{2\Lambda}{|\mu|}=\I Y_{\IR}\;{\rm Im}\left(2\Lambda\, e^{-2\pi Y}\right).
\end{split}
\ee
Thus we see that this term reproduces a shift\footnote{Clearly, \eqref{V_3} gives only a half of that contribution that we expect to get to derive \eqref{sup1}. We assume that this discrepancy can be eliminated by redefining ambiguous cut-off parameter $\Lambda$.} of the superpotential:
\be\label{sup1}
W=\mu Y-\frac{1}{\pi}\Lambda e^{-2\pi Y}\,,
\ee 
so that its critical points define exactly the vev of the defect field \eqref{defect_vev}.

We can redefine the field $Y$ by an overall shift so the superpotential has no additional parameters:
\be
W=\mu Y+\frac{1}{2\pi}e^{-2\pi Y}\,.
\ee

As well, let us note that the explicit K\"ahler metric \eqref{Kahler} is irrelevant for the purpose of soliton counting since the quantum numbers of solitonic configurations include the superpotential only. Therefore we assume that our Landau-Ginzburg model has a flat target space for simplicity.

As a result we conclude that the model with supercharge $\bar\CQ_B$ is equivalent to a Landau-Ginzburg model with the following supercharge operator by the means of Fourier transform:
\be\label{A-twist}
\begin{split}
	\CQ_A=\int\lm_0^L dx^1\left[-\I\chi_+\left(\delta_{Y}+\I \p_1\bar Y-\left(\mu-e^{-2\pi Y}\right)\right)-\I\bar\chi_-\left(\delta_{\bar Y}-\I \p_1Y-\left(\bar\mu-e^{-2\pi \bar Y}\right)\right)\right].
\end{split}
\ee

$\CQ_A$, in a complete analogy with $\bar\CQ_B$, defines BPS wave functions spannting the BPS Hilbert space:
$$
\CQ_A|\Psi_{\rm BPS}\rangle= \bar\CQ_A|\Psi_{\rm BPS}\rangle=0,
$$
we could  call a BPS Hilbert space in the theory of disorder parameter. 
It is natural to expect the following isomorphism:
\be\label{baby_mirror}
\fG_{\rm BPS}^{({\rm ord})}\cong\fG_{\rm BPS}^{({\rm disord})}.
\ee
Relation \eqref{baby_mirror} is a baby version of the \emph{Higgs-Coulomb duality} on a cylinder. 
This isomorphism is primarily given by the Fourier transform \eqref{Fourier}. 
However during our consideration we have made a set of modifications and simplifications to arrive from $\bar Q_B$ to $Q_A$. Therefore we would like to check explicitly if \eqref{baby_mirror} holds.

In the literature \cite{Aspinwall:2004jr} this pair of models is also referred to as A- and B-model 
by the type of the supersymmetry twist.

\subsection{Soliton condensate}
To describe the states of an A-model semi-classically it suffices to consider stationary field configurations preserving A-twist \eqref{A-twist}. 
Variation operators $\delta_Y$, $\delta_{\bar Y}$ correspond to momenta operators $\p_0 \bar Y$, $\p_0 Y$ and are zeroes on stationary field configurations. 
The semi-classical BPS field configuration satisfies the following differential equation:
\be\label{sol1}
\I \p_1\bar Y-\left(\mu-e^{-2\pi\, Y}\right)=0,
\ee
with a twisted periodic boundary conditions for field $Y$:
\be\label{bc}
Y(L)=Y(0)+\I k,\quad k\in \IZ.
\ee

Using the standard techniques we derive that along the solutions the derivative of the superpotential has a stationary phase:
\be
\p_1W=\I\,\left|\mu-e^{-2\pi\, Y}\right|^2.
\ee

And for a valid solution to \eqref{sol1}, \eqref{bc}  called a \emph{soliton} there is a constraint:
\be\label{soliton}
-\I\Delta W\Big|_0^L=\mu\, k\in \IR_{\geq 0}.
\ee

A soliton has a well-defined electric charge, indeed, dualizing \eqref{el_ch} we find:
\be
\CY=\Delta\, {\rm Im}\,Y\Big|_{0}^L=k.
\ee
Topological charge $k$ is called a soliton number.

Constraint \eqref{soliton} implies that for generic chiral mass $\mu_{\II}\neq 0$ the only possible topological solution corresponds to $k=0$. 
This is a simple solution when the field $Y$ takes a constant expectation value defined by the vacuum equation \eqref{defect_vev}. 
On the other hand when $\mu_{\II}=0$ depending on the sign of $\mu_{\IR}$ equation \eqref{soliton} admits solitonic solutions with either positive or negative topological charge:
$$
\begin{array}{ll}
	\mu_{\IR}>0: & k\in\IZ_{\geq 0};\\
	\mu_{\IR}<0: & k\in\IZ_{\leq 0}.
\end{array}
$$
\begin{figure}
	\begin{center}
		\begin{tikzpicture}[scale=0.8]
			\node at (-7,0) {\includegraphics[scale=0.48]{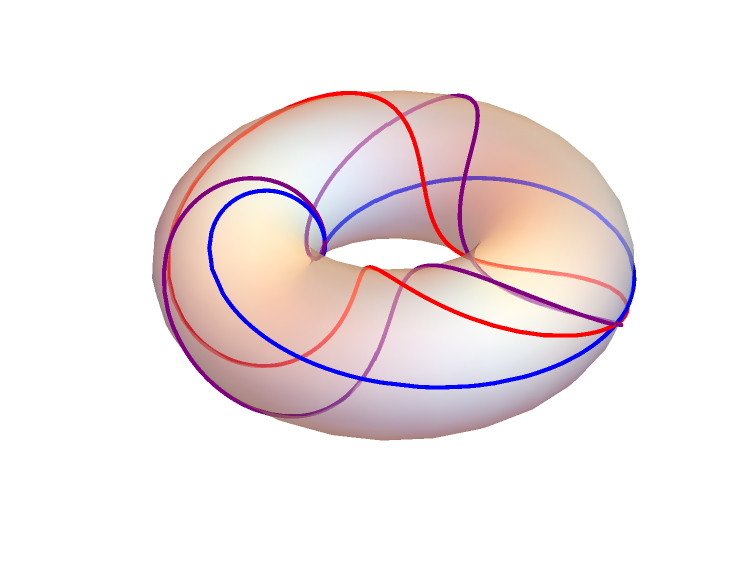}};
			\node at (0,0) {\includegraphics[scale=0.32]{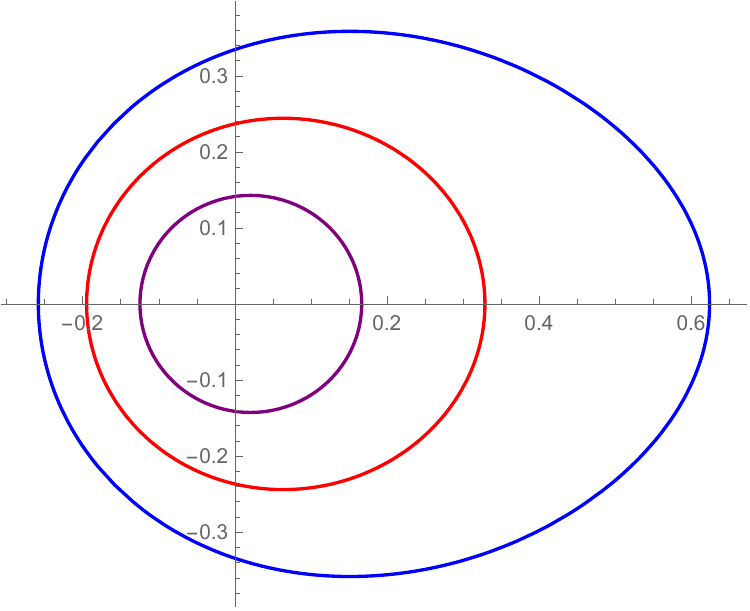}};
			\begin{scope}[shift={(4,1)}]
				\draw[fill=blue] (-0.1,-0.1) -- (-0.1,0.1) -- (0.1,0.1) -- (0.1,-0.1) -- cycle;
				\node[right] at (0.1,0) {$k=1$};
			\end{scope}
			\begin{scope}[shift={(4,0)}]
				\draw[fill=red] (-0.1,-0.1) -- (-0.1,0.1) -- (0.1,0.1) -- (0.1,-0.1) -- cycle;
				\node[right] at (0.1,0) {$k=2$};
			\end{scope}
			\begin{scope}[shift={(4,-1)}]
				\draw[fill=purple] (-0.1,-0.1) -- (-0.1,0.1) -- (0.1,0.1) -- (0.1,-0.1) -- cycle;
				\node[right] at (0.1,0) {$k=3$};
			\end{scope}
			\node at (0,-2.5) {(b)};
			\node at (-7,-2.5) {(a)};
		\end{tikzpicture}
		\caption{1-, 2- and 3-soliton solutions: a) soliton trajectories on torus $\CT$ with coordinates $(e^{2\pi\I \frac{x^1}{L}},e^{-2\pi\I Y_I})$, b) soliton trajectories on complex $e^{-2\pi \, Y}$-plane.\label{fig_solitons}}
	\end{center}
\end{figure}
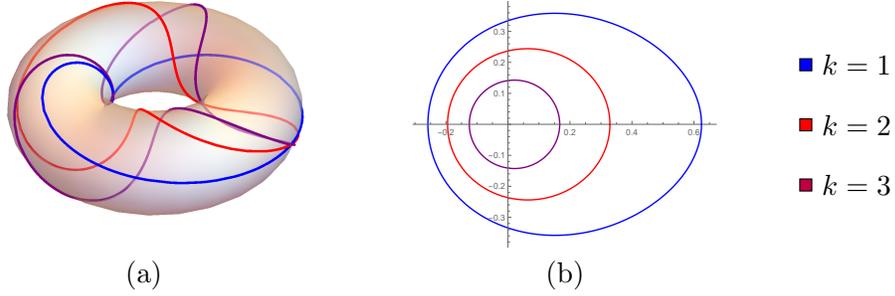
At this point it is clear that the order operators $\phi$ and $\bar \phi$ that shifted the topological charge of the BPS state and were perturbative in the previous picture may be identified with non-perturbative operators creating solitons and anti-solitons:
$$
\phi=\sum\lm_{k=0}^{\infty}|k+1\rangle\langle k|,\quad \bar\phi=\sum\lm_{k=0}^{\infty}|-k-1\rangle\langle -k|.
$$

The soliton equation \eqref{sol1} describes a motion of a particle in a 2d real phase space spanned by $({\rm Re}\,Y,{\rm Im}\,Y)$ induced by a Hamiltonian flow with a Hamiltonian given by ${\rm Re}\,W$. 
Since the number of Hamiltonians for this system is exactly a half of the phase space dimension the soliton equation is integrable, and one can explicitly write a solution in terms of action-angle variables. 
Since this procedure for solution derivation is rather standard and the resulting expression is rather involved we will not present it here. 
Rather we depict the results of numerical solutions for $\mu=1.0$ and $L=10.0$ (see Fig.\ref{fig_solitons}). 
We depict soliton trajectories in two ways. 
On one hand, trajectories are constructed as a result of applying a map $x^1\mapsto e^{-2\pi\, Y(x^1)}$ to the interval $x^1\in[0,L]$.
Notice all the solitons are represented by closed loops as expected. 
On the other hand, we consider a torus $\CT\cong S^1\times S^1$ spanned by coordinates $(e^{2\pi\I \frac{x^1}{L}},e^{-2\pi\I Y_I})$. 
On this torus $k$-solitons are again represented by closed loops winding $k$ times around $\CT$.

Finally, we should note that a soliton solution has a moduli space spanned by $u$ corresponding to translations in the spatial direction, so that if $Y_*(x^1)$ is a solution then $Y_*(x^1+u)$ is also a solution. 
This modulus is bounded to an interval $u\in[0,L)$ since solution $Y_*(x^1+L)$ is equivalent to $Y_*(x^1)$. 
Supersymmetry relates each modulus to fermionic zero mode $\chi$ in the soliton background. 
The corresponding spinor is given by $(\p_uY_*,\p_u\bar Y_*)$. 
Therefore to each $k$-soliton solution one associates the corresponding pair of quasi-classical wave-functions:
$$
\Psi_k,\quad \chi^{\dagger}\Psi_k.
$$

Summarizing the calculations of the soliton spectra we conclude that the BPS Hilbert spaces in theories of the order and disorder parameters are isomorphic as bi-graded vector spaces in all phases of the theory, therefore we confirm \eqref{baby_mirror}.

To conclude this section let us note that in calling $Y$ a disorder parameter we have admitted a certain abuse of notations, since a physical observable corresponding to the disorder parameter is an operator $e^{-2\pi Y}$. 
Nevertheless we would like to save this terminology and keep calling non-physical operator $Y$ disorder operator since it turns out to be more suitable for further considerations. 
A $U(1)$ rotation group of $e^{-2\pi Y}$ acting by imaginary shifts on $Y$ is a symmetry of the Landau-Ginzburg Lagrangian broken explicitly by the chiral mass operator $\mu$.
In the case $\mu_{\II}=0$ this symmetry is restored on the BPS solitons:
$$
\left\langle e^{-2\pi Y}\right\rangle=0.
$$ 
Indeed the value of $\left\langle e^{-2\pi Y}\right\rangle$ is averaged over the soliton trajectory depicted in Fig.\ref{fig_solitons}(b). 
After this we have to average over the position of the soliton core, or, in other words, to integrate over the soliton collective coordinate. 
The latter action rotates soliton trajectories  in Fig.\ref{fig_solitons}(b) around zero, therefore the average value of $e^{-2\pi Y}$ on the soliton trajectories is zero. 
We can summarize this information in the following table:
\be
\begingroup
\renewcommand*{\arraystretch}{1.4}
\begin{array}{c|c|c|c|c}
	\mu_{\II} & U(1)_{\rm ord} & \mbox{Order op.} & U(1)_{\rm disord} & \mbox{Disorder op.}\\
	\hline
	\mu_{\II}\neq 0 & \mbox{unbroken} & \langle\phi\rangle=0 & \mbox{broken} & \left\langle e^{-2\pi Y}\right\rangle\neq 0\\
	\mu_{\II}= 0 & \mbox{broken} & \langle\phi\rangle\neq 0 & \mbox{unbroken} & \left\langle e^{-2\pi Y}\right\rangle= 0
\end{array}
\endgroup
\ee

\section{Localization on a strip and BPS states}\label{sec:Localization}

\subsection{Localization and renormalization group}

\subsubsection{Localization in Schr\"odinger picture} \label{sec:Heis_loc}

A canonical generic approach to localization in a quantum mechanical system with a supersymmetry in the Schr\"odinger picture was proposed in a seminal paper \cite{Witten:1982im} (see also reviews in \cite{GMW,Behtash:2017rqj}). 
Here let us simply mention some basic steps we will apply in our consideration.

A quantum field theory can be treated as an ordinary quantum mechanical system describing the motion in the following space:
$$
\CT_{\rm QM}={\rm Map}\left(\CV_{d-1}\longrightarrow \CT\right).
$$
Supersymmetry produces supercharge operators $\CQ$ and $\CQ^{\dagger}$ on the Hilbert space of states satisfying the following superalgebra relation:
\be
\left\{\CQ,\CQ^{\dagger}\right\}=2\left(\CH-|Q_{\rm top}|\right)\geq 0,
\ee
where $\CH$ is the Hamiltonian of the system, and $Q_{\rm top}$ is a topological charge of the field configuration. The BPS bound
$$
\CH\geq |Q_{\rm top}|
$$
is saturated by BPS states annihilated by both supercharges:
\be\label{BPS_wave}
\CQ|\Psi_{\rm BPS}\rangle=\CQ^{\dagger}|\Psi_{\rm BPS}\rangle=0.
\ee

Suppose $x^i$ are coordinates on $\CT_{\rm QM}$. 
Supersymmetry mixes bosonic fields $x^i$ with fermionic fields $\psi_i$ and $\psi_i^{\dagger}$. 
A behavior of fermionic field operators is analogous to the behavior of differential forms, so one could identify:
\be
\psi_i^{\dagger}\rightsquigarrow dx^i\wedge,\quad \psi_i\rightsquigarrow g^{ij}\iota_{\p/\p x^j},
\ee
where $\iota$ is an interior product operator, and $g_{ij}$ is a metric tensor for $\CT_{\rm QM}$.
This allows one to identify the quantum mechanical quantities with geometric quantities.

In particular, the supercharges $\CQ$ and $\CQ^{\dagger}$ in the theory have a geometrical meaning of an extended differential and its Hodge dual. 
Depending on the initial geometric structures on $\CT_{\rm QM}$ -- Riemannian metric, complex structure, equivariant group action -- this differential will inherit its properties, so one will associate supercharge $\CQ$ with a de Rham differential, a Dolbeault differential, an equivariant Cartan model differential, etc. 
In addition the supercharge may acquire contributions from scalar functions characterizing a potential of the system. 
Under this treatment wave functions of BPS states are identified with harmonic forms on $\CT_{\rm QM}$.

Let us consider a system with concrete properties implying that a generic system will behave similarly. Suppose $\CT_{\rm QM}$ admits an action of a Lie group $G$, and we could pick a Morse height function $\fH$ on $\CT_{\rm QM}$. In this case the supercharge reads:
\be
\CQ=d+(d\fH)+\iota_V=\psi^{\dagger}_i\frac{\p}{\p x^i}+\psi^{\dagger}_i(\p_i \fH)+\psi_ig^{ij}V_j,
\ee
where $V$ is a Killing vector field created by the $G$-action on $\CT_{\rm QM}$. 

A localization mechanism follows four steps:
\begin{enumerate}
	\item Using Hodge decomposition one identifies harmonic forms annihilated by both $\CQ$ and $\CQ^{\dagger}$ with elements of the cohomology group:
	$$
	\fG_{\rm BPS}\cong H^*(\CQ).
	$$
	\item Consider an isomorphism of cohomologies:
	$$
	\phi^{(\hbar)}:\; H^*(\CQ)\longrightarrow H^*(\CQ(\hbar)),
	$$
	where the map $\phi^{(\hbar)}$ is a mere multiplication by an operator:
	$$
	\Phi_{\hbar}:=\exp\left(-\left(\hbar^{-1}-1\right)\fH+(\log\, \hbar)\sum\lm_i\psi^{\dagger}_i\psi_i\right).
	$$
	For a transformed supercharge we find:
	$$
	\CQ(\hbar):=\hbar\,\Phi_{\hbar}\CQ \Phi_{\hbar}^{-1}=\hbar\, d+(d\fH)+\iota_V.
	$$
	In principle the transformed supercharge belongs to a \emph{different} quantum system with a new Hamiltonian:
	\be\label{newHamiltonian}
	\begin{split}
		\CH(\hbar):=&|Q_{\rm top}(\hbar)|+\frac{1}{2}\left\{\CQ(\hbar),\CQ^{\dagger}(\hbar)\right\}=\\
		=&\hbar^2\Delta +\left|\vec \nabla \fH\right|^2+\left|\vec V\right|^2+{\rm fermions}.
	\end{split}
	\ee
	\item On one hand operator $\Phi_{\hbar}$ restores the Plank constant dependence often omitted in QFT calculations, on the other hand 
	it establishes an invariance of the BPS Hilbert spaces under variations of $\hbar$:
	$$
	\fG_{\rm BPS}(\hbar)\cong \fG_{\rm BPS}(\hbar').
	$$
	\item Using this invariance it is easier to compute corresponding BPS wave-functions in the semi-classical limit: 
	$$
	\hbar\longrightarrow 0.
	$$
\end{enumerate}

A calculation process in the semi-classical limit has its own known loopholes -- a necessity in certain situations to consider non-perturbative instanton corrections \cite{D-book_1}. We will not go into details widely presented in the literature, see e.g. \cite{GMW}. 
In our particular situation of supersymmetric quantum system the computation of $\fG_{\rm BPS}$ boils down to a calculation of cohomologies of a \emph{Morse-Smale-Witten} (MSW) complex $({\bf M}^*,{\bf Q})$. 

Zero locus of the potential term \eqref{newHamiltonian} corresponds to the classical vacua, therefore we call it a vacuum locus and denote it $\fV$. 
Assume for now $\fV$ is a set of isolated points. 
For each point $p\in\fV$ we change coordinates $x$ on $\CT_{\rm QM}$ as
$$
x\longrightarrow p+\hbar^{\frac{1}{2}}x.
$$
The Hamiltonian can be decomposed as:
$$
\CH=|Q_{\rm top}(\hbar)|+\hbar\; \CH_p^{(0)}+O(\hbar^2).
$$
The Hamiltonian $\CH_p^{(0)}$ is just a free particle Hamiltonian and its ground state wave function $\Psi_p$ can be easily calculated. 
$\Psi_p$ is the zeroth order perturbative approximation to the actual BPS wave-function. 
The MSW complex as a vector space is spanned by perturbative BPS wave-functions:\footnote{We have to warn the reader that the construction of the MSW complex has its own peculiarities depending on additional structures carried on by the quantum system. Here we review just the most basic one. For example, if $\CT_{\rm QM}$ has a complex structure compatible with the supercharge Hamiltonian $\CH_p^{(0)}$ would correspond to a model of a free particle on a plane put in a magnetic field perpendicular to that plane, therefore the ground states of such Hamiltonian will be described by a condensate of lowest Landau level wave functions \cite{landau1958quantum} with non-negative angular momenta corresponding to a structure sheaf of an affine complex line.}
\be\label{MSW}
{\bf M}:=\bigoplus\lm_{p\in\fV} \IC \; \Psi_p.
\ee
This vector space is graded by the fermion number $f$. The differential of the complex
$$
{\bf Q}:\quad {\bf M}^{f}\longrightarrow {\bf M}^{f+1}
$$
is defined through its matrix elements $\langle\Psi_p|{\bf Q}|\Psi_{p'}\rangle$ that up to a non-zero renormalization factor coincide with matrix elements of $\langle\Psi_p|\CQ^{\dagger}(\hbar)|\Psi_{p'}\rangle$:
\be\label{differential}
\langle\Psi_p|{\bf Q}|\Psi_{p'}\rangle=\delta_{f_p,f_{p'}+1}\times\sum\lm_{{\rm 1-instantons}\;(p\to p')}\Delta,
\ee
where $n$-instantons interpolating between points $p$ and $p'$ are solutions to the following differential equation boundary value problem:
\be\label{instanton}
\begin{split}
	&\p_{\tau}x^i(\tau)=-g^{ij}\p_{x^j}\fH(x(\tau)),\\
	&\lim\lm_{\tau\to-\infty}x^i(\tau)=p,\quad \lim\lm_{\tau\to-\infty}x^i(\tau)=p'.
\end{split}
\ee
1-instanton solutions to \eqref{instanton} have only a single translation modulus $m$ mapping a solution $x^i(\tau)$ to another solution $x^i(\tau+m)$. $\Delta$ in \eqref{differential} is a $\pm 1$-valued contribution
$$
\Delta=\frac{{\rm Det}'\hat D}{|{\rm Det}'\hat D|},
$$
where $\hat D$ is a Dirac operator in the instanton background:
$$
\hat D^i_k:=\p_{\tau}\delta^i_k+\p_{x^k}\left(g^{ij}\p_{x^j}\fH\right),
$$
and we subtract the contribution of the zero mode corresponding to the modulus $m$ from the determinant.

The BPS Hilbert space is defined as a cohomology:
\be\label{BPS_Hilb}
\fG_{\rm BPS}^*\cong H^*({\bf M},{\bf Q}),
\ee
where cohomological grading coincides with the fermion number grading of the Hilbert space.

\subsubsection{Wilsonian renormalization ``exact'' in one loop}

In the previous section we assumed that $\fV$ is a set of isolated points, now we rather assume $\fV$ to be a single component connected hypersurface in $\CT_{\rm QM}$. 
One can divide coordinates $x^i$ spanning $\CT_{\rm QM}$ in two groups: perpendicular to $\fV$, or ``fast'' variables $x_f$; and tangent to $\fV$, or ``slow'' variables $x_s$. Slow variables $x_s$ are also referred to as \emph{vacuum moduli}. 
Perturbative modes of $x_f$-fields around vacuum value $x_f=0$ have non-zero masses, therefore their temporal frequencies are rather high, at least greater or equal to corresponding masses in their absolute values, this is why we call these modes fast. 
The perturbative slow modes $x_s$ are analogs of Goldstone modes and have zero masses, they can form even stationary field configurations -- condensates.

Separation of slow and fast field modes is a usual prologue to the Wilsonian renormalization group \cite{ZinnJustin:2002ru}.
We easily calculate the action of the renormalization group on the supercharges. 
Let us just redefine variables:
$$
x_f\longrightarrow \hbar^{\frac{1}{2}}x_f.
$$
For the supercharge we have the following decomposition:
\be
\CQ(\hbar)=\sum\lm_{n=0}^{\infty} \hbar^{\frac{n+1}{2}}\CQ^{(n)},
\ee
where
\be
\CQ^{(0)}=d_f+\Omega(x_s)\cdot x_f.
\ee
Here $\Omega(x_s)\neq 0$ defines a linear functional in $x_f$, and differential $d_f$ defines differentiation only in fast variables $x_f$. We search for a BPS wave function having a suitable decomposition:
\be\label{expan}
\Psi_{\rm BPS}(\hbar)=\sum\lm_{n=0}^{\infty}\hbar^{\frac{n}{2}}\Psi^{(n)}.
\ee
Substituting such $\hbar$-expansions into \eqref{BPS_wave} we have at the zeroth order:
\be\label{0th}
\CQ^{(0)}\Psi^{(0)}=\bar\CQ^{(0)}\Psi^{(0)}=0.
\ee
These equations are simple linear differential equations in $x_f$, $\Psi^{(0)}$ has a natural form of a wave function localized at $x_f=0$ with quantum corrections suppressed by a Gaussian exponent. 
Solutions to the linear differential equations \eqref{0th} form a one-dimensional linear space $C\cdot \Psi^{(0)}$, where $C$ is usually referred to as an integration constant. 
Since fields $x_s$ in \eqref{0th} play a role of mere parameters we conclude that if  $\Psi^{(0)}(x_f,x_s)$ is a solution to \eqref{0th} then any solution  to \eqref{0th} has the following form:
$$
C(x_s)\cdot \Psi^{(0)}(x_f,x_s),
$$
where $C(x_s)$ is a generic functional of $x_s$ remaining undefined so far. 
At the first order of approximation we find the following equations:
\be
\begin{split}
	\CQ^{(1)}\cdot C\cdot\Psi^{(0)}+\CQ^{(0)}\cdot \Psi^{(1)}=0,\\
	\bar\CQ^{(1)}\cdot C\cdot\Psi^{(0)}+\bar\CQ^{(0)}\cdot \Psi^{(1)}=0.
\end{split}
\ee
Multiplying both equations from the left by $\Psi^{(0)\dagger}$ and integrating over $x_f$ we derive equations defining $C$:
\be
\CQ_{\rm eff}C=\bar\CQ_{\rm eff} C=0,
\ee
where
\be\label{Q_eff}
\begin{split}
	\CQ_{\rm eff}(x_s)=\int dx_f\; \Psi^{(0)\dagger}(x_f,x_s)\CQ^{(1)}(x_f,x_s)\Psi^{(0)}(x_f,x_s),\\
	\bar \CQ_{\rm eff}(x_s)=\int dx_f\; \Psi^{(0)\dagger}(x_f,x_s)\bar\CQ^{(1)}(x_f,x_s)\Psi^{(0)}(x_f,x_s).
\end{split}
\ee
Supercharges \eqref{Q_eff} represent a one-loop Wilsonian renormalization of initial supercharges, therefore we call them \emph{effective supercharges} defining the IR theory, and $C$ is an \emph{effective wave-function}.

Since we are working in the semi-classical limit $\hbar\to 0$ all the higher orders except the zeroth one in expansion \eqref{expan} can be neglected, and we have just described a procedure to calculate the latter. 
This procedure allows one to reformulate the problem of defining BPS wave-functions in terms of an effective theory, and since only one loop renormalization correction is taken into account we call it one-loop exact.

As well there is a non-trivial Jacobian modification to a Hilbert space norm of the  effective wave-functions:
\be
\begin{split}
	\langle C'|C\rangle&=\int dx_s \;J(x_s)\;\bar C'(x_s)C(x_s),\\
	\mbox{where }J(x_s)&=\int dx_f\; \Psi^{(0)\dagger}(x_f,x_s)\Psi^{(0)}(x_f,x_s).
\end{split}
\ee

\subsection{Localization in GLSM with an interface}\label{sec:loc_GLSM}
\subsubsection{Gauged linear sigma-model}
Now let us turn back to 2d $\CN=(2,2)$ gauged linear sigma models (GLSM). 
We consider a $U(1)$-theory with $n_f$ chiral multiplets with charges $Q_a$ and masses $\mu_a$ for $a=1,\ldots,n_f$ on an interval $[0,L]$. 
GLSM with a generic gauge group can be considered in a similar fashion.
At boundaries of the interval we choose Chan-Paton boundary conditions (see Appendix  \ref{sec:App_boundary}).
Chan-Paton factors contribute twice. 
The first contribution is to the supercharge, it is given by a holomorphic function $\bar \CQ_{\rm bdry}$ of chiral fields $\phi_a$ and boundary anti-fermion fields $\bar \chi_i$. 
Another contribution is to the boundary electric charge $\bf q$ of the brane. 
An expression for the B-twist in this theory with a restored $\hbar$-dependence reads:
\be\label{supercharge}
\begin{split}
	\bar\CQ_B=\int dx^1\;\Bigg[&\lambda_1\left(-\I\hbar\delta_{\sigma_{\II}}-\I\p_1\sigma_{\IR}-\hbar\delta_{A_1}+\I\hbar \theta\right)+\\
	+&\lambda_2\left(\hbar\delta_{\sigma_{\IR}}-\p_1\sigma_{\II}+\left(\sum\lm_a Q_a|\phi_a|^2-r\right)\right)-\\
	-&\I\sqrt{2}\sum\lm_{a}\bar\psi_{\dot 1,a}\left(\hbar\delta_{\bar\phi_a}+(Q_a\sigma_{\IR}-\mu_{\IR,a})\phi_a\right)-\\-&\sqrt{2}\sum\lm_{a}\bar\psi_{\dot 2,a}\left(\left(\p_1\phi_a+\I Q_a A_1\phi_a\right)-(Q_a\sigma_{\II}-\mu_{\II,a})\phi_a\right)\Bigg]+\bar\CQ_{\rm bdry},
\end{split}
\ee 

As well we will need the Gauss current operator expression:
\be\label{CJ}
\CJ= \sum\lm_{a}Q_a\left(\left(\bar\phi_a \delta_{\bar\phi_a}-\phi_a \delta_{\phi_a}\right)+\bar\psi_{\dot 1,a}\psi_{1,a}+\bar\psi_{\dot 2,a}\psi_{2,a}\right)+\I\p_1 \delta_{A_1}+{\bf q}\delta({\rm bdry}),
\ee
where $\bf q$ is an electric charge produced by the boundary branes.

Supercharge satisfies the following superalgebra:
\be
\CQ_B^2=\CZ,\quad \left\{\CQ_B,\bar{\CQ}_B\right\}=2\CH+2{\rm Re}\,(\zeta^{-1}\tilde \CZ),,
\ee
where $\zeta$ is an extra phase (see \eqref{AB-twists}), and $\CZ$ and $\tilde \CZ$ are supercharge and twisted supercharge operators respectively.

From this algebraic relations it is clear that the Hamiltonian spectrum is bounded from below.
Wave functions of physical BPS states are annihilated by the following operators:
\be
\CQ_B|\Psi_{\rm BPS}\rangle=\bar\CQ_B|\Psi_{\rm BPS}\rangle=\CJ|\Psi_{\rm BPS}\rangle=0.
\ee
On the BPS Hilbert space the Hamiltonian acquires an eigen value:
\be
\CH|\Psi_{\rm BPS}\rangle=-{\rm Re}(\zeta^{-1}\tilde \CZ) |\Psi_{\rm BPS}\rangle.
\ee

To introduce an interface dependence of parameters on the spatial coordinate $x^1$ it suffices to substitute constant parameters in the expression for the supercharge \eqref{supercharge} by functions:
\be
\theta\to\theta(x^1),\quad r\to r(x^1),\quad \mu_a\to \mu_a(x^1).
\ee
Using the inverse Lagrange transform on the Hamiltonian operator $\CH$ one can reconstruct the original action of the theory in the presence of an interface. 
We will restrict ourselves to an insertion of an interface defect only for FI parameters and $\theta$-angles. 
In this case it is easy to trace back the interface modification needed for the action to be invariant under the B-twist. 
It suffices to modify the corresponding term \eqref{S_FI}:
\be
\CS_{\rm FI,\,\theta}'=\int dx^0dx^1\,\left[-r\,{\bf D}+\theta\, F_{01}+\p_1r\,\sigma_{\II}-\p_1\theta\,\sigma_{\IR}\right].
\ee

The central charge in this model reads:
\be\label{cc}
\begin{split}
	\tilde \CZ=\Delta\I\left\{\hbar\sum\lm_a \bar\psi_{-,a}\psi_{+,a}+\hbar\bar\sigma\bar  t'-\sum\lm_a\left(Q_a\bar\sigma-\bar\mu_a\right)|\phi_a|^2-\hbar \bar\sigma\delta_{A_1}\right\}+\\
	+\hbar\sum\lm_{a}\bar\mu_a\int dx^1\,\left(\bar\phi_a\delta_{\bar\phi_a}-\phi_a\delta_{\phi_a}+\bar\psi_{1,a}\psi_{1,a}+\bar\psi_{2,a}\psi_{2,a}\right),
\end{split}
\ee
where $\Delta$ denotes the difference of values on interval boundaries, and $t'$ is an FI parameter complexified by the topological angle and shifted by the boundary charge:
\be
t'=\hbar^{-1}r-\I (\theta+{\bf q}_{\rm bdry})\,.
\ee

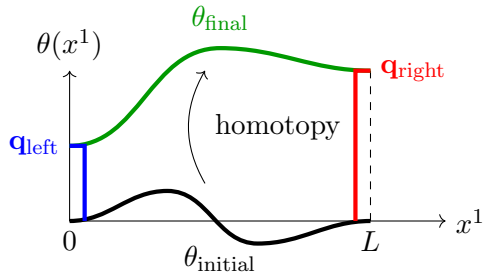
\begin{figure}[h!]
	\begin{center}
		\begin{tikzpicture}
			\draw[ultra thick] (0,0) to[out=0,in=180] (1.3,0.4) to[out=0,in=180] (2.5,-0.3) to[out=0,in=180] (4,0);
			\draw[<->] (0,2) -- (0,0) -- (5,0);
			\draw[dashed] (4,0) -- (4,2);
			\node[right] at (5,0) {$x^1$};
			\node[above] at (0,2) {$\theta(x^1)$};
			\node[below] at (0,0) {$0$};
			\node[below] at (4,0) {$L$};
			\draw[ultra thick, black!40!green] (0,1) to[out=0,in=180] (2,2.3) to[out=0,in=180] (4,2);
			\draw[ultra thick, blue] (0,1) -- (0.2,1) -- (0.2,0);
			\draw[ultra thick, red] (4,2) -- (3.8,2) -- (3.8,0);
			\node[left,blue] at (0,1) {${\bf q}_{\rm left}$};
			\node[right,red] at (4,2) {${\bf q}_{\rm right}$};
			\draw[->] (1.8,0.5) to[out=120,in=240] (1.8,2);
			\node[right] at (1.8,1.25) {homotopy};
			\node at (2,-0.5) {$\theta_{\rm initial}$};
			\node[black!40!green] at (2,2.7) {$\theta_{\rm final}$};
		\end{tikzpicture}
		\caption{Boundary brane charge contribution to the topological angle.}\label{fig:bdry_chg}
	\end{center}
\end{figure}

The gauge symmetry allows one to vary the phase of the wave-function:
\be
\Psi\to \exp\left(\I \int\varphi(x^1)A_1(x^1)\;dx^1\right)\, \Psi.
\ee
This variation does not change the integrability properties of the wave function, however it adds a shift to the Gauss charge operator and to the topological angle:
\be
\begin{split}
	\CJ(x^1)\to\CJ(x^1)-\p_1\varphi(x^1),\quad \theta(x^1)\to \theta(x^1)-\varphi(x^1).
\end{split}
\ee
We will use this shift to delete the contribution form the boundary charges:
\be
\varphi={\bf q}\;\Theta({\rm bdry}),
\ee
where $\Theta$ is the Heaviside step function. 
So we can move shifts due to the boundary brane electric charges to contribute as step-like shifts. 
We expect that the parallel transport is a homotopy invariant of interface path $\wp$.
One can smear step-like contributions at the boundaries  using this homotopy invariance (see Fig.\ref{fig:bdry_chg}). 
In what follows rather than considering the contribution of the boundary charges to the Gauss law we will imply that they perform appropriate shifts to the boundary values of the topological angle function $\theta(x^1)$ on the interface. 
Surely, the parameter space $\CP$ spanned by $t'$ will have singular loci where the theory is ill-defined. 
The homotopy of $\wp$ has to take these singularities into account.

To pursue our goal to study the IR physics of BPS states in this theory we follow the steps discussed in Section \ref{sec:Heis_loc} and construct the MSW complex. 
The first step in this paradigm is to define the vacuum locus $\fV$. $\fV$ is spanned by field configurations satisfying the following set of equations:
\be\label{sol_vac}
\begin{split}
	\p_1\sigma_{\IR}=O(\hbar),\\
	-\p_1\sigma_{\II}+\left(\sum\lm_aQ_a|\phi_a|^2-r\right)=O(\hbar),\\
	(Q_a\sigma_{\IR}-\mu_{\IR,a})\phi_a=O(\hbar),\\
	\left(\p_1\phi_a+\I Q_a A_1\phi_a\right)-(Q_a\sigma_{\II}-\mu_{\II,a})\phi_a=O(\hbar).
\end{split}
\ee
By choosing $O(\hbar)$ as the right hand side for those equations rather than just zeroes we would like to stress again that we allow quantum corrections to ``blur'' $\fV$.

Constant solutions to those equations satisfy the following constraints:
\be
\sum\lm_aQ_a|\phi_a|^2-r=O(\hbar),\quad (Q_a\sigma-\mu_a)\phi_a=O(\hbar).
\ee

Based on parameter values of these equations we could distinguish three phases:
\begin{enumerate}
	\item {\bf Coulomb branch.} It appears when $r=O(\hbar)$, $\mu_a=O(1)$. The solution is dominated by expectation values of the scalar in the gauge multiplet $\sigma=O(1)$. The chiral multiplets are considered to perform quantum corrections $\phi_a=O\left(\hbar^{\frac{1}{2}}\right)$.
	\item {\bf ``Soft'' Higgs branch.} It appears when $r=O(1)$, $\mu_a=O(\hbar)$. Here the situation is opposite to the Coulomb branch, the scalars from the chiral multiplet acquire expectation values $|\phi_a|=O(1)$, and the scalar from the gauge multiplet delivers corrections $\sigma=O(\hbar)$.
	\item {\bf Mixed/``rigid'' Higgs branch/crystal Coulomb branch.} It appears when $r=O(1)$, $\mu_a=O(1)$. The theory is localized in one of the fixed points on the soft Higgs branch: $|\phi_a|=r^{\frac{1}{2}}+O(\hbar)$, $\sigma=Q_a^{-1}\mu_a+O(\hbar)$.
\end{enumerate}

In the literature a more preferable name for the third option is a mixed branch since scalars belonging to both gauge and chiral multiplets acquire expectation values. 
We would like to save also both names ``rigid Higgs branch'' and ``crystal Coulomb branch'' and stress that a distinction between the rigid and the soft Higgs branches is rather contingent. 
In principle the Morse height functional can be split in two gauge-invariant parts separating $r$ and $\mu_a$. 
Those parts can be re-scaled also separately introducing two RG flow parameters $\hbar_1$ and $\hbar_2$.
The mass parameters in the resulting supercharge expression will scale as
$$
\mu_a\sim \hbar_1/\hbar_2.
$$
The general theory predicts that RG flows in $\hbar_1$ and $\hbar_2$ commute. The soft Higgs branch will appear naturally if we take limit $\hbar_1\to 0$ first, then limit $\hbar_2\to 0$ will perform a further flow to the rigid Higgs branch. 
Mathematically this corresponds to a further localization of coherent sheaves on the soft Higgs branch moduli space due to equivariant action with $\mu_a$ parameterizing equivariant tori.
We will comment on ``crystal Coulomb branch'' name in Section \ref{sec:quiver}.

So we would like to distinguish Higgs and Coulomb phases of the theory not by fields expectation values rather by orders of the parameter $r$ that we allow to vary along the interface. 
So a BPS state in our theory will look like a thick ``sandwich'' of various phases. 
Since these phases are just dual descriptions of the same physics we do not expect an appearance of domain walls between those phases.

It is clear that $\fV$ may have moduli. 
We would like to flow along $\hbar\to 0$ to integrate out degrees of freedom perpendicular to $\fV$ and derive an effective description. 
In what follows we will describe effective theories for both Higgs and Coulomb branches for constant values of parameters $\mu_a$ and $t$. 
In the remaining part we will briefly remind the web formalism technique \cite{GMW} allowing one to compute the MSW complex cohomologies for supersymmetric interfaces including non-trivial dependence of parameters on the spatial coordinate $x^1$ in a universal way.

\subsubsection{Brane boundary conditions}
To proceed we need to choose boundary conditions on the ends of interval $[0,L]$ by imposing constraints that are invariant under the RG flow. 
Only in this way we could guarantee that the actual Hilbert spaces of BPS states are dual to each other.

In general, the supercharge we have chosen $\CQ_B$ to localize with respect to is not nilpotent (see \eqref{AB-algebra}). 
For its nilpotency one has to impose constraints on the superpotential and charge $\CZ$ leading to boundary conditions for field modes. 
The boundary conditions for gauge multiplet naturally follow from manifestly RG invariant constraint:
$$
\CQ_B^2=0.
$$

To set RG invariant boundary conditions for the chiral fields we follow approach of \cite{Dimofte:2019zzj,Cordes:1994fc} and rewrite the chiral fields in terms of superfields of the B-type supersymmetry preserved by the boundary. 
A relation between the bulk and boundary supersymmetries is described in Appendix \ref{sec:brane_cond}. 

For chiral fields we choose Neumann type boundary conditions allowing the branes to cover the whole vacuum variety. If some subvariety is needed we will use Chan-Paton factors carrying a complex associated to the corresponding structure sheaf. Thus for scalars we have a gauge-invariant version of \eqref{chiral_brane}:
\be
D_1\phi_a-(Q_a\sigma_{\II}-\mu_{\II,a})\phi_a\big|_{\rm bdry}=0,\quad (Q_a\sigma_{\IR}-\mu_{\IR,a})\phi_a\big|_{\rm bdry}=0.
\ee
Boundary conditions for superpartners can be produced by B-twist actions.

Surely, proposed boundary conditions admit generalizations \cite{Hori:2013ika} by putting additional operators and charges on the brane, we will not consider these instances for keeping things simple.

\subsection{Coulomb branch: Landau-Ginzburg (LG) model}\label{sec:LG_model}
\subsubsection{Wilsonian renormalization}
According to the Coulomb branch localization prescription we expect that the Higgs fields $\phi_a$ do not acquire expectation values. For this situation to represent a valid classical vacuum it has to satisfy the D-term equation implying $r=0$. However this constraint has to be satisfied on the classical level only, we are able to admit a softer quantum limit $r=O(\hbar)$. Therefore the following redefinition is suitable: 
$$
r=\hbar \tilde r,\quad \tilde r=O(\hbar^0).
$$
We decompose remaining fields accordingly:
\be
\phi_a\to \hbar^{\frac{1}{2}}\phi_a,\quad \sigma\to \Sigma+\hbar^{\frac{1}{2}}\sigma,
\ee
where $\Sigma$ is a slow component of the scalar $\sigma$ satisfying 
$$
|\p_1\Sigma|= O(\hbar).
$$

After integration over the fast fields $\phi_a$ and $\sigma$ and rescaling the spatial coordinate $x^1 \to \hbar^{-1}x^1$ the supercharge and the Gauss law constraint have the following form:
\be
\begin{split}
	\bar\CQ_{\rm eff}=\int dx^1\;\Bigg[&\lambda_1\left(-\I\delta_{\Sigma_{\II}}-\I\p_1\Sigma_{\IR}-\delta_{A_1}+\I\theta\right)+\\
	+&\lambda_2\left(\delta_{\Sigma_{\IR}}-\p_1\Sigma_{\II}+\left(\sum\lm_a Q_a\;{\rm Re}\,Y_a-\tilde r\right)\right)\Bigg],\\
	\CJ=&\I\p_1 \delta_{A_1}-\sum\lm_aQ_a\;\p_1{\rm Im}\,Y_a,
\end{split}
\ee 
where the following notion for the disorder operators $Y_a$ is introduced:
\be
\begin{split}
	{\rm Re}\,Y_a&=\big\langle \Psi^{(0)}\big||\phi_a|^2\big|\Psi^{(0)}\big\rangle,\\
	\p_1{\rm Im}\,Y_a&=\big\langle \Psi^{(0)}\big|-\left(\bar\phi_a\delta_{\bar\phi_a}-\phi_a\delta_{\phi_a}+\bar\psi_{\dot 1,a}\psi_{1,a}+\bar\psi_{\dot 2, a}\psi_{2,a}\right)\big|\Psi^{(0)}\big\rangle.
\end{split}
\ee
Expectation values of $Y_a$ can be easily computed at one loop along the lines of Section \ref{sec:Preamble}, we put an explicit computation in Appendix \ref{sec:disorder}. 
The result reads:
\be
Y_a=-\frac{1}{2\pi}\log\frac{Q_a\Sigma-\mu_a}{\Lambda}.
\ee
As in Section \ref{sec:Preamble} logarithm multi-valuedness corresponds to a discrete ambiguity in a choice of a Dirac string contribution. 

The Gauss law constraint can be easily resolved by the phase dependence of the wave function on the gauge field:
$$
\exp\left(-\I\int dx^1A_1\sum\lm_a Q_a\;{\rm Im}\,Y_a\right)\Psi(\Sigma).
$$

The resulting effective supercharge reads:
\be\label{LG_eff_Q}
\begin{split}
	\bar\CQ_{\rm LG}=\int dx^1&\left[\zeta\left(\lambda^2-\lambda^1\right)\left(\delta_{\bar\Sigma}+\frac{\I}{2}\p_1\Sigma-\zeta^{-1}\frac{\bar W'}{2}\right)+\right.\\
	&\left.+\zeta^{-1}\left(\lambda^1+\lambda^2\right)\left(\delta_{\Sigma}-\frac{\I}{2}\p_1\bar\Sigma-\zeta\frac{W'}{2}\right)\right],
\end{split}
\ee
where we used an effective superpotential generated in this model:
\be
W(\Sigma)=t\,\Sigma+\sum\lm_a\frac{Q_a\Sigma-\mu_a}{2\pi}\left[\log\left(\frac{Q_a\Sigma-\mu_a}{\Lambda}\right)-1\right],
\ee
and the FI parameter is complexified with the use of a topological angle contribution:
$$
t=\tilde r-\I \theta.
$$
The effective central charge reads:
\be\label{ecc}
\tilde Z=\I\zeta^{-1}\,\Delta\bar W.
\ee

This supercharge defines a Landau-Ginzburg model on a complex plane spanned by $\Sigma$ with (twisted) superpotential $W$ (compare to \eqref{nlin_sup}).

\subsubsection{Landau-Ginzburg model}
Let us consider a more generic Landau-Ginzburg (LG) model on an $m$-dimensional K\"ahler manifold $X_{\rm LG}$ spanned by scalars $\Sigma^I$, $I=1,\ldots, m$, with a metric tensor given by $g_{I\bar J}$. Despite the case we derived above corresponds to a simple flat complex plane a more generic analysis will still go through. The superpotential is given by a holomorphic function $W$. The vacuum locus $\fV$ is spanned by solutions to the following equation:
\be\label{LG_soliton}
\I\p_1\Sigma^I=\zeta^{-1}g^{I\bar J}\overline{\p_{\Sigma^J}W}.
\ee
Constant field configurations satisfying \eqref{LG_soliton} correspond to roots of an algebraic system:
$$
\p_{\Sigma^J}W(\Sigma)=0.
$$
We call these roots LG vacua and denote by an index with an asterisk as $\Sigma_{*i}$.

To define appropriate boundary conditions apply the B-twist to $\CQ_B^2+\bar\CQ_B^2$. We will find that the result is equivalent to a sum of supercurrents through the boundary. Literally repeating arguments of \cite[Section 39.2.2]{D-book_1} we will arrive to a conclusion that the boundary D-brane has a form of a special Lagrangian. As a basis of such special Lagrangians we choose Lefshetz thimbles.

\begin{figure}[h!]
	\begin{center}
		\begin{tikzpicture}
			\draw[<-] (-5,0.25) -- (1.5,0.25);
			\draw[->] (1.5,0.25) -- (3,-1.25);
			\node[right] at (3,-1.25) {$-{\rm Im}\,\zeta W$};
			\node[left] at (-5,0.25) {$-{\rm Re}\,\zeta W$};
			\draw[->] (1.5,0.25) -- (1.5,2);
			\draw[ultra thick, red,postaction={decorate},decoration={markings, 
				mark= at position 0.65 with {\arrow{stealth}}}] (-2,0) -- (-1.3,-0.7) -- (-1,-1);
			\node[above right] at (1.5,2) {$\Sigma$-fiber};
			\node[right] at (0.1,0) {$\zeta W_{*i}$};
			\node[right] at (1.4,-1) {$\zeta W_{*j}$};
			\draw[ultra thick, -stealth]  (0,0) -- (-4,0);
			\draw[ultra thick] (-0.1,-0.1) -- (0.1,0.1) (0.1,-0.1) -- (-0.1,0.1);
			\begin{scope}[shift={(0,2)}]
				\begin{scope}[yscale=0.3]
					\draw[fill=white!40!gray] (-4,-1) -- (-2,-1) to[out=0,in=270] (0,0) to[out=90,in=0] (-2,1) -- (-4,1);
				\end{scope}
				\begin{scope}[shift={(-4,0)}]
					\begin{scope}[xscale=0.5]
						\draw[fill=white!40!gray] (0,0) circle (0.3);
					\end{scope}
				\end{scope}
			\end{scope}
			\draw[fill=black] (0,2) circle (0.05);
			\draw[thick, dashed] (0,0) -- (0,2);
			\draw[red,dashed] (-2,0) -- (-2,2);
			\begin{scope}[shift={(1.3,-1)}]
				\draw[ultra thick, -stealth]  (0,0) -- (-4,0);
				\draw[ultra thick] (-0.1,-0.1) -- (0.1,0.1) (0.1,-0.1) -- (-0.1,0.1);
				\begin{scope}[shift={(0,2)}]
					\begin{scope}[yscale=0.3]
						\draw[fill=white!40!gray] (-4,-1) -- (-2,-1) to[out=0,in=270] (0,0) to[out=90,in=0] (-2,1) -- (-4,1);
					\end{scope}
					\begin{scope}[shift={(-4,0)}]
						\begin{scope}[xscale=0.5]
							\draw[fill=white!40!gray] (0,0) circle (0.3);
						\end{scope}
					\end{scope}
				\end{scope}
				\draw[fill=black] (0,2) circle (0.05);
				\draw[thick, dashed] (0,0) -- (0,2);
			\end{scope}
			\draw[red,dashed] (-1,-1) -- (-1,1);
			\node[left] at (-4.25,2) {$\CL_i$};
			\node[left] at (-2.95,1) {$\CL_j$}; 
			\draw[ultra thick, red] (-2,2) to[out=350,in=150] (-1.3,2.5) to[out=330,in=60]  (-1,1);
			\draw[fill=red, red] (-2,0) circle (0.07) (-1,-1) circle (0.07) (-2,2) circle (0.07) (-1,1) circle (0.07);
			\node[right, red] at (-1.2,2.5) {BPS string};
		\end{tikzpicture}
		\caption{Lefshetz thimbles.}\label{fig:thimbles} 
	\end{center}
\end{figure}
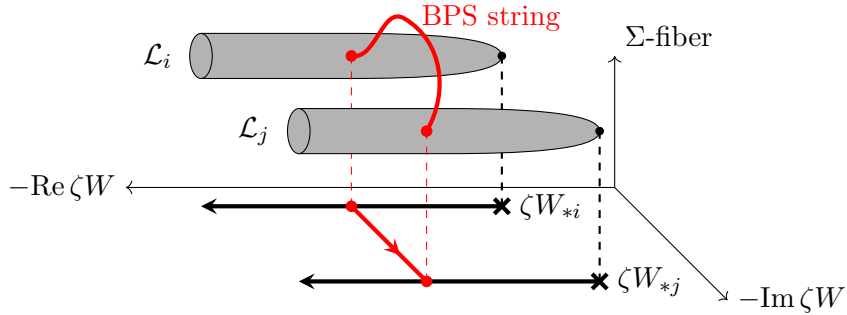

A Lefshetz thimble $\CL_i$ is labeled by LG vacuum $(*i)$ and as a manifold is given by a union of all trajectories satisfying the following asymptotic value problem:
\be\label{Lef_thi_eq}
\p_\tau\Sigma^I=-\zeta^{-1}g^{I\bar J}\overline{\p_{\Sigma^J}W},\quad \lim\lm_{\tau\to-\infty}\Sigma(\tau)=\Sigma_{*i}.
\ee
For equations defining $\fV$ and the Lefshetz thimble we easily derive:
\be
\p_1(\zeta W)=-\I g^{I\bar J}\p_{\Sigma^I}W\overline{\p_{\Sigma^J}W}\in-\I \IR_{\geq 0},\quad \p_\tau(\zeta W)=-g^{I\bar J}\p_{\Sigma^I}W\overline{\p_{\Sigma^J}W}\in- \IR_{\geq 0}.
\ee
Therefore, in the $\zeta W$-plane the thimbles are represented by rays starting in critical values $W_{*i}$ and flowing parallel to the real axis to the left, and solutions to \eqref{LG_soliton} are segments connecting those rays. In the field space $X_{LG}$ the Lefshetz thimbles are finger-looking Lagrangian fibrations of $S^{m-1}$ with the sphere shrinking (vanishing) at the thimble tip. Solutions to \eqref{LG_soliton} are BPS strings stretched between different thimbles or zero size strings concentrated at thimble tips (see Fig.\ref{fig:thimbles}).

Consider an auxiliary category $\CC$ where objects are Lefshetz thimbles and morphisms are defined as BPS Hilbert spaces \eqref{BPS_Hilb} for corresponding choices of boundary conditions:
$$
{\rm Hom}^*(\CL_i,\CL_j):=\fG_{\rm BPS}(\CL_i,\CL_j).
$$
A Fukaya-Seidel category is a category whose objects are Lagrangian submanifolds given by vanishing cycles \cite{Haydys:2010dv}. 
In our case these vanishing cycles are fibrations of spheres $S^{m-1}$ along rays in $\zeta W$-plane emanating from critical values of $W$. 
Having defined superpotential $W$ as well we can define the morphisms similarly as BPS Hilbert spaces for appropriate boundary conditions.\footnote{In principle, this definition requires a modification known as a wrapped Fukaya category \cite{Auroux,aganagic2021knot} since we allow Lagrangian submanifolds to approach singularities. In this case their asymptotic behavior has to be fixed by additional data.} We clearly see that morphism space for a pair of thimbles $\CL_i$ and $\CL_j$ is nonzero only if 
\be\label{Lef_ord}
{\rm Im}\,\zeta W_{*i}\geq {\rm Im}\,\zeta W_{*j}.
\ee
Therefore ordering thimbles according to ${\rm Im}\,\zeta W$ values one acquires naturally a sequence satisfying an exceptional collection constraint automatically. 
The Fukaya-Seidel category is a triangulated category, and it can be generated by acting on exceptional objects through shifts and exact triangles \cite{Seidel_book}. 
Thus for our purposes we accept as a definition for a Fukaya-Seidel category of brane boundary conditions a derived category of the auxiliary category $\CC$. 
Some details on a definition of a derived category are presented in Appendix \ref{sec:categories}.

Physically, this setup implies that as boundary conditions we accept Lefshetz thimbles with possible boundary operator insertions \cite{GMW} allowing a brane to carry a non-trivial complex. 
We will consider some simple examples of boundary operators when turn to the instantons in Section \ref{sec:Web}.

The very construction implies that we consider our theory in the limit of a large interval $L\to\infty$. Physical sizes of BPS strings in the spatial dimension and $\zeta W$-plane are related:
$$
L=-\int\frac{d\,{\rm Im}\,\zeta W}{\left|g^{I\bar J}\p_{\Sigma^I}W\overline{\p_{\Sigma^J}W}\right|}.
$$
To approach the limit $L\to\infty$ we need to move the BPS string segment in the $\zeta W$-plane towards the thimble tips -- LG vacua --  where the denominator diverges. 
Near LG vacuum $(*i)$ one can linearize equation \eqref{LG_soliton} for deviation $\delta\Sigma$ from the vacuum value:
\be\label{LG_sol_lin}
\p_1\delta\Sigma^I=-\I\zeta^{-1}\left(g^{I\bar J}\overline{\p^2_{JK}W}\right)_{*i}\overline{\delta\Sigma^K}
\ee
A solution of this linear equation decays exponentially fast, as a solution to \eqref{LG_soliton} approaches the vacuum value. 
This shrinks a core of a solution to \eqref{LG_soliton} where the solution differs from either constant vacuum solution to a narrow region of size $\lambda^{-1}$, where $\lambda$ is the minimal absolute value of eigen values of the linear operator in \eqref{LG_sol_lin}. 
Such a solution behaves as a quasi-particle -- a \emph{soliton} --  or a domain wall in $(x^0,x^1)$-space-time.  

Depending on a relation between ${\rm Re}\,\zeta W$ for boundary vacua a soliton can be either confined at one of the branes, or freely moving in the spatial $x^1$-direction:
\be\label{quasi_par_loc}
\begin{array}{c}
	\begin{tikzpicture}
		\draw[ultra thick] (-1,0) -- (1,0) (-1,-0.1) -- (-1,0.1) (1,-0.1) -- (1,0.1);
		\node[above] at (-1,0.2) {$0$};
		\node[above] at (1,0.2) {$L$};
		\shade[ball color = red] (-1,0) circle (0.15);
		\node[below] at (0,-0.2) {${\rm Re}\,\zeta W_{*i}>{\rm Re}\,\zeta W_{*j}$};
		\node[left] at (-1.2,0) {$*i$};
		\node[right] at (1.2,0) {$*j$};
		\begin{scope}[shift={(4,0)}]
			\draw[ultra thick] (-1,0) -- (1,0) (-1,-0.1) -- (-1,0.1) (1,-0.1) -- (1,0.1);
			\node[above] at (-1,0.2) {$0$};
			\node[above] at (1,0.2) {$L$};
			\shade[ball color = red] (0,0) circle (0.15);
			\node[below] at (0,-0.2) {${\rm Re}\,\zeta W_{*i}={\rm Re}\,\zeta W_{*j}$};
			\node[left] at (-1.2,0) {$*i$};
			\node[right] at (1.2,0) {$*j$};
			\node[above] at (0,0.15) {$x_c$};
		\end{scope}
		\begin{scope}[shift={(8,0)}]
			\draw[ultra thick] (-1,0) -- (1,0) (-1,-0.1) -- (-1,0.1) (1,-0.1) -- (1,0.1);
			\node[above] at (-1,0.2) {$0$};
			\node[above] at (1,0.2) {$L$};
			\shade[ball color = red] (1,0) circle (0.15);
			\node[below] at (0,-0.2) {${\rm Re}\,\zeta W_{*i}<{\rm Re}\,\zeta W_{*j}$};
			\node[left] at (-1.2,0) {$*i$};
			\node[right] at (1.2,0) {$*j$};
		\end{scope}
	\end{tikzpicture}
\end{array}
\ee
In the latter case the core position $x_c$ is a translation modulus of a soliton solution at $L\to\infty$. Depending on situation we name corresponding solutions left or right confined soliton or a free $ij$-soliton. We should stress that the modulus $x_c$ is fictitious and does not appear at finite $L$,  at finite $L$ there is still a small interaction between the quasi-particle and boundaries forcing $x_c$ to have a fixed equilibrium value.
\subsection{``Soft'' Higgs branch: coherent sheaves}

Let us start with the strict case $\mu_a=0$.

In this case vacuum equations \eqref{sol_vac} have a trivial almost stationary solution:
\be
\begin{split}
	\sigma_{\IR}=\sigma_{\II}=0,\\
	\phi_a=e^{\I Q_a \vartheta_0}\Phi_a,\\
	A_1=-\p_1\vartheta_0,
\end{split}
\ee
where $\Phi_a$ are Higgs branch moduli satisfying
\be\label{Hb}
\sum\lm_{a=1}^n Q_a|\Phi_a|^2=r,
\ee
and $\vartheta_0$ is an arbitrary function. 
Let us argue there are no other soliton solutions. 
We can deduce a secondary equation:
\be\label{no_soli}
\left(-\p_1^2+\sum\lm_a Q_a^2|\phi_a|^2\right)\sigma_{\II}=0.
\ee
Notice the operator in brackets in \eqref{no_soli} has only positive eigenvalues for constant vacuum boundary conditions, therefore this equation has no soliton solutions except the trivial one.

The equivariant term in the supercharge localizes field configurations to constant modes up to a gauge transformation. 
After integration over quickly oscillating modes we derive an effective supercharge depending only on constant modes of $\phi_a$ and $\sigma_{\IR}$.
We put details on the one-loop renormalization in this system in Appendix \ref{sec:Ren_soft}. 
The effective supercharge reads:
\be\label{eff_supercharge}
\begin{split}
	\bar\CQ_B^{(\rm eff)}=-\I\sqrt{2}\sum\lm_a\bar\eta_{a}\left(\hbar \p_{\bar \Phi_{a}}+Q_a\Sigma_{\IR}\Phi_{a}\right)+\\
	+\nu\left(\p_{\Sigma_{\IR}}+\left(\sum\lm_aQ_a|\Phi_{a}|^2-r\right)\right)+\bar{\CQ}_{\rm bdry}.
\end{split}
\ee
The boundary electric charge constraint also reduces solely to a contribution of constant modes:
\be\label{chrg}
\sum\lm_a Q_a\left(\bar \Phi_{a}\frac{\p}{\p\bar \Phi_{a}}- \Phi_{a}\frac{\p}{\p \Phi_{a}}+\bar\eta_{a}\eta_{a }\right)=\Delta{\bf q}.
\ee
Consider a projective variety $X$ given by tuples of complex numbers $z_a$ $a=1,\ldots,n$ modulo a transform:
$$
z_a\sim \lambda^{Q_a} z_a,\quad \lambda\in \IC^{\times}.
$$
Now redefining fields as:
$$
\Phi_{a}=e^{Q_a(\rho+\I\vartheta)}z_a, \quad\rho,\vartheta\in \IR.
$$
we will find that both fields $\rho$ and $\vartheta$ acquire expectation values in the IR. 
The expectation value of field $\rho$ is fixed by the D-term constraint, and the value of field $\vartheta$ is fixed by eigenvalue \eqref{chrg}. 
We derive that the effective theory describes a particle moving in $X$ so that the effective supercharge can be decomposed as:
\be
\bar \CQ_{\rm eff}=\bar\nabla+\bar\CQ_{\rm bdry},
\ee
where $\bar\nabla$ is a $\bar\p$-component of a Dolbeault differential on $X$ extended by the corresponding Levi-Civita connection and the $U(1)$ connection with charge $\Delta{\bf q}$. 
We will give a derivation of this fact in Appendix \ref{s:sheaf_coho_CP_n} for the case $X=\IC\IP^{n-1}$. $\bar\CQ_{\rm bdry}$ has contributions from two branes located at interval ends $x^1=0$ and $x^1=L$.
We call these contributions $\bar\CQ_\CA$ and $\bar\CQ_\CB$ correspondingly:
\be\label{superch}
\bar \CQ_{\rm eff}=\bar\nabla+\bar\CQ_{\CA}-\bar\CQ_{\CB}.
\ee
Both $\bar\CQ_\CA$ and $\bar\CQ_\CB$ define complexes of vector bundles, or more generally, coherent sheaves we call $\CA$ and $\CB$. 
Details on coherent sheaf properties can be found in Appendix \ref{s:DCoh}. Supercharge \eqref{superch} defines a differential on a complex with the following cochains:
\be
\CC^n:=\bigoplus\lm_{k=0}^n\Omega^{(0,k)}(X)\otimes {\rm Hom}_{\CO_X}(\CA,\CB[n-k]).
\ee
This complex gives an injective resolution of a Hom-sheaf of $\CA$ and $\CB$. Since we have identified the BPS Hilbert space with the cohomologies of $\bar \CQ_{\rm eff}$ we can rewrite our Hilbert space as cohomologies of a derived functor:
\be
\fG_{\rm BPS,\; soft\; Higgs}^{\CF}\cong {\bf R}^{\CF}{\rm Hom}_{D{\rm Coh}(X)}\left(\CA,\underline{\CB}\right),
\ee
where derivations are taken in the underlined argument, and homological degree $\CF$ corresponds to the fermion number eigenvalue.

A simple application of this setup to a textbook example of sheaf valued cohomologies of $\IC\IP^n$ can be found in Appendix \ref{s:sheaf_coho_CP_n}.

\subsection{``Rigid'' Higgs branch: equivariant action}\label{sec:RigidHiggs}

The roles of real and imaginary parts of the chiral masses $\mu_i$ appear to be different. 
Real parts $\mu_{\IR,a}$ contribute to the Morse height functional, imaginary ones $\mu_{\II,a}$ contribute to the equivariant action on the field space. 
Therefore we can refer to the rigid Higgs branch as an equivariant derived category as opposed to the soft Higgs branch.
Unfortunately, a definition of derived equivariant categories of coherent sheaves is rather involved \cite{Equivariant_sheves} and lacks a simple operational definition in the literature like a Cartan model or a BRST model for equivariant cohomologies \cite{Cordes:1994fc}.
So one could consider this physical model as a Cartan model-like version of equivariant cohomologies valued in sheaves.

Constant vacua of the theory correspond to the points on the vacuum manifold fixed with respect to the complexified gauge group:
\be
b^{\rm th}\;{\rm fixed \;point}:\quad \sigma=Q_b^{-1}\mu_b,\quad |\phi_{a=b}|=r^{\frac{1}{2}},\quad \phi_{a\neq b}=0,\quad\mbox{if}\;Q_b^{-1}r>0.
\ee

We call these fixed points $\sigma$-model vacua for simplicity. 
We would like to study the BPS wave function behavior in a presence of such a $\sigma$-model vacuum.
The following decomposition allows one to fix automatically the gauge by extracting overall phase $\vartheta$:
\be
\begin{split}
	\phi_{a=b}=e^{\I Q_b\vartheta}\left(r^{\frac{1}{2}}+\hbar^{\frac{1}{2}}\rho\right),\quad \phi_{a\neq b}=e^{\I Q_a\vartheta}\hbar^{\frac{1}{2}}f_a,\quad \psi_{\alpha,a}=e^{-\I Q_a\vartheta}\eta_{\alpha,a},\quad  A_1=-\p_1\vartheta.
\end{split}
\ee

The supercharge contribution from $f_a$-modes reads:
\be
\begin{split}
	\bar Q_B=\sum\lm_{a\neq b}\int dx^1\;\bigg(-\I\sqrt{2}\bar\eta_{\dot 1,a}\left(\delta_{\bar f_a}+\left[Q_aQ_b^{-1}\mu_{\IR,b}-\mu_{\IR,a}\right]f_a\right)-\\
	-\sqrt{2}\bar\eta_{\dot 2,a}\left(\p_1 f_a-\left[Q_aQ_b^{-1}\mu_{\II,b}-\mu_{\II,a}\right]f_a\right)\bigg).
\end{split}
\ee
The peculiarity of this supercharge is that the equivariant vector field term has a fixed point in maps from $[0,L]$:
$$
f_{a,0}\sqrt{\frac{2\mu'_{\II}}{e^{2\mu_{\II}'L}-1}}e^{\mu_{\II}'x^1},\quad \mu_{\II}'=Q_aQ_b^{-1}\mu_{\II,b}-\mu_{\II,a}.
$$
The supercharge for this mode is purely chiral and corresponds to a supercharge of a particle in a magnetic field perpendicular to the complex plane spanned by complex coordinate $f_{a,0}$:
$$
\bar Q_B=-\I\sqrt{2}\bar\eta_{\dot 1, a ,0}\left(\frac{\p}{\p\bar f_{a,0}}+\mu_{\IR}'f_{a,0}\right),\quad \mu_{\IR}'=Q_aQ_b^{-1}\mu_{\IR,b}-\mu_{\IR,a}.
$$
Therefore rather having just a single BPS state we have the whole lowest Landau level Hilbert space with states labeled by angular momentum number $\ell$:
\be\label{cond_wf}
\Psi=\left\{\begin{array}{lll}
	f_{a,0}^{\ell}e^{-|\mu_{\IR}'||f_{a,0}|^2}|0\rangle, & \mu_{\IR}'>0, & \ell\in\IZ_{\geq 0};\\
	\bar f_{a,0}^{\ell}e^{-|\mu_{\IR}'||f_{a,0}|^2}\bar\eta_{\dot 1,a,0}|0\rangle, & \mu_{\IR}'<0, & \ell\in\IZ_{\geq 0}.
\end{array}\right.
\ee
This Landau Hilbert space is a condensate of gauge invariant mesons
\be\label{meson}
f_a=\phi_a\phi_b^{-Q_a/Q_b},
\ee
analogous to one observed in the cylindrical case in Section \ref{sec:Preamble}, the wave functions form a sheaf of holomorphic or anti-holomorphic (depending on the sign of $\mu_{\IR}'$) functions in a local chart containing the fixed point. 

The angular momentum number $\ell$ is captured by the equivariant degree, or the central charge shift:
\be
\Delta\tilde Z=-\I\left(Q_aQ_b^{-1}\mu_b-\mu_a\right)\ell.
\ee

When the interval $L$ between branes becomes large zero mode condensate may be attracted to either left or right brane depending on the sign of $\mu_{\II}'$:
\be\label{condensate}
\begin{array}{c}
	\begin{tikzpicture}
		\node[left] at (0,0) {$\lim\lm_{L\to\infty}\sqrt{\dfrac{2\mu'_{\II}}{e^{2\mu_{\II}'L}-1}}e^{\mu_{\II}'x^1}$};
		\node[right] at (6,1) {$\delta(L-x)$}; 
		\node[right] at (6,0) {const};
		\node[right] at (6,-1) {$\delta(x)$};
		\draw[->] (0.1,0.2) to[out=45,in=180] (1,1) -- (6,1); 
		\draw[->] (0.1,0) -- (6,0); 
		\draw[->] (0.1,-0.2) to[out=315,in=180] (1,-1) -- (6,-1);
		\node[above right] at (1.2,1) {$\mu_{\II}'>0,\;{\rm Re}\Delta\tilde Z>0$};
		\node[above right] at (1.2,0) {$\mu_{\II}'=0,\;{\rm Re}\Delta\tilde Z=0$};
		\node[above right] at (1.2,-1) {$\mu_{\II}'<0,\;{\rm Re}\Delta\tilde Z<0$};
	\end{tikzpicture}
\end{array}
\ee

This behavior is a duality counterpart of localization \eqref{quasi_par_loc}.

For generic chiral masses $\mu_a$ vacuum equations \eqref{sol_vac} also admit solitonic solutions interpolating between two constant solutions in asymptotic similarly to the Landau-Ginzburg model.
Linearized equations in the neighborhood of a $\sigma$-model vacuum predict that a solitonic particle has a finite size:
\be\label{sigma_sol_size}
{\rm min}\left(r^{\frac{1}{2}},|\mu_a|\right)^{-1}.
\ee
Unfortunately, unlike the LG model case we are not aware about techniques that will allow one to construct such a solution in a general setting. 
We will present an example of explicit calculation of a soliton solution using purely $\sigma$-model means without reference to the dual LG theory in Section \ref{sec:sigma-model-sol} for the case of equivariant $\IC\IP^1$.

\subsection{Parallel transport and web formalism}\label{sec:Web}
\subsubsection{Interfaces}
Parameter space $\CP$ is spanned by FI parameters and topological angles for all gauge groups as well as chiral mass parameters. 
In our particular case of a single $U(1)$ gauge group we allow only corresponding FI parameter and topological angle to vary along the interface, and it is convenient to combine them in a single complex parameter:
$$
t=r-\I \theta.
$$
So in our particular case $\CP$ is just a complex plane of $t$. 
However we should note that the technology of web formalism of \cite{GMW} works for generic $\CN=(2,2)$ 2d theories with massive vacua.

An interface path $\wp$ is a map:
$$
\wp:\quad [0,L]\longrightarrow \CP.
$$

The vacuum locus $\fV$ corresponds to a solution of a the vacuum equations \eqref{sol_vac} for the $\sigma$-model and \eqref{LG_soliton} correspondingly with constant parameter $t$ substituted with the interface path:
$$
t=\wp\left(x^1\right).
$$
This problem can be easily solved in the limit of \emph{adiabatic} approximation. 
We assume that the variation of parameter $t$ along $\wp$ is \emph{adiabatic} if a derivative $|\p_1t|$ is much smaller than all the soliton masses.
In a theory with generically massive vacua we always can choose such a path since we consider a limit $L\to \infty$, so that quantity $|\p_1t|\geq|\Delta t|/L$ is allowed to be as small as necessary.
In this limit $t$ may be assumed to be almost a constant, therefore a solution for generic adiabatic function $t(x^1)$ can be ``glued'' from solutions for constant $t$ we considered in the previous part of this section using a diagrammatic technique.

All our diagrams are constructed in a line segment $[0,L]$ and consist of vertices and links. Vertices are divided into bulk ones and boundary ones.

Bulk vertices correspond to $ij$-solitons having a transition modulus, after inserting such a vertex the soliton modulus is no more a free parameter, rather it is fixed to satisfy a stability constraint:
\be\label{stability}
\begin{array}{c}
	\begin{tikzpicture}
		\draw[ultra thick] (-0.5,0) -- (0.5,0);
		\shade[ball color = red] (0,0) circle (0.15);
		\node[left] at (-0.5,0) {$*i$};
		\node[right] at (0.5,0) {$*j$};
		\node[above] at (0,0.15) {$x_c$};
	\end{tikzpicture}
\end{array},\quad \tilde Z^{(i\to j)}\left(t(x_c)\right)\in -\I \IR_{\geq 0},,
\ee
where $\tilde Z$ is effective central charge \eqref{ecc} for the corresponding soliton solution.

Correspondingly boundary vertices are represented by solitons confined at boundaries accompanied by stability constraints or just empty boundary with a choice of a constant vacuum $*i$:
\be
\begin{split}
	\begin{array}{c}
		\begin{tikzpicture}
			\draw[ultra thick] (0,0) -- (0.5,0) (0,-0.1) -- (0,0.1);
			\node[left] at (0,0) {$*i$};
			\node[right] at (0.5,0) {$*i$};
			\node[above] at (0,0.1) {$0$};
		\end{tikzpicture}
	\end{array}\quad \begin{array}{c}
		\begin{tikzpicture}
			\draw[ultra thick] (0,0) -- (0.5,0) (0.5,-0.1) -- (0.5,0.1);
			\node[left] at (0,0) {$*i$};
			\node[right] at (0.5,0) {$*i$};
			\node[above] at (0.5,0.1) {$L$};
		\end{tikzpicture}
	\end{array}\\
	\begin{array}{c}
		\begin{tikzpicture}
			\draw[ultra thick] (0,0) -- (0.5,0);
			\shade[ball color = red] (0,0) circle (0.15);
			\node[left] at (-0.15,0) {$*i$};
			\node[right] at (0.5,0) {$*j$};
			\node[above] at (0,0.15) {$0$};
		\end{tikzpicture}
	\end{array},\quad {\rm Re}\,\tilde Z^{(i\to j)}\left(t(0)\right)>0 \quad {\rm Im}\,\tilde Z^{(i\to j)}\left(t(0)\right)<0\\
	\begin{array}{c}
		\begin{tikzpicture}
			\draw[ultra thick] (0,0) -- (0.5,0);
			\shade[ball color = red] (0.5,0) circle (0.15);
			\node[left] at (0,0) {$*i$};
			\node[right] at (0.65,0) {$*j$};
			\node[above] at (0.5,0.15) {$L$};
		\end{tikzpicture}
	\end{array},\quad {\rm Re}\,\tilde Z^{(i\to j)}\left(t(L)\right)<0 \quad {\rm Im}\,\tilde Z^{(i\to j)}\left(t(L)\right)<0
\end{split}
\ee

Links connect vertices in such a way that linked vacua match. 
From a diagram we can produce a solution function on a segment $[0,L]$ to the soliton equation in the following way. 
One needs just to scan a diagram along $x^1\in[0,L]$ from the left to the right:
\begin{enumerate}
	\item For a generic point of a link the solution is the corresponding solution in vacuum $*i$ for parameter values $t(x^1)$.
	\item If $x^1$ hits a diagram vertex at some $x_c$ we implant corresponding soliton solution in a spatial interval $(x_c-m^{-1},x_c+m^{-1})$, where $m$ is the corresponding soliton mass. 
	This solution approaches corresponding vacua at the interval ends exponentially fast $O(e^{-m|x|})$.
\end{enumerate}

All such diagrams are points in the vacuum locus $\fV$. 
The MSW complex \eqref{MSW} as a vector space is spanned by wave functions corresponding to Gaussian fluctuations around $p\in\fV$. 
In our case it is bi-graded by eigen values of fermion number $\CF$ and twisted central charge $\tilde\CZ$ operators. 
A universal way to produce the diagrams and calculate corresponding quantum numbers could be given for the LG model in terms of spectral networks techniques. 
The duality allows one to extend this technique to $\sigma$-models. 

The resemblance between soliton behavior and that of quasi-particles on a 2d world-volume will become more transparent when we consider instantons. We will incorporate a specific notations for 2d quasi-particles:
$$
\fX_{ij},
$$
implying that the world-line of such a particle represents a domain wall between $*i$ and $*j$ constant vacua. Suppose on a diagram we have a sequence of bound quasi-particles:
$$
\fX_{i_1i_2},\fX_{i_2i_3},\ldots, \fX_{i_{n-1}i_n}.
$$
Using technique described in this section one extracts a vacuum field configuration, then it is a simple task to derive a  perturbative wave function capturing perturbative quantum fluctuations around vacuum field configuration, and then a state vector associated to this critical point. This state has a fixed fermion number $f$, and we will denote this wave function in the following way:
\be\label{bpsi}
\bpsi_{i_1i_n}^{(f)}\left[\fX_{i_1i_2},\fX_{i_2i_3},\ldots, \fX_{i_{n-1}i_n}|\wp\right].
\ee
We do not choose normalization for these vectors, so $\bpsi$ is a whole 1d state subspace in the Hilbert space.
To denote a wave function associated to quantum fluctuations around a constant vacuum $*i$ we will use an argument $1$, and what constant vacuum is in consideration will be seen from subscripts of $\bpsi$. Explicit path notation $\wp$ will be omitted in $\bpsi$ when it is obvious.

A concatenation rule for diagrams is reflected in a relation for wave functions (fermion numbers add due to their extensive property, see Appendix \ref{sec:App_FS_cat}):
\be\label{concatenation}
\begin{split}
	\bpsi_{i_1i_k}^{(f_1)}\left[\fX_{i_1i_2},\ldots, \fX_{i_{k-1}i_k}|\wp_1\right]\otimes& \bpsi_{i_ki_n}^{(f_2)}\left[\fX_{i_ki_{k+1}},\ldots, \fX_{i_{n-1}i_n}|\wp_2\right]=\\
	=&\bpsi_{i_1i_n}^{(f_1+f_2)}\left[\fX_{i_1i_2},\ldots, \fX_{i_{n-1}i_n}|\wp_1\circ \wp_2\right].
\end{split}
\ee

\subsubsection{Instantons}
To construct the differential in the MSW complex one needs to consider instantons \eqref{instanton}. The simplest way to reproduce an instanton equation is to consider the corresponding supersymmetry variation of fermionic fields and apply the Wick rotation to the temporal direction:
$$
x^0\longrightarrow -\I x^2.
$$
After this rotation it is convenient to parameterize the world-sheet by a complex coordinate:
$$
z=x^1+\I x^2.
$$
In this way we derive instanton equations for the GLSM:
\be\label{sigma-instanton}
\begin{split}
	F_{z\bar z}=0,\quad
	2\p_{\bar z}\sigma=-\I\zeta^{-1}\left(r-\sum\lm_{a}Q_a|\phi_a|^2\right),\\
	2D_{\bar z}\phi_a=\I\zeta^{-1}(Q_a\bar\sigma-\bar\mu_a)\phi_a,\quad 
	2D_{\bar z}\bar\phi_a=\I\zeta^{-1}(Q_a\bar\sigma-\bar\mu_a)\bar\phi_a,
\end{split}
\ee
and the Landau-Ginzburg model:
\be\label{LG-instanton}
2\p_{\bar z}\Sigma=-\I\zeta^{-1}\bar W'.
\ee

In field theories we can impose a BPS bound on the Euclidean action below by a topological term. Instantons as BPS solutions saturate this bound, so that the  instanton field configurations satisfy integral equations:
\be\label{BPS_inst}
\begin{split}
	\int\lm_A d^2x\Bigg[\sum\lm_a &\left(|D_1\phi_a|^2+|D_2\phi_a|^2+2|Q_a\sigma-\mu_a|^2|\phi_a|^2\right)+\\
	&+4|\p_{\bar z}\sigma|^2+\left(r-\sum\lm_a Q_a|\phi_a|^2\right)^2\Bigg]=\\
	&=-2{\rm Re}\;\zeta^{-1}\ointctrclockwise\lm_{\p A}d\bar z\;\left(\bar\sigma\left(r-\sum\lm_aQ_a|\phi_a|^2\right)+\sum\lm_a\mu_a|\phi_a|^2\right);\\
	&\int\lm_A d^2 x\left(4|\p_{\bar z}\Sigma|^2+|W'|^2\right)=-2{\rm Re}\ointctrclockwise\lm_{\p A}\zeta W\;dz.
\end{split}
\ee
Then we come to a generic argument of \cite{GMW} that instanton equations \eqref{sigma-instanton} and \eqref{LG-instanton} do not have point-like solutions approaching a single LG or $\sigma$-model vacuum at infinity. 
For a single vacuum at infinity the right hand side of both relations in \eqref{BPS_inst} is zero, then constraint \eqref{BPS_inst} admits only trivially constant (up to gauge transformations) field configurations, unlike the choice of an A-twist \cite{Hori:2000kt} in the $\sigma$-model, where point-like BPS vortex solutions appear. 
Instead in our situation solutions may localize to 1d soliton quasi-particle world-lines.

Any steady soliton particle gives an Euclidean $x^2$-time independent solution to \eqref{sigma-instanton} or \eqref{LG-instanton}. Moving soliton quasi-particles are also solutions to the instanton equations.
Notice that \eqref{sigma-instanton} and \eqref{LG-instanton} are invariant under the following change of variables:
\be
z\longrightarrow e^{\I\varphi}z,\quad \zeta \longrightarrow e^{-\I\varphi}\zeta.
\ee
This transformation acts as an Euclidean boost on $(x^1,x^2)$-space-time with rapidity $\varphi$. 
The soliton with a free translation modulus can be put in the bulk and form a core of such an instanton solution. 
For such a soliton to satisfy the stability condition \eqref{stability} and to preserve B-twist with chosen parameter $\zeta$ its boost rapidity should satisfy:
\be\label{rapidity}
e^{\I\varphi}=-\zeta^{-1}\frac{\tilde \CZ}{|\tilde \CZ|}.
\ee
Due to the presence of the interface background the free soliton central charge depends adiabatically on $x^1$, its world-line forms a curved trajectory $(x^1(\tau),x^2(\tau))$ where $\tau$ is a trajectory proper time:
\be
\dot x^1(\tau)\cdot {\rm Re}\left[\zeta^{-1}\tilde Z\left(x^1(\tau)\right)\right]+\dot x^2(\tau)\cdot{\rm Im}\left[\zeta^{-1}\tilde Z\left(x^1(\tau)\right)\right]=0.
\ee

Allowing solitons to move around we have to allow them to interact through scattering, both mutual and with the boundary branes. 
The scattering processes are localized in the Euclidean space-time in point-like defects we call scattering vertices. 
The corresponding scattering amplitude is given by a QFT path integral for fixed parameter value $t\in\CP$ and a fan of asymptotic constant vacua:
\be
\begin{array}{ll}
	\mbox{bulk:} & \fa\left[\{i_1,i_2,\ldots,i_n\}|t\right]\sim \begin{array}{c}
		\begin{tikzpicture}
			\draw[dashed, fill=blue, opacity=0.4] (0,0) circle (1);
			\draw[thick] (0,-1) -- (0,0) (-0.7071,-0.7071) -- (0,0)
			(-0.7071,0.7071) -- (0,0) (0,0) -- (1,0);
			\draw[fill=red] (0,0) circle (0.1);
			\node[below] at (-0.382683, -0.92388) {$\scriptstyle *i_1$};
			\node[left] at (-1,0) {$\scriptstyle *i_n$};
			\node[below right] at (0.7071,-0.7071) {$\scriptstyle *i_2$};
			\draw[fill=black]
			(1.20104, 0.497488) circle (0.02)
			(0.919239, 0.919239) circle (0.02)
			(0.497488, 1.20104) circle (0.02)
			(0,1.3) circle (0.02)
			(-0.497488, 1.20104) circle (0.02);
		\end{tikzpicture}
	\end{array};\\
	\mbox{boundary:} & \fb_l\left[\{i_0,i_2,\ldots,i_n\}|t\right]\sim\begin{array}{c}
		\begin{tikzpicture}
			\draw[ultra thick] (0,-1) -- (0,1);
			\draw[dashed,  fill=blue, opacity=0.4] ([shift=(-90:1)]0,0) arc (-90:90:1);
			\draw[thick] (0,0) -- (0.7071,-0.7071) (0,0) -- (0.7071,0.7071);
			\draw[fill=black] (1.20104, 0.497488) circle (0.02) (1.20104, -0.497488) circle (0.02) (1.3,0) circle (0.02);
			\node[left] at (0,-0.7) {$\scriptstyle *i_0$};
			\node[left] at (0,0.7) {$\scriptstyle *i_n$};
			\node[below right] at (0.382683, -0.92388) {$\scriptstyle *i_1$};
			\node[above right] at (0.382683, 0.92388) {$\scriptstyle *i_{n-1}$};
			\draw[fill=red] (0,0) circle(0.1);
		\end{tikzpicture}
	\end{array} ,\\ 
	&\fb_r\left[\{i_0,i_2,\ldots,i_n\}|t\right]\sim \begin{array}{c}
		\begin{tikzpicture}[xscale=-1]
			\draw[ultra thick] (0,-1) -- (0,1);
			\draw[dashed,  fill=blue, opacity=0.4] ([shift=(-90:1)]0,0) arc (-90:90:1);
			\draw[thick] (0,0) -- (0.7071,-0.7071) (0,0) -- (0.7071,0.7071);
			\draw[fill=black] (1.20104, 0.497488) circle (0.02) (1.20104, -0.497488) circle (0.02) (1.3,0) circle (0.02);
			\node[right] at (0,-0.7) {$\scriptstyle *i_n$};
			\node[right] at (0,0.7) {$\scriptstyle *i_0$};
			\node[below left] at (0.382683, -0.92388) {$\scriptstyle *i_{n-1}$};
			\node[above left] at (0.382683, 0.92388) {$\scriptstyle *i_{1}$};
			\draw[fill=red] (0,0) circle(0.1);
		\end{tikzpicture}
	\end{array}.
\end{array}
\ee
Asymptotic scattering soliton states are uniquely defined by this data as boosted solitons with rapidity defined by \eqref{rapidity}. 
Vacuum fans for the bulk vertices with cyclically permuted vacua are equivalent. 
On the boundaries we allow confined soliton states (see Sections \ref{sec:LG_model} and \ref{sec:RigidHiggs}) that can be also defined using a pair of vacua. 
If we would like to specify an empty boundary brane as an asymptotic state we just set $*i_0=*i_1$ or $*i_{n-1}=*i_n$.

Solely bulk vertices satisfy a set of $A_{\infty}$-relations as Lagrangian disks in the Fukaya-Seidel category \cite{Auroux}, and being mixed up with boundary vertices they satisfy $L_{\infty}$-relations \cite{GMW}.
We will not consider and use these relations in the paper, rather we calculate necessary vertices using other methods.

Eventually, the supercharge matrix element is given by a sum over instantons -- web diagrams consisting of disks containing vertices and glued together along the boosted soliton trajectories.
Web diagrams give a name to this formalism \cite{Gaiotto:2015zna}.

\subsubsection{Interface homotopy and bootstrapping scattering vertices.}\label{sec:bootstrap}

The role of instantons becomes extra crucial when invariance of the interface MSW complex  under interface path homotopy is in consideration. 
Let us introduce a homotopy parameter $h$ and a homotopy family of paths $\wp(h)$ interpolating between $\wp_0$ and $\wp_1$:
$$
\begin{array}{c}
	\begin{tikzpicture}
		\draw[->] (-0.5,0) -- (3.5,0);
		\draw[ultra thick] (0,-0.08) -- (0,0.08) (3,-0.08) -- (3,0.08);
		\node(A)[above] at (0,0) {$\wp_0$};
		\node(B)[above] at (3,0) {$\wp_1$};
		\path (A) edge[<->] node[above] {\scriptsize homotopy} (B);
		\node[right] at (3.5,0) {$h$};
		\node[below] at (0,0) {$0$};
		\node[below] at (3,0) {$1$};
	\end{tikzpicture}
\end{array}
$$
The  homotopy morphism produces therefore a two-parametric family of theory couplings and effective superpotentials:
$$
t=\wp(x^1,h),\quad W(\phi, x^1,h).
$$

Isomorphism of MSW complex cohomologies, or quasi-isomorphism of MSW complexes is guaranteed (see Section 10.7 of \cite{GMW}) by an existence of a pair of chain maps $U$ and $\tilde U$ between corresponding complexes satisfying:
\be\nn
\begin{split}
	\begin{array}{c}
		\begin{tikzpicture}
			\node[left] at (0,0) {$({\bf M}^*_0,{\bf Q}_0)$};
			\node[right] at (1.5,0) {$({\bf M}^*_1,{\bf Q}_1)$};
			\draw[->] (0,0.07) -- (1.5,0.07);
			\draw[<-] (0,-0.07) -- (1.5,-0.07);
			\node[above] at (0.75,0.07) {$U$};
			\node[below] at (0.75,-0.07) {$\tilde U$};
		\end{tikzpicture}
	\end{array}\!\!\!\!,\\
	U\cdot\tilde U={\bf Id}+T_1\cdot {\bf Q}_0+{\bf Q}_0\cdot T_2,\\
	\tilde U\cdot U={\bf Id}+T_3\cdot {\bf Q}_1+{\bf Q}_1\cdot T_4,
\end{split}
\ee
where $T_i$ are some maps. 
Analogously to the non-perturbative corrections to the supercharge the map $U$ is saturated by contributions from a ``forced'' instanton equation with appropriate boundary conditions:
\be\label{forced_instanton}
(\p_{1}+\I\p_h)\Sigma=-\I\zeta^{-1}\overline{\p_{\Sigma}W(\Sigma,x^1,h)}.
\ee
There is a similar analog for \eqref{sigma-instanton}.

Another map $\tilde U$ is constructed using the inverse homotopy map. One constructs solutions for this equation in the same fashion as we did for the forced soliton equation by gluing those from solution pieces for constant parameters. 
This strategy leads to a summation over web diagrams where instantons contribute as vertices.

Homotopic invariance of the BPS Hilbert space leads to a categorification of a 2d wall-crossing formula \cite{GMW,Khan:2020hir}. 
Due to wall-crossing solitons form new stable BPS bound-states from old ones, or some BPS states become unstable and decay. 
It is natural to describe these decay-recombination processes by scattering amplitudes. We also can mimic the decay/recombination processes by homotopic moves connecting different regions of the parameter space where different BPS spectra are established. 
The decay/recombination amplitudes will be saturated by soliton scattering vertices $\fa$ contributing to \eqref{forced_instanton}.

We conclude that one way to bypass the necessity to calculate a solution to instanton equations \eqref{sigma-instanton} and \eqref{LG-instanton} is to bootstrap the amplitudes using the wall-crossing. In quite many cases the soliton states in the model are not degenerate: for a given central charge eigen value the dimension of the graded component of the BPS Hilbert space is 0 or 1. 
To achieve such level of granulation of the graded BPS Hilbert space one might need to include extra gradings using other charges commuting with the supercharge. In this case the wall-crossing becomes simple and is described by primitive-like wall-crossing formulas \cite{Denef:2007vg}, from those we can restore certain scattering amplitude values. 
To illustrate this approach with an example let us assume that we have a theory with at least three constant vacua $*i$, $*j$ and $*k$. Consider a homotopy between two phases where the spectra are inhabited by two stable solitons $\fX_{ij}$ and $\fX_{jk}$, and in one of the phases there is a new stable bound soliton state $\fX_{ik}$ that is a recombination of $\fX_{ij}$ and $\fX_{jk}$ due to wall-crossing. 
This is the only scenario for $\fX_{ik}$ to appear in the spectrum therefore \eqref{forced_instanton} has a solution interpolating between interface asymptotic states containing a pair of bound $\fX_{ij}$ and $\fX_{jk}$ in the far past and $\fX_{ik}$ in the far future:
\be
\begin{array}{c}
	\begin{tikzpicture}
		\draw[->] (0,-0.5) -- (0,2.5);
		\draw[fill=blue, opacity=0.4] (0,0) -- (3,0) -- (3,2) -- (0,2) -- cycle;
		\draw[ultra thick] (1,0) to[out=90,in=225] (1.5,1) (2,0) to[out=90,in=315] (1.5,1) (1.5,1) -- (1.5,2);
		\draw[dashed, thick] (1.5,1) to[out=135,in=270] (0.5,2) (1.5,1) to[out=45,in=270] (2.5,2);
		\draw[dashed] (0,1) -- (1.5,1);
		\draw[ultra thick] (-0.05,0) -- (0.05,0) (-0.05,1) -- (0.05,1) (-0.05,2) -- (0.05,2);
		\node[left] at (0,2.5) {$h$};
		\node[left] at (0,2) {$1$};
		\node[left] at (0,1) {$h_{\rm crit}$};
		\node[left] at (0,0) {$0$};
		\shade[ball color = red] (1,0) circle (0.1);
		\shade[ball color = blue] (2,0) circle (0.1);
		\shade[ball color = orange] (1.5,2) circle (0.1);
		\draw[fill=red] (1.5,1) circle (0.13);
		\node[right, fill=white] at (2.1,1) {$\fa_{\rm crit}\left[\{i,j,k\}\right]=\pm 1$};
		\node[below] at (1,-0.1) {$\fX_{ij}$};
		\node[below] at (2,-0.1) {$\fX_{jk}$};
		\node[above] at (1.5,2.1) {$\fX_{ik}$};
	\end{tikzpicture}
\end{array},\quad \fa\left[\{i,j,k\}|h\right]=\left\{\begin{array}{ll}
	\fa_{\rm crit}, & h\geq h_{\rm crit};\\
	0, & h<h_{\rm crit}.
\end{array}\right.
\ee
The same solution contributes to scattering amplitude $\fa$ for vacuum fan $\{i,j,k\}$. 
In principle, for calculating the supercharge non-perturbative corrections in practice we need only a sign value of the amplitude.

Having some a priori information about the behavior of the soliton states induced by wall-crossing one is able to bootstrap some soliton scattering amplitude values without solving instanton equations.

\subsubsection{Categorical parallel transport}

As we discussed in Section \ref{sec:LG_model} the brane category of the IR Landau-Ginzburg theory corresponds to the Fukaya-Seidel category. 
The data of this category are defined by the K\"ahler manifold and a  holomorphic superpotential function $W$ on it. 
Transition between Fukaya-Seidel categories with close superpotentials $W$ and $W'$ is a transition functor described well in the literature \cite{Seidel_book}. 
A slow,  ``adiabatic'' variation of parameters along  an interface path $\wp$ is lifted to a path in a family of superpotentials. 
We used the adiabaticity property to extrapolate the physical system on the interface by its behavior at constant parameter values. 
The adiabaticity allows one to parallel transport a physical system along a path in a parameter space identifying states between close systems. 
It is natural to expect that a similar phenomenon is established by the brane categories, so that brane boundary conditions $\CB$ and an interface defect $\fJ_{\wp}$ could be substituted by a brane ``hologram'' $\beta_{\wp}(\CB)$ producing the same spectrum of solitons as $\CB$ concatenated with $\fJ_{\wp}$, so that we have the following relation:
\be
\begin{array}{c}
	\begin{tikzpicture}[scale=0.85]
		\draw[dashed,fill=blue, opacity=0.2] (0,0) -- (3,0) -- (3,2) -- (0,2) -- cycle;
		\draw[dashed,fill=blue, opacity=0.4] (3,0) -- (4.5,0) -- (4.75,0.25) -- (4.5,0.5) -- (4.75,0.75) -- (4.5,1) -- (4.75,1.25) -- (4.5,1.5)  -- (4.75,1.75) -- (4.5,2) -- (3,2) -- cycle;
		\draw[ultra thick] (0,0) -- (0,2) (3,0) -- (3,2);
		\node[left] at (0,1) {$\CB$};
		\node[right] at (3,1) {$\beta_{\wp}\left(\CB\right)$};
		\draw[<->] (-1,2) -- (-1,-0.5) -- (4.5,-0.5);
		\node[right] at (4.5,-0.5) {$x^1$};
		\node[above] at (-1,2) {$x^0$};
		\draw[ultra thick] (0,-0.4) -- (0,-0.6) (3,-0.4) -- (3,-0.6);
		\node[below] at (0,-0.6) {$0$};
		\node[below] at (3,-0.6) {$0'$};
		\node at (1.5,1) {$\fJ_{\wp}$};
		\draw[->] (0.25,0.5) -- (2.75,0.5);
	\end{tikzpicture}
\end{array},\; \begin{array}{c}
	\fG_{\rm BPS}(\fJ_{\wp}|\CB,X)\cong \fG_{\rm BPS}(\fJ_{\rm triv}|\beta_{\wp}(\CB),X),\\
	\forall X\in \fD_{L}.
\end{array}
\ee

Here $\beta_{\wp}$ is a parallel transport functor, and $\fJ_{\rm triv}$ is a simple interface associated with a trivial path $[0,L]\to\mbox{point}$. Moreover, applying adiabaticity we can claim that $\beta_{\wp}$ is a functor coinciding with the transition functor of the Fukaya-Seidel category through a family of superpotentials along $\wp$, it can be calculated explicitly using various techniques.

An actual proof of the proposal above is extremely involved, a large portion of \cite{GMW} is devoted to this issue, and we have no means to repeat the proof here.

The action of the transition functor of the Fukaya-Seidel category has the most transparent form for an exceptional collection of Lefshetz thimbles. 
We will discuss explicit relations in Appendix \ref{sec:App_FS_cat}. 
The derived category of coherent sheaves is also a triangulated category and admits a choice of an exceptional collection, say, a set of $\CO(k)$ bundles for $\IC\IP^n$. 
However in the latter case there is no naive universal function like a superpotential allowing one to identify exceptional objects a priori and order them.
We could have cooked up an effective GLSM potential $\mathscr{W}$, so that the central charge on a field configuration interpolating between sectors $a$ and $b$ takes a potential form $\tilde Z_{ab} =\mathscr{W}_b-\mathscr{W}_a$ like \eqref{ecc},
however in terms of GLSM fields, rather than dual LG ones, the central charge \eqref{cc} does not have a naive potential form.

Having constructed equivalences of categories of brane boundary conditions for various phases one extends the action of the brane parallel transport functor to an abstract category of brane boundary conditions:
\be\label{parallel_transport}
\beta_{\wp}:\quad \fD_{p_1}\longrightarrow \fD_{p_2}.
\ee
One can substitute this abstract category with any category suitable for each phase in corresponding chambers of the parameter space $p_{1,2}\in \CP$.

The parallel transport functor in the Fukaya-Seidel category satisfies a composition law
\be
\beta_{\wp_2}\circ\beta_{\wp_1}=\beta_{\wp_1\circ\wp_2}.
\ee
This statement is also extendable to the abstract brane category.

\section{Equivariant \texorpdfstring{$\IC\IP^1$}{CP1}: soliton spectrum}\label{sec:equiv_CP_1}

\subsection{Model description}

Field defects (kinks, quasi-particles, domain walls) in sigma-models with a target space given by $\IC\IP^{\kappa}$ in 1+1 and higher dimensions attract interests of many researchers \cite{Eto:2006pg,Eto:2007aw,Eto:2007uc,Shifman:2010id,Fujimori:2016ljw,Harland:2009mf} as a toy model to test ideas about instanton behavior of Yang-Mills theories in 3+1 dimensions \cite{Polyakov:1987ez}. 
In this section we will construct spectra of BPS domain wall solutions in a 2d $\CN=(2,2)$ $\IC\IP^1$-model in both Higgs and Coulomb branch descriptions of the IR physics and will find a complete agreement between these spectra.
Depending on a concrete model these defects have different names, we will use a unifying name ``soliton''  for these quasi-particles to follow notions of \cite{GMW} as these defects are supported on 1d world-lines.
Eventually we will discuss scattering of these soliton solutions contributing to instantons discussed in Section \ref{sec:Web}.

\bigskip

$\IC\IP^1$ turns out to be the first rather simple, yet non-trivial, model establishing physical effects we are after. 
Moreover, it turns out the $\IC\IP^1$ model could be considered as a primitive building block for other more involved models. 
Therefore we would like to consider this setup in great details. 

The key observation we are chasing in this section is a support of the duality relation we described in the previous section. 
We will calculate spectra of BPS states on both branches explicitly and from the first principles. 
A comparison of the spectra leads to a conclusion that the Hilbert spaces for two branches are isomorphic:
\be\label{CP_1_mirror}
\fG_{\rm BPS,\, Higgs}\left(\IC\IP^1\right)\cong \fG_{\rm BPS,\, Coulomb}\left(\IC\IP^1\right).
\ee

The $\IC\IP^1$ model is described by a $U(1)$ gauged linear sigma-model with 2 chiral multiplets of the same charge $Q_1=Q_2=1$ and complex masses $\mu_1$ and $\mu_2$. The matter content of this theory may be encoded in a simple quiver:
$$
\begin{array}{c}
	\begin{tikzpicture}
		\node at (-2,0) {$1$};
		\node at (0,0) {$1$};
		\node at (2,0) {$1$};
		\draw (0,0) circle (0.3);
		\begin{scope}[shift={(2,0)}]
			\draw (-0.3,-0.3) -- (-0.3,0.3) -- (0.3,0.3) -- (0.3,-0.3) -- cycle;
		\end{scope}
		\begin{scope}[shift={(-2,0)}]
			\draw (-0.3,-0.3) -- (-0.3,0.3) -- (0.3,0.3) -- (0.3,-0.3) -- cycle;
		\end{scope}
		\draw[->] (1.7,0) -- (0.3,0);
		\draw[->] (-1.7,0) -- (-0.3,0);
	\end{tikzpicture}
\end{array}
$$
We immediately write down an expression for the effective Landau-Ginzburg superpotential on the Coulomb branch of the theory following a recipe of Section \ref{sec:LG_model}:
\be\label{CP_1_sup}
W(\Sigma)=t\Sigma+\frac{1}{2\pi}\sum\lm_{a=1}^2(\Sigma-\mu_a)\left(\log\frac{\Sigma-\mu_a}{\Lambda}-1\right),
\ee
and corresponding spectral curve (see Appendix \ref{sec:App_SN} for details):
\be
(\lambda-\mu_1)(\lambda-\mu_2)=z,
\ee
where we identify the spectral parameters with the complexified FI parameter and superpotential expectation value derivative:
\be
z=e^{-2\pi t},\quad \lambda=-2\pi \,z\frac{d}{dz}W.
\ee

Properties of this spectral curve are simpler to describe using a relative complex mass parameter:
$$
\mu:=\mu_1-\mu_2.
$$

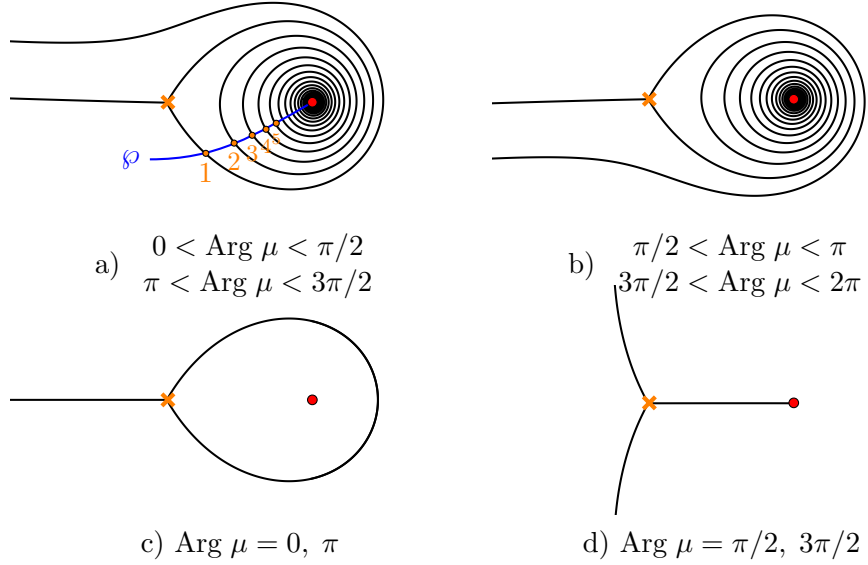
\begin{figure}[h!]
	\begin{center}
		\begin{tikzpicture}[scale=0.8]
			\begin{scope}
				\node at (0,0) {\includegraphics[scale=0.32]{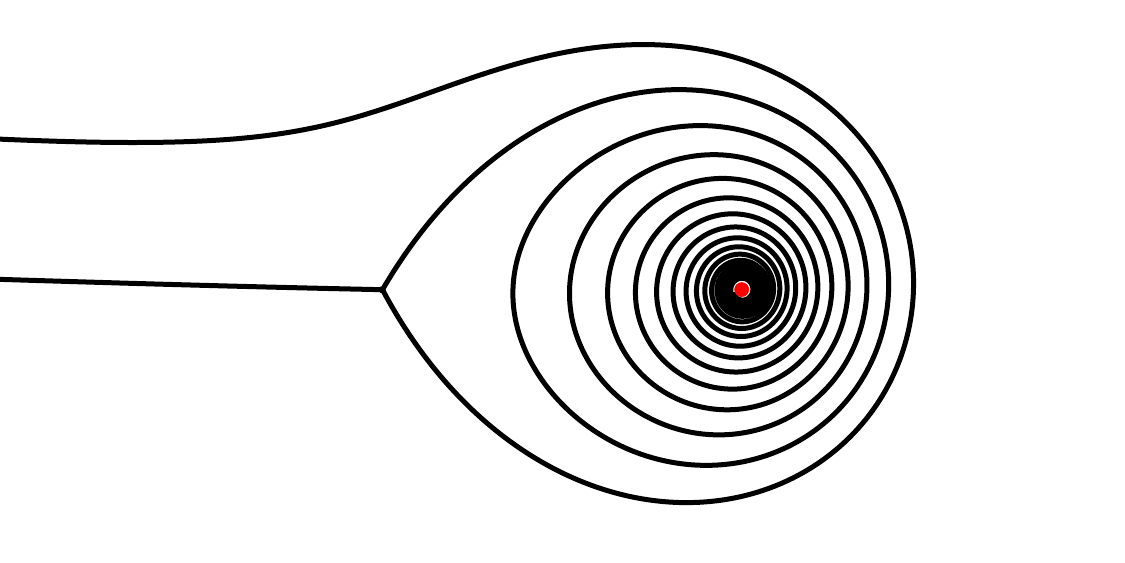}};
				\draw[thick, blue] (-1.5,-1) to[out=0,in=210] (1.2,-0.05);
				\node[left,blue] at (-1.5,-1) {$\wp$};
				\draw[fill=orange] (-0.57,-0.9) circle (0.05) (-0.1,-0.73) circle (0.05) (0.2,-0.6) circle (0.05) (0.43,-0.5) circle (0.05) (0.6,-0.4) circle (0.05);
				\node[below,orange] at (-0.57,-0.9) {1};
				\node[below,orange] at (-0.1,-0.73) { \small 2};
				\node[below,orange] at (0.2,-0.6) { \footnotesize 3};
				\node[below,orange] at (0.43,-0.5) { \scriptsize 4};
				\node[below,orange] at (0.6,-0.4) { \tiny 5};
				\begin{scope}[shift={(-1.2,-0.05)}]
					\draw[ultra thick, orange] (-0.1,-0.1) -- (0.1,0.1) (0.1,-0.1) -- (-0.1,0.1);
				\end{scope}
				\draw[fill=red] (1.2,-0.05) circle (0.08);
				\node[below] at (0,-2) {a) $\begin{array}{c}0<{\rm Arg}\;\mu<\pi/2\\
						\pi<{\rm Arg}\;\mu<3\pi/2
					\end{array}$}; 
			\end{scope}
			\begin{scope}[shift={(8,0)}]
				\node at (0,0) {\includegraphics[scale=0.32]{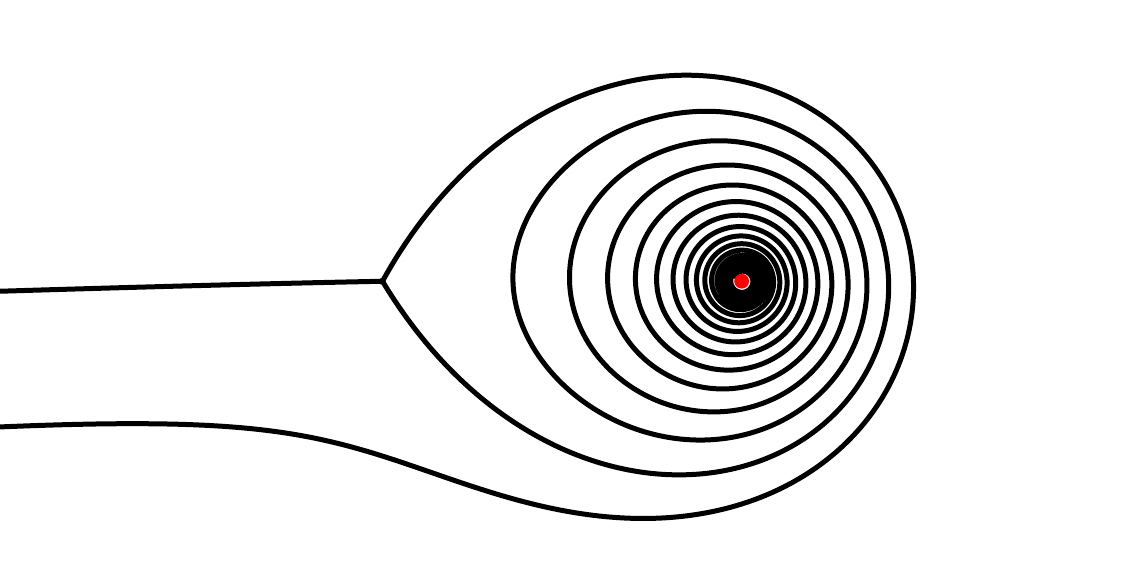}};
				\begin{scope}[shift={(-1.2,0)}]
					\draw[ultra thick, orange] (-0.1,-0.1) -- (0.1,0.1) (0.1,-0.1) -- (-0.1,0.1);
				\end{scope}
				\draw[fill=red] (1.2,0) circle (0.08);
				\node[below] at (0,-2) {b) $\begin{array}{c}\pi/2<{\rm Arg}\;\mu<\pi\\
						3\pi/2<{\rm Arg}\;\mu<2\pi
					\end{array}$};
			\end{scope}
			\begin{scope}[shift={(0,-5)}]
				\node at (0,0) {\includegraphics[scale=0.32]{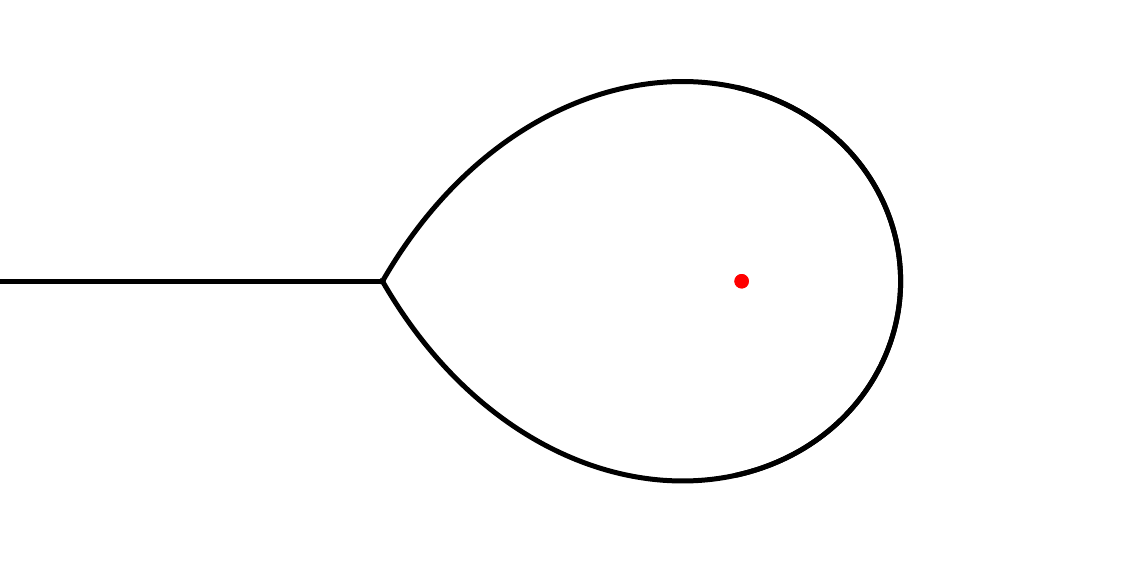}};
				\begin{scope}[shift={(-1.2,0)}]
					\draw[ultra thick, orange] (-0.1,-0.1) -- (0.1,0.1) (0.1,-0.1) -- (-0.1,0.1);
				\end{scope}
				\draw[fill=red] (1.2,0) circle (0.08);
				\node[below] at (0,-2) {c) ${\rm Arg}\;\mu=0,\;\pi$}; 
			\end{scope}
			\begin{scope}[shift={(8,-5)}]
				\node at (0,0) {\includegraphics[scale=0.32]{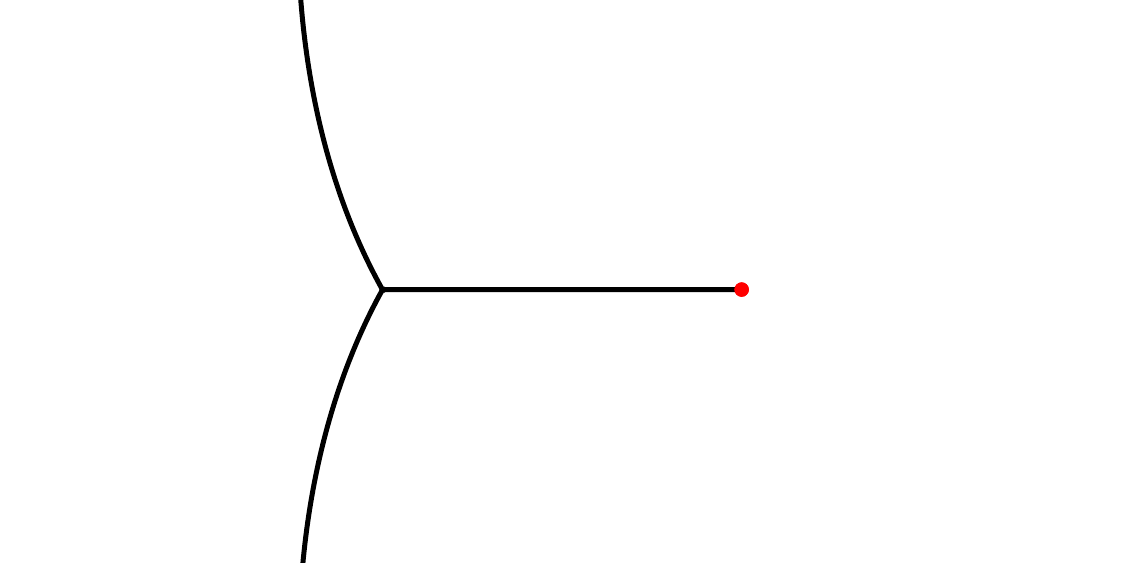}};
				\begin{scope}[shift={(-1.2,-0.05)}]
					\draw[ultra thick, orange] (-0.1,-0.1) -- (0.1,0.1) (0.1,-0.1) -- (-0.1,0.1);
				\end{scope}
				\draw[fill=red] (1.2,-0.05) circle (0.08);
				\node[below] at (0,-2) {d) ${\rm Arg}\;\mu=\pi/2,\;3\pi/2$}; 
			\end{scope}
		\end{tikzpicture}
		\caption{Spectral network topologies in $z$-plane for equivariant $\IC\IP^1$.}\label{fig:SN_CP_1}
	\end{center}
\end{figure}

There is a single ramification point:
$$
z_{r}=-\mu^2/4,
$$
and the topology of the spectral network depends on the argument of $\mu$. The topology experiences jumps at half-integer portions of $\pi$ (see Fig.\ref{fig:SN_CP_1}).

\subsection{Coulomb branch soliton spectrum}

\begin{figure}
	\begin{center}
		\begin{tikzpicture}[xscale=1.5, yscale=0.8]
			\begin{scope}
				\foreach \i in {1,...,19}
				{
					\draw[gray] (0.05 * \i - 1,0) -- (0.05 * \i - 1.5,2);
					\draw[gray] (-1 - 0.025 * \i,0.1 * \i) -- (- 0.025 * \i ,0.1 * \i);
				}
				\draw[ultra thick] (-1,0) -- (0,0) -- (-0.5,2) -- (-1.5,2) -- cycle (-1,0) -- (-0.5,2);
				\draw[fill=blue] (-1,0) ellipse (0.05333 and 0.1) (1,0) ellipse (0.05333 and 0.1)  (0,0) ellipse (0.05333 and 0.1)  (1,0) ellipse (0.05333 and 0.1)  (2,0) ellipse (0.05333 and 0.1)  (3,0) ellipse (0.05333 and 0.1)  (-1.5,2) ellipse (0.05333 and 0.1)  (-0.5,2) ellipse (0.05333 and 0.1)  (0.5,2) ellipse (0.05333 and 0.1)  (1.5,2) ellipse (0.05333 and 0.1)  (2.5,2) ellipse (0.05333 and 0.1)  (3.5,2) ellipse (0.05333 and 0.1) ; 
				\draw[->] (1,0) to node[above] {$a$} (1.92,0);
				\begin{scope}[shift={(0.5,2)}]
					\draw[->] (1,0) to node[below] {$a$} (1.92,0);
				\end{scope}
				\draw[->] (1,0) to node[right]{$b$} (1.4806, 1.92239);
				\node[right] at (5,1) {$\begin{array}{l}
						a=\I(\mu_1-\mu_2)\\
						b=t(\mu_1-\mu_2)
					\end{array}$};
				\node[right] at (3.2,0) {$W_{*1,\IZ}$};
				\node[right] at (3.7,2) {$W_{*2,\IZ}$};
				\node[below left] at (-1,-0.08) {$-1$};
				\node[below right] at (0,-0.08) {$0$};
				\node[below] at (1,-0.08) {$+1$};
				\node[below] at (2,-0.08) {$+2$};
				\node[below] at (3,-0.08) {$+3$};
				\node[above left] at (-1.5,2.08) {$-2$};
				\node[above right] at (-0.5,2.08) {$-1$};
				\node[above] at (0.5,2.08) {$0$};
				\node[above] at (1.5,2.08) {$+1$};
				\node[above] at (2.5,2.08) {$+2$};
				\node[above] at (3.5,2.08) {$+3$};
				\node[below] at (-0.5,0) {$\fX_{A,-1}$};
				\node[above] at (-1,2) {$\fX_{B,-1}$};
				\node[left] at (-1.25,1) {$\fX_{C,-1}$};
				\node[right] at (-0.25,1) {$\fX_{C,-1}$};
				\draw[thick,->] (-2.5,0.2) to[out=0,in=180] (-1.5,0.2) to[out=0,in=150] (-0.75,1);
				\node[left] at (-2.5,0.2) {$\fX_{C,0}$};
			\end{scope}
		\end{tikzpicture}
		\caption{Equivariant $\IC\IP^1$ vacua positions in $W$-plane}\label{fig:CP_1_vacua}
	\end{center}
\end{figure}
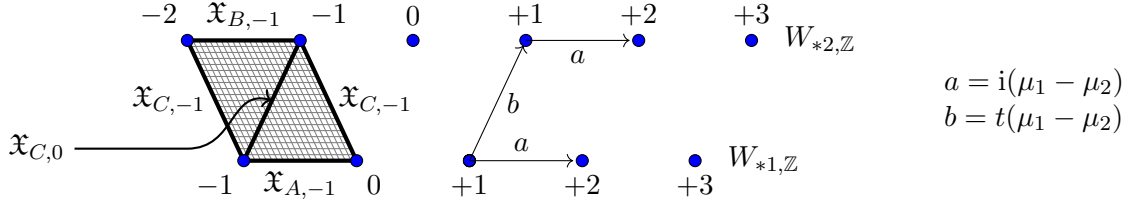

The vacua on the Coulomb branch of the theory are defined by solutions of the vacuum equation:
\be\label{CP_1_vacuum}
(\Sigma-\mu_1)(\Sigma-\mu_2)=\Lambda^2e^{-2\pi t}.
\ee
We are interested in solving it in a regime when Coulomb and Higgs branch overlap:
$$
{\rm Re}\;t\gg 0.
$$
In this case vacuum expectation value of field $\Sigma$ is given by a small deviation from either of complex mass values:
\be
\Sigma_{*1,2}=\mu_{1,2}\pm \frac{\Lambda^2}{\mu_1-\mu_2}e^{-2\pi t}+O\left(e^{-4\pi t}\right).
\ee
Despite the complex scalar expectation value is fixed by the vacuum choice the log function in the superpotential contributes with a multi-valuedness ambiguity. Therefore for the superpotential we acquire two infinite series of vacuum values labeled by integers:
\be
\begin{split}
	W_{*1,k}=t\mu_1+\frac{\mu_1-\mu_2}{2\pi}\left(\log\frac{\mu_1-\mu_2}{\Lambda}-1\right)+\I (\mu_1-\mu_2)k+O\left(e^{-2\pi t}\right),\quad k\in\IZ;\\
	W_{*2,k}=t\mu_2+\frac{\mu_2-\mu_1}{2\pi}\left(\log\frac{\mu_2-\mu_1}{\Lambda}-1\right)+\I (\mu_1-\mu_2)k+O\left(e^{-2\pi t}\right),\quad k\in\IZ.
\end{split}
\ee
To describe the soliton spectrum on the Coulomb branch we will depict vacua positions in the $W$-plane in Fig.\ref{fig:CP_1_vacua}.\footnote{A measurable value of $t$ gets renormalized in this picture: $$t_{\rm mes}:=\frac{W_{*1,0}-W_{*2,0}}{\mu_1-\mu_2}=t\pm\frac{\I}{2}-\frac{1}{\pi}+O\left((\Lambda/\mu) e^{-2\pi t}\right).$$ We neglect this shift in our consideration.}

The soliton equation \eqref{LG_soliton} has a solution interpolating from vacuum $(*i)$ to vacuum $(*j)$ if and only if the separation between vacua satisfies constraint $W_{*j}-W_{*i}\in \IR_{\geq 0}$. 

Solutions to these constraint are summarized in the following table:\footnote{By the superscript we will denote the soliton flow direction between vacua.}
\begingroup
\renewcommand*{\arraystretch}{1.7}
\be\label{CP_1_solitons}
\begin{array}{c|c|c}
	\mbox{Name} & \mbox{Constraint} & \mbox{Central charge}\\
	\hline
	\fX_{A,k}& \begin{array}{c}
		\mu_{\II,1}=\mu_{\II,2}\\
	\end{array}& \CZ_{\rm Clmb}^{(1\to 1)}=-|\mu_{\IR,1}-\mu_{\IR,2}|k,\;k\in\IZ_{\geq 0}\\
	\hline
	\fX_{B,k}& \begin{array}{c}
		\mu_{\II,1}=\mu_{\II,2}\\
	\end{array}& \CZ_{\rm Clmb}^{(2\to 2)}=-|\mu_{\IR,1}-\mu_{\IR,2}|k,\;k\in\IZ_{\geq 0}\\
	\hline
	\fX_{C,k}& \begin{array}{c}{\rm Re}\, t\,\frac{\mu_{\IR,1}-\mu_{\IR,2}}{\mu_{\II,1}-\mu_{\II,2}}-{\rm Im}\,t=:k\in\IZ\\
		\mu_{\II,1}>\mu_{\II,2}
	\end{array} & \CZ^{(1\to 2)}_{\rm Clmb}=-\frac{{\rm Re}\,t}{\mu_{\II,1}-\mu_{\II,2}}|\mu_1-\mu_2|^2\\
	\hline
	\fX_{D,k}& \begin{array}{c}{\rm Re}\, t\,\frac{\mu_{\IR,1}-\mu_{\IR,2}}{\mu_{\II,1}-\mu_{\II,2}}-{\rm Im}\,t=:k\in\IZ\\
		\mu_{\II,1}<\mu_{\II,2}
	\end{array} & \CZ^{(2\to 1)}_{\rm Clmb}=-\frac{{\rm Re}\,t}{\mu_{\II,1}-\mu_{\II,2}}|\mu_1-\mu_2|^2
\end{array}
\ee
\endgroup

Consider a path $\wp$ depicted in Fig.\ref{fig:SN_CP_1} (a). 
This path flows in the direction  ${\rm Re}\,t\to+\infty$ in the $t$-plane and intersects the spectral network in a sequence of points. 
Clearly this sequence is associated to either $\fX_{C,k}$ or $\fX_{D,k}$ soliton family depending on a sign of $\mu_{\II,1}- \mu_{\II,2}$. 
So that as soon $\wp$ intersects the spectral network in a point the expression
$$
{\rm Re}\, t\,\frac{\mu_{\IR,1}-\mu_{\IR,2}}{\mu_{\II,1}-\mu_{\II,2}}-{\rm Im}\,t
$$
hits an integer value. 

Solitons from $\fX_{A,k}$ and $\fX_{B,k}$ series have fixed masses independent of the $t$-parameter, therefore they do not produce in the $z$-plane a branching point where masses of a corresponding effective particle will flow to zero. 
The only way they produce a contribution in this setting is a critical behavior at ${\rm Arg}\,\mu=0,\,\pi$ (Fig.\ref{fig:SN_CP_1} (c)). 
At this critical value solitons from all families have co-directed central charges as vectors in a complex plane, therefore the solitons experience the wall-crossing phenomenon when some composite solitons decay to or recombine from more elementary ones. 
The cluster coordinates corresponding to the soliton partition function  experience a cluster transformation when the partition function associated with a soliton from $\fX_{C,k}$ or $\fX_{D,k}$ series is conjugated by a partition function associated to a gas of solitons from $\fX_{A,k}$ or $\fX_{B,k}$ series \cite{Manschot:2010qz,Gaiotto:2012rg,Galakhov:2014xba}. 
We will consider some details of this wall-crossing process when discuss instantons in this model.

\subsection{Higgs branch soliton spectrum} \label{sec:sigma-model-sol}

On the Higgs branch the theory has also two classical vacua:
\be
\begin{array}{lll}
	\mbox{1)}&\sigma_{*1}=\mu_1,& \vec \phi_{*1}=\left(r^{\frac{1}{2}},0\right);\\
	\mbox{2)}&\sigma_{*2}=\mu_2,& \vec \phi_{*2}=\left(0,r^{\frac{1}{2}}\right).
\end{array}
\ee

As it was discussed in Section \ref{sec:RigidHiggs} in the case $\mu_{\II,1}=\mu_{\II,2}$ BPS wave functions in the corresponding vacua are given by the following expressions (see \eqref{cond_wf}):
\be\label{CP_1_const_wave}
\Psi_1\sim\left\{\begin{array}{ll}
	(\phi_2/\phi_1)^k|0\rangle,& \mu_{\IR,1}>\mu_{\IR,2}\\
	(\bar\phi_2/\bar\phi_1)^k\bar\psi_{2,\dot 1}|0\rangle,& \mu_{\IR,1}<\mu_{\IR,2}
\end{array}\right.,\quad \Psi_2\sim\left\{\begin{array}{ll}
	(\phi_1/\phi_2)^k|0\rangle,& \mu_{\IR,1}<\mu_{\IR,2}\\
	(\bar\phi_1/\bar\phi_2)^k\bar\psi_{1,\dot 1}|0\rangle,& \mu_{\IR,1}>\mu_{\IR,2}
\end{array}\right.,
\ee
where $\phi$ and $\psi$ are constant field modes. 

Corresponding supercharges are:
\be\label{Higgs_spec_1}
\begin{split}
	\CZ_{\rm Higgs}^{(1\to 1)}=-|\mu_{\IR,1}-\mu_{\IR,2}|k,\;k\in\IZ_{\geq 0},\\
	\CZ_{\rm Higgs}^{(2\to 2)}=-|\mu_{\IR,1}-\mu_{\IR,2}|k,\;k\in\IZ_{\geq 0}.
\end{split}
\ee
So these states resemble solitonic states $\fX_{A,k}$ and $\fX_{B,k}$ on the Coulomb branch.

In the case $\mu_{\II,1}\neq \mu_{\II,2}$ we have to solve a system of equations describing the critical field configuration:
\be\label{complex_vortex}
\begin{split}
	-\p_1\sigma_{\II}+\left(\sum\lm_{a=1}^2|\phi_a|^2-r\right)=0,\\
	D_1\phi_a-(\sigma_{\II}-\mu_{\II,a})\phi_a=0,\; a=1,2.
\end{split}
\ee
Remaining equations of system \eqref{sol_vac} can have a non-trivial solution only if 
\be\label{mass_const}
\sigma_{\IR}=\mu_{\IR,1}=\mu_{\IR,2}.
\ee

Let us perform the following change of variables:
\be
\begin{split}
	\sigma_{\II}=\frac{\mu_{\II,1}+\mu_{\II,2}}{2}+s,\quad \Delta\mu=\frac{\mu_{\II,1}-\mu_{\II,2}}{2},\quad A_1=\p_1\vartheta,\\ \phi_1=e^{-\I\theta}r^{\frac{1}{2}}e^{\Phi}\cos\tau,\quad \phi_2=e^{-\I\theta} r^{\frac{1}{2}}e^{\Phi}\sin\tau.
\end{split}
\ee
In terms of the new variables equations \eqref{complex_vortex} are simplified:
\be\label{compl_v_simp}
\begin{split}
	\p_1\tau&=\Delta \mu \,\sin 2\tau,\\
	s&=\p_1\Phi+\Delta \mu\,\cos2\tau,\\
	\p_1s&=r\left(e^{2\Phi}-1\right).
\end{split}
\ee
The first equation in the column \eqref{compl_v_simp} can be solved analytically: 
\be\label{Higgs_sol}
\cos\tau=\frac{1}{\sqrt{1+e^{4\Delta \mu \left(x^1+c\right)}}},\quad \sin\tau=\frac{e^{2\Delta \mu \left(x^1+c\right)}}{\sqrt{1+e^{4\Delta \mu \left(x^1+c\right)}}},
\ee
This solution represents a domain wall of thickness\footnote{Solutions to the Liouville equation introduce another suppressing exponent $r^{-\frac{1}{2}}$. 
	So the actual width of the soliton core is given by \eqref{sigma_sol_size}.}
$|\Delta \mu|^{-1}$ and located at $x^1=c$. 
We call this solution a \emph{polar soliton} since it interpolates between poles of $\IC\IP^1$.
Depending on the sign of $\Delta \mu$ the soliton flows from vacuum 1 to vacuum 2 or in the inverse direction:
\be
\begin{array}{c}
	\begin{tikzpicture}
		\node[left] at (0,0) {$	{\vec \phi}_{*1}=\left(r^{\frac{1}{2}},0\right)$};
		\node[right] at (3,0) {${\vec \phi}_{*2}=\left(0,r^{\frac{1}{2}}\right)$};
		\draw[-stealth] (0,0.05) -- (3,0.05);
		\draw[stealth-] (0,-0.05) -- (3,-0.05);
		\node[above] at (1.5,0.05) {\footnotesize$\Delta\mu>0$};
		\node[below] at (1.5,-0.05) {\footnotesize$\Delta\mu<0$};
	\end{tikzpicture}
\end{array}
\ee
Unfortunately, we are unable to solve remaining equations in system \eqref{compl_v_simp} analytically since the resulting equation for $\Phi$ is a Liouville equation with a non-trivial external source. 
Rather we could perform perturbative expansion in a regime $\Delta\mu/\sqrt{r}\ll 1$ confirming each soliton solution \eqref{Higgs_sol} is accompanied by a single solution to the complete system. 
The resulting expressions turn out to be rather bulky and we will never use their explicit form, therefore we omit them in our consideration.

As usual the low energy dynamics of this solutions is governed by zero modes  -- variations of fields $\delta \phi_a$, $\delta\sigma_{\II}$, $\delta A_1$ along the soliton moduli we will calculate momentarily. 
We would like to accompany the usual zero mode equations by an additional constraint for field variation following from the Gauss law \eqref{Gauss}:
$$
\p_1\delta A_1=\I\sum\lm_{a=1}^2\left(\bar\phi_a\delta\phi_a-\phi_a\delta\bar\phi_a\right).
$$

Chosen B-twist \eqref{AB-twists} relates fermionic and bosonic zero modes in the usual way:
\be
\delta\phi_a=\sqrt{2}\epsilon\,\psi_{1,a},\quad \delta\sigma_{\II}-\I\delta A_1=2\epsilon\,\bar\lambda_1,
\ee
allowing one to re-assemble the operators annihilating zero modes in a Dirac operator:
\be
\nabla\left(\begin{array}{c}
	\psi_{1,1}\\
	\psi_{1,2}\\
	\bar\lambda_1
\end{array}\right)=0,\quad \nabla=\left(\begin{array}{ccc}
	D_1-(\sigma_{\II}-\mu_{\II,1}) & 0 & -\sqrt{2}\phi_1\\
	0& D_1-(\sigma_{\II}-\mu_{\II,2}) & -\sqrt{2}\phi_2\\
	-\sqrt{2}\bar\phi_1 & -\sqrt{2}\bar\phi_2 & \p_1
\end{array}\right).
\ee
Again we would like to redefine variables in the following way:
\be
\psi_{1,a}:=\phi_a \alpha_a,\quad \bar\lambda_1:=\beta.
\ee
In terms of these new variables the zero mode equations read (we suppress the superscript $1$ for the spatial coordinate $x^1$):
\be\label{system}
\begin{split}
	\p_x\alpha_1-\sqrt{2}\beta=0,\\
	\p_x\alpha_2-\sqrt{2}\beta=0,\\
	\p_x\beta-\sqrt{2}|\phi_1|^2\alpha_1-\sqrt{2}|\phi_2|^2\alpha_2=0.
\end{split}
\ee
The first two equations of this system can be solved easily:
\be
\alpha_1(x)=\alpha_0(x)+c_0,\quad\alpha_2(x)=\alpha_0(x)-c_0,\quad \beta=\frac{1}{\sqrt{2}}\p_x\alpha_0(x).
\ee
where $c_0$ is an integration constant.
The system \eqref{system} reduces to the following equation for $\alpha_0$:
\be
-\p_x^2\alpha_0+2(|\phi_1|^2+|\phi_2|^2)\alpha_0+2c_0\left(|\phi_1|^2-|\phi_2|^2\right)=0.
\ee
It is simpler to solve this equation in an approximation regime $\Delta\mu/\sqrt{r}\ll 1$ when we have:
\be\label{approx}
|\phi_1|^2\approx r\left(\cos\tau\right)^2,\quad|\phi_2|^2\approx r\left(\sin\tau\right)^2.
\ee
In this case a generic solution reads:\footnote{Green function for a massive 1d Laplace operator reads:
	$$
	\left(-\p_x^2+m^2\right)G_m(x)=\delta(x),\quad G_m(x)=\frac{1}{2m}e^{-m|x|},\quad m>0.
	$$}
\be
\alpha_0=c_1e^{-\sqrt{2r}x}+c_2e^{\sqrt{2r}x}-\frac{c_0 r}{\sqrt{2r}}\int\lm_{-\infty}^{+\infty}dy\,e^{-\sqrt{2r}|x-y|}\frac{1-e^{2(\mu_{\II,1}-\mu_{\II,2}) y}}{1+e^{2(\mu_{\II,1}-\mu_{\II,2}) y}}.
\ee
The latter integral can be rewritten in terms of hypergeometric series, however the resulting expression is rather bulky, moreover applying assumption $\Delta\mu/\sqrt{r}\ll 1$ we could substitute the integration kernel by a kernel supported in a single point:
$$
e^{-\sqrt{2r}|x|}\to \frac{2}{\sqrt{2r}}\delta(x).
$$
Thus one derives:
\be\label{alpha_0}
\alpha_0\approx c_1e^{-\sqrt{2r}x}+c_2e^{\sqrt{2r}x}-c_0\frac{1-e^{2(\mu_{\II,1}-\mu_{\II,2}) x}}{1+e^{2(\mu_{\II,1}-\mu_{\II,2}) x}}.
\ee
Zero modes governed by integration constants $c_1$ and $c_2$ are concentrated near interval boundaries and shift the boundary brane electric charges (see \eqref{condensate}). 
The zero mode governed by $c_0$ is a mode bound to the soliton core. 

The mass constraint \eqref{mass_const} for the vacuum field configuration needs to be satisfied up to the quantum $\hbar$ order. Therefore we could put:
$$
\mu_{\IR,1}=\hbar \tilde\mu_{\IR,1},\quad \mu_{\IR,2}=\hbar \tilde\mu_{\IR,2},
$$
where $\tilde\mu_{\IR,a}$ are of order 1.
The central charge acquires a constant correction to the equilibrium zero value due to this core zero mode -- ``vacuum supercharge'':
\be
\bar Q_0=\int\left(\bar\lambda_1\theta+\sqrt{2}\tilde\mu_{\IR,1}\bar\phi_1\psi_{1,1}+\sqrt{2}\tilde\mu_{\IR,2}\bar\phi_2\psi_{1,2}\right)\,dx
\ee
The energy of the state acquires additional  gap:
$$
E_0\sim|\tilde\CZ|+|\bar Q_0|^2.
$$
For the state to be BPS it should satisfy  constraint $|Q_0|^2=0$.
After massaging the vacuum supercharge expression a little and applying $\beta(x\to\pm \infty)=0$ we derive:
\be
\bar Q_0=\int\left(\beta\theta-\frac{\tilde\mu_{\IR,1}-\tilde\mu_{\IR,2}}{2}\left(\p_x\frac{|\phi_1|^2-|\phi_2|^2}{|\phi_1|^2+|\phi_2|^2}\right)\beta+\frac{\tilde\mu_{\IR,1}-\tilde\mu_{\IR,2}}{\sqrt{2}}\frac{4|\phi_1|^2|\phi_2|^2}{|\phi_1|^2+|\phi_2|^2}c_0\right)\,dx.
\ee
For the approximate expression \eqref{alpha_0} we derive another BPS constraint:
\be
\bar Q_0\approx\frac{2c_0}{\sqrt{2}|\mu_{\II,1}-\mu_{\II,2}|}\left[\theta(\mu_{\II,1}-\mu_{\II,2})+r(\tilde\mu_{\IR,1}-\tilde\mu_{\IR,2})\right]=0.
\ee

In principle the zero mode $\phi$ has its own monomial contribution $\phi^k$ to the wave-function (as in \eqref{CP_1_const_wave}), this contribution shifts the effective topological angle $\theta$ by $k$, therefore the BPS constraint reads (compare to \eqref{CP_1_solitons}):
\be
k={\rm Re}\, t\,\frac{\mu_{\IR,1}-\mu_{\IR,2}}{\mu_{\II,1}-\mu_{\II,2}}-{\rm Im}\,t\in\IZ.
\ee
If this constraint is satisfied the complex vortex soliton flows in a direction defined by the difference between masses $\mu_{\II,1}$ and $\mu_{\II,2}$:
\begingroup
\renewcommand*{\arraystretch}{1.7}
\be\label{Higgs_spec_2}
\begin{array}{ll}
	\CZ^{(1\to 2)}_{\rm Higgs}=-\dfrac{{\rm Re}\,t}{\mu_{\II,1}-\mu_{\II,2}}|\mu_1-\mu_2|^2,& \mu_{\II,1}>\mu_{\II,2};\\
	\CZ^{(2\to 1)}_{\rm Higgs}=-\dfrac{{\rm Re}\,t}{\mu_{\II,1}-\mu_{\II,2}}|\mu_1-\mu_2|^2,& \mu_{\II,1}<\mu_{\II,2}.
\end{array}
\ee
\endgroup
Comparing the eventual Higgs branch BPS soliton spectrum \eqref{Higgs_spec_1} and \eqref{Higgs_spec_2} with the Coulomb branch soliton spectrum \eqref{CP_1_solitons} we find them to be {\bf identical}. This identity is a manifestation of the \emph{cHC duality} \eqref{HC_du} in this simple case of equivariant $\IC\IP^1$ \eqref{CP_1_mirror}.

\subsection{Instantons and boundary operators}
As we discussed have in Section \ref{sec:bootstrap} instanton contributions become crucial when homotopy relations between interface paths $\wp$ are in consideration.
One type of instantons appearing in this model is a standard instanton contributing to the homotopy of the interface path around the branching point \cite{Galakhov:2017pod}. 
We will not consider it here.

Rather we concentrate on another type of instanton vertices delivering a homotopy invariance depicted in Fig.\ref{fig:bdry_chg} through the corresponding boundary operator. 
The same instanton vertex governs soliton recombination during critical wall-crossing between soliton spectra (a) and (b) depicted in Fig.\ref{fig:SN_CP_1}. 
This process could be described as a cluster transformation of soliton partition functions in coupled  2d-4d systems due to ``flavor'' solitons $\fX_{A,k}$ and $\fX_{B,k}$ that are effective vector multiplet 4d quasi-particles coupled to a 2d defect \cite{Gaiotto:2009hg,Klemm:1996bj}.
These solitons interpolate between the same type LG vacua \eqref{CP_1_vacuum} on different sheets of $W$-cover (see Fig.\ref{fig:CP_1_vacua}). 
The instanton vertices represent the following reactions:
\be\label{sol_scatt}
\begin{split}
	\fX_{A,-1}+\fX_{C,-1} \longleftrightarrow \fX_{C,0},\\
	\fX_{B,-1}+\fX_{C,-1} \longleftrightarrow \fX_{C,0}.
\end{split}
\ee
To prove that corresponding scattering vertices are non-zero we apply the bootstrap method of Section \ref{sec:bootstrap}. 

\begin{figure}[h!]
	\begin{center}
		\begin{tikzpicture}
			\node {\includegraphics[scale=0.4]{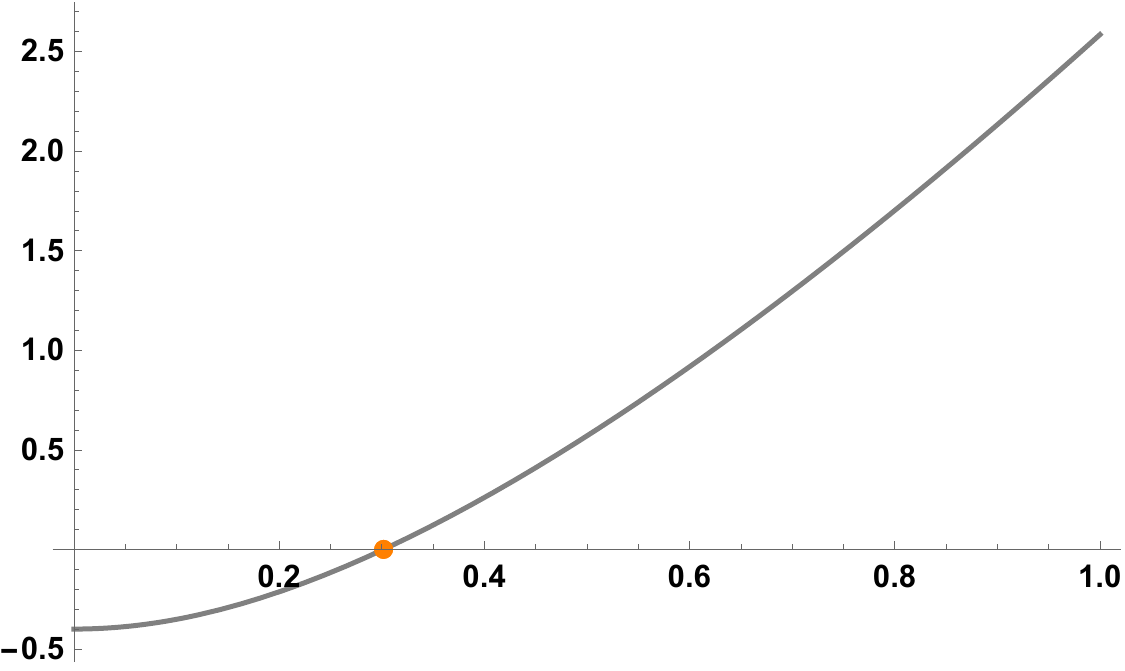}};
			\node[black!10!orange] at (-2,-1) {$\mu_{\rm crit}=0.3018$};
			\node at (4.5,-1.7) {$\scriptstyle\mu$};
			\node at (-5,1.8) {$\scriptstyle{\rm Re}\,\Delta W$};
		\end{tikzpicture}
		\caption{Plot ${\rm Re}\,\Delta W(\mu).$}\label{fig:plot}
	\end{center}
\end{figure}
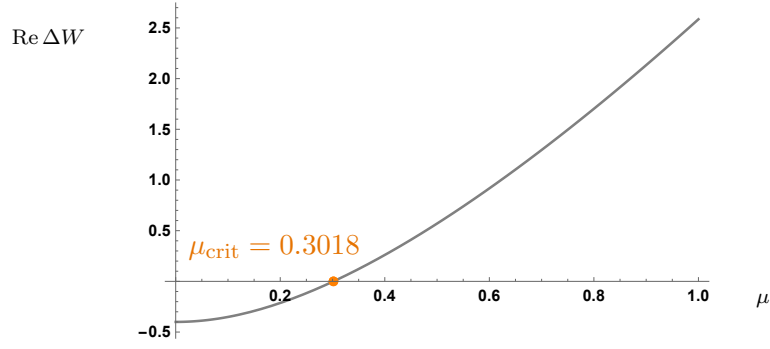

It will be more spectacular if we consider a dual model. 
First we restore disorder operator $Y$ in the superpotential expression \eqref{CP_1_sup} then integrate over field $\Sigma$. 
The resulting class of models can be described by the following superpotential:
$$
W=\mu Y-R \,e^{Y}-e^{-Y},\quad R\ll 1\,,
$$
where $\mu$ is some combination of mass parameters and $R\sim e^{-2\pi t}\ll 1$. 
This model has two LG vacuum series corresponding to two vacua series in the $W$-plane (Fig.\ref{fig:CP_1_vacua}):
$$
Y_{*1}=\log\frac{\mu+\sqrt{\mu^2+4R}}{2R}+2\pi\I\IZ,\quad Y_{*2}=\log\frac{\mu-\sqrt{\mu^2+4R}}{2R}+2\pi\I\IZ\,.
$$
The wall-crossing decay/recombination occur when the central charges of solitons are co-directed in the complex plane. For flavor solitons $\Delta W$ is purely imaginary for real parameter $\mu$. 
The marginal stability wall separating chambers with different BPS spectra is located in the point where ${\rm Re}\,(\Delta W)$ for $\fX_{C,k}$ changes sign:
$$
{\rm Re}\,(\Delta W)=\mu\;\log\frac{\sqrt{\mu^2+4R}+\mu}{\sqrt{\mu^2+4R}-\mu}-2\sqrt{\mu^2+4R}\,.
$$
The behavior of this function is depicted in a plot in Fig.\ref{fig:plot}.

\begin{figure}[h!]
	\begin{center}
		\begin{tikzpicture}
			\begin{scope}
				\node {\includegraphics[scale=0.18]{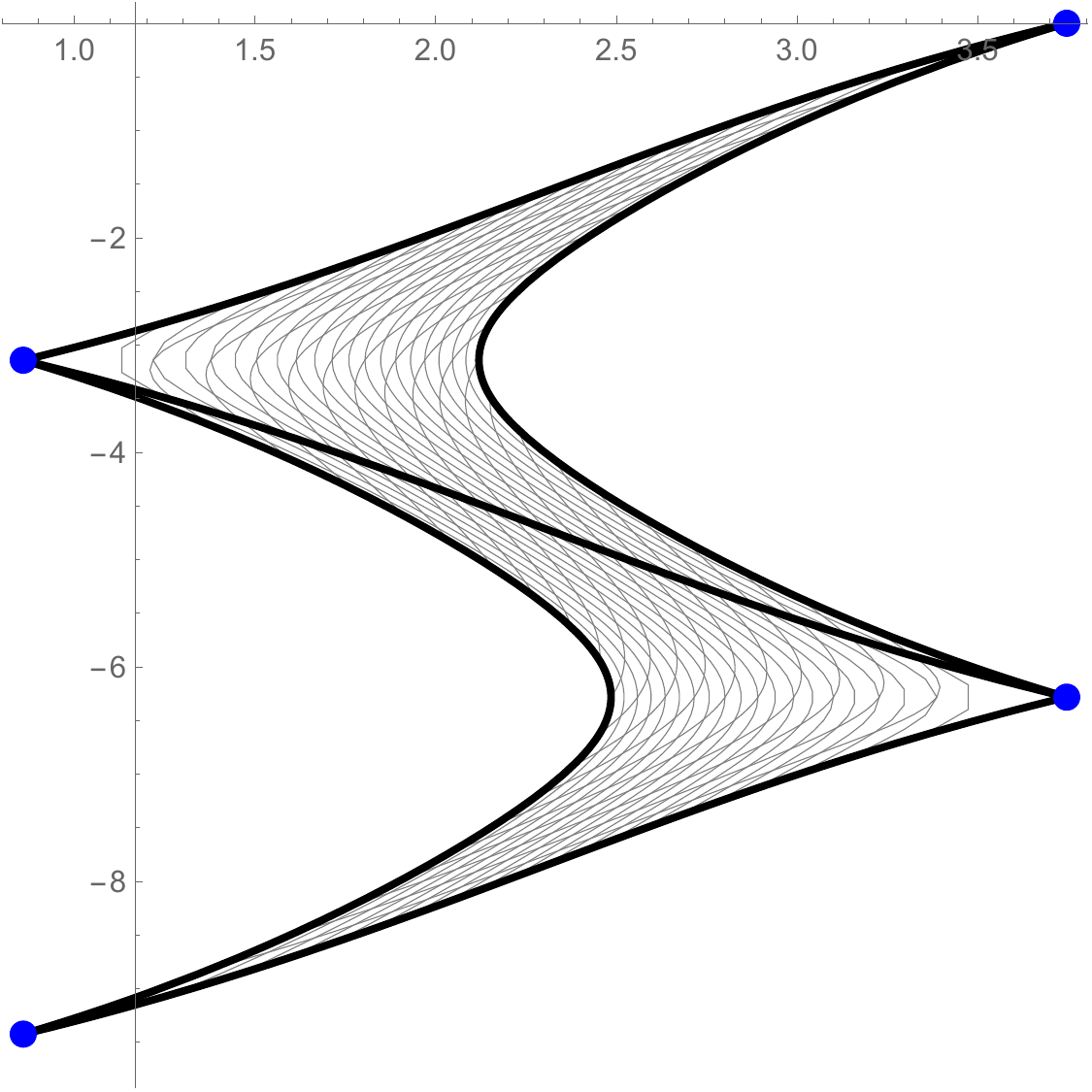}};
				\node at (0,2) {$\scriptstyle{\rm Re}\,Y$};
				\node at (-2,0) {$\scriptstyle{\rm Im}\,Y$};
				\node[black!10!blue] at (1.8,1.4) {$\scriptstyle Y_{\rm I}$};
				\node[black!10!blue] at (1.8,-0.8) {$\scriptstyle Y_{\rm II}$};
				\node[black!10!blue] at (-2,0.8) {$\scriptstyle Y_{\rm III}$};
				\node[black!10!blue] at (-2,-1.8) {$\scriptstyle Y_{\rm IV}$};
			\end{scope}
			\begin{scope}[shift={(5,0)}]
				\node {\includegraphics[scale=0.18]{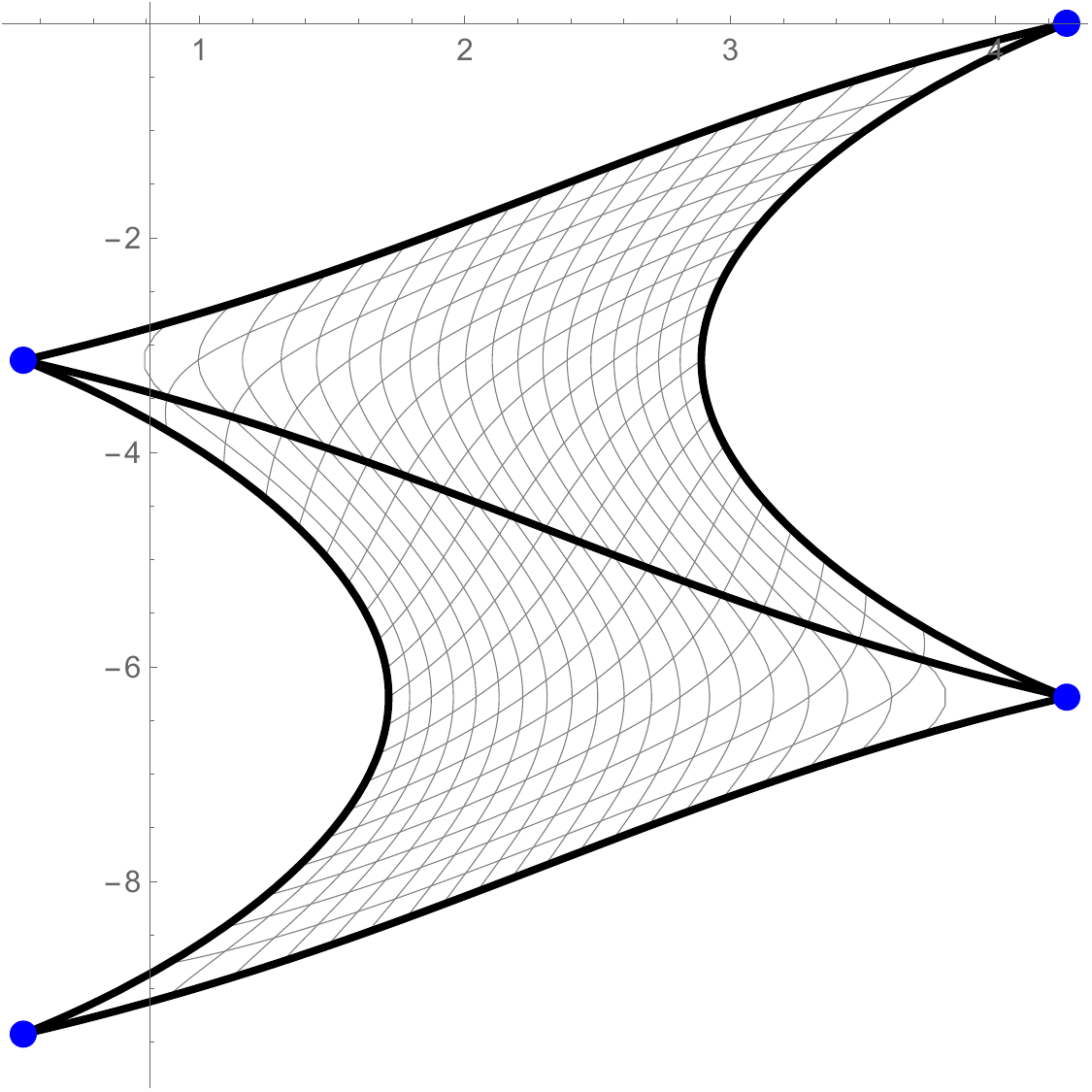}};
				\node at (0,2) {$\scriptstyle{\rm Re}\,Y$};
				\node at (-2,0) {$\scriptstyle{\rm Im}\,Y$};
				\node[black!10!blue] at (1.8,1.4) {$\scriptstyle Y_{\rm I}$};
				\node[black!10!blue] at (1.8,-0.8) {$\scriptstyle Y_{\rm II}$};
				\node[black!10!blue] at (-2,0.8) {$\scriptstyle Y_{\rm III}$};
				\node[black!10!blue] at (-2,-1.8) {$\scriptstyle Y_{\rm IV}$};
			\end{scope}
			\begin{scope}[shift={(10,0)}]
				\node {\includegraphics[scale=0.18]{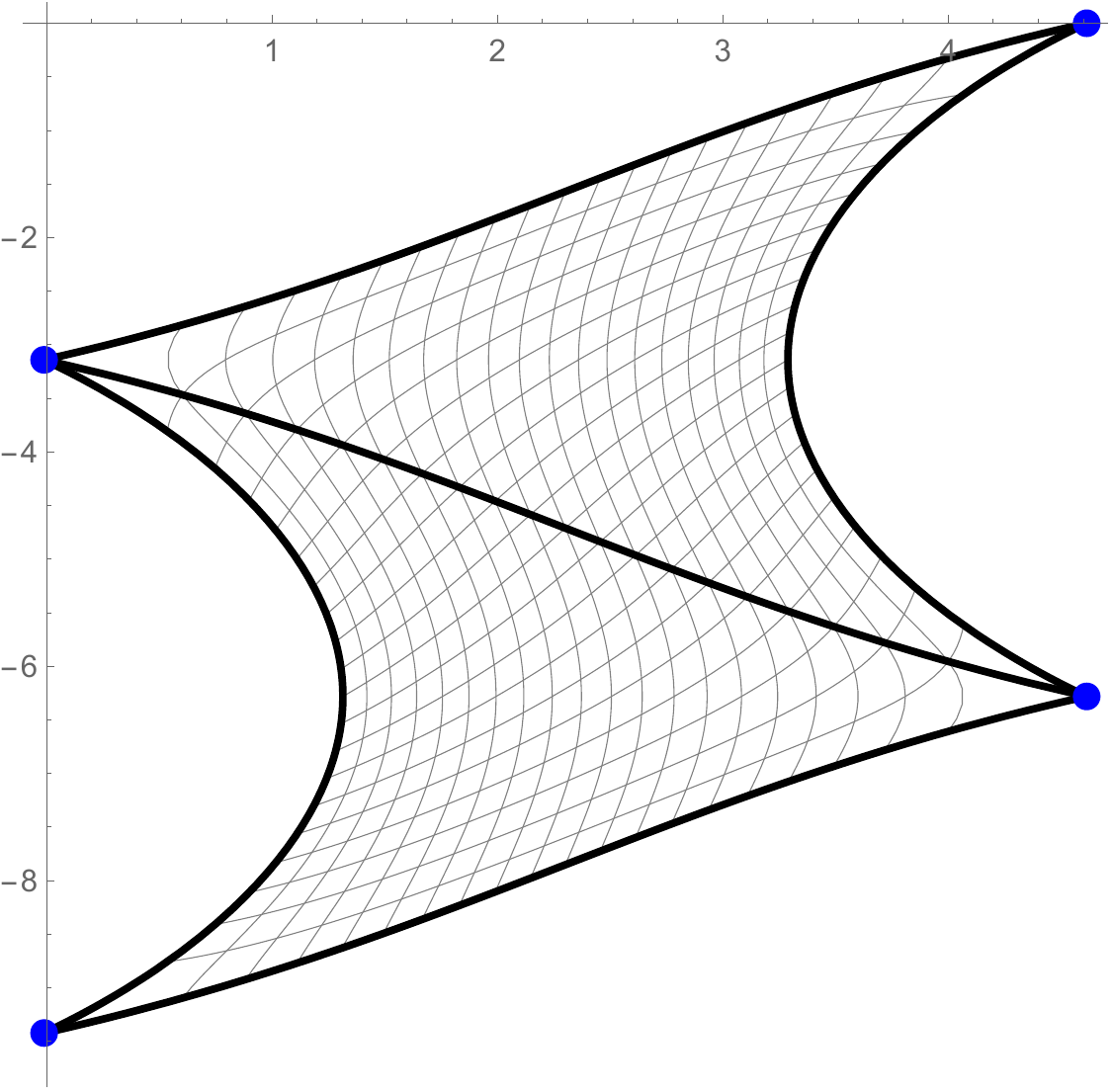}};
				\node at (0,2) {$\scriptstyle{\rm Re}\,Y$};
				\node at (-2,0) {$\scriptstyle{\rm Im}\,Y$};
				\node[black!10!blue] at (1.8,1.4) {$\scriptstyle Y_{\rm I}$};
				\node[black!10!blue] at (1.8,-0.8) {$\scriptstyle Y_{\rm II}$};
				\node[black!10!blue] at (-2,0.8) {$\scriptstyle Y_{\rm III}$};
				\node[black!10!blue] at (-2,-1.8) {$\scriptstyle Y_{\rm IV}$};
			\end{scope}
			\begin{scope}[shift={(0,-2.5)}]
				\draw[->] (-2,0) -- (11,0);
				\node[right] at (11,0) {$\mu$};
				\draw[ultra thick] (-1.8,-0.1) -- (-1.8,0.1) (0,-0.1) -- (0,0.1) (5,-0.1) -- (5,0.1) (10,-0.1) -- (10,0.1);
				\node[below] at (-1.8,-0.1) {$\mu_{\rm crit}$};
				\node[below] at (0,-0.1) {$0.4$};
				\node[below] at (5,-0.1) {$0.7$};
				\node[below] at (10,-0.1) {$1.0$};
				\draw (-1.5,0.4) to[out=330,in=120] (0,0) to[out=60,in=210] (1.5,0.4);
				\begin{scope}[shift={(5,0)}]
					\draw (-1.5,0.4) to[out=330,in=120] (0,0) to[out=60,in=210] (1.5,0.4);
				\end{scope}
				\begin{scope}[shift={(10,0)}]
					\draw (-1.5,0.4) to[out=330,in=120] (0,0) to[out=60,in=210] (1.5,0.4);
				\end{scope}
			\end{scope}
		\end{tikzpicture}
		\caption{Shrinking holomorphic disk movie in $\IC\IP^1$ model.}\label{fig:holo_disk}
	\end{center}
\end{figure}

The soliton equation in this model is integrable, therefore $\zeta$-solitons in Fig.\ref{fig:CP_1_vacua} are stable as long as the inverse map $W^{-1}$ to $Y$-plane is holomorphic on a disk inside the rectangle generated by soliton paths $\fX_{C,-1}$, $\fX_{A,-1}$, $\fX_{B,-1}$. We can construct a family of conformal $W^{-1}$-maps to $Y$-plane for this disk as a function of parameter $\mu$. In Fig.\ref{fig:holo_disk} we present this family for a numerical value $R=0.01$, clearly the holomorphic disk shrinks as one approaches to and cease to exist after
$$
\mu_{\rm crit}=0.3018\ldots\,.
$$

Thus we conclude that there is a \emph{non-zero} soliton scattering amplitude for processes like \eqref{sol_scatt}. In principle, we conclude that any amplitude for a triplet of solitons forming a triangle with vertices positioned in local vacua superpotential values are non-zero if all three are stable:
\be\label{amplitudes}
\begin{split}
	\fa\left[W_{*1,k_1},W_{*1,k_2},W_{*2,k_3}||{\rm Re}\,t|\gg 1\right]=1,\quad \forall\;k_1,k_2,k_3;\\
	\fa\left[W_{*1,k_1},W_{*2,k_2},W_{*2,k_3}||{\rm Re}\,t|\gg 1\right]=1,\quad \forall\;k_1,k_2,k_3.
\end{split}
\ee

In such a simple model of a single field $Y$ we can perform even some simple numerical estimates, in particular, the instanton equation \eqref{LG-instanton} can be solved numerically. 

\begin{figure}[h!]
	\begin{center}
		\begin{tikzpicture}[scale=0.9]
			\node at (0,0) {\includegraphics[scale=0.36]{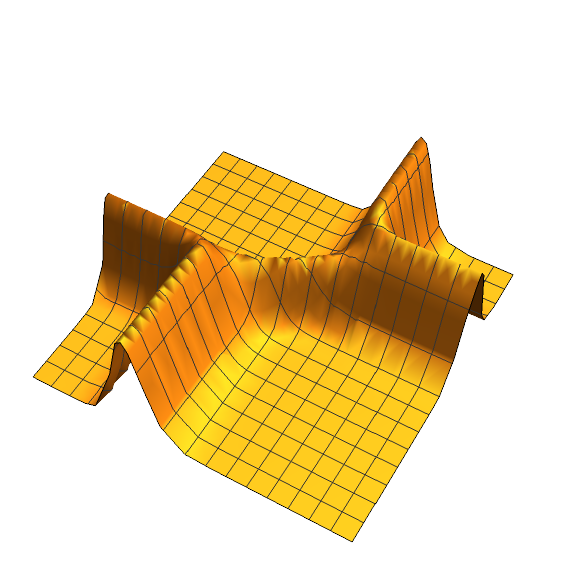}};
			\node at (4.5,0) {\includegraphics[scale=0.18]{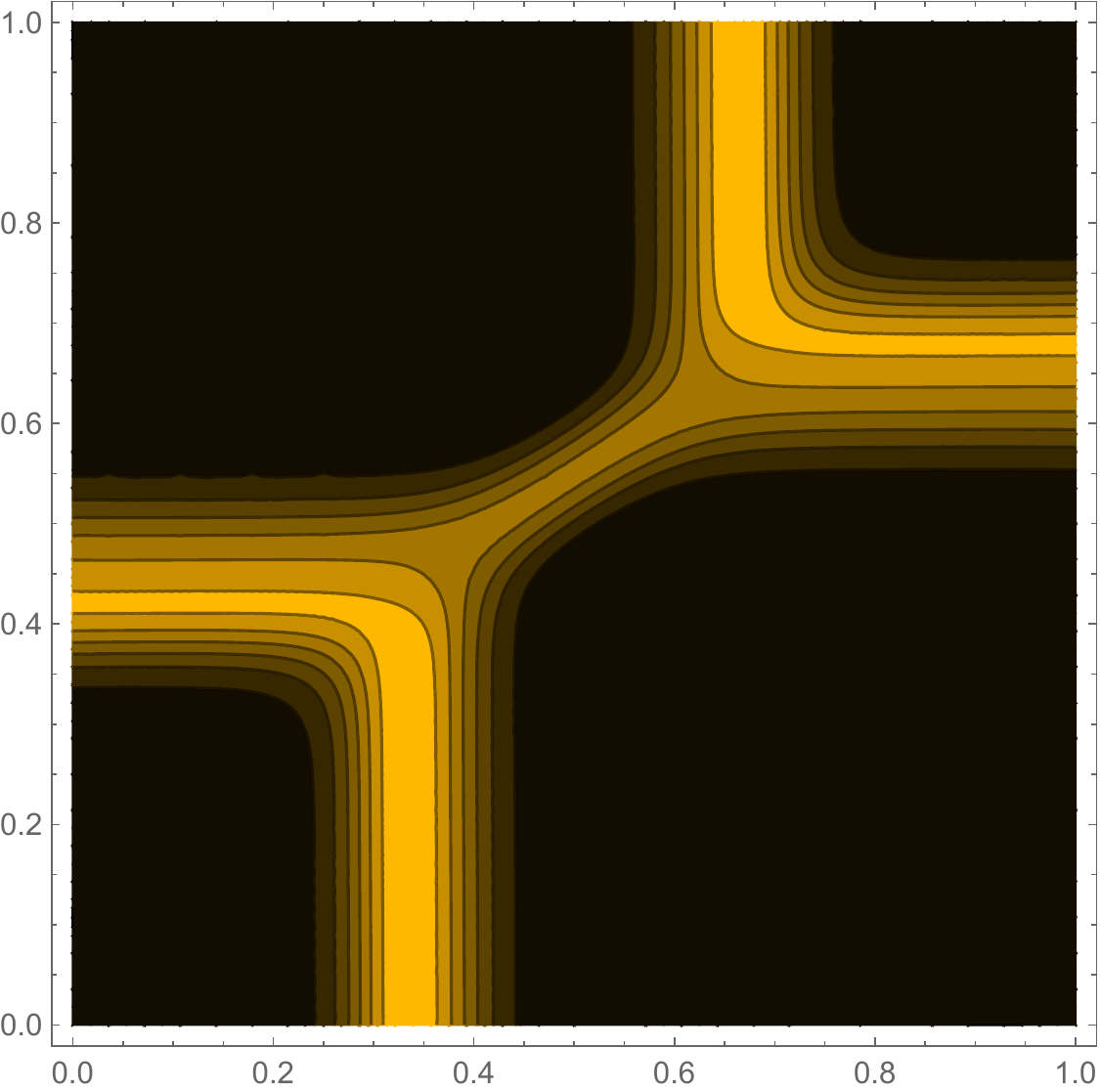}};
			\node at (7,0) {\includegraphics[scale=0.315]{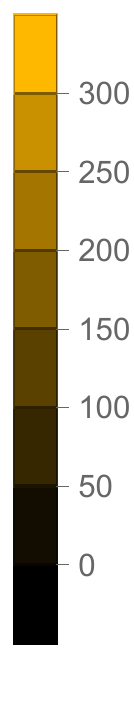}};
			\node[below] at (3.5,-2.3) {(a)};
			\begin{scope}[shift={(10.5,0)}]
				\draw[dashed,fill=blue,opacity=0.2] (-1.5,-1.5) -- (-1.5,1.5) -- (1.5,1.5) -- (1.5,-1.5) -- cycle;
				\draw[ultra thick] (-1.5,-0.5) -- (-0.5,-0.5) -- (-0.5,-1.5) (1.5,0.5) -- (0.5,0.5) -- (0.5,1.5) (-0.5,-0.5) -- (0.5,0.5); 
				\draw[fill=black] (-0.5,-0.5) circle (0.08) (0.5,0.5) circle (0.08);
				\node[below] at (-0.5,-1.5) {$\fX_{A,-1}$};
				\node[left] at (-1.5,-0.5) {$\fX_{C,-1}$};
				\node[above] at (0.5,1.5) {$\fX_{B,-1}$};
				\node[right] at (1.5,0.5) {$\fX_{C,-1}$};
				\node[above left] at (0,0) {$\fX_{C,0}$};
				\node[below left] at (-1.5,-1.5) {$W_{\rm I}$};
				\node[below right] at (1.5,-1.5) {$W_{\rm II}$};
				\node[above left] at (-1.5,1.5) {$W_{\rm III}$};
				\node[above right] at (1.5,1.5) {$W_{\rm IV}$};
			\end{scope}
			\node[below] at (10.5,-2.3) {(b)};
		\end{tikzpicture}
		\caption{ (a) Density $\rho(t,s)$ for numerically simulated soliton scattering process in the model $W(Y)=1.0\times Y-0.01\times e^Y-e^{-Y}$;  and (b) diagrammatic depiction of the soliton scattering process.}\label{fig:inst}
	\end{center}
\end{figure}
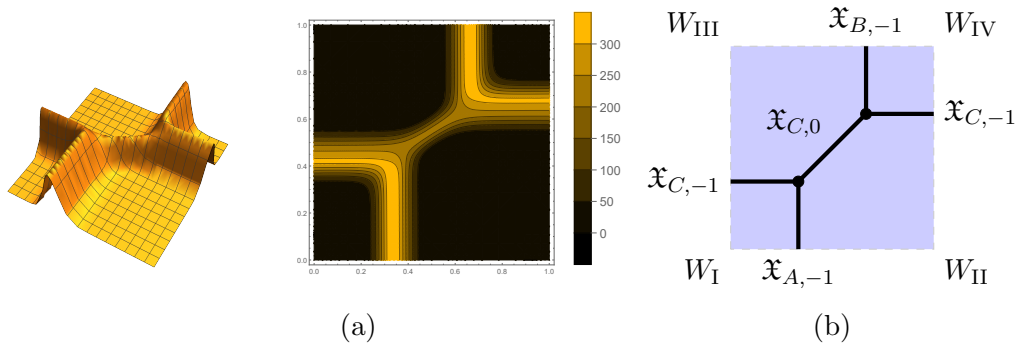

To do so let us make few preparations. First of all, let us choose notations for vacua:
\be\nn
\begin{split}
	Y_{\rm I}=Y_{*1},\quad Y_{\rm II}=Y_{*1}-2\pi \I,\quad Y_{\rm III}=Y_{*2}-2\pi\I,\quad Y_{\rm IV}=Y_{*2}-4\pi \I
\end{split}
\ee
Further let us choose the phase $\zeta=\I$ and coordinates in $z$-plane:
\be\nn
z=\alpha\tilde x^1+\beta\tilde x^2,\quad \alpha=\frac{\bar W_2-\bar W_1}{|W_2-W_1|},\; \beta=\frac{\bar W_3-\bar W_1}{|W_3-W_1|}.
\ee
So that the instanton equation reads:
\be
(\alpha\p_{\tilde 1}+\beta\p_{\tilde 2})Y=h\,\overline{\p_Y W},
\ee
where $h$ is an overall scaling factor for the world-sheet plane to control a fast convergence of the soliton solution to a LG vacuum.
We take $h=20.0$. 
If such a coordinate transformation is chosen trajectories of asymptotic soliton quasi-particles will be parallel to axes $\tilde x^1$ and $\tilde x^2$. 
We depict the result of such numerical simulation in Fig.\ref{fig:inst}(a) in terms of a density function:
$$
\rho\left(\tilde x^1,\tilde x^2\right):=|Y\left(\tilde x^1,\tilde x^2\right)-Y_{\rm I}|\cdot |Y\left(\tilde x^1,\tilde x^2\right)-Y_{\rm II}|\cdot |Y\left(\tilde x^1,\tilde x^2\right)-Y_{\rm III}|\cdot |Y\left(\tilde x^1,\tilde x^2\right)-Y_{\rm IV}|.
$$
When density $\rho$ is zero it means that the field $Y$ is in one of the LG vacuum states. 
Clearly, the simulation in Fig.\ref{fig:inst} depicts as two asymptotic soliton states are scattered through a s-channel process to a new pair of asymptotic soliton states. 
The intermediate state in s-channel is also a soliton, so we can depict this process diagrammatically as in Fig.\ref{fig:inst} (b). 
This process incorporates two scattering vertices \eqref{sol_scatt}.

\section{Categorified analytic continuation of hypergeometric series}\label{sec:cat_contin_hyper}

\subsection{Model description}

In this section we discuss the categorification of the analytic continuation for hypergeometric series following the program proposed in Section \ref{sec:Atiyah_intro}.
first of all let us survey the corresponding GLSM.

As a model for the conifold transition we pick a $U(1)$ B-twisted GLSM with $N_f=4$ chiral multiplets with charges and masses. 
This matter content can be formulated in the following quiver diagram:
\be
\begin{array}{c|c|c|c|c}
	{\rm field} & \phi_1 & \phi_2 & \phi_3 & \phi_4\\
	\hline
	Q_a & +1 & +1 & -1 & -1\\
	\hline
	\mu_a & \mu_1 & \mu_2 & -\mu_3 & -\mu_4\\
\end{array},\quad \begin{array}{c}
	\begin{tikzpicture}
		\node at (-2,0) {$1$};
		\node at (0,0) {$1$};
		\node at (2,0) {$1$};
		\draw (0,0) circle (0.3);
		\begin{scope}[shift={(2,0)}]
			\draw (-0.3,-0.3) -- (-0.3,0.3) -- (0.3,0.3) -- (0.3,-0.3) -- cycle;
		\end{scope}
		\begin{scope}[shift={(-2,0)}]
			\draw (-0.3,-0.3) -- (-0.3,0.3) -- (0.3,0.3) -- (0.3,-0.3) -- cycle;
		\end{scope}
		\draw[<-] (1.7,0.1) -- (0.3,0.1);
		\draw[->] (1.7,-0.1) -- (0.3,-0.1);
		\draw[->] (-1.7,0.1) -- (-0.3,0.1);
		\draw[<-] (-1.7,-0.1) -- (-0.3,-0.1);
		\node[above] at (-1,0.1) {$\mu_1$};
		\node[below] at (-1,-0.1) {$-\mu_3$};
		\node[above] at (1,0.1) {$-\mu_4$};
		\node[below] at (1,-0.1) {$\mu_2$};
	\end{tikzpicture}
\end{array}.
\ee
We assume that an arrangement of complex values $\mu_a$ are of general position, in particular, a difference of imaginary parts of any pair of $\mu_a$ never acquire an integer value, otherwise some of classical vacua might end up being degenerate.

Depending on the sign of FI parameter $r$ the Higgs branch is a conifold resolution:
\be
\begin{split}
	X_+:=X_{\rm Higgs}(r>0)=\left[\CO(-1)_{(\phi_3,\phi_4)}^{\oplus 2}\longrightarrow \IC\IP^1_{(\phi_1:\phi_2)}\right];\\
	X_-:=X_{\rm Higgs}(r<0)=\left[\CO(-1)_{(\phi_1,\phi_2)}^{\oplus 2}\longrightarrow \IC\IP^1_{(\phi_3:\phi_4)}\right].
\end{split}
\ee
where pairs of field $(\phi_1,\phi_2)$ and $(\phi_3,\phi_4)$ play roles of a base and a fiber. 

The disk partition function for a boundary condition given by the structure sheaf reads \cite{Hori:2013ika}:
\be\label{conifold_pf}
Z(t)=\int\lm_{-\I\infty}^{\I\infty}d\sigma\; e^{2\pi\, t\, \sigma}\,\Gamma(\sigma-\I\mu_1)\Gamma(\sigma-\I\mu_2)\Gamma(-\sigma+\I\mu_3)\Gamma(-\sigma+\I\mu_4)
\ee

We can calculate this integral by the standard Cauchy formula method after closing the integration contour at infinity. 
The way we close the contour depends on the sign of ${\rm Re}\;t$. 
Rather than considering different operator insertions for the bundle on the boundary brane we can consider various integration cycles $\gamma_a$, $a=1,\ldots,4$ encircling different series of poles from the gamma-functions. 
Clearly, these four choices correspond to constant vacua choices $*a$ when the corresponding field acquires expectation value $|\phi_a|=r^{\frac{1}{2}}$:
\be
\begin{array}{c}
	\begin{tikzpicture}
		\begin{scope}
			\draw[fill=black] (0,0) circle (0.05) (-0.5,0) circle (0.05) (-1,0) circle (0.05) (-1.5,0) circle (0.05) (-2,0) circle (0.05);
			\draw[->,thick,red] (-2.3,-0.4) to[out=0,in=180] (0,-0.4) to[out=0,in=270] (0.2,0) to[out=90,in=0] (0,0.4) to[out=180,in=0] (-1,0.4);
			\draw[thick,red] (-2.3,0.4) -- (-1,0.4);
			\node[left,red] at (-2.3,0.4) {$\gamma_1$};
			\node[right] at (0.2,0) {$\I\mu_1$}; 
		\end{scope}
		\begin{scope}[shift={(0,-1.5)}]
			\draw[fill=black] (0,0) circle (0.05) (-0.5,0) circle (0.05) (-1,0) circle (0.05) (-1.5,0) circle (0.05) (-2,0) circle (0.05);
			\draw[->,thick,blue] (-2.3,-0.4) to[out=0,in=180] (0,-0.4) to[out=0,in=270] (0.2,0) to[out=90,in=0] (0,0.4) to[out=180,in=0] (-1,0.4);
			\draw[thick,blue] (-2.3,0.4) -- (-1,0.4);
			\node[left,blue] at (-2.3,0.4) {$\gamma_2$};
			\node[right] at (0.2,0) {$\I\mu_2$}; 
		\end{scope}
		\begin{scope}[shift={(3,0)}]
			\begin{scope}[scale=-1]
				\draw[fill=black] (0,0) circle (0.05) (-0.5,0) circle (0.05) (-1,0) circle (0.05) (-1.5,0) circle (0.05) (-2,0) circle (0.05);
				\draw[->,thick,black!40!green] (-2.3,-0.4) to[out=0,in=180] (0,-0.4) to[out=0,in=270] (0.2,0) to[out=90,in=0] (0,0.4) to[out=180,in=0] (-1,0.4);
				\draw[thick,black!40!green] (-2.3,0.4) -- (-1,0.4);
				\node[right,black!40!green] at (-2.3,0.4) {$\gamma_3$};
				\node[left] at (0.2,0) {$\I\mu_3$}; 
			\end{scope}
		\end{scope}
		\begin{scope}[shift={(3,-1.5)}]
			\begin{scope}[scale=-1]
				\draw[fill=black] (0,0) circle (0.05) (-0.5,0) circle (0.05) (-1,0) circle (0.05) (-1.5,0) circle (0.05) (-2,0) circle (0.05);
				\draw[->,thick,purple] (-2.3,-0.4) to[out=0,in=180] (0,-0.4) to[out=0,in=270] (0.2,0) to[out=90,in=0] (0,0.4) to[out=180,in=0] (-1,0.4);
				\draw[thick,purple] (-2.3,0.4) -- (-1,0.4);
				\node[right,purple] at (-2.3,0.4) {$\gamma_4$};
				\node[left] at (0.2,0) {$\I\mu_4$}; 
			\end{scope}
		\end{scope}
	\end{tikzpicture}
\end{array}
\ee

For each of the basis integration cycle we can easily calculate the integral as a sum over pole contributions in terms of hypergeometric series (see Appendix \ref{sec:hypergeom} for reference):
\be
\begin{split}
	&Z_{\gamma_1}=-2\pi\I \,e^{2\pi\I t \mu_1}\,\left[\Gamma(\I\mu_{12})\Gamma(-\I\mu_{13})\Gamma(-\I\mu_{14})\right]\;{}_2F_1\left[\begin{array}{c}
		-\I\mu_{13};\;-\I\mu_{14}\\
		1-\I\mu_{12}
	\end{array}\right]\left(e^{-2\pi t}\right),\\
	&Z_{\gamma_2}=-2\pi\I \,e^{2\pi\I t \mu_2}\,\left[\Gamma(\I\mu_{21})\Gamma(-\I\mu_{23})\Gamma(-\I\mu_{24})\right]\;{}_2F_1\left[\begin{array}{c}
		-\I\mu_{23};\;-\I\mu_{24}\\
		1-\I\mu_{21}
	\end{array}\right]\left(e^{-2\pi t}\right),\\
	&Z_{\gamma_3}=2\pi\I\,e^{2\pi\I t \mu_3}\,\left[\Gamma(\I\mu_{43})\Gamma(-\I\mu_{13})\Gamma(-\I\mu_{23})\right]\;{}_2F_1\left[\begin{array}{c}
		-\I\mu_{13};\;-\I\mu_{23}\\
		1-\I\mu_{43}
	\end{array}\right]\left(e^{2\pi t}\right),\\
	&Z_{\gamma_4}=2\pi\I\,e^{2\pi\I t \mu_4}\,\left[\Gamma(\I\mu_{34})\Gamma(-\I\mu_{14})\Gamma(-\I\mu_{24})\right]\;{}_2F_1\left[\begin{array}{c}
		-\I\mu_{14};\;-\I\mu_{24}\\
		1-\I\mu_{34}
	\end{array}\right]\left(e^{2\pi t}\right),
\end{split}
\ee
where $\mu_{ab}=\mu_a-\mu_b$. 
A unit circle $|e^{2\pi t}|=1$ corresponding to value $r=0$ divides the Riemann sphere spanned by $e^{2\pi t}$ in two hemispheres. 
Correspondingly, series for $\gamma_{1,2}$ converge absolutely in the hemisphere containing point $e^{2\pi t}=\infty$, and series for $\gamma_{3,4}$ converge in the opposite hemisphere containing point $e^{2\pi t}=0$ (see Fig.\ref{fig:wp}(c)).

The calculated series may be continued from one hemisphere to the other using the fact that the hypergeometric function is a solution to the hypergeometric equation,  and a pair of solutions are flat sections of a holomorphic $SL(2,\IC)$ connection. 
This connection allows one to parallel transport a hypergeometric solution across the unit circle boundary and re-decompose it on the other side over basic convergent solutions. 
This linear map between hypergeometric series bases can be summarized in the following form:
\be
\begingroup
\renewcommand*{\arraystretch}{1.7}
\left(\begin{array}{c}
	Z_{\gamma_1}\\
	Z_{\gamma_2}
\end{array}\right)={\bf A}
\left(\begin{array}{c}
	Z_{\gamma_3}\\
	Z_{\gamma_4}
\end{array}\right),\quad {\bf A}=\left(\begin{array}{cc}
	\dfrac{\chi_3}{\chi_1}\cdot\dfrac{\lambda_{23}}{\lambda_{12}} & \dfrac{\chi_4}{\chi_1}\cdot\dfrac{\lambda_{24}}{\lambda_{12}}\\
	\dfrac{\chi_3}{\chi_2}\cdot\dfrac{\lambda_{13}}{\lambda_{21}} & \dfrac{\chi_4}{\chi_2}\cdot\dfrac{\lambda_{14}}{\lambda_{21}}
\end{array}\right),
\endgroup
\ee
where
\be
\chi_a:=e^{\pi\mu_a},\quad \lambda_{ab}:=\frac{\chi_a}{\chi_b}-\frac{\chi_b}{\chi_a}.
\ee
Matrix $\bf A$ follows directly from \eqref{analytic}, it is induced by a parallel transport along path $\wp$ in Fig.\ref{fig:wp}(b), we will call it an analytic continuation matrix.

\subsection{Analytic continuation in the language of the Grothendieck group}

As we mentioned in Section \ref{sec:dec_cat_par} the disk partition function represents a map from  the category to its Grothendieck group given by a vector space. 
In this section we will discuss a simpler behavior of the Grothendieck groups under the parallel transport.
This will allow us to capture static behavior of solitons confined to the defect. 
Here we apply the methods of spectral networks (see Appendix~\ref{sec:App_SN}). 

We consider the superpotential of this theory with the contributions of the disorder operator fields $Y_a$ restored:
\be\label{Conif_sp}
W(\Sigma,Y_a)=t\Sigma+\frac{1}{2\pi}\sum\lm_{a=1}^2\left[(\Sigma-\mu_a)Y_a-\Lambda e^{Y_a}\right]+\frac{1}{2\pi}\sum\lm_{b=3}^4\left[(\mu_b-\Sigma)Y_b-\Lambda e^{Y_b}\right].
\ee
Clearly, substituting this expression in the formula for the disk partition function \eqref{Clmb_br_pf} and integrating over disorder operator fields $Y_a$ we will get an expression of type \eqref{conifold_pf}.
Unfortunately, the topology of Lefschetz thimbles in terms of the original twisted field $\Sigma$ is rather messy, we would rather use a dual model where it will be more transparent.
Following this purpose we first integrate over $\Sigma$ in \eqref{Conif_sp}. 
It plays a role of a Lagrange multiplier imposing a constraint:
\begin{equation}\label{delta_constr}
	2\pi t +Y_1+Y_2-Y_3-Y_4=0,
\end{equation}
realized in terms of the partition function as a delta-function in the integrand. 
This constraint\footnote{The author would like to thank Mina Aganagic for a suggestion to apply here this variable redefinition.} can be resolved in terms of new variables $y_1$, $y_2$, $x$ and $w$:
\be\label{Ys}
\begin{split}
	Y_1&=y_2-\log\left(1-x^{-1}e^{-2\pi t}\right)-2\pi t-\log(-x)+w,\\
	Y_2&=y_1-\log(1-x)+\log(-x),\\
	Y_3&=y_1-\log(1-x),\\
	Y_4&=y_2-\log\left(1-x^{-1}e^{-2\pi t}\right).
\end{split}
\ee
In terms of these variables the delta function constraint reduces to simply $w=0$.
Fields $y_1$ and $y_2$ decouple: corresponding effective theories describe two free neutral massive scalars with masses $\mu_{32}$ and $\mu_{41}$ having a single constant vacuum field configuration. 
In the partition function integral these variables are also separated giving simple contributions:
\be\label{extra_PF}
\begin{split}
	\int dy_1\,e^{y_1(\I \mu_{32})-\Lambda e^{y_1}}\sim\Gamma(\I\mu_{32}),\quad 	\int dy_2\,e^{y_2(\I \mu_{41})-\Lambda e^{y_2}}\sim\Gamma(\I\mu_{41}).
\end{split}
\ee

We denote the resulting partition function as $\Xi$ to stress it is a LG partition function as opposed to the $\sigma$-model partition function denoted by $Z$. 
The resulting expression depends on a choice of Lagrangian integration cycle $\CL$: 
\be
\Xi_{\CL}(t)=e^{2\pi\I t\,\mu_4}\;\Gamma(\I\mu_{32})\Gamma(\I\mu_{41})\;\int\lm_{\CL} \frac{dx}{x}\;x^{\I\mu_{42}}(1-x)^{\I\mu_{23}}\left(1-x\,e^{2\pi t}\right)^{\I\mu_{14}}.
\ee
This partition function corresponds (up to an overall multiplier) to a LG model of field $x$ with a superpotential:
\be\label{sup_0}
W(x)=\frac{1}{2\pi}\left[\mu_{42}\,\log \,x+\mu_{23}\,\log(1-x)+\mu_{14}\log\left(1-x \,e^{2\pi t}\right)\right].
\ee

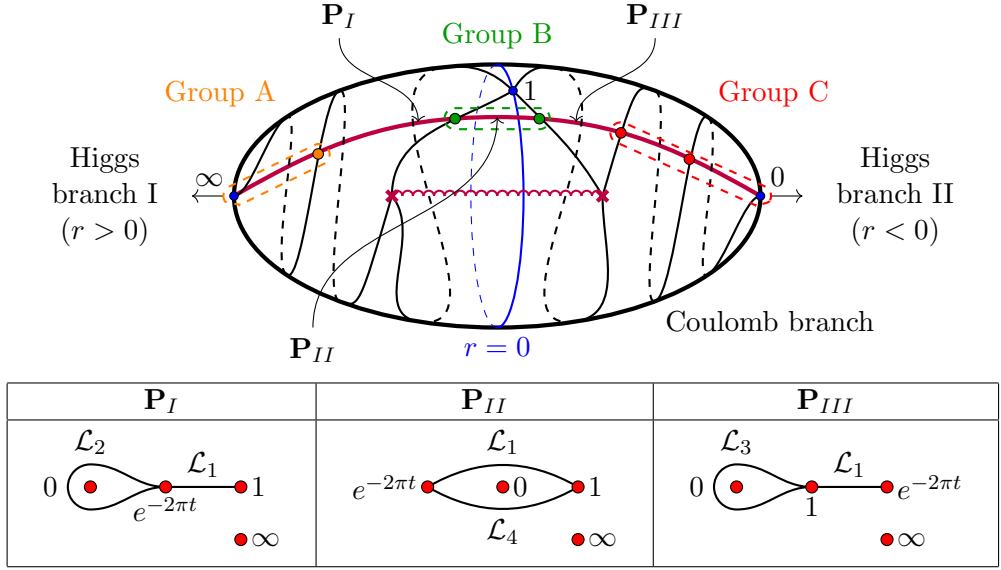
\begin{figure}
	\begin{center}
		$\begin{array}{c}
			\begin{tikzpicture}[scale=0.7]
				\begin{scope}[yscale=0.5]
					\draw[ultra thick] (0,0) circle (5);
				\end{scope}
				\begin{scope}[xscale=0.2]
					\draw[dashed, blue] (0,2.5) arc (90:270:2.5);
					\begin{scope}[xscale=-1]
						\draw[thick, blue] (0,2.5) arc (90:270:2.5);
					\end{scope}
				\end{scope}
				\draw[ultra thick, color=purple] (-5,0) to[out=30,in=180] (0,1.5) to[out=0,in=150] (5,0);
				\draw[thick] (-2,0) to[out=90,in=210] (0.3,2) (2,0) to[out=120,in=315] (0.3,2);
				\draw[thick] (-2,0) to[out=310,in=171.064] (-1.5,-2.38485) (-1,2.44949) to[out=5.82755,in=135] (0.3,2);
				\draw[thick, dashed] (-1.5,-2.38485) to[out=351.064,in=185.828] (-1,2.44949);
				\begin{scope}[xscale=-1]
					\draw[thick] (-2,0) to[out=270,in=171.064] (-1.5,-2.38485) (-1,2.44949) to[out=5.82755,in=120] (-0.3,2);
					\draw[thick, dashed] (-1.5,-2.38485) to[out=351.064,in=185.828] (-1,2.44949);
				\end{scope}
				\begin{scope}[yscale=-1]
					\draw[thick] (-2,0) .. controls (-2.62547, 2.14045) .. (-3,2) (-3,-2) .. controls (-3.37453, -1.85955) and (-3.66718, 1.72188) .. (-4, 1.5) (-4,-1.5) .. controls (-4.33282, -1.27812) and (-4.7,0) .. (-5,0);
					\draw[thick, dashed] (-3,2) .. controls (-3.37453, 1.85955) and (-2.62547, -2.14045) .. (-3,-2) (-4,1.5) .. controls (-4.33282, 1.27812) and (-3.66718, -1.72188) ..  (-4,-1.5);
				\end{scope}
				\begin{scope}[xscale=-1]
					\draw[thick] (-2,0) .. controls (-2.62547, 2.14045) .. (-3,2) (-3,-2) .. controls (-3.37453, -1.85955) and (-3.66718, 1.72188) .. (-4, 1.5) (-4,-1.5) .. controls (-4.33282, -1.27812) and (-4.7,0) .. (-5,0);
					\draw[thick, dashed] (-3,2) .. controls (-3.37453, 1.85955) and (-2.62547, -2.14045) .. (-3,-2) (-4,1.5) .. controls (-4.33282, 1.27812) and (-3.66718, -1.72188) ..  (-4,-1.5);
				\end{scope}
				\draw[fill=blue] (-5,0) circle (0.08) (0.3,2) circle (0.08) (5,0) circle (0.08);
				\node[above left] at (-5,0) {$\infty$};
				\node[above right] at (5,0) {$0$};
				\node[right] at (0.3,2) {$1$};
				\foreach \i in {0,1,...,19}
				{
					\draw[thick, purple] (\i / 5 - 2, 0) to[out=90,in=180] (\i / 5 + 0.1 - 2, 0.1) to[out=0,in=90] (\i / 5 + 0.2 - 2, 0);
				}
				\begin{scope}[shift={(-2,0)}]
					\draw[ultra thick, purple] (-0.1,-0.1) -- (0.1,0.1) (-0.1,0.1) -- (0.1,-0.1);
				\end{scope}
				\begin{scope}[shift={(2,0)}]
					\draw[ultra thick, purple] (-0.1,-0.1) -- (0.1,0.1) (-0.1,0.1) -- (0.1,-0.1);
				\end{scope}
				\draw[->] (5.2,0) -- (5.8,0);
				\draw[->] (-5.2,0) -- (-5.8,0);
				\node[left] at (-6,0) {$\begin{array}{c} \mbox{Higgs} \\ \mbox{branch I}\\ (r>0) \end{array}$};
				\node[right] at (6,0) {$\begin{array}{c} \mbox{Higgs} \\ \mbox{branch II}\\ (r<0)\end{array} $};
				\draw[fill=green!60!black] (-0.8,1.47) circle (0.1) (0.8,1.47) circle (0.1);
				\draw[thick, green!60!black, dashed] (-1,1.47) to[out=90,in=180] (-0.8,1.67) -- (0.8,1.67) to[out=0,in=90] (1,1.47) to[out=270,in=0] (0.8,1.27) -- (-0.8,1.27) to[out=180,in=270] (-1,1.47);
				\node[above] at (0,2.6) {\color{green!60!black} Group B};
				\draw[fill=orange] (-3.4,0.8) circle (0.1);
				\begin{scope}[shift={(-5,0)}]
					\begin{scope}[rotate=26]
						\draw[dashed, thick, orange] (-0.2,0) to[out=90,in=180] (0,0.2) -- (1.8,0.2) to[out=0,in=90] (2,0) to[out=270,in=0] (1.8,-0.2) -- (0,-0.2) to[out=0,in=270] (-0.2,0);
					\end{scope}
				\end{scope}
				\node[above left] at (-4,1.5) {\color{orange} Group A};
				\begin{scope}[xscale=-1]
					\draw[fill=red] (-2.35,1.2) circle (0.1) (-3.65,0.7) circle (0.1);
					\begin{scope}[shift={(-5,0)}]
						\begin{scope}[rotate=24.36]
							\draw[dashed, thick, red] (-0.2,0) to[out=90,in=180] (0,0.2) -- (2.91,0.2) to[out=0,in=90] (3.11,0) to[out=270,in=0] (2.91,-0.2) -- (0,-0.2) to[out=0,in=270] (-0.2,0);
						\end{scope}
					\end{scope}
					\node[above right] at (-4,1.5) {\color{red} Group C};
				\end{scope}
				\draw[->] (-3,3) to[out=270,in=90] (-1.5,1.4);
				\draw[->] (3,3) to[out=270,in=90] (1.5,1.4);
				\draw[->] (-3.5,-2.5) to[out=90,in=270] (0,1.5);
				\node[above] at (-3,3) {${\bf P}_{I}$};
				\node[above] at (3,3) {${\bf P}_{III}$};
				\node[below] at (-3.5,-2.5) {${\bf P}_{II}$};
				\node[below] at (0,-2.5) {\color{blue} $r=0$};
				\node[below right] at (3,-2) {Coulomb branch};
			\end{tikzpicture}\\
			\begin{array}{|c|c|c|}
				\hline
				{\bf P}_{I} & {\bf P}_{II} & {\bf P}_{III}\\
				\hline
				\begin{array}{c}
					\begin{tikzpicture}
						\draw[thick] (1,0) to[out=180,in=0] (0,0.3) to[out=180,in=90] (-0.3,0) to[out=270,in=180] (0,-0.3) to[out=0,in=180] (1,0) (1,0) -- (2,0);
						\draw[fill=red] (0,0) circle (0.08) (1,0) circle (0.08) (2,0) circle (0.08) (2,-0.7) circle (0.08);
						\node[left] at (-0.3,0) {$0$};
						\node[right] at (2,0) {$1$};
						\node[below] at (1,0) {$e^{-2\pi t}$};
						\node[right] at (2,-0.7) {$\infty$};
						\node[above] at (0,0.3) {$\CL_2$};
						\node[above] at (1.5,0) {$\CL_1$};
					\end{tikzpicture}
				\end{array}&\begin{array}{c}\begin{tikzpicture}
						\draw[thick] (0,0) to[out=30,in=150] (2,0) (0,0) to[out=330,in=210] (2,0);
						\draw[fill=red] (0,0) circle (0.08) (1,0) circle (0.08) (2,0) circle (0.08) (2,-0.7) circle (0.08);
						\node[left] at (0,0) {$e^{-2\pi t}$};
						\node[right] at (2,0) {$1$};
						\node[right] at (1,0) {$0$};
						\node[right] at (2,-0.7) {$\infty$};
						\node[above] at (1,0.3) {$\CL_1$};
						\node[below] at (1,-0.3) {$\CL_4$};
				\end{tikzpicture}\end{array}&\begin{array}{c}
					\begin{tikzpicture}
						\draw[thick] (1,0) to[out=180,in=0] (0,0.3) to[out=180,in=90] (-0.3,0) to[out=270,in=180] (0,-0.3) to[out=0,in=180] (1,0) (1,0) -- (2,0);
						\draw[fill=red] (0,0) circle (0.08) (1,0) circle (0.08) (2,0) circle (0.08) (2,-0.7) circle (0.08);
						\node[left] at (-0.3,0) {$0$};
						\node[below] at (1,0) {$1$};
						\node[right] at (2,0) {$e^{-2\pi t}$};
						\node[right] at (2,-0.7) {$\infty$};
						\node[above] at (0,0.3) {$\CL_3$};
						\node[above] at (1.5,0) {$\CL_1$};
					\end{tikzpicture}
				\end{array}\\
				\hline
			\end{array}
		\end{array}$
	\end{center}
	\caption{Conifold Coulomb branch parameterized by $z=e^{2\pi t}$.}\label{fig:phase}
\end{figure}

Now it is easy to describe the topology of the Lefschetz thimbles. 
Those are curves in the $x$-plane that spiral towards singularities at $0$, $1$, $\infty$ and $e^{2\pi t}$ similarly to the situation depicted in Fig.\ref{fig:SN_CP_1} for $\IC\IP^1$. 
The topology of Lefshetz thimbles jumps across S-walls.
A spectral cover equation determined by superpotential \eqref{sup_0} reads:
\be\label{cover}
z(z\lambda-\mu_1)(z\lambda-\mu_2)=(z\lambda-\mu_3)(z\lambda-\mu_4),
\ee
where we introduced a parameter:
$$
z:=e^{2\pi t},
$$
and the meromorphic SW differential reads (see Appendix \ref{sec:App_SN} for definitions):
$$
\Omega =2\pi \I\; \lambda\; dz.
$$
\begin{figure}
	\begin{center}
		\begin{tikzpicture}
			\node {\includegraphics[scale=0.7]{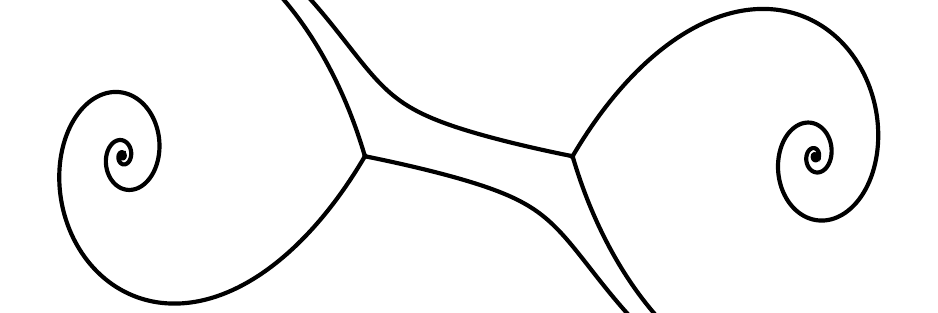}};
			\draw[gray, thick, postaction={decorate},decoration={markings, 
				mark= at position 0.65 with {\arrow{stealth}}}] (-4.1,0) to[out=10,in=170] node[pos=0.65,above]{$\Gamma_1$} (4.1,0) ;
			\draw[gray, thick, postaction={decorate},decoration={markings, 
				mark= at position 0.65 with {\arrow{stealth}}}] (-4.1,0) to[out=350,in=190] node[pos=0.35,below]{$\Gamma_2$} (4.1,0);
			\draw[fill=blue] (-4.1,0) circle (0.07);
			\node[blue] at (-4,-0.7) {$-1$};
			\draw[fill=blue] (4.1,0) circle (0.07);
			\node[blue] at (4,0.7) {$+1$};
		\end{tikzpicture}
		\caption{Spectral networks for a conifold with the following choice of parameters:
			$\mu_1=1.2\times e^{0.07\pi \I}$, $\mu_2=2.2\times e^{0.07\pi \I}$, $\mu_3=3.248\times e^{0.07\pi \I}$, $\mu_4=2.248\times e^{0.07\pi \I}$. The spectral networks is drawn in the plane $z=e^{2\pi t}$ under a conformal map $z\mapsto \frac{1+z}{1-z}$, so that $0\mapsto 1$, $\infty\mapsto -1$.}\label{fig:Conif_SN}
	\end{center}
\end{figure}
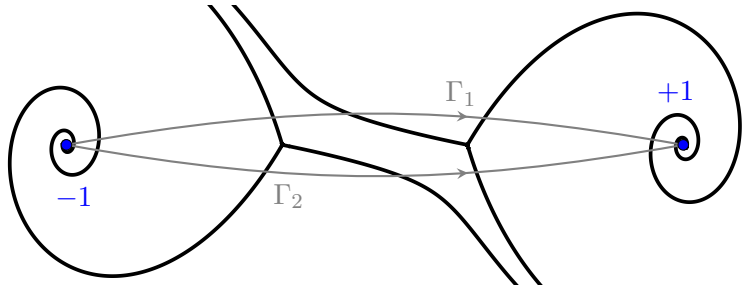
\indent Using this differential we construct a collection of S-walls on the phase diagram spanned by $z=e^{2\pi t}$ in  Fig.\ref{fig:Conif_SN} and schematize an interface path and corresponding Lefschetz thimble topologies in different chambers in Fig.\ref{fig:phase}.
In practice, from the Lefschetz thimbles at the current stage of consideration only their homology classes are needed, and in Fig.\ref{fig:phase} we restrict to illustrating solely this information.
To determine the form of these curves we solved numerically equation \eqref{Lef_thi_eq} for parameters specified in  Fig.\ref{fig:Conif_SN}.
We should warn the reader that a different choice of twisted mass parameters $\mu_i$ would reveal a different topology for the thimbles.
This is a contribution from the wall-crossing processes accompanying braiding in $\mu_i$ not considered in this paper.
Nevertheless, as it would be transparent in the following, the resulting parallel transport functor can be written in universal terms without a reference to the twisted masses.
Eventually we distinguish four cycles denoted $\CL_{i=1,2,3,4}$ respectively.

Comparing S-wall patterns of Fig.\ref{fig:SN_CP_1} and Fig.\ref{fig:phase} we observe that the conifold Coulomb branch is actually glued out of two $\IC\IP^1$ Coulomb branches. The cut line goes along the unit circle $r=0$. At singularities $0$ and $\infty$ the Coulomb branch is sewed with Higgs branches as in the $\IC\IP^1$ case.

S-walls divide the Coulomb branch in three chambers ${\bf P}_I$, ${\bf P}_{II}$ and ${\bf P}_{III}$ with corresponding topologies of Lefschetz thimbles depicted schematically in a table in Fig.\ref{fig:phase}. Corresponding expressions for the partition functions are:
\be\nn
\begin{split}
	\Xi_{\CL_1}=&\frac{\Gamma(\I\mu_{32})\Gamma(1+\I\mu_{23})\Gamma(\I\mu_{41})\Gamma(1+\I\mu_{14})}{\Gamma(2+\I\mu_{14}+\I\mu_{23})}e^{2\pi\I t\,\mu_3}\left(1-e^{2\pi t}\right)^{1+\I\mu_{23}+\I\mu_{14}}\times\\
	&\times {}_2F_1\left[\begin{array}{c}
		1+\I\mu_{14};\;1+\I\mu_{24}\\
		2+\I\mu_{14}+\I\mu_{23}
	\end{array}\right]\left(1-e^{2\pi t}\right),\\
	\Xi_{\CL_2}=&\frac{\sinh \pi\mu_{42}\sinh\pi\mu_{21}}{\sinh\pi\mu_{41}}e^{2\pi\I t\,\mu_2}\left[\Gamma(\I\mu_{32})\Gamma(\I\mu_{42})\Gamma(\I\mu_{21})\right]\;{}_2F_1\left[\begin{array}{c}
		\I\mu_{42};\;\I\mu_{32}\\
		1+\I\mu_{12}
	\end{array}\right]\left(e^{-2\pi t}\right),\\
	\Xi_{\CL_3}=&\frac{\sinh \pi\mu_{42}\sinh\pi\mu_{34}}{\sinh\pi\mu_{32}}e^{2\pi\I t\,\mu_4}\left[\Gamma(\I\mu_{41})\Gamma(\I\mu_{42})\Gamma(\I\mu_{34})\right]\;{}_2F_1\left[\begin{array}{c}
		\I\mu_{42};\;\I\mu_{41}\\
		1+\I\mu_{43}
	\end{array}\right]\left(e^{2\pi t}\right),\\
	\Xi_{\CL_4}=&e^{\pi(\mu_{32}+\mu_{42})}\;\Xi_1+\Xi_2=e^{\pi \mu_{34}}\;\Xi_{\CL_1}-\Xi_{\CL_3}.
\end{split}
\ee

We perform a parallel transport along path $\wp$ connecting Higgs branches at $z=\infty$ and $0$. 
This path intersect S-walls in a selection of points we divide in three groups (see Fig.\ref{fig:phase}). 
In groups A and C there are infinitely many intersection points from the spirals, these intersection points are in one-to-one correspondence with a tower of solitons obtained from intersection points at Fig.\ref{fig:SN_CP_1}(a). 
There are only two intersection points in group B.

A pair of hypergeometric functions form a basis in the space of solutions to a hypergeometric equation. 
Transition between bases $Z_{\gamma}$ and $\Xi_{\CL}$ is given by Stokes matrices\footnote{At this point a clarification should be made. 
	We have chosen to use a calculation of the transition matrices via genuine solution bases to the hypergeometric equation due to a simplicity reason.
	The properties of hypergeometric functions are well understood, therefore we are able to find all the necessary relations in canonical textbooks on this subject.
	The jumps of asymptotic behavior we would construct by applying machinery of Appendix \ref{sec:App_SN} would coincide with asymptotic of matrices ${\bf S}_{A,B,C}$.
	Moreover it turns out that re-structuring of Lefschetz thimbles for the model in question \eqref{sup_0} leads to coefficients that are quasi-classically exact (see also discussion in \cite[Section 6.3]{Gaiotto:2011nm}).
	Indeed comparing expressions \eqref{StokesM} for exact matrices $S_{A,B,C}$ and Stokes matrices \eqref{Stokes} we find an agreement -- the matrices are of upper(lower) triangular form, and off-diagonal elements are series in monomials $\exp\left(\sum_i c_i \mu_i\right)$ --  provided that the soliton action in this model is a linear function in complex masses $\mu_i$ in the large volume limit.
	Following this reasoning we call matrices ${\bf S}_{A,B,C}$ Stokes matrices.
} ${\bf S}_A$, ${\bf S}_B$, ${\bf S}_C$ that are given by supersymmetric indices of solitons contributing to intersection points of $\wp$ with the spectral network. 
Transitions between different bases can be summarized in the following form:
\be\label{sequence}
\begin{array}{c}
	\begin{tikzpicture}
		\node (A) at (0,0) {$
			\left(\begin{array}{c}
				Z_{\gamma_1}\\ Z_{\gamma_2}
			\end{array}\right)
			$};
		\node (B) at (3,0) {$
			\left(\begin{array}{c}
				\Xi_{\CL_1}\\ \Xi_{\CL_2}
			\end{array}\right)
			$};
		\node (C) at (6,0) {$
			\left(\begin{array}{c}
				\Xi_{\CL_1}\\ \Xi_{\CL_4}
			\end{array}\right)
			$};
		\node (D) at (9,0) {$
			\left(\begin{array}{c}
				\Xi_{\CL_1}\\ \Xi_{\CL_3}
			\end{array}\right)
			$};
		\node (E) at (12,0) {$
			\left(\begin{array}{c}
				Z_{\gamma_3}\\ Z_{\gamma_4}
			\end{array}\right)
			$};
		\path (A) edge[->] node[above] {$G_I{\bf \color{orange} S}_{\color{orange} A}$} (B)  (B) edge[->] node[above] {${\bf \color{green!60!black} S}_{\color{green!60!black} B,1}$} (C) (C) edge[->] node[above] {${\bf \color{green!60!black} S}_{\color{green!60!black} B,2}$} (D) (D) edge[->] node[above] {${\bf \color{red} S}_{\color{red} C}G_{II}$} (E);
	\end{tikzpicture}
\end{array},
\ee
so that
$$
{\bf A}=G_I{\bf \color{orange} S}_{\color{orange} A} {\bf \color{green!60!black} S}_{\color{green!60!black} B,1}{\bf \color{green!60!black} S}_{\color{green!60!black} B,2} {\bf \color{red} S}_{\color{red} C}G_{II}\,,
$$
where we marked the matrices corresponding to different groups by the same colors as they appear in Fig.\ref{fig:phase}, and we distinguish contributions of the left and right points in group B as ${\bf S}_{B,1}$ and ${\bf S}_{B,2}$.

Here $G_{I}$ and $G_{II}$ are diagonal matrices of normalization coefficients:
$$
G_I={\rm diag}\left(\begin{array}{cc}
	\frac{\chi_3\chi_4}{\chi_1\chi_2}\frac{\lambda_{23}}{\lambda_{13}},&\frac{\lambda_{14}}{\lambda_{21}\lambda_{24}}
\end{array}\right),\quad G_{II}={\rm diag}\left(\begin{array}{cc}
	\frac{\lambda_{13}}{\lambda_{14}},&\frac{\lambda_{24}\lambda_{34}}{\lambda_{23}}
\end{array}\right);
$$ 
and the Stokes matrices are:
\begingroup
\renewcommand*{\arraystretch}{1.7}
\be\label{StokesM}
{\bf S}_A=\left(\begin{array}{cc}
	1 & \frac{h_1}{\lambda_{12}}\\
	0 & 1 \\
\end{array}\right),\quad {\bf S}_{B,1}=\left(\begin{array}{cc}
	1 & 0 \\
	-\frac{\chi_1}{h_1\chi_2}& 1
\end{array}\right),\quad 
{\bf S}_{B,2}=\left(\begin{array}{cc}
	1 & 0 \\
	\frac{\chi_3}{\chi_4} & -1
\end{array}\right),\quad {\bf S}_C=\left(\begin{array}{cc}
	1 & \frac{1}{\lambda_{34}}\\
	0 & 1 \\
\end{array}\right)\,,
\ee
\endgroup
where $h_1$ is related to the residue of $\Omega$ at puncture $e^{2\pi t}=1$:
\begin{equation}
	h_1:=\frac{\chi_1\chi_2}{\chi_3\chi_4}=\exp\left(-\frac{1}{2}\oint\lm_1\Omega^{(1)}\right),\quad \oint\lm_1\Omega^{(2)}=0\,.
\end{equation}
Here we assume that $\Omega^{(i)}$ are values of SW differential on sheets 1 and 2 respectively,
and enumeration of sheets of cover \eqref{cover} is chosen in such a way that integrals of $\Omega^{(i)}$ describe asymptotics of the first and of the second entry correspondingly in the column vectors of \eqref{sequence}.
The branch cut is chosen to connect two ramification points.
It is easy to calculate these values also from monodromies of a local basis of functions $\Xi_{\CL_1}$ and $\Xi_{\CL_4}$ around the singularity $e^{2\pi t}=1$.

\subsection{Herbst-Hori-Page invariant branes and parallel transport}

We could reorder matrices \eqref{StokesM} to derive the following relations between solution bases:
\begin{equation}\label{HHP-O}
	\left(\begin{array}{c}
		\Xi_{\CL_1}\\
		\chi_3\chi_4\Xi_{\CL_4}
	\end{array}\right)=\left(\begin{array}{cc}
		1 & 1\\
		{\color{blue}\chi_3^2} &{\color{blue}\chi_4^2}\\
	\end{array}\right)\left(\begin{array}{c}
		\tilde Z_{\gamma_3}\\
		\tilde Z_{\gamma_4}\\
	\end{array}\right),\quad 
	\left(\begin{array}{c}
		\Xi_{\CL_1}\\
		{\color{black!40!green}h_1^2}\chi_3\chi_4\Xi_{\CL_4}
	\end{array}\right)=\left(\begin{array}{cc}
		1 & 1\\
		{\color{blue}\chi_1^2} &{\color{blue}\chi_2^2}\\
	\end{array}\right)\left(\begin{array}{c}
		\tilde Z_{\gamma_1}\\
		\tilde Z_{\gamma_2}\\
	\end{array}\right)\,,
\end{equation}
where\footnote{
	A discrepancy between normalizations of $Z_k$ and $\tilde Z_k$ is induced by our rather inaccurate treatment of gamma function integrals like \eqref{extra_PF}.
	In practice, these relations depend on a particular choice of the integration contour dictated by the QFT, for instance, relation
	$$
	\int\lm_{-\infty}^{\infty} dy\; e^{\mu \;y-e^{y}} =\Gamma(\mu)
	$$
	holds only for ${\rm Re}\;\mu>0$. 
	If the situation is different we have to choose a different integration cycle. 
	Similarly, in the definition of the disk partition function \eqref{conifold_pf} we have chosen simple brane boundary conditions.
	As we will see in what follows this produces condensates of neutral mesons \eqref{cond_wf} that have to be dualized to the LG phase.
	To cancel contributions of these condensates one may choose a zero section sheaf following recipe of \eqref{Weil}.
	Such a sheaf produces a non-trivial D-brane operator \eqref{Z_B} $f(\sigma)\sim\sin\pi\sigma$ shifting normalizations of disk partition functions.
	See also \cite[Section 2.1.1]{Zenkevich:2017ylb}.
}
$$
\tilde Z_{\gamma_1}=h_1\frac{\lambda_{13}\lambda_{14}}{\lambda_{32}\lambda_{41}}Z_{\gamma_1},\quad \tilde Z_{\gamma_2}=h_1\frac{\lambda_{23}\lambda_{24}}{\lambda_{32}\lambda_{41}}Z_{\gamma_2},\quad \tilde Z_{\gamma_3}=\frac{\lambda_{31}\lambda_{32}}{\lambda_{32}\lambda_{41}}Z_{\gamma_3},\quad \tilde Z_{\gamma_4}=\frac{\lambda_{41}\lambda_{42}}{\lambda_{32}\lambda_{41}}Z_{\gamma_4}.
$$

Rather than considering various Lagrangian cycles as integration contours we can produce various solutions to the hypergeometric equations by modifying brane boundary conditions. 
There is a single brane on the disk boundary and we can insert a supersymmetric Wilson line of charge $k$.
According to \cite{Hori:2013ika} it produces a shift of the complexified FI parameter in the partition function argument:
$$
t\longrightarrow t-\I k.
$$
It will be simpler to calculate the result of this shift in the basis $Z_{\gamma}$. Let us denote the corresponding shift operator by $\CW_k$:
\be\label{Wilson_line}
\CW_k\,Z_{\gamma_a}=\chi_a^{2k}Z_{\gamma_a}.
\ee

The space of hypergeometric solutions is two-dimensional, therefore the basis in the Wilson loop operators is also two-dimensional. 
As basis elements one could choose, say, $\CW_0$ and $\CW_1$ that correspond to pullbacks of an exceptional set of bundles $\CO$ and $\CO(1)$ from base $\IC\IP^1$ to the resolved conifold. 
More precisely, the fields spanning the resolution $X_{\pm}$ base $\IC\IP^1$ are $\phi_{1,2}$ or $\phi_{3,4}$ depending on the sign of $r$, therefore since $\phi_{1,2}$ and $\phi_{3,4}$ are charged oppositely with respect to the gauge $U(1)$ Wilson line insertions produce bundles with opposite Chern classes (for definition of bundles on conifold resolutions see Appendix \ref{s:BKS}):
$$
\begin{array}{c}
	\begin{tikzpicture}
		\node(A) {$\CW_k$};
		\node(B) at (3,0) {$\CY_+(k)$}; 
		\node(C) at (-3,0) {$\CY_-(-k)$};
		\path (A) edge[->] node[above] {$\scriptstyle r>0$} (B) (A) edge[->] node[above] {$\scriptstyle r<0$} (C); 
	\end{tikzpicture}
\end{array}
$$

Using this prospective we analyze solutions \eqref{HHP-O}.
The rows of matrices in \eqref{HHP-O} correspond to the action of the $\CW$-operators, in particular, the Lagrangian cycle $\CL_1$ is associated with simply structure sheaves on both Higgs branches, whereas $\CL_4$ is associated with twists $\CY_{\pm}(\pm 1)$ respectively.
An additional multiplier $h_1^2$ appearing in the second relation in \eqref{HHP-O} corresponds to an equivariant morphism of multiplying by a global section $\phi_1\phi_2\phi_3\phi_4$, we will discuss these details in Section \ref{sec:ideal}.

The pair of K-theory classes $\left(\left[\CY_+(0)\right],\left[\CY_+(1)\right]\right)$ as we start our journey along $\wp$ at the Higgs branch $r>0$ becomes a pair of Lagrangian branes $\left(\left[\CL_1\right],\left[\CL_4\right]\right)$ on the dual Coulomb branch opening near $r\sim 0$, and then further becomes $\left(\left[\CY_-(0)\right],\left[\CY_-(-1)\right]\right)$ at the second Higgs branch when $r<0$.
We could summarize this transformation in the following schematic diagram of boundary condition class bases fibered over $r$ as one moves along $\wp$:
\begin{equation}\label{parallel_K-thy}
	\begin{array}{c}
		\begin{tikzpicture}
			\draw(-2,0) -- (10,0);
			\draw[ultra thick] (0,-0.1) -- (0,0.1) (4,-0.1) -- (4,0.1) (8,-0.1) -- (8,0.1);
			\node[below] at (0,-0.1) {$r>0$};
			\node[below] at (4,-0.1) {$r=0$};
			\node[below] at (8,-0.1) {$r<0$};
			\node[right] at (10,0) {$\wp$};
			\node[below] at (2,0) {$\wp_+$};
			\node[below] at (6,0) {$\wp_-$};
			\node[above] at (0,0.1) {$\left(\left[\CY_+(0)\right],\left[\CY_+(1)\right]\right)$};
			\node[above] at (4,0.1) {$\left(\left[\CL_1\right],\left[\CL_4\right]\right)$};
			\node[above] at (8,0.1) {$\left(\left[\CY_-(0)\right],\left[\CY_-(-1)\right]\right)$};
		\end{tikzpicture}
	\end{array}\,.
\end{equation}
The branes $[\CL_1]$ and $[\CY_{\pm}(0)]$ ($[\CL_4]$ and $[\CY_{\pm}(\pm 1)]$ respectively) are not different, rather they are dual -- describing the IR behavior of the same physical system in dual frames, on the level of categories these branes will be equivalent objects of boundary condition categories under the mirror map.
Therefore it is natural to think of branes $\CL_1$ and $\CL_4$ as non-perturbative invariant branes that do not vary during the transition along $\wp$, rather they have different representations if different local descriptions are chosen.

The result of the brane parallel transport can be calculated constructing invariant branes using the brane grade restriction rule of \cite{HHP}. 
We will call these branes Herbst-Hori-Page branes, or HHP branes for short.
It is easy to check that HHP branes coincide with the invariant branes $\CL_1$ and $\CL_4$ we have just described.

A lift of this construction from the level of K-theory classes to the level of categories, and a calculation of the corresponding Fourier-Mukai kernel requires a further discussion of pecularities of these systems we devote to the next subsection.

\subsection{Fourier-Mukai transform}

The lift of the parallel transport \eqref{parallel_K-thy} to the level of categories gives the following transformation of branes along $\wp$:
\begin{equation}\label{parallel_cat}
	\begin{split}
		&\CY_+(0)\mathop{\longrightarrow}\lm^{\scriptstyle \wp_+}\CL_1\mathop{\longrightarrow}\lm^{\scriptstyle \wp_-}\CY_-(0)\,,\\
		&\CY_+(1)\mathop{\longrightarrow}\lm^{\scriptstyle \wp_+}\CL_4\mathop{\longrightarrow}\lm^{\scriptstyle \wp_-}\CY_-(-1)\otimes \CJ\,,
	\end{split}
\end{equation}
where $\CJ$ is an ideal sheaf of sections of $\CO(-1)^{\oplus 2}$-bundle in $X_-$.

As it was shown in \cite{Bondal_Kapranov_Schecntman} (see also a details on this calculation in Appendix \ref{s:BKS}) this transformation of sheaves is delivered by a Fourier-Mukai kernel:
\begin{equation}\label{kernel}
	\CK=\CO_{\{S_{(+)}=S_{(-)}\}}\,,
\end{equation}
where $S_{(\pm)}$ are matrices of fiber coordinates \eqref{S-mat} over $X_{\pm}$ respectively.

In this section we will supply these statements with calculations in three steps:
\begin{enumerate}
	\item First of all we will argue that relation \eqref{HHP-O} can be lifted to the level of categories unspoiled, so that branes $\CL_1$ and $\CL_4$ are indeed invariant branes for the parallel transport along $\wp$.
	\item Further we would propose a reasoning of how kernel \eqref{kernel} could have been guessed a priori without performing calculations of Appendix \ref{s:BKS}.
	\item Finally, we will explain details of appearance of sheaf $\CJ$ in \eqref{parallel_cat}.
\end{enumerate}

\subsubsection{Categorical parallel transport}

Varieties $X_+$ and $X_-$ are isomorphic and the set of vacuum equations possess a reflection symmetry mapping $r$ to $-r$.
Due to this symmetry there is no need to consider categorified parallel transport along the whole $\wp$, either half, $\wp_+$ or $\wp_-$, would be enough.
Let us choose, say, $\wp_-$.
Path $\wp_-$ intersects the spectral network (see Fig.\ref{fig:phase}) in a single point in group B and and an infinite family of poles in group C.
The parallel transport of the Lefshcetz thimble basis is given by a product of corresponding Stokes matrices:
\begin{equation}\label{SB2SC}
	{\bf S}_{B,2}{\bf S}_C=\left(\begin{array}{cc}
		1 & 0 \\
		\frac{\chi_3}{\chi_4} & -1
	\end{array}\right)\cdot\left(\begin{array}{cc}
		1 & \frac{1}{\lambda_{34}}\\
		0 & 1 \\
	\end{array}\right)=\left(\begin{array}{cc}
		1 & \lambda_{34}^{-1} \\
		\frac{\chi_3}{\chi_4} & \frac{\chi_4}{\chi_3}\lambda_{34}^{-1}
	\end{array}\right).
\end{equation}

Stokes coefficient $\lambda_{34}^{-1}$ in matrix ${\bf S}_C$ is a partition function contribution from an infinite series of poles in group C, in particular, the very coefficient is a trivial consequence of series summation:
$$
\lambda_{34}^{-1}=\sum\lm_{k=0}^{\infty}\left(\chi_3^{-1}\chi_4\right)^{2k+1}.
$$
Each such solitonic contribution to the partition function is reflected in wave functions spanning the BPS Hilbert space. 
So we write easily an expression for the MSW complex vector space associated to an interface parameterized by path $\wp_-$. 
In this section to denote component vectors of the MSW complex we use notations \eqref{bpsi}. 
The $\IC\IP^1$-solitonic particles of \eqref{CP_1_solitons} in this section will be denoted as $\fX_{\alpha\beta;k}$, where $\alpha$ and $\beta$ are corresponding fixed points (vacua) of $\IC\IP^1$ base implying that the soliton flows along direction $\alpha\to\beta$, and $k$ is defined in \eqref{CP_1_solitons}.
Homological degrees -- fermion numbers -- of distinct vectors are defined by homological degrees of complex \eqref{subcomplex} and can be calculated alternatively using Maslov index \eqref{Maslov} (see also \cite[Appendix A]{Galakhov:2016cji}):
\be\label{MSW1}
\begin{split}
	{\bf M}^*=\Psi_L\cdot\left(\begin{array}{cc}
		\bpsi_{33}^{(0)}[1] & 0\\
		\bpsi_{43}^{(0)}[\fX_{43;0}] & \bpsi_{44}^{(-1)}[1]\\
	\end{array}\right)\cdot \left(\begin{array}{cc}
		\bpsi_{33}^{(0)}[1] & \bigoplus\lm_{k\geq 0}\bpsi_{34}^{(0)}[\fX_{34;k}]\\
		0 & \bpsi_{44}^{(0)}[1]\\
	\end{array}\right)\cdot \Psi_R=\\
	=\Psi_L\cdot\left(\begin{array}{cc}
		\bpsi_{33}^{(0)}[1] & \bigoplus\lm_{k\geq 0}\bpsi_{34}^{(0)}[\fX_{34;k}]\\
		\bpsi_{43}^{(0)}[\fX_{43;0}] & \bigoplus\lm_{k\geq 0}\bpsi_{44}^{(0)}[\fX_{43;0},\fX_{34;k}]\oplus\bpsi_{44}^{(-1)}[1] \\
	\end{array}\right)\cdot \Psi_R\,.
\end{split}
\ee
\eqref{MSW1} ``categorifies'' \eqref{SB2SC} in a naive way -- when instanton contributions are ignored.
Multiplication of matrices corresponds to a concatenation of soliton diagrams discussed in Section \ref{sec:Web}, where the first element in the basis corresponds to a vacuum represented by thimble $\CL_1$ on the Coulomb branch, and the second element corresponds to a vacuum represented by $\CL_4$ respectively.

So far we have not specified boundary conditions yet, therefore we use wave functions $\Psi_L$ and $\Psi_R$ to encode these data. 
$\Psi_L$ corresponds to a brane boundary condition on the left. 
We consider it to be given by either Lagrangian brane $\CL_1$ or $\CL_4$. 
Solutions for $\CL_1$ and $\CL_4$ approach asymptotically constant vacua $*3$ and $*4$ correspondingly. 
We can represent these two choices in a form of vectors of complexes:\footnote{Here notation \eqref{bpsi} and concatenation rule \eqref{concatenation} are naturally extended to include simple boundary condition choices. 
	So that $\CL$ corresponding to a specific vacuum choice are allowed to terminate a ``string'' of quasi-particle letters $\fX_{ij}$.}
$$
\left(\begin{array}{cc}
	\psi_{33}^{(0)}[\CL_1]& 0\\
\end{array}\right),\quad  \left(\begin{array}{cc}
	0& \psi_{44}^{(0)}[\CL_4]\\
\end{array}\right).
$$
$\Psi_R$ and boundary conditions at the right end of the interval will remain unspecified.

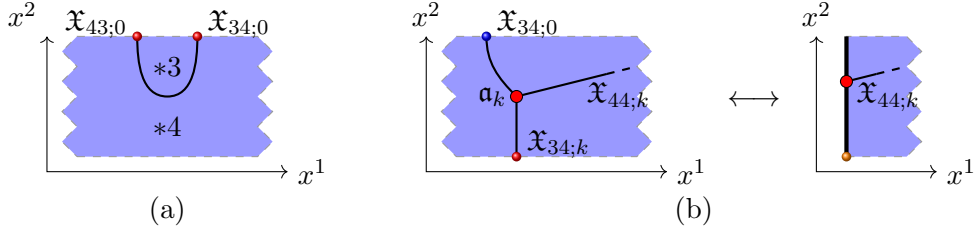
\begin{figure}[h!]
	\begin{center}
		\begin{tikzpicture}
			\node (A) at (0,0) {$\begin{array}{c}
					\begin{tikzpicture}[scale=0.8]
						\draw[dashed,fill=blue, opacity=0.4] (0,0) -- (3,0) -- (3.25,0.25) -- (3,0.5) -- (3.25,0.75) -- (3,1) -- (3.25,1.25) -- (3,1.5) -- (3.25,1.75) -- (3,2) -- (0,2) -- (-0.25,1.75) -- (0,1.5) -- (-0.25,1.25) -- (0,1) -- (-0.25,0.75) -- (0,0.5)  -- (-0.25,0.25) -- (0,0);
						\draw[<->] (-0.5,2) -- (-0.5,-0.25) -- (3.5,-0.25);
						\node[above left] at (-0.5,2) {$x^2$};
						\node[right] at (3.5,-0.25) {$x^1$};
						\draw[thick] (1,2) to[out=270,in=180] (1.5,1) to[out=0,in=270] (2,2);
						\shade[ball color = red] (1,2) circle (0.08);
						\shade[ball color = red] (2,2) circle (0.08);
						\node[above left] at  (1,1.8) {$\fX_{43;0}$};
						\node[above right] at  (2,1.8) {$\fX_{34;0}$};
						\node at (1.5,1.5) {$*3$};
						\node at (1.5,0.5) {$*4$};
					\end{tikzpicture}
				\end{array}$};
			\node (B) at (7,0) {$\begin{array}{c}
					\begin{tikzpicture}[scale=0.8]
						\draw[dashed,fill=blue, opacity=0.4] (0,0) -- (3,0) -- (3.25,0.25) -- (3,0.5) -- (3.25,0.75) -- (3,1) -- (3.25,1.25) -- (3,1.5) -- (3.25,1.75) -- (3,2) -- (0,2) -- (-0.25,1.75) -- (0,1.5) -- (-0.25,1.25) -- (0,1) -- (-0.25,0.75) -- (0,0.5)  -- (-0.25,0.25) -- (0,0);
						\draw[<->] (-0.5,2) -- (-0.5,-0.25) -- (3.5,-0.25);
						\node[above] at (-0.5,2) {$x^2$};
						\node[right] at (3.5,-0.25) {$x^1$};
						\draw[thick] (1,0) -- (1,1) to[out=135,in=270] (0.5,2) (1,1) -- (2.5,1.375);
						\draw[thick, dashed] (2.5,1.375) -- (3,1.5);
						\shade[ball color = red] (1,0) circle (0.08);
						\shade[ball color = blue] (0.5,2) circle (0.08);
						\draw[fill=red] (1,1) circle (0.1);
						\node[left] at (1,1) {$\fa_k$};
						\node[above right] at (1,-0.2) {$\fX_{34;k}$};
						\node[above right] at (0.5,1.8) {$\fX_{34;0}$};
						\node[right] at (2,1) {$\fX_{44;k}$};
					\end{tikzpicture}
				\end{array}\!\!\longleftrightarrow\!\! \begin{array}{c}
					\begin{tikzpicture}[scale=0.8]
						\draw[dashed,fill=blue, opacity=0.4] (2,0) -- (3,0) -- (3.25,0.25) -- (3,0.5) -- (3.25,0.75) -- (3,1) -- (3.25,1.25) -- (3,1.5) -- (3.25,1.75) -- (3,2) -- (2,2) -- cycle;
						\draw[<->] (1.5,2) -- (1.5,-0.25) -- (3.5,-0.25);
						\node[above] at (1.5,2) {$x^2$};
						\node[right] at (3.5,-0.25) {$x^1$};
						\draw[thick] (2,1.25) -- (2.5,1.375);
						\draw[thick, dashed] (2.5,1.375) -- (3,1.5);
						\draw[ultra thick] (2,0) -- (2,2);
						\node[right] at (2,1) {$\fX_{44;k}$};
						\draw[fill=red] (2,1.25) circle (0.1);
						\shade[ball color = orange] (2,0) circle (0.08);
					\end{tikzpicture}
				\end{array}$};
			\node at (0,-1.5) {(a)};
			\node at (7,-1.5) {(b)};
		\end{tikzpicture}
		\caption{Web diagrams}\label{fig:web}
	\end{center}
\end{figure}

Let us reshuffle some terms of an expression for this complex and start with the bottom right element in matrix \eqref{MSW1}:
$$
\bigoplus\lm_{k\geq 0}\bpsi_{44}^{(0)}[\fX_{43;0},\fX_{34;k}]\oplus\bpsi_{44}^{(-1)}[1] =\underline{\left(\bpsi_{44}^{(0)}[\fX_{43;0},\fX_{34;0}]\oplus\bpsi_{44}^{(-1)}[1]\right)}\oplus \bigoplus\lm_{k\geq 1}\bpsi_{44}^{(0)}[\fX_{43;0},\fX_{34;k}].
$$
The underlined term in this expression is a subcomplex:
\be\label{subcomplex}
\begin{array}{c}
	\begin{tikzcd}
		0 \ar[r]& \bpsi_{44}^{(-1)}[1] \ar[r,"{\bf Q}"]& \bpsi_{44}^{(0)}[\fX_{43;0},\fX_{34;0}] \ar[r]& 0
	\end{tikzcd}
\end{array},
\ee
that is quasi-isomorphic to zero. This is a simple conclusion following from the fact that a homotopy move of an interface path forcing it to go through the ramification point of the spectral network gives a family of quasi-isomorphic MSW complexes (see \cite{Galakhov:2017pod}). 
In particular, all the states in MSW complex will contain either both cochains of this two-term complex or none. 
The instanton map saturating the supercharge differential $\bf Q$ in this complex is given by a soliton web diagram depicted in Fig.\ref{fig:web}(a). 

In a similar way we treat a tower of solitons contribution $\fX_{34;k}$ as condensate \eqref{condensate}.
First notice that soliton scattering amplitudes $\fa_k$ \eqref{amplitudes} justify chemical transitions of type \eqref{sol_scatt} between soliton particles:
\begin{equation}
	\fX_{34;k}\mathop{\longleftrightarrow}\lm^{\fa_k} \fX_{34;0}+\fX_{44;k}\,.
\end{equation}
On the level of complexes this scattering process is promoted to the following equivalence:
\begin{equation}
	\bpsi_{34}^{(0)}[\fX_{34;k}] \cong \bpsi_{34}^{(0)}[\fX_{34;0}]\otimes \IC \left(\phi_3\phi_4^{-1}\right)^k\,,
\end{equation}
where a soliton particle $\fX_{44;k}$ is identified with a monomial sheaf section $(\phi_3\phi_4^{-1})^k$ (cf. \eqref{meson}) condensing at the boundary.
We have constructed an explicit example of such condensation in \eqref{alpha_0}.
Indeed, scattering vertex $\fa_k$ and a boundary vertex when soliton $\fX_{44;k}$ slides off the boundary create equivalent left boundary conditions for web diagrams (see Fig.\ref{fig:web}(b)).
Moreover, this process incorporates only a transition in the $\IC\IP^1$ base of $X_-$, therefore one is able to use duality in spectra of Higgs and Coulomb branch solitons in $\IC\IP^1$ model discussed in Section \ref{sec:equiv_CP_1}.

If $\CL_1$ is chosen as a boundary condition the resulting MSW complex is homotopic equivalent to simply:
\be\label{complex}
\left(\bpsi^{(0)}_{33}\left[\CL_1\right]\otimes \Psi_R\right)\oplus \left(\bpsi^{(0)}_{34}\left[\CL_1;\fX_{34;0}\right]\otimes \bigoplus\lm_{k\geq 0}\IC \left(\phi_3\phi_4^{-1}\right)^k\otimes \Psi_R\right)=: \Psi_L[\CY(0)]\otimes \Psi_R\,.
\ee
This complex  is equivalent to a choice of sheaf $\CY(0)$ as a brane boundary condition in the GLSM, therefore we have denoted the corresponding wave function as $\Psi_L[\CY(0)]$.
This sheaf is naturally expected from HHP arguments.
Indeed if such a sheaf is chosen due to equivariant localization it reduces to either constant vacuum $*3$ or $*4$. 
In the case of vacuum $*4$ there is a tower of condensed mesons \eqref{meson} for our choice of numerical values of complex masses $\mu_a$. 
These are exactly two terms in complex \eqref{complex}.

Similarly, for $\CL_4$ choice as a left boundary brane we have the following result for ${\bf M}^*$:
\be
\begin{split}
	\scalebox{0.86}{$\left(\bpsi^{(0)}_{43}\left[\CL_4;\fX_{43;0}\right]\otimes \Psi_R\right)\oplus \left(\bpsi^{(0)}_{44}\left[\CL_4;\fX_{43;0};\fX_{34;0}\right]\otimes \bigoplus\lm_{k\geq 1}\IC \left(\phi_3\phi_4^{-1}\right)^k\right)\otimes \Psi_R=:\Psi_L[\CY(-1)\otimes \CJ]\otimes \Psi_R.$}
\end{split}
\ee
Due to a shift in the electric charges $k$ this boundary condition is equivalent to a sheaf $\CY(-1)$.
The sheaf $\CY(-1)$ is expected from HHP arguments, however we have to modify it by an additional sheaf $\CJ$ having a trivial class (see Appendix \ref{Groth_groups}).
The ideal sheaf $\CJ$ appears since the currents for each chiral $\phi_i$ are preserved during the soliton scattering individually, and solely $\CY(-1)$ would not be enough to preserve those charges. 
We will explain this phenomenon in Section \ref{sec:ideal}.

In a similar way we analyze $\wp_+$.

We conclude that the parallel transport formula \eqref{parallel_cat} is valid. 
The theory at the end of the interval $[0,L]$ is in the Higgs phase, so the brane boundary conditions are represented by derived categories of coherent sheaves and the parallel transport is a functor:
\be
\beta_{\wp}:\quad D^{(b)}{\rm Coh}(X_+)\longrightarrow D^{(b)}{\rm Coh}(X_-).
\ee
According to Orlov's theorem \cite{Huybrechts,orlov2003derived} this functor can be represented as a Fourier-Mukai (FM) transform with some kernel $\CK$ (see Appendix \ref{sec:App_FM} for our conventions):
\be
\beta_{\wp}=\bPhi_{\CK}.
\ee

Fortunately, sheaves $\CY(k)$ and $\CJ$ are equivariant, therefore the action of the parallel transport functor on those sheaves will be preserved even if the equivariant action is lifted (see a definition of equivariant derived category in \cite{Equivariant_sheves}). 
It can be explicitly shown that for kernel \eqref{kernel} the FM transform action on branes $\CY(0)$ and $\CY(1)$ will coincide with the action of parallel transport \eqref{parallel_cat}. 
This check is performed in \cite{Kapranov:2014uwa} and in Appendix \ref{s:BKS} using slightly different tools in algebraic geometry. 

\subsubsection{Fourier-Mukai kernel}

In this section we will consider a proposition on how the kernel in the form \eqref{kernel} could have been guessed a priori without performing the algebraic geometry calculation.

Let us consider the theory in the neighborhood of the left brane in either constant vacuum $*1$ or $*2$ when field $\phi_1$, or field $\phi_2$ respectively, acquires an expectation value.
According to our calculations in Section \ref{sec:RigidHiggs} the IR description will contain electrically neutral effective meson scalars \eqref{meson}. 
We can combine chiral fields in meson operators in different ways, however the only holomorphic combinations are the following four we arrange in a matrix:
\be\label{S}
S=\left(\begin{array}{cc}
	M_{11} & M_{12}\\
	M_{21} & M_{22}
\end{array}\right)=\left(\begin{array}{cc}
	\phi_2\phi_4& -\phi_1\phi_4\\ 
	-\phi_2\phi_3 & \phi_1\phi_3\\
\end{array}\right).
\ee
Effective theory for the right brane with two other choices -- $*3$ and $*4$ --  as constant vacua will have the same set of neutral holomorphic meson fields. 
According to Section \ref{sec:RigidHiggs} the localization position of meson degrees of freedom depends on effective masses:
\be
\mu_{\rm eff}(M_{ij})=\left(\begin{array}{cc}
	\mu_4-\mu_2 & \mu_4-\mu_1\\
	\mu_3-\mu_2 & \mu_3-\mu_1\\
\end{array}\right).
\ee
A meson condensate wave function is a generic holomorphic or anti-holomorphic function of the meson field -- structure sheaf of an affine plane spanned by this field. 
A condensate of neutral meson chiral field $\fm$ corresponds to a ring of lowest Landau level wave functions $\IC[\fm]$ and reproduces a condensate variety as points in 
$$
{\rm Spec}\; \IC[\fm].
$$
In our case the whole meson matrix $S$ condenses and parameterizes a fiber of $X_{\pm}$ (see such parameterization in Appendix \ref{s:BKS}). 
We could even distinguish values of $S$ on two ends of spatial interval $[0,L]$:
$$
S_{(\pm)}\subset X_{\pm}.
$$
$M_{ij}$ localized at the left or right brane depending on the sign of ${\rm Im}\,\mu(M_{ij})$. In the case ${\rm Im}\,\mu(M_{ij})=0$ meson $M_{ij}$ is delocalized and is represented by a constant mode for the whole interface. We can even decouple a half of them from the IR theory explicitly, see decoupling of partition functions in \eqref{extra_PF}. 
These observations indicate that even if we have distinguished subvarieties $S_{(+)}$ and $S_{(-)}$ for the left and the right branes the BPS state wave function depends on common meson degrees of freedom $M_{ij}(x^1)$ so that:\footnote{We could probe this constraint by applying such a morphism with a constant chiral mode to skyscraper sheaves.
	Suppose on the left and on the right branes one puts stalks supported in points $z_L$ and $z_R$ correspondingly.
	The resulting boundary supercharge contribution reads (see \eqref{stalk}):
	$$
	\bar\CQ_{\rm bdry}=\chi_L\left(\phi(0)-z_L\right)+\chi_R\left(\phi(L)-z_R\right).
	$$
	The effective supercharge for the constant mode $\phi_0:=\phi(0)=\phi(L)$ has the following form:
	$$
	\bar\CQ\sim \bar\psi_{\dot 1}\p_{\bar\phi_0}+\left(\chi_L+\chi_R\right)\left(\phi_0-\frac{z_L+z_R}{2}\right)-\frac{\chi_L-\chi_R}{2}(z_L-z_R).
	$$
	As a result the Laplacian is bounded from below $\left\{\bar\CQ,\CQ\right\}\geq \frac{1}{2}|z_L-z_R|^2$, and there is no BPS state in the spectrum unless $z_L=z_R$.
}
$$
S_{(+)}=M_{ij}(0)=M_{ij}(L)=S_{(-)}.
$$
An alternative way to impose this constraint is to include the corresponding $\delta$-function in the IR BPS wave-function:
\be\label{Psi}
\Psi_L(S_{(+)})\otimes \Psi_{\rm bulk}\otimes \Psi_R(S_{(+)})\;\cdot\;\delta\left(\{S_{(+)}=S_{(-)}\}\subset X_+\times X_-\right).
\ee
Geometrically the $\delta$-function contribution implies that kernel $\CK$ is supported on the corresponding locus, and we checked that $\CK$ itself is just a structure sheaf of this locus \eqref{kernel}.

\subsubsection{Ideal sheaf $\CJ$}\label{sec:ideal}

We would like to start this subsection with putting forward an argument that the highlighted multiplier $h_1^2$ in \eqref{HHP-O} corresponds to a morphism of multiplication by a global section:
\begin{equation}\label{global_sec}
	\phi_1\phi_2\phi_3\phi_4\otimes \ldots\,.
\end{equation}

The passing process from vacuum $*1$ to vacuum $*3$ has two channels through $\CL_1$ and $\CL_4$ respectively:
$$
*1\mathop{\longrightarrow}\lm^{\CL_1}*3,\quad *1\mathop{\longrightarrow}\lm^{\CL_4}*3\,.
$$
In terms of detours of Appendix \ref{sec:App_FS_cat} $(*1\mathop{\longrightarrow}\lm^{\CL_1}*3)$ corresponds to a lift of path $\wp$ to sheet 1 (see Fig.\ref{fig:phase}) of cover \eqref{cover}.
Whereas channel $(*1\mathop{\longrightarrow}\lm^{\CL_4}*3)$ is going along a lift of $\wp$ to sheet 1 till the first intersection point in group B, then it makes a detour along the spectral network to sheet 2 and further another detour at the second point of group B back to sheet 1.
One could depict relative homology classes of these detour paths as $\Gamma_1$ and $\Gamma_2$ on sheet 1 in Fig.\ref{fig:Conif_SN} respectively.
The resulting shift of gradings is due to a classical soliton central charge difference along these paths, and it is given by a holonomy around singularity $e^{2\pi t}=1$:
\begin{equation}
	\int\lm_{\Gamma_1}\Omega^{(1)}-	\int\lm_{\Gamma_2}\Omega^{(1)}=\oint\lm_{e^{2\pi t}=1}\Omega^{(1)}=-\log\, h_1^2\,.
\end{equation}

Factor $h_1^2$ is an equivariant degree of a monomial $\phi_1\phi_2\phi_3\phi_4$. 
The appearance of this monomial in morphism \eqref{global_sec} can be argued by passing to the dual point of view.
The above holonomy around $e^{2\pi t}=1$ can be translated into hysteresis shifts of disorder fields \eqref{Ys}:
\begin{equation}\label{mono}
	\begin{split}
		&\frac{\Delta^{(1)}}{2\pi\I}Y_1=\frac{\Delta^{(1)}}{2\pi\I}Y_2=\frac{\Delta^{(1)}}{2\pi\I}Y_3=\frac{\Delta^{(1)}}{2\pi\I}Y_4=-1\,,\\
		&\frac{\Delta^{(2)}}{2\pi\I}Y_1=\frac{\Delta^{(2)}}{2\pi\I}Y_2=\frac{\Delta^{(2)}}{2\pi\I}Y_3=\frac{\Delta^{(2)}}{2\pi\I}Y_4=0\,.
	\end{split}
\end{equation}
where the superscript in $\Delta^{(i)}$ denotes vacuum sheet that is followed during parallel transport.
Shift of an imaginary part of $Y_a$ by $2\pi$-quanta are shifts in corresponding electrical currents \eqref{disord} of $\phi_a$:
\begin{equation}\label{curr_du}
	\begin{array}{c}
		\begin{tikzpicture}
			\node(A) at (0,0) {$\frac{\Delta}{2\pi \I}Y_a=1$};
			\node(B) at (4,0) {$\phi_a\otimes$};
			\draw[<->] (A.east) -- (B.west) node[pos=0.5,above] {\tiny duality};
		\end{tikzpicture}
	\end{array}\,.
\end{equation}
Thus for \eqref{mono} we arrive again to morphism \eqref{global_sec}.

To observe an appearance of sheaf $\CJ$ we have to compare as sheaf sections transform along $\wp$.
This calculation is analogous to one in Appendix \ref{s:BKS}.
To deliver a bundle twist $\CY(1)$ in \eqref{delta_constr} one has to consider a net shift $+1$ of $Y_a$ that corresponds in vacua $*1$ and $*2$ to the following monomials respectively:
\begin{equation}\label{polys}
	\phi_1(\phi_1\phi_3)^{k_1}(\phi_1\phi_4)^{k_2}(\phi_2/\phi_1)^{k_3},\;k_{1,2,3}\in\IZ_{\geq 0};\quad \phi_2(\phi_2\phi_3)^{k'_1}(\phi_2\phi_4)^{k'_2}(\phi_1/\phi_2)^{k'_3},\;k'_{1,2,3}\in\IZ_{\geq 0}\,.
\end{equation}

If these monomials are transported through the channel of invariant brane $\CL_4$ they are both multiplied by global section \eqref{global_sec} corresponding to multiplier $h_1^2$ in \eqref{HHP-O} that does not change the classes of monomials \eqref{polys}, this is simply a multiplication by a global section of the structure sheaf -- the identity element of the Picard group. 
To transform these sections to sections of sheaves in vacua $*3$ and $*4$ one should transform respective charged multipliers as follows (compare to integration \eqref{integral_1} and \eqref{integral_2}):
\begin{equation}
	\phi_1= (\phi_1\phi_{3,4})\cdot\phi_{3,4}^{-1},\quad \phi_2= (\phi_2\phi_{3,4})\cdot\phi_{3,4}^{-1}\,.
\end{equation}
Monomials $\phi_{3,4}^{-1}$ in the vacua $*3$ and $*4$ correspond to sections of $\CY(-1)$, whereas monomials $\phi_{1,2}\phi_{3,4}$ are neutral mesons $M_{ij}$.
These neutral meson functions are coordinates on the fiber of the bundle $\CO(-1)^{\oplus 2}$, therefore these functions are sections of the ideal sheaf $\CJ$ of zero sections of $\CO(-1)^{\oplus 2}$.
We conclude that $\CL_4$ transforms to sheaf $\CY_-(-1)\otimes \CJ$ for $r<0$.

\section{Categorification of braid group action}\label{sec:quiver}

\subsection{Braid group in Wess-Zumino-Witten model}

\subsubsection{Braid group}

Consider a generic configuration of $N$ points in a complex plain $z_i\in\IC$, $i=1,\ldots,N$. 
Suppose this set is ordered by any ordering function, say, a real part, an imaginary part or an absolute value. 
So if we choose the latter variant we say that inequality $i<j$ holds for indices when
$$
|z_i|<|z_j|.
$$
Consider a counterclockwise permutation move of two neighbor points with indices $i$ and $i+1$. 
Let us parameterize this move by ``time'' interval $I_t$. 
World-lines of  these points in ``space-time'' $\IC\times I_t$ form a simple braid of $N$ strands where only two neighbor strands are braided:
\be\label{bii+1_+}
b_{i,i+1}=\begin{array}{c}
	\begin{tikzpicture}
		\draw (-2.1,-0.3) -- (-2.5,-0.7) -- (1.6,-0.7) -- (2,-0.3) -- cycle;
		\draw[thick] (-0.5,-0.5) to[out=90,in=210] (0,0) to[out=30,in=270] (0.5,0.5) (0.5,-0.5) to[out=90,in=330] (0.1,-0.07) (-0.1,0.07) to[out=150,in=270] (-0.5,0.5);
		\draw[thick] (-1.5,-0.5) -- (-1.5,0.5) (-2,-0.5) -- (-2,0.5) (1.5,-0.5) -- (1.5,0.5);
		\draw[fill=black] (-2,-0.5) circle (0.07) (-1.5,-0.5) circle (0.07) (-0.5,-0.5) circle (0.07) (0.5,-0.5) circle (0.07) (1.5,-0.5) circle (0.07)
		(-2,0.5) circle (0.07) (-1.5,0.5) circle (0.07) (-0.5,0.5) circle (0.07) (0.5,0.5) circle (0.07) (1.5,0.5) circle (0.07)
		(-1.25,-0.5) circle (0.03) (-1,-0.5) circle (0.03) (-0.75,-0.5) circle (0.03)
		(1.25,-0.5) circle (0.03) (1,-0.5) circle (0.03) (0.75,-0.5) circle (0.03)
		(-1.25,0.5) circle (0.03) (-1,0.5) circle (0.03) (-0.75,0.5) circle (0.03)
		(1.25,0.5) circle (0.03) (1,0.5) circle (0.03) (0.75,0.5) circle (0.03);
		\begin{scope}[shift={(0,1)}]
			\draw (-2.1,-0.3) -- (-2.5,-0.7) -- (1.6,-0.7) -- (2,-0.3) -- cycle;
		\end{scope}
		\node[right] at (2,-0.5) {$\IC$};
		\node[right] at (2,0.5) {$\IC$};
		\node[below] at (-2,-0.7) {$1$};
		\node[below] at (-1.5,-0.7) {$2$};
		\node[below] at (-0.5,-0.7) {$i$};
		\node[below] at (0.5,-0.7) {$i+1$};
		\node[below] at (1.5,-0.7) {$N$};
		\node[above] at (-2,0.7) {$1$};
		\node[above] at (-1.5,0.7) {$2$};
		\node[above] at (-0.5,0.7) {$i$};
		\node[above] at (0.5,0.7) {$i+1$};
		\node[above] at (1.5,0.7) {$N$};
	\end{tikzpicture}
\end{array}\,.
\ee
The inverse element is generated by a clockwise permutation move:
\be\label{bii+1_-}
b_{i,i+1}^{-1}=\begin{array}{c}
	\begin{tikzpicture}
		\draw (-2.1,-0.3) -- (-2.5,-0.7) -- (1.6,-0.7) -- (2,-0.3) -- cycle;
		\begin{scope}[xscale=-1]
			\draw[thick] (-0.5,-0.5) to[out=90,in=210] (0,0) to[out=30,in=270] (0.5,0.5) (0.5,-0.5) to[out=90,in=330] (0.1,-0.07) (-0.1,0.07) to[out=150,in=270] (-0.5,0.5);
		\end{scope}
		\draw[thick] (-1.5,-0.5) -- (-1.5,0.5) (-2,-0.5) -- (-2,0.5) (1.5,-0.5) -- (1.5,0.5);
		\draw[fill=black] (-2,-0.5) circle (0.07) (-1.5,-0.5) circle (0.07) (-0.5,-0.5) circle (0.07) (0.5,-0.5) circle (0.07) (1.5,-0.5) circle (0.07)
		(-2,0.5) circle (0.07) (-1.5,0.5) circle (0.07) (-0.5,0.5) circle (0.07) (0.5,0.5) circle (0.07) (1.5,0.5) circle (0.07)
		(-1.25,-0.5) circle (0.03) (-1,-0.5) circle (0.03) (-0.75,-0.5) circle (0.03)
		(1.25,-0.5) circle (0.03) (1,-0.5) circle (0.03) (0.75,-0.5) circle (0.03)
		(-1.25,0.5) circle (0.03) (-1,0.5) circle (0.03) (-0.75,0.5) circle (0.03)
		(1.25,0.5) circle (0.03) (1,0.5) circle (0.03) (0.75,0.5) circle (0.03);
		\begin{scope}[shift={(0,1)}]
			\draw (-2.1,-0.3) -- (-2.5,-0.7) -- (1.6,-0.7) -- (2,-0.3) -- cycle;
		\end{scope}
		\node[right] at (2,-0.5) {$\IC$};
		\node[right] at (2,0.5) {$\IC$};
		\node[below] at (-2,-0.7) {$1$};
		\node[below] at (-1.5,-0.7) {$2$};
		\node[below] at (-0.5,-0.7) {$i$};
		\node[below] at (0.5,-0.7) {$i+1$};
		\node[below] at (1.5,-0.7) {$N$};
		\node[above] at (-2,0.7) {$1$};
		\node[above] at (-1.5,0.7) {$2$};
		\node[above] at (-0.5,0.7) {$i$};
		\node[above] at (0.5,0.7) {$i+1$};
		\node[above] at (1.5,0.7) {$N$};
	\end{tikzpicture}
\end{array}\,.
\ee
We will treat such braids being equivalent up to ambient isotopy. 
Also we endow them with a group composition law reflected in a natural braid concatenation. 
Under these circumstances braids $b_{i,i+1}$ and $b^{-1}_{i,i+1}$ for different indices $i$ generate braid group $\fB\fr_N$ of $N$ strands and satisfy the following set of relations:
\be\label{braid_group}
\begin{split}
	& b_{i,i+1}b_{i,i+1}^{-1}=b_{i,i+1}^{-1}b_{i,i+1}={\bf Id};\\
	& b_{i,i+1}b_{j,j+1}=b_{j,j+1}b_{i,i+1},\quad {\rm if}\; |i-j|>1;\\
	& b_{i,i+1}b_{i+1,i+2}b_{i,i+1}=b_{i+1,i+2}b_{i,i+1}b_{i+1,i+2};
\end{split}
\ee
where by $\bf Id$ we imply a collection of $N$ unbraided strands.

\subsubsection{Wess-Zumino-Witten model}\label{sec:WZW_model}

Wess-Zumino-Witten(WZW) 2d conformal model is characterized by a level $k$ and a Lie group $G$. 
A holomorphic primary field $g_{\lambda}(z)$ depends on a complex coordinate $z$ on a 2d surface and transforms as a representation of Lie algebra $\fg$ of the highest weight $\lambda$ \cite{DiFrancesco:639405}. 
$N$-point conformal blocks are holomorphic functions of positions of punctures:
\be
\CB_{\vec\lambda}^{\vec\alpha}(\vec z)=\left\langle g_{\lambda_1}^{\alpha_1}(z_1)g_{\lambda_2}^{\alpha_2}(z_2)\ldots g_{\lambda_N}^{\alpha_N}(z_N)\right\rangle,
\ee
where $\alpha_i$ parameterize vectors of corresponding representations of weights $\lambda_i$. 
Conformal blocks for various choices of $\vec \lambda$ and $\vec\alpha$ form a vector space of $N$-point conformal blocks $\fB_N$. 
$\fB_N$ is a space of solutions to Knizhnik-Zamolodchikov equations. 
In other words, a vector
\be
\Psi(\vec z):=\sum\lm_{\vec \lambda,\vec \alpha}c_{\vec\alpha}^{\vec\lambda}\CB_{\vec\lambda}^{\vec\alpha}(\vec z)
\ee
is annihilated by a Knizhnik-Zamolodchikov (KZ) connection \cite{Knizhnik:1984nr}:
\be\label{KZ}
\nabla_a^{\rm KZ}\Psi=\left(\p_{z_a}-A_a\right)\Psi=\left(\p_{z_a}+\epsilon\sum\lm_{b\neq a}\frac{\eta_{ij}T^i_aT_b^j}{z_a-z_b}\right)\Psi=0,\quad a=1,\ldots,N,
\ee
where $\eta$ is the Killing form, and $\epsilon=(k+c_2(\fg))^{-1}$, where $c_2(\fg)$ is the dual Coxeter number. 
Generators $T$ act as generators $t^i$ of corresponding Lie algebra $\fg$ on fields:
\be
T_a^i\left\langle g_{\lambda_1}^{\alpha_1}(z_1)\ldots g_{\lambda_N}^{\alpha_N}(z_N)\right\rangle=\left\langle g_{\lambda_1}^{\alpha_1}(z_1)\ldots \left(\left(t^i\right)_{\alpha_a'}^{\alpha_a}g_{\lambda_a}^{\alpha_a'}(z_a)\right)\ldots g_{\lambda_N}^{\alpha_N}(z_N)\right\rangle.
\ee

The KZ connection is flat on the configuration space of $N$ points outside the singular locus where points $z_a$ collide. 
Corresponding parallel transport map will be an invariant of the parallel transport path homotopy class. Consider paths associated to simple tangles $b_{i,i+1}^{\pm}$ in \eqref{bii+1_+}, \eqref{bii+1_-} and denote corresponding paths projected to $\IC$ as $\wp_{i,i+1}^{\pm}$. 
Parallel transport operators:
\be
\begin{split}
	&U_{i,i+1}:\quad \fB_N\longrightarrow\fB_N,\\
	&U_{i,i+1}={\rm P}\exp\int\lm_{\wp_{i,i+1}}A_adz^a.
\end{split}
\ee
satisfy braid group relations \eqref{braid_group} due to parallel transport flatness and form a representation of the braid group $\fB\fr_N$ on conformal blocks $\fB_N$.

A categorification of the braid group representation for generic Lie groups and representations is a hard problem \cite{Aganagic:2020olg,aganagic2021knot}, therefore we restrict ourselves to the simplest case of $G=SU(2)$ and its spin-1/2 representations. 
We put the WZW model on a cylinder, for this purpose two primary fields are located in $0$ and $\infty$, corresponding (not necessarily integral) weights are $\lambda_0$ and $\lambda_{\infty}$, remaining fields are in a spin-1/2 representation of $SU(2)$:
\be\label{cylinder}
\begin{array}{c}
	\begin{tikzpicture}
		\draw[thick] (-2,0.3) -- (2,0.3) (-2,-0.3) -- (2,-0.3);
		\begin{scope}[shift={(-2,0)}]
			\begin{scope}[xscale=0.5]
				\draw[thick] (0,0) circle (0.3);
			\end{scope}
		\end{scope}
		\begin{scope}[shift={(2,0)}]
			\begin{scope}[xscale=0.5]
				\draw[thick] (0,0) circle (0.3);
			\end{scope}
		\end{scope}
		\node[left] at (-2.2,0) {$\langle \lambda_0|$};
		\node[right] at (2.2,0) {$|\lambda_{\infty}\rangle$};
		\draw[fill=black] (-1.5,0) circle (0.06) (-1,0) circle (0.06) (1.5,0) circle (0.06) (0.25,0) circle (0.03) (0.5,0) circle (0.03) (0,0) circle (0.03);
		\node[below] at (-1.5,-0.3) {$z_0$};
		\node[below] at (-1,-0.3) {$z_1$};
		\node[below] at (1.5,-0.3) {$z_{m-1}$};
	\end{tikzpicture}
\end{array} 
\ee

Index $\alpha$ for primary fields $g$ in spin-1/2 representation takes two values corresponding to +1/2 and -1/2 spin projections. We will denote these indices as ``$+$'' and ``$-$'' correspondingly for the sake of brevity. 
The action of the parallel transport maps splits the space of conformal blocks in invariant subspaces of fixed weights. 
Therefore the relative numbers of positive and negative spins are invariants of the braid group action. Suppose among our $m$ spins $n$ spins are positive and $m-n$ are negative. 
We denote the set of all such spin configurations as 
$$
\CY_{n,m-n}.
$$

In what follows we will exploit an isomorphism between the set of spin configurations $\CY_{n,m-n}$ and all Young diagrams that can be embedded in a $(m-n)\times n$ rectangular field.
This isomorphism is constructed as follows. 
Consider a rectangular $(m-n)\times n$ field of unit cells. 
We construct a path going from the top left corner of this field towards the bottom right corner.
This path goes along the borders of cells. At each crossroad one is able to turn either downwards or to the right, the downward flow corresponds to spin ``$+$'' and the right flow corresponds to spin ``$-$''.
A road-map describing the route and written as a sequence of ``$+$'''s and ``$-$'''s is in a one-to-one correspondence with a spin configuration in $\CY_{n,m-n}$. 
The corresponding Young diagram is given by a box profile under the path.
For example, there are two following representations of partition $\{3,2,1,1\}$ in $\CY_{3,5}$
\be\label{Xi_iso}
(-,+,-,+,-,-,+,-)\quad\leftrightarrow \quad
\begin{array}{c}
	\begin{tikzpicture}[scale=0.5]
		\begin{scope}[shift={(0,0)}]
			\draw[fill=gray] (0,0) -- (0,1) -- (1,1) -- (1,0) -- cycle;
		\end{scope}
		\begin{scope}[shift={(1,0)}]
			\draw[fill=gray] (0,0) -- (0,1) -- (1,1) -- (1,0) -- cycle;
		\end{scope}
		\begin{scope}[shift={(2,0)}]
			\draw[fill=gray] (0,0) -- (0,1) -- (1,1) -- (1,0) -- cycle;
		\end{scope}
		\begin{scope}[shift={(3,0)}]
			\draw[fill=gray] (0,0) -- (0,1) -- (1,1) -- (1,0) -- cycle;
		\end{scope}
		\begin{scope}[shift={(4,0)}]
			\draw[fill=white!40!blue] (0,0) -- (0,1) -- (1,1) -- (1,0) -- cycle;
		\end{scope}
		\begin{scope}[shift={(0,1)}]
			\begin{scope}[shift={(0,0)}]
				\draw[fill=gray] (0,0) -- (0,1) -- (1,1) -- (1,0) -- cycle;
			\end{scope}
			\begin{scope}[shift={(1,0)}]
				\draw[fill=gray] (0,0) -- (0,1) -- (1,1) -- (1,0) -- cycle;
			\end{scope}
			\begin{scope}[shift={(2,0)}]
				\draw[fill=white!40!blue] (0,0) -- (0,1) -- (1,1) -- (1,0) -- cycle;
			\end{scope}
			\begin{scope}[shift={(3,0)}]
				\draw[fill=white!40!blue] (0,0) -- (0,1) -- (1,1) -- (1,0) -- cycle;
			\end{scope}
			\begin{scope}[shift={(4,0)}]
				\draw[fill=white!40!blue] (0,0) -- (0,1) -- (1,1) -- (1,0) -- cycle;
			\end{scope}
		\end{scope}
		\begin{scope}[shift={(0,2)}]
			\begin{scope}[shift={(0,0)}]
				\draw[fill=gray] (0,0) -- (0,1) -- (1,1) -- (1,0) -- cycle;
			\end{scope}
			\begin{scope}[shift={(1,0)}]
				\draw[fill=white!40!blue] (0,0) -- (0,1) -- (1,1) -- (1,0) -- cycle;
			\end{scope}
			\begin{scope}[shift={(2,0)}]
				\draw[fill=white!40!blue] (0,0) -- (0,1) -- (1,1) -- (1,0) -- cycle;
			\end{scope}
			\begin{scope}[shift={(3,0)}]
				\draw[fill=white!40!blue] (0,0) -- (0,1) -- (1,1) -- (1,0) -- cycle;
			\end{scope}
			\begin{scope}[shift={(4,0)}]
				\draw[fill=white!40!blue] (0,0) -- (0,1) -- (1,1) -- (1,0) -- cycle;
			\end{scope}
		\end{scope}
		\path[thick,>=stealth'] (0,3) edge[->] (1,3) (1,3) edge[->] (1,2) (1,2) edge[->] (2,2) (2,2) edge[->] (2,1) (2,1) edge[->] (3,1) (3,1) edge[->] (4,1) (4,1) edge[->] (4,0) (4,0) edge[->] (5,0);
		\draw[fill=orange] (0,3) circle (0.08) (1,3) circle (0.08) (1,2) circle (0.08) (2,2) circle (0.08) (2,1) circle (0.08) (3,1) circle (0.08) (4,1) circle (0.08) (4,0) circle (0.08) (5,0) circle (0.08);
	\end{tikzpicture}
\end{array}
\ee
Compare to homological projective duality of \cite{Chen:2020iyo}.

\subsubsection{Picard-Lefschetz parallel transport}

Conformal block expressions can be constructed using so called free field representation \cite{Dotsenko:1984nm,1990IJMPA...5.2495G}. 
Resulting expressions take the form of integral representation of solution to the KZ equations \eqref{KZ}. 
In our particular case, this solution reads \cite{schechtman1991arrangements,etingof1998lectures}:
\be\label{free_field}
\begin{split}
	&\Psi_{\CL}(z_1,\ldots,z_m)=\prod\lm_{a=0}^{m-1}z_a^{-\epsilon\lambda_0/2}\prod\lm_{0\leq a<b\leq m-1}(z_a-z_b)^{-\epsilon/2}\times\\
	&\times\int\lm_{\CL}\prod\lm_{i=1}^n dx_i \;\prod\lm_{i=1}^{n}x_i^{\epsilon\lambda_0}\;\prod\lm_{i=1}^n\prod\lm_{a=0}^{m-1} (x_i-z_a)^{\epsilon}\;\prod\lm_{1\leq i<j\leq n} (x_i-x_j)^{-2\epsilon}.
\end{split}
\ee

The integration contour $\CL$ is chosen in such a way that the integral is absolutely convergent. 
The basis of integration cycles is isomorphic to the basis of conformal blocks.
However a direct identification of a cycle with fixed values of $\vec\lambda$ and $\vec\alpha$ is rather involved.
In the limit $|\lambda_0|\gg 1$ it is easy to show \cite{Galakhov:2017pod} that the basis of conformal blocks is identical to the basis of Lefschetz thimbles. 

The form of Knizhnik-Zamolodchikov equations suggests a natural WKB approximation \cite{Reshetikhin:1994qw} to the parallel transport problem of calculating $U_{i,i+1}$ \cite{Galakhov:2014aha}. 
That is analogous to Picard-Lefschetz parallel transport discussed in Appendix \ref{sec:App_FS_cat}.

Eventually we can categorify the Picard-Lefschetz parallel transport as a Fukaya-Seidel category of a Landau-Ginzburg model with a superpotential:
\be\label{check_sup}
\check{W}_{\rm LG}(\vec x)=\epsilon\lambda_0\sum\lm_{i=1}^{n}\log\;x_i+\epsilon\sum\lm_{i=1}^n\sum\lm_{a=0}^{m-1} \log\;(x_i-z_a)-2\epsilon\sum\lm_{1\leq i<j\leq n} \log\;(x_i-x_j).
\ee
This superpotential is a generalization of the superpotential we used to construct the categorification of the hypergeometric series analytic continuation in \eqref{sup_0}.
Analogous  LG model can be derived as an effective model on collective coordinates of monopole-like solutions in a 5d theory compactified on a cigar \cite{Gaiotto:2011nm,GMW}.

\subsection{Assets of quiver varieties}

\subsubsection{Nakajima quiver varieties}

Nakajima quiver varieties parameterize classical Higgs branches of super Yang-Mills-Higgs theories. 
The gauge-matter content of these theories can be also summarized in the form of a quiver according to the rules listed in Section \ref{sec:Intro_mirror}. 
We will consider quiver varieties of so called ${\rm A}_k$-type:
\be
\begin{array}{c}
	\begin{tikzpicture}
		\node at (0,0) {$v_1$};
		\draw (0,0) circle (0.3);
		\begin{scope}[shift={(0,-2)}]
			\node at (0,0) {$w_1$};
			\draw (-0.3,-0.3) -- (-0.3,0.3) -- (0.3,0.3) -- (0.3,-0.3) -- cycle;
		\end{scope}
		\draw[->] (0.15, -0.259808) -- (0.15, -1.7);
		\draw[<-] (-0.15, -0.259808) -- (-0.15, -1.7);
		\draw[->] (0.15, 0.259808) to[out=60,in=0] (0,1) to[out=180,in=120] (-0.15, 0.259808);
		\node[above] at (0,1) {$P_1$};
		\node[left] at (-0.15,-1) {$\Gamma_1$};
		\node[right] at (0.15,-1) {$\Delta_1$};
		\draw[->] (0.259808, 0.15) -- (1.74019, 0.15);
		\draw[<-] (0.259808, -0.15) -- (1.74019, -0.15);
		\node[above] at (1,0.15) {$A_1$};
		\node[below] at (1,-0.15) {$B_1$};
		\begin{scope}[shift={(2,0)}]
			\node at (0,0) {$v_2$};
			\draw (0,0) circle (0.3);
			\begin{scope}[shift={(0,-2)}]
				\node at (0,0) {$w_2$};
				\draw (-0.3,-0.3) -- (-0.3,0.3) -- (0.3,0.3) -- (0.3,-0.3) -- cycle;
			\end{scope}
			\draw[->] (0.15, -0.259808) -- (0.15, -1.7);
			\draw[<-] (-0.15, -0.259808) -- (-0.15, -1.7);
			\draw[->] (0.15, 0.259808) to[out=60,in=0] (0,1) to[out=180,in=120] (-0.15, 0.259808);
			\node[above] at (0,1) {$P_2$};
			\node[left] at (-0.15,-1) {$\Gamma_2$};
			\node[right] at (0.15,-1) {$\Delta_2$};
			\draw[->] (0.259808, 0.15) -- (1.74019, 0.15);
			\draw[<-] (0.259808, -0.15) -- (1.74019, -0.15);
			\node[above] at (1,0.15) {$A_2$};
			\node[below] at (1,-0.15) {$B_2$};	
		\end{scope}
		\begin{scope}[shift={(8,0)}]
			\node at (0,0) {$v_k$};
			\draw (0,0) circle (0.3);
			\begin{scope}[shift={(0,-2)}]
				\node at (0,0) {$w_k$};
				\draw (-0.3,-0.3) -- (-0.3,0.3) -- (0.3,0.3) -- (0.3,-0.3) -- cycle;
			\end{scope}
			\draw[->] (0.15, -0.259808) -- (0.15, -1.7);
			\draw[<-] (-0.15, -0.259808) -- (-0.15, -1.7);
			\draw[->] (0.15, 0.259808) to[out=60,in=0] (0,1) to[out=180,in=120] (-0.15, 0.259808);
			\node[above] at (0,1) {$P_k$};
			\node[left] at (-0.15,-1) {$\Gamma_k$};
			\node[right] at (0.15,-1) {$\Delta_k$};	
		\end{scope}
		\begin{scope}[shift={(6,0)}]
			\draw[->] (0.259808, 0.15) -- (1.74019, 0.15);
			\draw[<-] (0.259808, -0.15) -- (1.74019, -0.15);
			\node[above] at (1,0.15) {$A_{k-1}$};
			\node[below] at (1,-0.15) {$B_{k-1}$};
		\end{scope}
		\draw[fill=black] (4.5,0) circle (0.05) (5,0) circle (0.05) (5.5,0) circle (0.05);
	\end{tikzpicture}
\end{array}
\ee

The theory is accompanied by a canonical superpotential:
\be\label{Naka_sup}
W=\sum\lm_{i=1}^{k}\Tr\; P_i\left(B_iA_i-A_{i+1}B_{i+1}-\Gamma_i\Delta_i\right).
\ee
Here and in what follows we imply that fields with indices $i$ outside the interval $1,\ldots,k$ are zero.

Masses of chiral fields -- equivariant weights --  are compatible with the superpotential, so that the overall equivariant weight of the superpotential is 0. 
We denote mass parameters $\epsilon_1$, $\epsilon_2$ as a reference to the canonical parameterization of the $\Omega$-background deformation \cite{Nekrasov:2002qd,Nekrasov:2003rj,Moore:1998et}:
\be
\mu(A_i)=\epsilon_1,\;\;\;\mu(B_i)=\epsilon_2,\;\;\;\mu(\Gamma_i)=0,\;\;\; \mu(\Delta_i)=\epsilon_1+\epsilon_2,\;\;\;\mu(P_i)=-\epsilon_1-\epsilon_2.
\ee
These masses are introduced as additional twisted masses --  flavor $U(1)$ framing multiplets associated with quiver arrows \cite{Galakhov:2020vyb}.

As before we denote FI parameters and topological angles associated to  gauge groups of quiver nodes correspondingly:
$$
r_i,\quad \theta_i,\quad i=1,\ldots,k.
$$

The classical Higgs branch is described by a locus of chiral fields satisfying D-term and F-term constraints:
\be\label{Nakajima}
\begin{split}
	{\bf D}\mbox{-term:}\quad& A_{i-1}
	A_{i-1}^{\dagger}+B_iB_i^{\dagger}+\Gamma_i\Gamma_i^{\dagger}-\\
	&-A_i^{\dagger}A_i-B_{i-1}^{\dagger}B_{i-1} -\Delta_{i}^{\dagger}\Delta_i=r_i,\quad i=1,\ldots,k;\\
	{\bf F}\mbox{-term:}\quad& A_{i}
	B_{i}=A_{i+1}B_{i+1}-\Gamma_i\Delta_i,\quad i=1,\ldots,k;\\
	&P_i=0,\quad i=1,\ldots,k.
\end{split}
\ee
The latter constraint $P_i=0$ is very simple and maps $P_i$ do not appear anywhere else, so in a canonical description of Nakajima quiver varieties these constraints and fields $P_i$ are usually omitted, we will proceed in a similar fashion.
Also we admit the following usual simplification of notations.
Since all the nodes are connected by a doublet of arrows we substitute this doublet by a single link:
$$
\begin{array}{c}
	\begin{tikzpicture}[scale=0.6]
		\draw (0,0) circle (0.3) (2,0) circle (0.3);
		\draw[->] (0.259808, 0.15) -- (1.74019, 0.15);
		\draw[<-] (0.259808, -0.15) -- (1.74019, -0.15);
	\end{tikzpicture}
\end{array}\leftrightsquigarrow \begin{array}{c}
	\begin{tikzpicture}[scale=0.6]
		\draw (0,0) circle (0.3) (2,0) circle (0.3);
		\draw (0.3,0) -- (1.7,0);
	\end{tikzpicture}
\end{array}\,.
$$

Classical vacua satisfying \eqref{Nakajima} connected by the action of the gauge group are equivalent.
Therefore the vacuum manifold -- the classical Higgs branch -- in this case is described by the following quotient called quiver representation moduli space:
\be
\CR(\vec v,\vec w):=\{{\bf D}\mbox{-term},{\bf F}\mbox{-term}\}\Big/\prod\lm_{i=1}^kU(v_i).
\ee
This manifold is smooth for values of $r_i$ belonging to certain stability chambers. 
In what follows we will exploit mostly a stability chamber called a cyclic chamber:
$$
r_i>0,\quad i=1,\ldots,k.
$$
As we mentioned in Section \ref{sec:Intro_mirror} the $\bf D$-term constraint can be traded for a stability condition and enlargement of the gauge group to its complexified analog, so the quiver representation moduli space admits an isomorphic description:
\be
\CR(\vec v,\vec w)=\{\mbox{stability},{\bf F}\mbox{-term}\}\Big/\prod\lm_{i=1}^kGL(v_i,\IC).
\ee
For Nakajima quiver varieties this theorem is proven in \cite{nakajima1994instantons}. 
We will denote a stable Nakajima quiver variety representation corresponding to the cyclic chamber as $\CR_+(\vec v,\vec w)$.  
It is a smooth algebraic variety of complex dimension:
\be\label{Nakajima_dim}
{\rm dim}_{\IC}\;\CR(\vec v,\vec w)=2(\vec w,\vec v)-2(\vec v,\vec v)+(\vec v,A\vec v),
\ee
where $A$ is the quiver adjacency matrix:
$$
A_{i,j}=\delta_{i,j+1}+\delta_{i+1,j}.
$$

\subsubsection{Spectral duality}
The spectral duality establishes isomorphism of vacuum moduli spaces for certain theories (see e.g. \cite{Mironov:2012uh}).
We will be interested in the case of 3d theories \cite{Gaiotto:2013bwa,Bullimore:2014awa}.
This duality is also tightly related to a 3d mirror symmetry (see e.g.  \cite{Gaiotto:2013bwa,Bullimore:2016nji,Bullimore:2014awa,Intriligator:1996ex,deBoer:1996ck}) and $q$-Langlands correspondence (see e.g. \cite{Kapustin:2006pk,Aganagic:2016jmx,Aganagic:2017smx,Okounkov:2016sya,Rimanyi:2019zyi,2020arXiv201108603D}).

One could extract an equivalence of solutions to Bethe ansatz equations describing vacuum moduli spaces for theories with target spaces associated with a dual pair of quivers $\fQ$ and $\fQ^!$. This duality is well established when both $\fQ$ and $\fQ^!$ describe so called $T[SU(k)]$ theories \cite{Bullimore:2014awa,Aprile:2018oau}:
\be\label{T(SU(k))}
\begin{array}{c}
	\begin{tikzpicture}[scale=0.4]
		\draw[thick] (0,0) -- (7,0) (11,0) -- (15,0);
		\draw[thick,dashed] (7,0) -- (11,0);
		\draw[fill=white] (3,0) circle (0.3) (6,0) circle (0.3) (12,0) circle (0.3) (15,0) circle (0.3);
		\draw[fill=white] (-0.3,-0.3) -- (-0.3,0.3) -- (0.3,0.3) -- (0.3,-0.3) -- cycle;
		\node[above] at (0,0.3) {$k$};
		\node[above] at (3,0.3) {$k-1$};
		\node[above] at (6,0.3) {$k-2$};
		\node[above] at (12,0.3) {$2$};
		\node[above] at (15,0.3) {$1$};
		\node[left] at (-0.5,0) {$T[SU(k)]=$};
	\end{tikzpicture}
\end{array}
\ee

The duality exchanges the roles of complex mass parameters $\mu$ associated with the framing nodes and complexified FI parameters $r_i-\I\theta_i$ associated with the gauge nodes.

Similarly, the spectral duality \cite{Gaiotto:2013bwa} connects the theory with the target space given by the cotangent bundle to a Grassmannian:
\be
\begin{array}{c}
	\begin{tikzpicture}[scale=0.4]
		\draw[thick] (0,0) -- (3,0);
		\draw[fill=white] (3,0) circle (0.3);
		\draw[fill=white] (-0.3,-0.3) -- (-0.3,0.3) -- (0.3,0.3) -- (0.3,-0.3) -- cycle;
		\node[above] at (0,0.3) {$m$};
		\node[above] at (3,0.3) {$n$};
		\node[left] at (-0.5,0) {$T^*{\rm Gr}(n,m)=$};
	\end{tikzpicture}
\end{array}
\ee
and a theory of type we denote as $\CS_{n,m-n}$:
\be\label{S_{n,m-n}}
\begin{array}{c}
	\begin{tikzpicture}[scale=0.4]
		\draw[thick] (0,0) -- (3,0) (6,0) -- (12,0) (15,0) -- (21,0) (24,0) -- (27,0) (9,0) -- (9,-1.3) (18,0) -- (18,-1.3);
		\draw[thick,dashed] (3,0) -- (6,0) (12,0) -- (15,0) (21,0) -- (24,0);
		\draw[fill=white] (0,0) circle (0.3) (3,0) circle (0.3) (6,0) circle (0.3) (9,0) circle (0.3) (12,0) circle (0.3) (15,0) circle (0.3) (18,0) circle (0.3) (21,0) circle (0.3) (24,0) circle (0.3) (27,0) circle (0.3);
		\node[above] at (0,0.3) {$1$};
		\node[above] at (3,0.3) {$2$};
		\node[above] at (6,0.3) {$n-1$};
		\node[above] at (9,0.3) {$n$};
		\node[above] at (12,0.3) {$n$};
		\node[above] at (15,0.3) {$n$};
		\node[above] at (18,0.3) {$n$};
		\node[above] at (21,0.3) {$n-1$};
		\node[above] at (24,0.3) {$2$};
		\node[above] at (27,0.3) {$1$};
		\begin{scope}[shift={(9,-1.3)}]
			\draw[fill=white] (-0.3,-0.3) -- (-0.3,0.3) -- (0.3,0.3) -- (0.3,-0.3) -- cycle;
		\end{scope}
		\begin{scope}[shift={(18,-1.3)}]
			\draw[fill=white] (-0.3,-0.3) -- (-0.3,0.3) -- (0.3,0.3) -- (0.3,-0.3) -- cycle;
		\end{scope}
		\node[below] at (9,-1.6) {$1$};
		\node[below] at (18,-1.6) {$1$};
		\node[left] at (-0.5,0) {$\CS_{n,m-n}=$};
		\begin{scope}[shift={(0,1.5)}]
			\draw (-0.3,0) to[out=90,in=180] (0,0.3) -- (13.2,0.3) to[out=0,in=270] (13.5,0.6) to[out=270,in=180] (13.8,0.3) -- (27,0.3) to[out=0,in=90] (27.3,0);
			\node[above] at (13.5,0.6) {\small $m-1$ nodes};
		\end{scope}
	\end{tikzpicture}
\end{array}
\ee

In the IR these theories experience a symmetry breaking phenomenon. 
Effective fields remaining present in the IR description on the Coulomb branch are eigen values of the scalars $\sigma$ in the gauge multiplet. 
We can denote those fields by two indices $\Sigma^{(a)}_{\alpha}$ where index $a$ runs over nodes $a=1,\ldots, m-1$ and $\alpha=1,\ldots, v_a$.

A procedure to generate an effective superpotential is analogous to that discussed in Section \ref{sec:Preamble} (see also \cite{Aganagic:2001uw}).
Analogously we could reduce dimensionally the 3d superpotential of \cite{Gaiotto:2013bwa} or take a logarithm of disk partition function integrand \cite[Section 10]{Hori:2013ika} substituting gamma-functions by their Stirling's approximations. 
The result for a Nakajima quiver variety reads:
\be\label{uncheck_sup}
\begin{split}
	&W_{\rm LG}=\sum\lm_{a=1}^Nt_a \Sigma_{\alpha}^{(i)}+\sum\lm_{a=1}^N\sum\lm_{\substack{\alpha,\beta=1\\ \alpha\neq \beta}}^{N}{\bf w}\left(\Sigma_{\alpha}^{(a)}-\Sigma_{\beta}^{(a)}-\epsilon_1-\epsilon_2\right)+\\
	&+\sum\lm_{a=1}^{N-1}\sum\lm_{\alpha=1}^{v_a}\sum\lm_{\beta=1}^{v_{a+1}}{\bf w}\left(\Sigma_{\beta}^{(a+1)}-\Sigma_{\alpha}^{(a)}-\epsilon_1\right)+\sum\lm_{a=1}^{N-1}\sum\lm_{\alpha=1}^{v_a}\sum\lm_{\beta=1}^{v_{a+1}}{\bf w}\left(\Sigma_{\alpha}^{(a)}-\Sigma_{\beta}^{(a+1)}-\epsilon_2\right)+\\
	&+\sum\lm_{a=1}^{N}\sum\lm_{p=1}^{w_a}{\bf w}\left(\Sigma_{\alpha}^{(a)}-\mu_p^{(a)}\right)+\sum\lm_{a=1}^{N}\sum\lm_{p=1}^{w_a}{\bf w}\left(\mu_p^{(a)}-\Sigma_{\alpha}^{(a)}-\epsilon_1-\epsilon_2\right)\,,
\end{split}
\ee
where $\mu_p^{(a)}$, $p=1,\ldots,w_a$ are masses associated with the framing nodes, $t_a$ are complexified FI parameters:
$$
t_a=r_a-\I\theta_a,
$$
and $\bf w$ is an elementary single chiral superpotential \eqref{small_superpotential}.

Analogous dimensional reduction of $T^*{\rm Gr}(n,m)$ will produce an effective theory with superpotential \eqref{check_sup}.
A spectral duality between $T^*{\rm Gr}(n,m)$ and $\CS_{n,m-n}$ indicates that corresponding vacuum moduli spaces are isomorphic, in particular, in the Landau-Ginzburg phase critical point sets of $\check W_{\rm LG}$ \eqref{check_sup} and $W_{\rm LG}$ \eqref{uncheck_sup} are isomorphic under the following identification of parameters:
\be\label{reparm}
z_a=:e^{2\pi \tau_a},\quad t_a=\tau_a-\tau_{a-1}.
\ee

Here let us review a quick two line proof of this fact for the simplest case $n=1$.
A complete analytic proof of this duality relation for generic $m$ and $n$ is rather involved  technically, it is described in \cite{Koroteev:2021lvp}.

In the case $n=1$ theory $\CS_{1,m-1}$ has the following quiver depiction:
\be
\begin{array}{c}
	\begin{tikzpicture}[scale=0.4]
		\draw[thick] (-3,0) -- (3,0) (6,0) -- (12,0);
		\draw[thick,dashed] (3,0) -- (6,0);
		\draw[fill=white] (0,0) circle (0.3) (3,0) circle (0.3) (6,0) circle (0.3) (9,0) circle (0.3);
		\node[above] at (-3,0.3) {$1$};
		\node[above] at (0,0.3) {$1$};
		\node[above] at (3,0.3) {$1$};
		\node[above] at (6,0.3) {$1$};
		\node[above] at (9,0.3) {$1$};
		\node[above] at (12,0.3) {$1$};
		\begin{scope}[shift={(12,0)}]
			\draw[fill=white] (-0.3,-0.3) -- (-0.3,0.3) -- (0.3,0.3) -- (0.3,-0.3) -- cycle;
		\end{scope}
		\begin{scope}[shift={(-3,0)}]
			\draw[fill=white] (-0.3,-0.3) -- (-0.3,0.3) -- (0.3,0.3) -- (0.3,-0.3) -- cycle;
		\end{scope}
		\node[left] at (-3.5,0) {$\CS_{1,m-1}=$};
	\end{tikzpicture}
\end{array}\;.
\ee

Corresponding superpotential reads:
\be
W_{\rm LG}=\sum\lm_{a=0}^{m-1}{\bf w}\left(\Sigma^{(a)}-\Sigma^{(a+1)}\right)+\sum\lm_{a=0}^{m-1}{\bf w}\left(\Sigma^{(a+1)}-\Sigma^{(a)}-\epsilon\right)+\sum\lm_{a=1}^{m-1}t_a\Sigma^{(a)},
\ee
where we shifted fields $\Sigma^{(a)}\to\Sigma^{(a)}+a\epsilon_1$,
\be
\epsilon=\epsilon_1+\epsilon_2,
\ee
and we assume boundary conditions for framing nodes: 
$$
\Sigma^{(0)}=\mu_+,\quad \Sigma^{(m)}=\mu_-,
$$
where complex masses $\mu_+$ and $\mu_-$ are associated with framing nodes $w_1$ and $w_{m-1}$ correspondingly.

In our notations vacuum equations read:
\be
e^{2\pi\tau_{a-1}}\left(1-\frac{\epsilon}{\Sigma^{(a-1)}-\Sigma^{(a)}}\right)=e^{2\pi\tau_{a}}\left(1-\frac{\epsilon}{\Sigma^{(a)}-\Sigma^{(a+1)}}\right).
\ee
The right hand side of this equation is equivalent to the left hand side with indices shifted by one.
We conclude that both expressions for the l.h.s. and the r.h.s. are independent of index $a$.
Introduce a new variable:
\be\label{xx}
x:=e^{2\pi\tau_{a}}\left(1-\frac{\epsilon}{\Sigma^{(a)}-\Sigma^{(a+1)}}\right).
\ee

It is simple to solve \eqref{xx} for $\Sigma^{(a)}$ variables, then one arrives to a single constraint for $\Sigma^{(a)}$ to match the boundary conditions:
\be
\frac{\mu_+-\mu_--m\epsilon}{x}+\sum\lm_{a=0}^{m-1}\frac{\epsilon}{x-z_a}=0.
\ee

The latter equation describes critical points of the dual superpotential of $T^*{\rm Gr(1,m)}$ theory (compare to \eqref{check_sup}):
\be
{\check W}_{\rm LG}=\epsilon\left(\frac{\mu_+-\mu_-}{\epsilon}-m\right)\log\; x+\epsilon \sum\lm_{a=0}^{m-1}\log(x-z_a).
\ee

\subsubsection{Maffei's isomorphism}

Consider an $N$-step flag variety of hyperplanes $F_a$ in a $d$-dimensional complex space (see \cite{Affleck:2021ypq}):
\be\label{flag_def}
\CF^d_{\vec q}:=\left\{\begin{array}{l}
	\IC^d=F_0\supseteq F_1 \supseteq F_2\supseteq \ldots \supseteq F_N \supseteq F_{N+1}=\{0\}\\
	{\rm dim}F_a-{\rm dim}F_{a+1}=q_a\geq 0,\quad a=0,\ldots,N
\end{array}\right\}.
\ee

A cotangent bundle to a flag variety is produced by adding a nilpotent element $z$:
\be\label{cotang}
\CN^{d}_{\vec q}:=\left\{(z,F_{\bullet})\in {\rm End}(\IC^d)\times \CF_{\vec q}^d\;\big|\; z(F_a)\subseteq F_{a+1}\right\}.
\ee

We say that $z$ is an element of type $\lambda$, where $\lambda=\{\lambda_1\geq \lambda_2\geq \ldots\geq \lambda_p\}$ is a partition of $d$, if nilpotent matrix  $z$ has blocks of sizes $\lambda_1,\ldots, \lambda_p$ in its Jordan decomposition. A slice in $\CN^{d}_{\vec q}$ where $z$ is an element of type $\lambda$ we denote as
$$
\CN^{d}_{\vec q,\lambda}
$$

Cotangent bundle $\CN^{d}_{\vec q}$ is isomorphic to a Nakajima quiver variety with dimensional vectors
$$
\vec v=\left({\rm dim}\; F_1,\;{\rm dim}\; F_2,\ldots, {\rm dim} \;F_N\right),\quad \vec w=(d,0,0,\ldots).
$$
in the cyclic chamber \cite{Rimanyi:2019zyi,nakajima1994instantons}. To establish this isomorphism it is useful to identify $W_1$ with $V_0$ and maps:
$$
A_0:=\Gamma_1,\quad B_0:=\Delta_1.
$$
If the vector space $\IC^d$ is identified with $V_0$ then hyperplanes $F_a$ are derived as
\be
F_a={\rm Im}\;H_a,
\ee
where maps $H_a$ read:
\be\label{embedding}
H_a=A_0^\dagger\cdot\ldots\cdot A_{a-1}^\dagger:\quad V_a\longrightarrow V_0.
\ee
Nilpotent element $z$ is defined in this framework as:
\be
z=(B_0A_0)^\dagger.
\ee

We should note that according to our prescriptions element $z$ has equivariant weight $\epsilon=\epsilon_1+\epsilon_2$. Vectors in $V_0$ belonging to the same Jordan subspace form a module generated by the cyclic vector $\vec u$:
$$
\vec u,\; z {\vec u},\; z^2 {\vec u},\; z^3 {\vec u},\ldots.
$$
This imposes a constraint on non-dynamical expectation values of field $\sigma$ associated with the framing node $W_1$. To produce a nilpotent element of type $\lambda$ it has to have the following form:
\be
\sigma={\rm diag}\left(\mu_1,\mu_1-\epsilon,\ldots,\mu_1-\epsilon \lambda_1,\;\;\mu_2,\ldots,\mu_2-\epsilon \lambda_2,\;\;\ldots,\mu_p-\epsilon\lambda_p\right).
\ee 

Maffei's isomorphism  \cite{maffei2005quiver} relates Nakajima quiver varieties and slices in the cotangent bundle to a flag variety:
\be
\CR_+(\vec v, \vec w)\cong \CN_{\vec q, \lambda}^d,
\ee
where 
$$
q_a=-v_a+v_{a+1}+\sum\lm_{j=1}^aw_j,
$$
and $\lambda$ is a partition of type $1^{w_N}2^{w_N-1}\ldots N^{w_1}$, $d=\sum_a q_a$.
In general Maffei's isomorphism is rather involved, nevertheless it can be simplified a lot for equivariant fixed points.

Variety $\CN_{\vec q, \lambda}^d$ describes a transverse slice in a point defined by nilpotent element $z$ in a convolution space representation of orbit closure resolution $\tilde{\rm Gr}_{\vec \chi}$ in an affine Grassmannian of $PGL_2$ \cite{CautisKamnitzerI,CautisKamnitzerII,Mirkovic_Vybornov,Aganagic:2020olg,aganagic2021knot}.

\subsubsection{Fixed points}
For the GLSM with a quiver variety target space we have introduced a non-trivial superpotential \eqref{Naka_sup}. 
On one hand superpotential adds a twist to the supercharge differential operator \cite{Galakhov:2020vyb}, nevertheless localization procedure is still applicable.
The theory localizes to the critical locus of the height function and superpotential fixed with respect to gauge symmetry.
On the other hand superpotential $W$ complicates consideration of the GLSM model, it introduces a non-trivial boundary condition for the boundary supercharge leading to an effect of matrix factorization \cite{Khovanov:2004bc,Hori:2013ika,HHP}:
$$
\bar\CQ_{\rm bdry}^2=W\cdot{\bf Id}\;.
$$
We will try to avoid this difficulty using an equivariant localization trick. 
We split localization flows as we proposed in Section~\ref{sec:loc_GLSM} so that first we localize to the locus \eqref{Nakajima} so that the IR target space is a cotangent bundle to \eqref{Nakajima} (similarly to an example in Appendix \ref{s:sheaf_coho_CP_n}), then we continue localization to the equivariant fixed points.
Fortunately, on the critical locus \eqref{Nakajima} $W\equiv0$ and there is no need to consider matrix factorization for coherent sheaves on the IR target space.
For other types of quiver varieties with singular moduli spaces this trick will not work as higher quantum corrections will contribute \cite{Galakhov:2020vyb}.

\bigskip

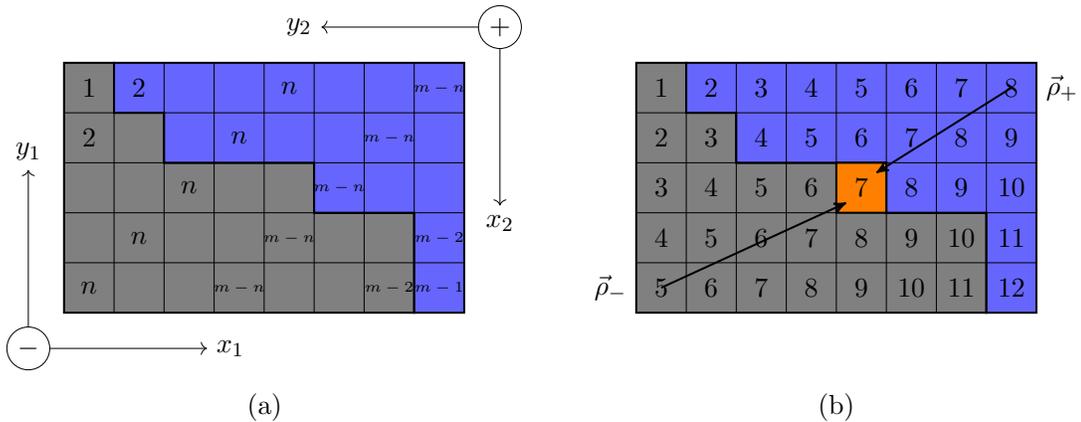
\begin{figure}
	\begin{center}
		\scalebox{0.95}{$\begin{array}{c}
				\begin{tikzpicture}
					\node { 
						$\begin{array}{c}
							\begin{tikzpicture}
								\begin{scope}[scale=0.7]
									\draw[fill=gray,thick] (0,0) -- (0,5) -- (1,5) -- (1,4) -- (2,4) -- (2,3) -- (5,3) -- (5,2) -- (7,2) -- (7,0) -- cycle;
									\draw[fill=white!40!blue,thick] (8,5) --  (1,5) -- (1,4) -- (2,4) -- (2,3) -- (5,3) -- (5,2) -- (7,2) -- (7,0) -- (8,0) -- cycle;
									\foreach \i in {1,...,7}
									{
										\draw (\i,0) -- (\i,5);
									}
									\foreach \i in {1,...,4}
									{
										\draw (0,\i) -- (8,\i);
									}
									\node at (0.5,4.5) {$1$};
									\node at (0.5,3.5) {$2$};
									\node at (1.5,4.5) {$2$};
									\foreach \i in {0,...,4}
									{
										\node at (\i+0.5,\i+0.5) {$n$};
									}
									\foreach \i in {0,...,4}
									{
										\node at (-\i+7.5,-\i+4.5) {\tiny $m-n$};
									}
									\node at (7.5,0.5) {\tiny $m-1$};
									\node at (6.5,0.5) {\tiny $m-2$};
									\node at (7.5,1.5) {\tiny $m-2$};
								\end{scope}
								\node at (-0.5,-0.5) {$-$};
								\node at (6.1, 4.) {$+$};
								\draw (-0.5,-0.5) circle (0.3) (6.1, 4.) circle (0.3);
								\draw[->] (-0.2,-0.5) -- (2,-0.5);
								\draw[->] (-0.5,-0.2) -- (-0.5,2);
								\node[right] at (2,-0.5) {$x_1$};
								\node[above] at (-0.5,2) {$y_1$};
								\draw[->] (6.1, 3.7) -- (6.1, 1.5);
								\draw[->] (5.8, 4.) -- (3.6, 4.);
								\node[below] at (6.1, 1.5) {$x_2$};
								\node[left] at (3.6, 4.) {$y_2$};
							\end{tikzpicture}
						\end{array}$
					};
					\node at (0,-3) {(a)};
					\begin{scope}[shift={(8,0)}]
						\node { 
							$\begin{array}{c}
								\begin{tikzpicture}
									\begin{scope}[scale=0.7]
										\draw[fill=gray,thick] (0,0) -- (0,5) -- (1,5) -- (1,4) -- (2,4) -- (2,3) -- (5,3) -- (5,2) -- (7,2) -- (7,0) -- cycle;
										\draw[fill=white!40!blue,thick] (8,5) --  (1,5) -- (1,4) -- (2,4) -- (2,3) -- (5,3) -- (5,2) -- (7,2) -- (7,0) -- (8,0) -- cycle;
										\foreach \i in {1,...,7}
										{
											\draw (\i,0) -- (\i,5);
										}
										\foreach \i in {1,...,4}
										{
											\draw (0,\i) -- (8,\i);
										}
										\draw[thick,fill=orange] (4,2) -- (4,3) -- (5,3) -- (5,2) -- cycle;
										\foreach \i in {1,...,5}
										\foreach \j in {1,...,5}
										{
											\pgfmathparse{int(\j-\i+1)}
											\let\r\pgfmathresult
											\ifnum \r > 0
											\breakforeach
											\fi
											\node at (\j-0.5,-\i+4.5+\j) {\i};
										}
										\foreach \i in {6,...,7}
										\foreach \j in {1,...,5}
										{
											\node at (\j+\i-5.5,-0.5+\j) {\i};
										}
										\foreach \i in {1,...,5}
										\foreach \j in {1,...,5}
										{
											\pgfmathparse{int(\j-\i+1)}
											\let\r\pgfmathresult
											\ifnum \r > 0
											\breakforeach
											\fi
											\pgfmathparse{int(\i-\j+8)}
											\let\s\pgfmathresult
											\node at (8.5-\j,-\i+5.5) {\s};
										}
										\draw[->,thick,>=stealth'] (0.5,0.5) -- (4.2,2.2);
										\draw[->,thick,>=stealth'] (7.5,4.5) -- (4.8,2.8);
										\node[left] at (0,0.5) {${\vec \rho}_-$};
										\node[right] at (8,4.5) {${\vec \rho}_+$};
									\end{scope}
								\end{tikzpicture}
							\end{array}$
						};
						\node at (0,-3) {(b)};
					\end{scope}
				\end{tikzpicture}
			\end{array}$}
		\caption{Diagrammatic depiction of equivariant fixed points on $\CS_{n,m-n}$.}\label{fig:S_fixed}
	\end{center}
\end{figure}

Equivariant fixed points on a Nakajima quiver variety in the cyclic chamber are enumerated by $\sum\lm_i w_i$-tuples of $N$-colored Young diagrams.
In the case of theory $\CS_{n,m-n}$ this tuple is simply a doublet of Young diagrams $\vec Y=\{Y_1,Y_2\}$.
Moreover, for specified quiver dimensions two Young diagrams complement each other, so that an actual fixed point is labeled by a single Young diagram that can be embedded in a $(m-n)\times n$ cell field.
If $Y_1$ fills the gray area in diagram depicted in Fig.\ref{fig:S_fixed}(a), then $Y_2$ transposed and reflected afterwards along both axes with respect to the origin is filling the complementary blue area.
There is an additional coloring of the diagrams associated with quiver nodes. 
We enumerate/color nodes of diagram \eqref{S_{n,m-n}} from left to right starting with color 1, so that nodes linked to the framing nodes have colors $n$ and $m-n$ respectively.
Coloring of field cells goes in diagonal rows as it is depicted in Fig.\ref{fig:S_fixed}(a) starting with color 1 in the top left corner and ending with color $m-1$ in the right bottom corner.

We denote the left bottom corner as ``$-$'' and along the gray area use coordinates $x_1$ and $y_1$ so that the most bottom left cell has coordinates $(0,0)$. 
Also the top right corner is denoted as ``$+$'', along the blue area coordinates $x_2$ and $y_2$ are used, so that the most top right cell has coordinates $(0,0)$.

In Section \ref{sec:WZW_model} we have already identified a set of such Young diagrams with spin arrangements $\CY_{n,m-n}$.

On the 3d mirror dual side diagrams $\CY_{n,m-n}$ define Schubert cells of Grassmannian ${\rm Gr}(n,m)$ (see, for example,  \cite{2015arXiv150803010S}).

There is a simple procedure to present field vevs in the corresponding constant vacuum having a diagrammatic labeling of the equivariant fixed point.
In the IR vacuum the gauge group is broken to its Weyl subgroup and scalar field $\sigma$ in the gauge multiplets acquire diagonal expectation values of order of complex masses \cite{Galakhov:2020vyb}:
$$
\left\langle\sigma^{(a)}\right\rangle={\rm diag}\left(\sigma^{(a)}_{1},\sigma^{(a)}_{2},\ldots,\sigma^{(a)}_{v_a}\right),\quad a=1,\ldots,m-1.
$$
The unbroken remnant Weyl subgroup of the gauge group acts on these expectation values by permutations:
$$
\prod\lm_{i=1}^{m-1}S_{v_i}.
$$
Therefore vacuum values $\left\langle\sigma^{(a)}\right\rangle$ different by reordering $\sigma^{(a)}_{\alpha}$ inside each $a$-group are not distinguished.
These expectation values can be extracted directly from diagrams like in Fig.\ref{fig:S_fixed}(a).
Each cell in the diagram corresponds to a particular expectation value.
Denote the complex mass associated with the framing node $w_n$ as $\mu_+$ and the mass associated with node $w_{m-n}$ as $\mu_-$. 
An association between cells and $\sigma^{(a)}$-vevs depends on the area type the cell belongs to and also its $x$, $y$ and color $c$ coordinates:
\be\label{eigen}
\begin{array}{rcl}
	\mbox{{\color{gray} Gray} cell with }(x_1,y_1,c) & \longleftrightarrow& \sigma^{(c)}=\mu_-+x_1\epsilon_1+y_1\epsilon_2;\\
	\mbox{{\color{white!40!blue} Blue} cell with }(x_2,y_2,c') & \longleftrightarrow& \sigma^{(c')}=\mu_++x_2\epsilon_1+y_2\epsilon_2.
\end{array}
\ee

After symmetry breaking components of chiral fields behave as independent effective IR fields. We could introduce the following notations for their indices:
$$
\phi_{(b,\beta)(a,\alpha)}.
$$
This field is a component of the chiral field represented by an arrow $a\to b$ linking nodes $a$ and $b$. 
Indices $\alpha$ and $\beta$ are just matrix indices of the corresponding linear map associated with arrow $a\to b$.
Effective IR mass of such field reads:
\be\label{Nakajima_meson}
\mu\left(\phi_{(b,\beta)(a,\alpha)}\right)=\sigma_\beta^{(b)}-\sigma_\alpha^{(a)}-\mu(\mbox{arrow }a\to b).
\ee

Only chiral fields that turn out to be effectively massless acquire IR expectation values of order $\sim\sqrt{r_i}$.
We provide an explicit example of effective theory calculation in Appendix \ref{sec:app_quiver}. 
A nice geometric way to enumerate chiral fields acquiring expectation values is in terms of composite meson operators.
As in the the case of fields $\sigma$ such meson operators acquiring an expectation value in the IR are defined by cells of Young diagrams.
So a cell in the gray area with coordinates and color $(x_1,y_1,c)$ corresponds to the following operator:
\be\label{quiver_meson}
\left( e_{(x_1,y_1,c)},\left( B_{n+y_1-x_1}\cdot\ldots \cdot B_{n+x_1-1}\cdot A_{n+x_1}\cdot\ldots\cdot A_{n+1} \right)\Gamma_n\right),
\ee
where $e_{(x_1,y_1,c)}$ is a unit eigenvector of matrix $\langle\sigma^{c}\rangle$ with eigenvalue $\sigma^{(c)}$ corresponding to cell $(x_1,y_1,c)$ \eqref{eigen} and $( *,*)$ is a simple scalar product.
An expression for the meson operator associated with a cell in the blue area is given by an expression similar to \eqref{quiver_meson} starting with $\Gamma_{m-n}$.
Clearly operators $A_a$ and $B_a$ in brackets in \eqref{quiver_meson} can be re-arranged freely due to $\bf F$-term constraints \eqref{Nakajima} on maps.

\bigskip

Young diagram depiction represents a part of a $\IZ^2$-lattice. 
Expectation values $\langle\sigma^{(c)}\rangle$ acquire positions in nodes of this lattice embedded in $\IC$.
In a complete analogy with a 3d crystal description of toric fixed points on Calabi-Yau threefolds \cite{Ooguri:2008yb,Okounkov:2003sp,Yamazaki:2010fz,Li:2020rij,Aganagic:2010qr} Young diagram labeling for fixed points on a Nakajima variety can be associated with 2d crystals.
A finite 2d crystal is formed by vevs of the scalar in the gauge multiplet so it is also natural to call this phase we called ``rigid Higgs branch'' in Section \ref{sec:RigidHiggs} a crystal Coulomb branch.

\bigskip

Now we are in a position to describe an inverse isomorphism map $\Xi$ mapping a fixed point on $\CS_{n,m-n}$ to a spin configuration. This construction goes as follows.

Maffei's isomorphism maps $\CS_{n,m-n}$ to a flag variety with $q_a=1$ (see \eqref{flag_def}), in other words to a slice in a complete flag $T[SU(m)]$ moduli space.
Let us denote the corresponding flag cotangent bundle as $\CN_{n,m-n}$  and vector spaces associated with gauge and framing nodes for original $\CS_{n,m-n}$ and its Maffei's image as $V$,  $W$ and $\tilde V$, $\tilde W$ correspondingly. For the largest space we have:
\be
\tilde V_0=\tilde W_1=\IC^m=W_n^{\oplus n}\oplus W_{m-n}^{\oplus (m-n)},
\ee
where $W_n^{\oplus n}$ is just $n$ copies of 1d space $W_n$. 
Operator $z$ in this basis has the following form:
$$
z=\left(\begin{array}{cc}
	J_{n} & 0\\
	0 & J_{m-n}\\
\end{array}\right)
$$
where $J_k$ are Jordan blocks of appropriate sizes.

We assign to $W_n$ and $W_{m-n}$ 1d spaces spanned by 1/2-spin projection vectors:
\be
W_n={\rm Span}\left\{ \left|+\frac{1}{2},\frac{1}{2}\right\rangle\right\},\quad W_{m-n}={\rm Span}\left\{ \left|-\frac{1}{2},\frac{1}{2}\right\rangle\right\},
\ee
or simply $|+\rangle$ and $|-\rangle$ for brevity. 
Then vectors of $W_n^{\oplus n}$ and  $W_{m-n}^{\oplus (m-n)}$ can be denoted as $|+,a\rangle$, $a=1,\ldots,n$ and $|-,b\rangle$, $b=1,\ldots,m-n$ correspondingly. 
Also introduce a forgetful morphism:
\be
\xi:\quad |+,i\rangle\mapsto |+\rangle,\quad |-,i\rangle\mapsto |-\rangle.
\ee

Flag cotangent bundle $\CN_{n,m-n}$ admits an orthogonal decomposition in lines with a natural norm on $\tilde V_0$:
\be\label{ortho}
\begin{split}
	&\tilde V_0=\IC^m=\ell_0\oplus\ell_1\oplus\ldots\oplus\ell_{m-1},\quad {\rm dim}\;\ell_a=1,\quad \ell_a \perp \ell_b \;\mbox{for}\;a\neq b;\\
	&F_a=\bigoplus\lm_{k=a}^{m-1}\ell_k.
\end{split}
\ee

It turns out that for fixed points $\xi(\ell_i)$ is parallel to either $|+\rangle$ and $|-\rangle$. 
Thus we construct the following map on Young diagrams $Y\in\CY_{n,m-n}$. 
First we construct a quiver representation of $\CS_{n,m-n}$ based on fixed point labeling, then we apply to it Maffei's isomorphism to derive a point on flag cotangent bundle $\CN_{n,m-n}$, and, finally we apply the forgetful morphism to the flag's orthogonal decomposition:
\be
\Xi:\quad Y\mapsto \left(\xi(\ell_0),\xi(\ell_1),\ldots,\xi(\ell_{m-1})\right).
\ee
The letter word is an arrangement of spins with $n$ ``$+$''-spins and $m-n$ ``$-$''-spins, and $\Xi$ is an isomorphism. As it might have been expected $\Xi$ is equivalent to isomorphism \eqref{Xi_iso} constructed in Section \ref{sec:WZW_model}.

\subsection{Categorified tangles as interfaces}

\subsubsection{Braid group action on quiver varieties}
Let us assume that parameters $z_a$ of the WZW conformal block are arranged on the cylinder as depicted in diagram \eqref{cylinder}:
\be
|z_0|<|z_1|<|z_2|<\ldots<|z_{m-1}|.
\ee
In this case parameters $\tau_a$ \eqref{reparm} are ordered according to their real parts:
\be
{\rm Re}\;\tau_0<{\rm Re}\;\tau_1<{\rm Re}\;\tau_2<\ldots<{\rm Re}\;\tau_{m-1}.
\ee
Then the dual $\CS_{n,m-n}$ model is in a cyclic chamber of the parameter space:
\be
r_1>0,\;r_2>0,\;\ldots,\;r_{m-1}>0.
\ee
Path $\wp_{a,a+1}$ defined by braid element $b_{a,a+1}$ permutes $\tau_a$ and $\tau_{a+1}$ and flows outside the cyclic chamber of the stability parameters to a new chamber:
\be\label{new_chamb}
r_1>0,\;\ldots,\;r_a>0,\;r_{a+1}<0,\;r_{a+2}>0,\;\ldots,\;r_{m-1}>0.
\ee
On a wall in the parameter space separating two chambers the quiver variety may become \emph{singular}.
Physically, we may observe a phenomenon similar to one discussed in Section \ref{sec:cat_contin_hyper} when Higgs branch description fails to mimic the effective behavior of the theory, and one has to switch to the Coulomb branch description via a LG model.
We would like to calculate categorified parallel transport induced by $\wp_{a,a+1}$, however it is inconvenient to work with chamber \eqref{new_chamb}.
We are not aware if there is a nice combinatorial way to count fixed points in that chamber as we did with the cyclic one.
Fortunately, there is an isomorphism $\varphi$ of the original quiver variety $\CS_{n,m-n}$ moduli space in chamber \eqref{new_chamb} to another variety $\CS'_{n,m-n}$ moduli space in a cyclic chamber:
\be\label{Phi_{i,i+1}}
\begin{array}{c}
	\begin{tikzpicture}
		\node[shape          
		= rectangle,
		rounded corners=10pt,
		draw,                  
		fill           = black!20!white,
		align          = center,
		thick	](A) at (0,0) {$\begin{array}{c}
				\begin{tikzpicture}[scale=0.4]
					\draw[thick] (-1,0) -- (11,0);
					\draw[fill=black!20!white] (0,0) circle (0.3) (5,0) circle (0.3) (10,0) circle (0.3);
					\node[above] at (0,0.3) {$r_{a}>0$};
					\node[above] at (5,0.3) {$r_{a+1}>0$};
					\node[above] at (10,0.3) {$r_{a+2}>0$};
				\end{tikzpicture}
			\end{array}$};
		\node[shape          
		= rectangle,
		rounded corners=5pt,
		draw,                  
		fill           = black!20!white,
		align          = center,
		thick	](B) at (9,0) {$\begin{array}{c}
				\begin{tikzpicture}[scale=0.4]
					\draw[thick] (-1,0) -- (11,0);
					\draw[fill=black!20!white] (0,0) circle (0.3) (5,0) circle (0.3) (10,0) circle (0.3);
					\node[above] at (0,0.3) {$r_{a}>0$};
					\node[above] at (5,0.3) {$r_{a+1}<0$};
					\node[above] at (10,0.3) {$r_{a+2}>0$};
				\end{tikzpicture}
			\end{array}$};
		\node[shape          
		= rectangle,
		minimum height=1cm,
		rounded corners=10pt,
		draw,                  
		fill           = black!20!white,
		align          = center,
		thick	](C) at (0,-2.5) {LG model};
		\node[shape          
		= rectangle,
		rounded corners=10pt,
		draw,                  
		fill           = black!20!white,
		align          = center,
		thick	](D) at (9,-2.5) {$\begin{array}{c}
				\begin{tikzpicture}[scale=0.4]
					\draw[thick] (-1,0) -- (11,0);
					\draw[fill=black!20!white] (0,0) circle (0.3) (5,0) circle (0.3) (10,0) circle (0.3);
					\node[above] at (0,0.3) {$r'_{a}>0$};
					\node[above] at (5,0.3) {$r'_{a+1}>0$};
					\node[above] at (10,0.3) {$r'_{a+2}>0$};
				\end{tikzpicture}
			\end{array}$};
		\path (A) edge[->,thick] node[above] {$\wp_{a,a+1}$} (B) (A) edge[->,thick] node[left] {\tiny melting}(C) (C) edge[->,thick] node[below] {\tiny solidifying}(D) ([shift={(-2.92893pt,2.92893pt)}]A.south east) edge[->,thick] node[below left] {$\bPhi_{a,a+1}$} ([shift={(2.92893pt,-2.92893pt)}]D.north west) (B) edge[draw=none] node[right] {\rotatebox[origin=c]{-90}{$\cong$}} node [right=0.5cm] {$\varphi$} (D);
	\end{tikzpicture}
\end{array}
\ee
Indeed both theories are dual to the same LG model with only reshuffled indices of punctures $z_a$.
In the construction of the parallel transport if it is along a path going through a singular variety one is unable to proceed directly and has to dualize theory observables to the LG model and then map them back.
Along this route rather than dualizing directly back the LG model to a new chamber of $\CS_{n,m-n}$ we simply reshuffle subscripts of punctures $z_a$ and map theory to $S'_{n,m-n}$ as it it is depicted in diagram \eqref{Phi_{i,i+1}}.
This identification gives a precise map between parameters of $\CS_{n,m-n}$ and $\CS'_{n,m-n}$:
\be
\begin{split}
	&r'_a=r_a+r_{a+1},\quad r'_{a+1}=-r_{a+1},\quad r'_{a+2}=r_{a+2}+r_{a+1},\\
	&\mbox{if }b<a\mbox{ or }b>a+2\quad r'_b=r_b.
\end{split}
\ee

For a LG model, in general, a deviation of effective IR fields $\Sigma$ from vacuum values are suppressed by the effective superpotential behavior, this suppression is much ``softer'' than quantum suppression of deviations of fields $\sigma$ from crystal nodes in GLSMs.
Borrowing an analogy from condensed matter physics we could call the Coulomb phase where the LG model description is incorporated a ``\emph{liquid}''\footnote{Similarly one could have borrowed as a physical analogy a resemblance between conformal blocks given by disk amplitudes in a theory with superpotential ${\check W}_{LG}$ and averages of Penner type matrix models (or $\beta$-ensembles) \cite{Dijkgraaf:2009pc,Itoyama:2009sc,Eguchi:2009gf,Schiappa:2009cc,Mironov:2010ym,Mironov:2011jn}.
	In the canonical large $N$ limit effective particles -- matrix eigen values -- form Wigner-like \emph{droplets} confined in potential extrema.
}, the transition from the GLSM phase to the LG phase we could call ``melting'' and the inverse process ``solidifying''.
Using this terminology one arrives to a picturesque image of the parallel transition process: as one varies parameters along a path through a liquid phase a crystal state first melts then solidifies to, in principle, a new crystal.

\bigskip

To get a slight glimpse of how this isomorphism could work on original varieties let us consider the simple case of $\CS_{1,m-1}$.
If we are searching for constant vacua the dominant part of the height function defining these vacua is one without derivatives:
\be\label{height_red}
\fH_{\rm red}=-\int dx^1\;\sum\lm_{b}\Tr\;\sigma^{(b)}_{\II}\left(|A_{b-1}|^2+|B_b|^2-|A_b|^2-|B_{b-1}|^2-{\rm Re}\;t_b\right).
\ee

Clearly, \eqref{height_red} is invariant with respect to the following change of coordinates:
\be
\begin{array}{c}
	\begin{array}{ccc}
		t_{a-1}\to t_{a-1}+t_a, & t_a\to -t_a, & t_{a+1}\to t_{a+1}+t_a;\\
		\sigma^{(a-1)} \to \sigma^{(a-1)}, & \sigma^{(a)}\to \sigma^{(a-1)}+\sigma^{(a+1)}-\sigma^{(a)}, &\sigma^{(a+1)}\to \sigma^{(a+1)};
	\end{array}\\
	\begin{array}{cccc}
		A_a\to A_{a+1}, & B_a\to B_{a+1}, & A_{a+1}\to A_{a}, & B_{a+1}\to B_{a}.
	\end{array}
\end{array}
\ee

We denote a complete move from $\CS_{n,m-n}$ to $\CS'_{n,m-n}$ as $\bPhi_{a,a+1}$ as depicted in diagram \eqref{Phi_{i,i+1}}:
\be
\bPhi_{a,a+1}=\varphi\circ\beta_{\wp_{a,a,+1}}.
\ee

$\bPhi_{a,a+1}$ will be also represented by a Fourier-Mukai transform with a specific kernel. In what follows we will calculate this kernel.

\subsubsection{Fourier-Mukai transform}

In this subsection we will study properties of the parallel transport along the path $\wp_{a,a+1}$ and calculate morphism $\bPhi_{a,a+1}$ in a form of a Fourier-Mukai transform on the flag cotangent bundle $\CN_{\vec q,\lambda}^d$ associated with theory $\CS_{n,m-n}$.
Along the path $\wp_{a,a+1}$ taking $r_{a+1}>0$ to $r_{a+1}<0$ the variety becomes singular at $r_{a+1}=0$, however  this may not affect equivariant fixed points.
Equivariant localization may prevent field values from hitting the singularity and the effective description from collapsing. 
As a matter of fact the situation depends on the spin arrangement in the fixed point and precisely on two spins located at positions $a$ and $a+1$.

Let us introduce the following space (were we use orthogonal decomposition \eqref{ortho}):
\be
E_{a,a+1}:=F_{a+1}/F_{a-1}=\ell_a\oplus\ell_{a+1}.
\ee

There are 4 possible spin configurations divided in two groups when spins are parallel and anti-parallel. 
Jordan decomposition of operator $z$ restricted to $E_{a,a+1}$ is different for these two groups:
\be
\begin{array}{c|c}
	\begin{array}{ccc}
		(\underset{a}{-},\underset{a+1}{-}) & \quad &	(\underset{a}{+},\underset{a+1}{+})
	\end{array} & 
	\begin{array}{ccc}
		(\underset{a}{-},\underset{a+1}{+}) & \quad &	(\underset{a}{+},\underset{a+1}{-})
	\end{array}\\
	z\big|_{E_{a,a+1}}=J_2=\left(\begin{array}{cc}
		0 & 1\\
		0 & 0\\
	\end{array}\right) & z\big|_{E_{a,a+1}}=J_1\oplus J_1=\left(\begin{array}{cc}
		0 & 0\\
		0 & 0\\
	\end{array}\right)
\end{array}
\ee

Notice that in the former case when spins are co-aligned the information about the very flag can be restored from the information about $E_{a,a+1}$ and operator $z$, one can construct both lines $\ell_a$ and $\ell_{a+1}$ as a kernel of $z$ and its orthogonal complement:
$$
\ell_a={\rm Ker}\; z\big|_{E_{a,a+1}},\quad \ell_{a+1}=\left({\rm Ker}\; z\big|_{E_{a,a+1}}\right)^{\perp}.
$$
In the latter case when spins are opposite the information about embedding $\ell_a\hookrightarrow E_{a,a+1}$ is stored neither in $E_{a,a+1}$ nor in $z$.

To observe that in one case the variety becomes singular notice that  all $\ell_a$ are lines, so they can be considered as elements of $\IC\IP^1$.
In the physical theory all these $\IC\IP^1$'s have finite volumes controlled by expectation values of chiral condensates defining embedding of the lines in $\IC^m$ \eqref{embedding}.
Denote corresponding volumes $U_a$, eventually they are functions of FI stability parameters $r_a$:
$$
\vec U({\vec r}).
$$

As $r_{a+1}$ goes to zero along $\wp_{a,a+1}$ volume vectors in the cases of co-aligned and opposite spins behave in a different way:
\be
\begin{split}
	&{\vec U}_{J_2}=(O(1),\ldots,O(1),\ldots,O(1)),\\
	&{\vec U}_{J_1\oplus J_1}=(O(1),\ldots,\underset{a}{O(1)},\underset{a+1}{\sim{\color{red}\sqrt{r_{a+1}}}},\underset{a+2}{O(1)},\ldots,O(1)).
\end{split}
\ee

In the first case the classical vacuum stays away from the singularity along the whole path $\wp_{a,a+1}$, so the transition is smooth and does not require gluing in a Coulomb branch resolution. 
In the case when spins are opposite some classical expectation values of fields disappear in the point $r_{a+1}=0$, as in the the case of the conifold discussed in Section \ref{sec:cat_contin_hyper} the Higgs branch description fails to reflect the IR behavior and we have to switch to the Coulomb branch description.
Let us consider the situation of opposite spins in more details.

If two spin arrangements differ by a permutation:
$$
(\underset{a}{-},\underset{a+1}{+}) \longrightarrow	(\underset{a}{+},\underset{a+1}{-}),
$$
corresponding Young diagrams differ by a single cell of color $a+1$ at the boundary of the Young diagram \emph{migrating} from the gray area to the blue area, see an example in Fig.\ref{fig:S_fixed}(b) for a pair of the following fixed points where the \emph{migrating cell} is denoted by the orange color marker:
\be
\begin{array}{c}
	\begin{tikzpicture}
		\node(0) at (0,0) {$-$};
		\node[left] at (0) {$($};
		\node[below right] at (0) {$,$};
		\node(1) at (0.5,0) {$+$};
		\node[below right] at (1) {$,$};
		\node(2) at (1,0) {$-$};
		\node[below right] at (2) {$,$};
		\node(3) at (1.5,0) {$+$};
		\node[below right] at (3) {$,$};
		\node(4) at (2,0) {$-$};
		\node[below right] at (4) {$,$};
		\node(5) at (2.5,0) {$-$};
		\node[below right] at (5) {$,$};
		\node(6) at (3,0) {$-$};
		\node[below right] at (6) {$,$};
		\node(7) at (3.5,0) {$+$};
		\node[below right] at (7) {$,$};
		\node(8) at (4,0) {$-$};
		\node[below right] at (8) {$,$};
		\node(9) at (4.5,0) {$-$};
		\node[below right] at (9) {$,$};
		\node(10) at (5,0) {$+$};
		\node[below right] at (10) {$,$};
		\node(11) at (5.5,0) {$+$};
		\node[below right] at (11) {$,$};
		\node(12) at (6,0) {$-$};
		\node[right] at (12) {$)$};
		\begin{scope}[shift={(0,-1.5)}]
			\node(0) at (0,0) {$-$};
			\node[left] at (0) {$($};
			\node[below right] at (0) {$,$};
			\node(1) at (0.5,0) {$+$};
			\node[below right] at (1) {$,$};
			\node(2) at (1,0) {$-$};
			\node[below right] at (2) {$,$};
			\node(3) at (1.5,0) {$+$};
			\node[below right] at (3) {$,$};
			\node(4) at (2,0) {$-$};
			\node[below right] at (4) {$,$};
			\node(5) at (2.5,0) {$-$};
			\node[below right] at (5) {$,$};
			\node(6) at (3,0) {$+$};
			\node[below right] at (6) {$,$};
			\node(7) at (3.5,0) {$-$};
			\node[below right] at (7) {$,$};
			\node(8) at (4,0) {$-$};
			\node[below right] at (8) {$,$};
			\node(9) at (4.5,0) {$-$};
			\node[below right] at (9) {$,$};
			\node(10) at (5,0) {$+$};
			\node[below right] at (10) {$,$};
			\node(11) at (5.5,0) {$+$};
			\node[below right] at (11) {$,$};
			\node(12) at (6,0) {$-$};
			\node[right] at (12) {$)$};
		\end{scope}
		\begin{scope}[shift={(0,-2)}]
			\node(0) at (0,0) {\tiny 0};
			\node(1) at (0.5,0) {\tiny 1};
			\node(2) at (1,0) {\tiny 2};
			\node(3) at (1.5,0) {\tiny 3};
			\node(4) at (2,0) {\tiny 4};
			\node(5) at (2.5,0) {\tiny 5};
			\node(6) at (3,0) {\tiny 6};
			\node(7) at (3.5,0) {\tiny\color{orange} 7};
			\node(8) at (4,0) {\tiny 8};
			\node(9) at (4.5,0) {\tiny 9};
			\node(10) at (5,0) {\tiny 10};
			\node(11) at (5.5,0) {\tiny 11};
			\node(12) at (6,0) {\tiny 12};
		\end{scope}
		\draw[thick] (3,-0.3) -- (3,-0.5) -- (3.5,-0.5) -- (3.5,-0.3) (3,-1.2) -- (3,-1) -- (3.5,-1) -- (3.5,-1.2) (3.25,-0.5) -- (3.25,-1);
	\end{tikzpicture}
\end{array}
\ee

Let us denote vectors pointing to the migrating cell from the $+$ and $-$ corners as ${\vec\rho}_+$ and ${\vec\rho}_-$ correspondingly. 
According to association \eqref{Nakajima_meson} these vertices correspond to meson operators, we denote these operators as $y_+$ and $y_-$ respectively.
These are exactly the operators whose expectation values go to zero as one approaches singularity at $r_{a+1}=0$:
\be
\left\langle|y_+|^2\right\rangle\sim r_{a+1},\quad \left\langle|y_-|^2\right\rangle\sim r_{a+1}.
\ee 
As expectation values of these fields approach $\hbar$ in orders of magnitude the fields are no longer considered as classical.

Deviations of gauge multiplets from classical values $\langle\sigma\rangle$ located in cell positions of the Young diagrams are suppressed by masses of the fields corresponding to the tangent bundle and generated through the Higgs mechanism by chiral fields expectation values.
As fields $y_{\pm}$ become quantum the Higgs mass of the gauge field corresponding to the migrating cell becomes null.
We denote corresponding scalar in this gauge multiplet as $\sigma_0$.

\bigskip

Effective masses, or equivariant weights, of fields $y_{\pm}$ read:
\be\label{y_masses}
\mu(y_{\pm})=\sigma_0-\tilde \mu_{\pm},
\ee
where
$$
\tilde\mu_{\pm}:=\mu_{\pm}+{\vec \rho}_{\pm}\cdot\vec\epsilon,
$$
where we introduced the following notations:
$$
{\vec \rho}_{-}\cdot\vec\epsilon=x_1\epsilon_1+y_1\epsilon_2,\quad {\vec \rho}_{+}\cdot\vec\epsilon=x_2\epsilon_1+y_2\epsilon_2.
$$

When the migrating cell is in the gray area of diagram in Fig\ref{fig:S_fixed}(b) $\sigma_0=\tilde\mu_-$, so that mass of field $y_-$ is zero according to \eqref{y_masses}, so $y_-$ condensates and produces the $(-,+)$ equivariant point. 
Similarly, if the migrating cell is in the blue area $\sigma_0=\tilde\mu_+$, $y_+$ condensates and produces the $(+,-)$ point.

A tangent space to the quiver variety produces other quantum fields we divide in two groups.
The first group of two fields has masses dependent on $\sigma_0$, we call them $z_+$ and $z_-$:
\be
\mu(z_{\pm})=\epsilon+\tilde\mu_{\pm}-\sigma_0.
\ee

Also there is a group of tangent fields with masses independent of $\sigma_0$, there are $2(n-1)$ of them:
\be
u_1,\ldots,u_{2(n-1)}.
\ee

$\bf D$-term and $\bf F$-term constraints for these fields read:
\be
\begin{split}
	& |y_+|^2+|y_-|^2-|z_+|^2-|z_-|^2=r_{a+1},\\
	& y_+z_+-y_-z_-=0.
\end{split}
\ee

Analyzing the field spectra and constraints we conclude that near singularity $r_{a+1}\to 0$ our $\CS_{n,m-n}$ theory flows to simply $\CS_{1,1}$ theory and a collection of $2(n-1)$ free neutral chiral scalars:
\be
\begin{array}{c}
	\begin{tikzpicture}
		\node (A) {$\begin{array}{c}
				\begin{array}{c}
					\begin{tikzpicture}
						\node at (-2,0) {$1$};
						\node at (0,0) {$1$};
						\node at (2,0) {$1$};
						\draw (0,0) circle (0.3);
						\begin{scope}[shift={(2,0)}]
							\draw (-0.3,-0.3) -- (-0.3,0.3) -- (0.3,0.3) -- (0.3,-0.3) -- cycle;
						\end{scope}
						\begin{scope}[shift={(-2,0)}]
							\draw (-0.3,-0.3) -- (-0.3,0.3) -- (0.3,0.3) -- (0.3,-0.3) -- cycle;
						\end{scope}
						\draw[<-] (1.7,0.1) -- (0.3,0.1);
						\draw[->] (1.7,-0.1) -- (0.3,-0.1);
						\draw[->] (-1.7,0.1) -- (-0.3,0.1);
						\draw[<-] (-1.7,-0.1) -- (-0.3,-0.1);
						\node[above] at (-1,0.1) {$y_+$};
						\node[below] at (-1,-0.1) {$z_+$};
						\node[above] at (1,0.1) {$z_-$};
						\node[below] at (1,-0.1) {$y_-$};
					\end{tikzpicture}
				\end{array}\oplus\mbox{Free } \left(u_1,\ldots,u_{2(n-1)}\right)
			\end{array}$};
		\node(B) at ([shift={(-3,0)}]A.west) {$\CS_{n,m-n}$};
		\path (B) edge[->] node[above] {$r_{a+1}\to 0$} (A);
		\node at ([shift={(-1.95,-0.3)}]A.north) {$\sigma_0$};
	\end{tikzpicture}
\end{array}
\ee

Spontaneously broken initial gauge symmetry is restored in the IR to $U(1)$, effective phonons with Higgs masses
$$
m_{\rm Higgs}\sim\langle|y_+|^2+|y_-|^2\rangle
$$
become massless and destroy crystal links fixing position of the migrating cell $\sigma_0$ in the lattice.
However as we see there is no need to melt the whole crystal, rather it suffices to dislocate a single migrating cell.

\bigskip

We should note that the number of actual tangent fields matches the quiver variety dimension:
\be
2\mbox{ of }y_{\pm}\;+\;2\mbox{ of }z_{\pm}\;+\;2(n-1)\mbox{ of }u_a\;-\;\mbox{({\bf D}-term)}\;-\;\mbox{({\bf F}-term)}\;=\;2n\;=\;{\rm dim}\;S_{n,m-n},
\ee
according to \eqref{Nakajima_dim}.

Effective theory $\CS_{1,1}$ describes a conifold transition we discussed in Section \ref{sec:cat_contin_hyper}. 
Path $\wp_{a,a+1}$ bringing $r_{a+1}$ from the positive values to the negative values coincides with the analytic continuation of hypergeometric functions we have constructed categorification of in terms of a Fourier-Mukai transform.
The kernel of the Fourier-Mukai transform was defined by a support of the delta-function contribution to the BPS state wave-function \eqref{Psi}.
Briefly speaking we had to identify all the neutral mesons on the ends of the interface segment $[0,L]$.
In this case in addition to the usual conifold neutral mesons there are new mesons $u_1,\ldots,u_{2(n-1)}$ that are neutral with respect to $U(1)$ of effective $\CS_{1,1}$.
Therefore, effective wave functions of the BPS states in the presence of an interface generated by a parallel transport along $\wp_{a,a+1}$ have the following multiplier (compare to \eqref{Psi}):
\be\label{Psi_Naka}
\begin{split}
	&\Psi_{\rm BPS}\sim \delta\left(y_+z_+-(y_+z_+)'\right) \delta\left(y_+z_--(y_+z_-)'\right)\times\\
	&\times\delta\left(y_-z_+-(y_-z_+)'\right)\delta\left(y_-z_--(y_-z_-)'\right)\times\\
	&\times\delta(u_1-u_1')\ldots\delta(u_{2(n-1)}-u_{2(n-1)}'),
\end{split}
\ee
where unprimed variables correspond to coordinates on initial $\CS_{n,m-m}$ at $x^1=0$ and primed variables correspond to coordinates on final $\CS_{n,m-n}'$ at $x^1=L$ in diagram \eqref{Phi_{i,i+1}}.

To give a geometric interpretation of this transform we would like to note that the following orthogonal decomposition over subspaces:
\be
\begin{split}
	&\tilde V_0=\ell_0\oplus\ldots\oplus \ell_{a-1}\oplus E_{a,a+1}\oplus \ell_{a+2}\oplus \ldots\oplus \ell_{m-1},
\end{split}
\ee
and operator $z$ are functions of neutral mesons only. 
Moreover, knowing this decomposition is sufficient to define values of all the neutral mesons.
Information about values of $y_\pm$ remains unknown since they control embedding $\ell_a\hookrightarrow E_{a,a+1}$ that can not be reconstructed from presented data.
From this data we can reconstruct all hyperplanes $F_b$ except $F_{a+1}$:
\be
\begin{split}
	&F_b=\bigoplus\lm_{k=b}^{m-1}\ell_k,\quad \mbox{if }b\geq a+2,\\
	&F_b=\left(\bigoplus\lm_{k=b}^{a-1}\ell_k\right)\oplus E_{a,a+1}\oplus \left(\bigoplus\lm_{j=a+2}^{m-1}\ell_j\right),\quad \mbox{if }b\leq a.
\end{split}
\ee

Summarizing, we conclude that the delta-function in \eqref{Psi_Naka} is supported on the following variety:
\be
\Upsilon_{a,a+1}:=\left\{(F_*,F_*'):\;F_b=F_b',\;\mbox{if }b\neq a+1 \right\}.
\ee

The resulting parallel transport functor $\bPhi_{a,a+1}$ can be constructed as Fourier-Mukai transform \eqref{FM_transform} with respect to the following kernel:
\be
\CO_{\Upsilon_{a,a+1}}.
\ee

Functors $\bPhi_{a,a+1}$ are in agreement with crossing functors constructed in \cite[Section 4.2.2]{CautisKamnitzerI} for $SL(2)$ and generalized in \cite{CautisKamnitzerII,CautisKamnitzerIV} -- our variety $\Upsilon_{i,i+1}$ corresponds to $Z_n^i$ in notations of \cite{CautisKamnitzerI}. 
As well we should note that we have constructed only direct braid functors (crossings \#1 and \#3 in \cite{CautisKamnitzerI}) that do not have twists by line bundles like $F_a/F_{a-1}$ that will appear for inverse functors (crossings \#2 and \#4 in \cite{CautisKamnitzerI}). 
Those additional twists by line bundles will be captured by equivariant degree shifts.

In \cite{CautisKamnitzerI} $\bPhi_{a,a+1}$ are proven to satisfy relations \eqref{braid_group}, whereas in our interpretation of $\bPhi_{a,a+1}$ as interfaces in GLSM braid group relations \eqref{braid_group} follow naturally from the fact that the categorified parallel transport is locally flat and transport paths in relations \eqref{braid_group} are homotopic.

We present an explicit example of cotangent bundle construction $\CN_{n,m-n}$ near a singular  point and counting meson degrees of freedom in Appendix \ref{sec:app_quiver}.

The cHC duality maps this calculation to a calculation on the LG model side \cite{Galakhov:2016cji,Galakhov:2017pod} and may be considered as another physical argument that the link cohomology proposed in \cite{GMW} is isomorphic to Khovanov link homology \cite{Khovanov:1999qla}.

\subsubsection{Remarks on decategorification and quantum groups}\label{sec:decat}

To conclude this section let us give few comments on decategorification of functors $\bPhi_{a,a+1}$.
In \cite{CautisKamnitzerI} it is shown that the K-theoretic reduction of functor $\bPhi_{a,a+1}$ gives the usual braid group element in a modular tensor category of $\CU_{\fq}(sl_2)$ constructed from a permutation operator and $U_{\fq}(sl_2)$ R-matrix in representation $\Box\otimes\Box$:
$$
R=\fq^{\frac{h\otimes h}{4}}\left(1+(\fq-\fq^{-1})\;e\otimes f\right)\fq^{\frac{h\otimes h}{4}},
$$
where $e$, $f$ and $h$ are the standard Chevalley generators of $U_{\fq}(sl_2)$ \cite{MR1492989}. The parameter of the quantum group is defined by the equivariant weight of operator $z$:
\be\label{quant_q}
\fq=e^{\pi\I\left(\mu(A_i)+\mu(B_i)\right)}=e^{\pi \I\epsilon}.
\ee

On the other hand, a flat section of KZ connections \eqref{free_field} is a free field representation of conformal blocks in a 2d Liouville CFT with a coupling constant $b=\sqrt{\epsilon}$ \cite{Mironov:2010su}.
Corresponding parallel transport operator $U_{i,i+1}$ also produces an R-matrix in $\CU_{\fq}(sl_2)$ \cite{Galakhov:2015fna} with a parameter defined by \eqref{quant_q}.

We conclude that indeed functor $\bPhi_{i,i+1}$ categorifies Berry parallel transport operator $U_{i,i+1}$:
$$
\begin{array}{c}
	\begin{tikzpicture}
		\node(A)[left] at (0,0) {$\bPhi_{i,i+1}$};
		\node(B)[right] at (4,0) {$U_{i,i+1}$};
		\draw[-stealth] (0.2,0.06) -- (3.8,0.06);
		\draw[stealth-] (0.2,-0.06) -- (3.8,-0.06);
		\node[above] at (2,0.06) {\tiny decategorification};
		\node[below] at (2,-0.06) {\tiny categorification};
	\end{tikzpicture}
\end{array}.
$$

\section{Open problems and future directions} \label{sec:future}

In conclusion we would like to mention some open problems and possible directions worth further investigation:
\begin{itemize}
	\item To construct brane parallel transport we used as a tool the algebra of the infrared of \cite{GMW}.
	Despite the soliton scattering vertices deliver algebraic structures and satisfy $L_{\infty}$-relations an effective general mechanism to calculate or bootstrap scattering amplitudes from the first principles is still lacking in the literature.
	This is not surprising since in the heart of this calculation lies a solution to a non-linear differential instanton boundary value problem.
	On the other hand the Higgs-Coulomb duality makes this problem dual to a problem in algebraic geometry where one expects to use algebraic means to solve it.
	Therefore it is natural to guess that the problem could be reversed, and computations of brane parallel transport in GLSM using alternative methods like \cite{HHP,GMW,Khan:2020hir,Galakhov:2017pod,Kerr:2017usa,Clingempeel:2018iub,Brunner:2020miu,Brunner:2021cga,Chen:2020iyo,Brunner:2009zt,Brunner:2008fa,Aspinwall:2004jr,Knapp:2016rec,Erkinger:2017aaa,Knapp:2020oba} can be used to benefit solving the instanton equation in LG models, or, at least, counting such solutions with signs.
	\item Another possible direction follows from the previously discussed one.
	$L_{\infty}$-relations for amplitudes appear in \cite{GMW} as structures on 2d polygons dual to web diagrams and may be extended to higher dimensional polytopes \cite{Kapranov:2014uwa}.
	A physical theory representing structures of \cite{Kapranov:2014uwa} has not been presented in the literature yet.
	Structures of \cite{GMW} are universal for 2d $\CN=(2,2)$ theories with massive vacua, and equivariant GLSMs in particular.
	So a good candidate for such a physical theory may be a higher dimensional Yang-Mills theory from what a 2d GLSM is derived by dimensional reduction.
	Some indications that this is the case are given in \cite{Costello:2020ndc}.
	\item As we have seen the models discussed in the paper have a direct application to a categorification of a braid group action and, therefore, the knot theory. 
	As it is explained in \cite{Aganagic:2020olg,aganagic2021knot} the string theory predicts a categorification scheme for links in generic representation and an arbitrary simple Lie group, however an algorithmic construction of link homologies for this general parameterization is absent in the literature as far as we can tell.
	Ideally, using physical approaches and intuition one expects to have an algorithmic program translatable to any computer language to calculate a link homology for arbitrary group and representations along the lines of \cite{Carqueville:2011zea}.
	Moreover, a closer work with representations higher than minuscule ones is expected to be related to a more detailed investigation of a monopole bubbling phenomenon \cite{Kapustin:2006pk,Dedushenko:2018icp}.
	\item Knot invariants, in particular, HOMFLY polynomials are used in Witten-Reshetikhin-Turaev-Viro construction \cite{reshetikhin1991invariants} of Chern-Simons invariants of 3d manifolds.
	So it is a natural tendency to try to extend categorified knot invariants to categorified invariants of 3d manifolds.
	Moreover physical theories (see e.g. \cite{Haydys:2010dv,Gukov:2017kmk,Wang:2020lqp}) suggest that a geometrical self-consistent definition for such invariants exists.
	\item Another possible application of a categorified representation of conformal blocks in WZW models is a categorification of the mapping class group that will probably deliver a new refinement to the standard representation of the mapping class group similarly to \cite{Arthamonov:2017oxw}.
	\item Appearance of Young diagram crystals in the problem about categorified braid group action colored with $m-1$ colors is not surprising and accounts to the skew Howe duality (see e.g. \cite{2015arXiv150206011M,Chun:2015gda}) where braid functors $\Phi_{i,i+1}$ are associated with a categorified action of raising/lowering operators in corresponding $\fs\fl(m)$.
	A similarity of the braid action with the crystal melting problem suggests that the action of a BPS algebra on molten crystals of Calabi-Yau 3-folds (see e.g. \cite{Li:2020rij,Galakhov:2020vyb,Rapcak:2020ueh}) and, hopefully, 4-folds (see e.g. \cite{Nekrasov:2018xsb,Kanno:2020ybd,Cao:2019tvv,Oh:2020rnj}) will acquire a physical categorification as well.
	
\end{itemize}

\section*{Acknowledgements}
The author is indebted to Mina Aganagic, Aleksandra Anokhina, Semeon Artamonov, Alexei Bondal, Tudor Dimofte, Eugene Gorsky, Kentaro Hori, Mikhail Kapranov, Petr Koroteev, Andrey Mironov,  Gregory W. Moore, Alexei Morozov, Hiraku Nakajima, Vivek Schende, Alexei Sleptsov, Andrey Smirnov, Lev Soukhanov, Masahito Yamazaki, Taizan Watari, Yegor Zenkevich for stimulating and illuminating discussions at various stages of this project.  This work is supported in part by WPI Research Center Initiative, MEXT, Japan.
The author would like to thank Eidgen\"ossische Technische Hochschule Z\"urich, Switzerland; Laboratoire de Physique Th\'eorique et Hautes Energies, Sorbonne Universit\'e et CNRS, France; Moscow Institute of Physics and Technology (MIPT), Russia, for generous hospitality during the work on different parts of this project. Section 4 was completed during the author's visit to NRC ``Kurchatov institute'', Moscow, Russia, this work was supported by the Russian Science Foundation (Grant No. 20-12-00195).

\appendix

\section{Hypergeometric series}\label{sec:hypergeom}

Here we collect some basic facts about hypergeometric functions used in the text, most of them can be found in canonical textbooks on mathematical physics \cite{landau1958quantum,Gradshteyn:1702455}.

The hypergeometric function is defined inside a unit disk $|z|<1$ as absolutely convergent hypergeometric series:
$$
{}_2F_1\left[\begin{array}{c}
	a; \; b\\
	c
\end{array}\right](z)=\sum\lm_{n=0}^{\infty}\left(\prod\lm_{k=0}^{n-1}\frac{(a+k)(b+k)}{c+k}\right)\frac{z^n}{n!}.
$$

It is a solution $f$ to the hypergeometric ODE analytic in a neighborhood of point $z=0$:
$$
z(1-z)f''+\left[c-(a+b+1)z\right]f'-(ab)\, f=0.
$$

This function admits two integral representations.

We can give an Euler integral representation of the hypergeometric series due to Riemann used for analytic continuation of the hypergeometric function. 
If ${\rm Re}\;a>0$, ${\rm Re}\;c>{\rm Re}\;a$ we have:
\be\label{Riemann_hyp}
{}_2F_1\left[\begin{array}{c}
	a;\; b\\
	c
\end{array}\right](z)=\frac{\Gamma(c)}{\Gamma(a)\Gamma(c-a)}\int\lm_{0}^1 t^{a-1}(1-t)^{c-a-1}(1-t z)^{-b}dt,
\ee
where one assumes that ${\rm arg}\;t={\rm arg}\;(1-t)=0$.

Another integral representation for hypergeometric series is given by Barnes \cite{Gradshteyn:1702455} in the form of summation over Cauchy residues:
\be\label{Barnes}
{}_2F_1\left[\begin{array}{c}
	a;\; b\\
	c
\end{array}\right](z)=\frac{\Gamma(c)}{\Gamma(a)\Gamma(b)}\frac{1}{2\pi \I}\int\lm_{-\I\infty}^{\I\infty}\frac{\Gamma(a+s)\Gamma(b+s)\Gamma(-s)}{\Gamma(c+s)}(-z)^s\,ds,
\ee
where $|{\rm arg}(-z)|<\pi$ and the integration contour separates poles of $\Gamma(-s)$ from the poles of $\Gamma(a+s)$ and $\Gamma(b+s)$.

The hypergeometric equation has regular singularities at points $0$, $1$ and $\infty$. 
Therefore the hypergeometric series as a solution to the hypergeometric equation can be extended to the disk $|z|>1$ and re-expanded in terms of series convergent in that area (in  applying this relation we imply that $c$, $\pm(a-b)$ are not integer so that both hand sides do not have poles in parameters):
\be\label{analytic}
\begin{split}
	{}_2F_1\left[\begin{array}{c}
		a;\; b\\
		c
	\end{array}\right](z)=\frac{\Gamma(c)\Gamma(b-a)}{\Gamma(b)\Gamma(c-a)}\;(-z)^{-a}{}_2F_1\left[\begin{array}{c}
		a;\; a+1-c\\
		a+1-b
	\end{array}\right](z^{-1})+\\
	+ \frac{\Gamma(c)\Gamma(a-b)}{\Gamma(a)\Gamma(c-b)}\; (-z)^{-b}{}_2F_1\left[\begin{array}{c}
		b;\; b+1-c\\
		b+1-a
	\end{array}\right](z^{-1}).
\end{split}
\ee
This relation is an analytic continuation of the hypergeometric series form $|z|<1$ to $|z|>1$. 
A value of multi-valued function $(-z)^a$ and similar ones is defined by the following condition: $${\rm Im}\;\log (-z)\in (-\pi,\pi].$$

\section{2d \texorpdfstring{$\CN=(2,2)$}{N=(2,2)} supersymmetry}\label{sec:App_SUSY}

\subsection{2d \texorpdfstring{$\CN=(2,2)$}{N=(2,2)} supersymmetric gauged linear sigma model}

The action of the $\CN=(2,2)$ two-dimensional gauged linear sigma model can be derived by a dimensional reduction of  four-dimensional $\CN=1$ supersymmetric Yang-Mills-Higgs theory \cite{Wess:1992cp,D-book_1}.

The vector multiplet describing the gauge field consists of gauge vector field $A_{\mu}$, complex scalar $\sigma$, auxiliary field $\bf D$ and fermions $\lambda$. Here for simplicity we consider only $U(1)$ gauge theory, the corresponding action for the gauge vector multiplet is given by the following expression:
\be
\begin{split}
	\CS_g
	=\int dx^0dx^1\;\frac{1}{2}\Bigg[|\p_0\sigma|^2-|\p_1\sigma|^2+&\I\bar\lambda_-\left(\overset{\leftrightarrow}{\p}_0+\overset{\leftrightarrow}{\p}_1\right)\lambda_- +\\+ &\I\bar\lambda_+\left(\overset{\leftrightarrow}{\p}_0-\overset{\leftrightarrow}{\p}_1\right)\lambda_++F_{01}^2+{\bf D}^2\Bigg].
\end{split}
\ee
Here $F_{01}$ is the corresponding gauge field curvature:
$$
F_{01}=\p_0A_1-\p_1A_0.
$$

SUSY transforms for the vector multiplet read ($A_{\pm}:=\frac{1}{2}(A_0\pm A_1)$):
\be\label{SUSY_vec}
\begin{split}
	&\delta A_{\pm}=\frac{\I}{2}\bar\epsilon_{\pm}\lambda_{\pm}+\frac{\I}{2}\epsilon_{\pm}\bar\lambda_{\pm},\\
	&\delta \sigma=-\I \bar\epsilon_+\lambda_--\I\epsilon_-\bar\lambda_+\\
	&\delta {\bf D}=\frac{1}{2}\left( -\bar\epsilon_+(\p_0-\p_1)\lambda_+-\bar\epsilon_-(\p_0+\p_1)\lambda_-+\epsilon_+(\p_0-\p_1)\bar\lambda_++\epsilon_-(\p_0+\p_1)\bar\lambda_- \right),\\
	&\delta \lambda_+=\I \epsilon_+({\bf D}+\I F_{01})+\epsilon_-(\p_0+\p_1)\bar\sigma,\\
	&\delta \lambda_-=\I\epsilon_-({\bf D}-\I F_{01})+\epsilon_+(\p_0-\p_1)\sigma.
\end{split}
\ee

One can introduce chiral multiples consisting of complex scalar fields $\phi$, fermions $\psi$ and auxiliary complex scalars $F$, all charged in some representation of the corresponding gauge group. 
Here we consider a field with electric charge $Q$:
\be
\begin{split}
	\CS_{\chi}
	=\int dx^0dx^1\,\Bigg[&|D_0\phi|^2-|D_1\phi|^2+\I \bar\psi_-\left( \overset{\leftrightarrow}{D}_0+\overset{\leftrightarrow}{D}_1 \right)\psi_-+\I \bar\psi_+\left( \overset{\leftrightarrow}{D}_0-\overset{\leftrightarrow}{D}_1 \right)\psi_++\\
	+&Q {\bf D}|\phi|^2+|{\bf F}|^2
	-Q^2|\sigma\phi|^2-Q\bar\psi_-\sigma\psi_+-Q\bar\psi_+\bar\sigma\psi_-+\\
	+&\I Q\bar\phi\left(\lambda_+\psi_--\lambda_-\psi_+\right)+\I Q\phi\left(\bar\psi_+\bar\lambda_--\bar\psi_-\bar\lambda_+ \right)\Bigg],
\end{split}
\ee
where
$$
D_{\mu}\phi=\p_{\mu}\phi+\I Q A_{\mu}\phi.
$$
SUSY transforms for the chiral multiplet read:
\be\label{SUSY_chir}
\begin{split}
	&\delta \phi=\epsilon_+\psi_--\epsilon_-\psi_+,\\
	&\delta \psi_+=\I\bar\epsilon_-(D_0+D_1)\phi+\epsilon_+ {\bf F}-Q\bar\epsilon_+\bar\sigma \phi,\\
	&\delta \psi_-=-\I\bar\epsilon_+(D_0-D_1)\phi+\epsilon_- {\bf F}+Q\bar\epsilon_-\sigma \phi,\\
	&\delta {\bf F}=-\I\bar\epsilon_+(D_0-D_1)\psi_+-\I\bar\epsilon_-(D_0+D_1)\psi_-+\\
	&\qquad\qquad\qquad\qquad\qquad\;\;+(\bar\epsilon_+\bar\sigma\psi_-+\bar\epsilon_-\sigma\psi_+)+\I(\bar\epsilon_-\bar\lambda_+-\bar\epsilon_+\bar\lambda_-)\phi.
\end{split}
\ee

Additional parameters can be introduced through topological and Fayet-Illiopolous terms:
\be\label{S_FI}
\CS_{\rm FI,\,\theta}=\int dx^0dx^1\,\left[-r{\bf D}+\theta F_{01}\right].
\ee
As well we can consider a superpotential term defined by a holomorphic function of chiral fields $W(\phi)$:
\be
\CS_{W}=\int dx^0dx^1\,{\rm Re}\left({\bf F} W'-W''\psi_+\psi_-\right).
\ee

Now we would like to quantize this theory. Fields ${\bf D}$ and ${\bf F}$ are non-dynamical, therefore they acquire only expectation values:
\be\label{const}
{\bf D}=r-Q|\phi|^2,\quad {\bf F}=-\frac{1}{2}\bar W'.
\ee
$A_0$ is also non-dynamical and produces a secondary constraint -- the Gauss law:
\be\label{Gauss}
\CJ= Q\left(i(\bar\phi D_0\phi-\phi D_0\bar\phi)+\bar\psi_+\psi_++\bar\psi_-\psi_-\right)-\p_1 F_{01}+\rho_{\rm bdry},
\ee
where $\rho_{\rm bdry}$ is an electric charge due to the boundary.

The fields are understood as operators on the Hilbert space, momentum operators are defined as corresponding variations:
$$
F_{01}(x^1)=-\I\frac{\delta}{\delta A_1(x^1)}-\theta,\quad \p_0\sigma(x^1)=-2\I\frac{\delta}{\delta \bar{\sigma}(x^1)},\quad D_0\phi(x^1)=-\I\frac{\delta}{\delta\bar{\phi}(x^1)}.
$$
The physical Hilbert space in this quantization scheme is constrained:
$$
\CJ|{\rm phys}\rangle=0.
$$

We derive the supercharges as spacial integrals of corresponding Noether supercurrents:\footnote{Here we used the standard relation for symmetry generators on the Hilbert space $\delta_{\varphi}\CO=\I\left[Q_{\varphi},\CO\right]$, and a definition of supersymmetry generators:
	$$
	\delta=\epsilon_+\CQ_--\epsilon_-\CQ_+-\bar\epsilon_+\bar \CQ_-+\bar\epsilon_-\bar\CQ_+.
	$$}
\be
\begin{split}
	\CQ_+=\int dx^1\Big[&\frac{1}{2}\Big(-\bar\lambda_-({\bf D}-\I F_{01})+\I\bar\lambda_+(\p_0+\p_1)\bar\sigma\Big)+\\
	&+\Big(\psi_+(D_0+D_1)\bar\phi+\I\bar\psi_- {\bf F}+\I Q\psi_-\bar\sigma\bar\phi\Big)\Big],\\
	\CQ_-=\int dx^1\Big[&\frac{1}{2}\Big(\bar\lambda_+({\bf D}+\I F_{01})-\I\bar\lambda_-(\p_0-\p_1)\sigma\Big)+\\
	&+\Big(\psi_-(D_0-D_1)\bar\phi-\I\bar\psi_+ {\bf F}+\I Q\psi_+\sigma\bar\phi\Big)\Big],\\
	\bar \CQ_+=\int dx^1\Big[&\frac{1}{2}\Big(-\lambda_-({\bf D}+\I F_{01})-\I\lambda_+(\p_0+\p_1)\sigma\Big)+\\
	&+\Big(\bar\psi_+(D_0+D_1)\phi-\I\psi_- \bar {\bf F}-\I Q\bar\psi_-\sigma\phi\Big)\Big],\\
	\bar \CQ_-=\int dx^1\Big[&\frac{1}{2}\Big(\lambda_+({\bf D}-\I F_{01})+\I\lambda_-(\p_0-\p_1)\bar\sigma\Big)+\\
	&+\Big(\bar\psi_-(D_0-D_1)\phi+\I\psi_+ \bar{\bf F}-\I Q\bar\psi_+\bar\sigma\phi\Big)\Big].
\end{split}
\ee

We calculate these generators satisfy the following algebra:
\be
\begin{split}
	\CQ_+^2=\int dx^1\;\frac{Q}{2}\bar \sigma\bar \phi\bar W',\quad \CQ_-^2=-\int dx^1\;\frac{Q}{2}\sigma\bar \phi\bar W',\\
	\left\{\CQ_{\pm},\bar\CQ_{\pm} \right\}=\CH\pm \CP,\\
	\left\{\CQ_{+},\CQ_{-} \right\}=\CZ,\\
	\left\{\CQ_{+},\bar\CQ_{-} \right\}=\tilde\CZ+\int dx^1\; \bar\sigma\CJ.
\end{split}
\ee
where $\CH$ and $\CP$ are Hamiltonian and momentum operator corresponding to the action
\be\label{action}
S=\CS_g+\CS_{\chi}+\CS_{\rm FI,\, \theta}+\CS_{W}.
\ee
$\CZ$ and $\tilde \CZ$ are central charges given by the following expressions:
\be
\CZ=\int dx^1\;\left[\I \bar W'D_1\bar\phi+\frac{\I}{2}\p_1(\bar\lambda_-\bar\lambda_+)\right],\\
\tilde\CZ=\int dx^1\;\left[\I\p_1(\bar\psi_-\psi_++\bar\sigma({\bf D}-\I F_{01}))\right].
\ee

These expressions admit an immediate generalization to arbitrary gauge groups and representations. 

As well we will need to consider multiple flavors of chiral multiplets. 
From this point of view this means that we charge our chiral fields with respect to non-dynamical flavor group $U(N_f)$, where $N_f$ is a number of flavors. 
We assign complex masses $\mu_i$, $i=1,\ldots,N_f$ to chiral fields by simply assuming that $U(N_f)$ is gauged, and by giving a non-dynamical expectation value to the scalar field in the flavor symmetry gauge vector multiplet:
$$
\sigma_f={\rm diag}(\mu_1,\ldots,\mu_{N_f}),
$$
all the remaining fields in the flavor symmetry gauge vector multiplet remain zero.

\subsection{A-twist and B-twist}
By A-twist and B-twist following the common terminology \cite{D-book_1} we call the following supercharge families ($\zeta$ is a complex phase):
\be\label{AB-twists}
\begin{split}
	\CQ_A=\zeta^{\frac{1}{2}}\CQ_-+\zeta^{-\frac{1}{2}}\bar{\CQ}_+,\quad \CQ_A^{\dagger}=\bar\CQ_A=\zeta^{-\frac{1}{2}}\bar\CQ_-+\zeta^{\frac{1}{2}}{\CQ}_+,\\
	\CQ_B=\zeta^{\frac{1}{2}}\CQ_-+\zeta^{-\frac{1}{2}}\CQ_+,\quad \CQ_B^{\dagger}=\bar\CQ_B=\zeta^{-\frac{1}{2}}\bar\CQ_-+\zeta^{\frac{1}{2}}\bar\CQ_+.
\end{split}
\ee

It is easy to calculate corresponding subalgebras generated by these supercharge (here we have already applied the Gauss law constraint $\CJ=0$):
\be\label{AB-algebra}
\begin{split}
	\CQ_A^2=-\I \int dx^1\;Q\sigma\,{\rm Im}(\zeta^{-1}\phi W')+\tilde \CZ^*,\quad \left\{\CQ_A,\bar{\CQ}_A\right\}=2\CH+2{\rm Re}(\zeta\CZ),\\
	\CQ_B^2=\I \int dx^1\; Q\,{\rm Im}(\zeta\sigma)\bar{\phi}\bar W'+\CZ,\quad \left\{\CQ_B,\bar{\CQ}_B\right\}=2\CH+2{\rm Re}(\zeta^{-1}\tilde \CZ).
\end{split}
\ee

We construct half-BPS ground states as cohomologies of the corresponding supercharge. For this procedure to work we require the corresponding supercharge to be nilpotent. 
In other words, when we consider a Landau-Ginzburg model preserving A-twist we assume $\CQ_A^2=0$, on the other hand when we consider GLSM preserving B-twist we take $\CQ_B^2=0$. 
In the first case we assume that the chiral fields are not charged and there is no gauge symmetry, the superpotential can be an arbitrary holomorphic function; in the second case we consider the superpotential to be a gauge invariant function of the chiral fields. 
In the both cases the term containing superpotential is annihilated in the supercharge square and the remaining part describes branes as boundary conditions on fields. 
Under these assumptions the supercharges in these models can be represented by the following expressions:
\begin{subequations}
	\be
	\begin{split}
		\CQ_A=e^{-\mathfrak{H}_A}\int dx^1 \left(-\I\sqrt{2}\zeta^{\frac{1}{2}}\psi_-\frac{\delta}{\delta\phi}-\I\sqrt{2}\zeta^{-\frac{1}{2}}\bar\psi_+\frac{\delta}{\delta\bar\phi}\right)\;e^{\mathfrak{H}_A},\\
		\mathfrak{H}_A=\int dx^1\left[\I\,\bar\phi\p_1\phi-{\rm Re}\left(\zeta^{-1}W\right)\right];
	\end{split}
	\ee
	\be
	\begin{split}
		\bar\CQ_B=e^{-\mathfrak{H}_B}\int dx^1\Bigg[&-\lambda_1\left(\frac{\delta}{\delta A_1}+\I\frac{\delta}{\delta\sigma_\II}\right)+\lambda_2\frac{\delta}{\delta \sigma_\IR}-\\
		&-\I\sqrt{2}\bar\psi_1\frac{\delta}{\delta\bar\phi}-\sqrt{2}\bar\psi_2 \left(\hat V\cdot \phi\right)+\frac{\I}{\sqrt{2}}\psi_2 W'\Bigg]e^{\mathfrak{H}_B},\\
		&\mathfrak{H}_B=\int dx^1\;\left(\sigma_\II\p_1\sigma_\IR-\sigma_\IR\left(r-Q|\phi|^2\right)-\I\theta A_1\right),\\
		&\hat V\cdot \phi=D_1\phi-Q\sigma_\II\phi.
	\end{split}
	\ee
\end{subequations}

Here we introduced the following notations:
\be
\begin{split}
	\lambda_1=\frac{1}{2}\left(\zeta^{\frac{1}{2}}\lambda_-+\zeta^{-\frac{1}{2}}\lambda_+\right),\quad \lambda_2=\frac{1}{2}\left(\zeta^{\frac{1}{2}}\lambda_--\zeta^{-\frac{1}{2}}\lambda_+\right),\\
	\psi_1=\frac{1}{\sqrt{2}}\left(\zeta^{\frac{1}{2}}\psi_-+\zeta^{-\frac{1}{2}}\psi_+\right),\quad \psi_2=\frac{1}{\sqrt{2}}\left(\zeta^{\frac{1}{2}}\psi_--\zeta^{-\frac{1}{2}}\psi_+\right),\\
	\sigma_{\IR}=\frac{\zeta\sigma+\zeta^{-1}\bar\sigma}{2},\quad \sigma_{\II}=\frac{\zeta\sigma-\zeta^{-1}\bar\sigma}{2\I}.
\end{split}
\ee
Notice that in these variables the norm of fermionic fields is changed:
\be
\frac{1}{2}\bar\lambda_-\lambda_-+\frac{1}{2}\bar\lambda_+\lambda_++\bar\psi_-\psi_-+\bar\psi_+\psi_+=\bar\lambda_1\lambda_1+\bar\lambda_2\lambda_2+\bar\psi_1\psi_1+\bar\psi_2\psi_2.
\ee

\subsection{Landau-Ginzburg sigma-model}\label{sec:App_LG}

As well we will need a sigma model with a target space given by a K\"ahler manifold $X$. 
Consider $n$ chiral fields $\phi^i$ defining coordinates on $X$.  
The metric is defined by a K\"ahler potential:
$$
g_{i\bar j}=\p_i\p_{\bar j}K(\phi,\bar\phi)
$$

The Lagrangian for this model reads:\footnote{$\p_{\pm}=\frac{1}{2}(\p_0\pm \p_1)$}
\be
\begin{split}
	\CS=\int dx^0dx^1\Bigg[-g_{i\bar j}\p^{\mu}\phi^i\p_{\mu}\overline{\phi^j}+2\I g_{i\bar j}\overline{\psi_-^j}D_+\psi_-^i+2\I g_{i\bar j}\overline{\psi_+^j}D_-\psi_+^i+\\
	+R_{i\bar j k\bar l}\psi_+^i\psi_-^k\overline{\psi_-^j}\overline{\psi_+^l}+g_{i\bar j}\left(F^i-\Gamma^i_{jk}\psi_+^j\psi_-^k\right)\left(\overline{F^i}-\Gamma^{\bar i}_{\bar j\bar k}\overline{\psi_+^j}\overline{\psi_-^k}\right)+\\
	+\frac{1}{2}F^i\p_i W-\frac{1}{2}\p_{ij}^2W\psi_+^i\psi_-^j+\frac{1}{2}\overline{F^i}\overline{\p_i W}-\frac{1}{2}\overline{\p_{ij}^2W}\overline{\psi_-^i}\overline{\psi_+^j}\Bigg].
\end{split}
\ee

Corresponding supercharges read:
\be
\begin{split}
	\CQ_{\pm}=\int dx^1\left(g_{i\bar j}(\p_0\pm\p_1)\overline{\phi^j}\psi_{\pm}^i\mp \frac{\I}{2}\overline{\psi_{\mp}^i}\overline{\p_iW}\right).
\end{split}
\ee

For the corresponding A-twist we have:
\be\label{nlin_sup}
\CQ_A=\int dx^1\left[-\I\psi_+^i\left(\delta_{\phi^i}+\I g_{i\bar j}\p_1\overline{\phi^j}-\frac{1}{2}\p_iW\right)-\I\overline{\psi_-^i}\left(\delta_{\overline{\phi^i}}-\I g_{\bar i j}\p_1\phi^j-\frac{1}{2}\overline{\p_iW}\right)\right].
\ee

\section{B-type brane boundary conditions}\label{sec:App_boundary}
\subsection{Boundary fermion}
Let us specify the B-twist action for the SUSY transformations of the GLSM:
$$
\epsilon_+=\zeta^{\frac{1}{2}}\nu,\quad \epsilon_-=-\zeta^{-\frac{1}{2}}\nu,\quad \bar\epsilon_+=\zeta^{-\frac{1}{2}}\bar\nu,\quad \epsilon_-=-\zeta^{\frac{1}{2}}\bar\nu
$$

The expressions \eqref{SUSY_vec} and \eqref{SUSY_chir} under this substitution are modified as:
\be
\begin{split}
	\begin{array}{l}
		\delta A_0=-\I\left(\bar\nu \lambda_2+\nu\bar\lambda_2\right),\\
		\delta A_1=\I\left(\bar\nu \lambda_1+\nu\bar\lambda_1\right),\\
		\delta\sigma_1=-\I\left(\bar\nu\lambda_2+\nu\bar\lambda_2\right),\\
		\delta\sigma_2=\nu\bar\lambda_1-\bar\nu\lambda_1,\\
	\end{array}\quad
	\begin{array}{l}
		\delta \lambda_1=\nu\left(-\p_1 \sigma _1+\I \p_0 \sigma _2-F_{01}\right),\\
		\delta \lambda_2=\nu\left(\p_0 \sigma _1-\I \p_1 \sigma _2-\I \Delta\right),\\
		\delta \phi=\sqrt{2}\nu\psi_1,\\
		\delta \psi_1=-\I\sqrt{2}\bar\nu\left(D_0\phi-\I Q\,\sigma_1\phi\right),\\
	\end{array}\\
	\delta \psi_2=-\I\sqrt{2}\bar\nu\left(-D_1\phi+Q\sigma_2\phi\right)-\sqrt{2}F\nu,\\
	\delta \Delta=\bar{\nu }\p_1 \lambda _1 +\bar{\nu }\p_0 \lambda _2 -\nu \p_1  \bar{\lambda }_1- \nu \p_0 \bar{\lambda }_2,\\
	\delta F=\I\sqrt{2}\bar\nu\left(D_1\psi_1-Q\sigma_2\psi_1\right)+\I\sqrt{2}\bar\nu\left(D_0\psi_2-\I Q \sigma_1\psi_2 \right)-2\I Q\bar\nu\bar\lambda_1\phi.\\ 
\end{split}
\ee

One observes immediately that the supersymmetry allows one to introduce a Fermi multiplet \cite{Hori:2000ic} by considering the following supersymmetric field configuration:
$$
\phi=0,\quad \psi_1=0.
$$
It will turn out that spacial derivative $D_1$ disappears from the action. 
Therefore we treat it as a boundary Fermi multiplet. 
It is allowed to interact with the bulk fields via superpotential, however from that superpotential only first derivative survives. We call it a fermion superpotential $V$. 

We introduce a Fermi boundary multiplet as a pair of fermion $\chi$ and auxiliary field $U$:
\be
\begin{split}
	\delta \chi=-\sqrt{2}U\nu,\\
	\delta U=\I\sqrt{2}\left(D_0\chi-\I Q \sigma_\IR\chi\right).
\end{split}
\ee

The action for this boundary fermion reads:
\be
\begin{split}
	S_{\rm b.f.}=\int dx^0\,\Bigg[\I\bar\chi\overset{\leftrightarrow}{D}_0\chi+|U|^2+Q\sigma_1\bar\chi\chi\Bigg]+\int dx^0\,{\rm Re}\left(U V(\phi)-V'(\phi)\psi_1\chi\right).
\end{split}
\ee

\subsection{Boundary charge}

We assume that at the boundary a Wilson loop is inserted in a representation $\kappa$ and a boundary fermion, in other words, we modify the action as follows:
\be
\CS_{\rm 2d}\to\CS_{\rm 2d}+\kappa\int dx^0 (A_0-\sigma_\IR)+S_{\rm b.f.}\;.
\ee
This additional term produces a non-trivial electric charge modifying the Gauss law constraint \eqref{Gauss}:
\be
\rho_{\rm bdry}=\left(F_{01}+Q_{\chi}\bar{\chi}\chi-\kappa+\theta\right)\delta\left(x^1-x^1_{\rm bdry}\right).
\ee
The supercharge is modified accordingly by a shift of the boundary value:
\be\label{stalk}
\bar \CQ_B\to \bar \CQ_B+\bar\CQ_{\rm bdry}, \quad \bar\CQ_{\rm bdry}=\frac{\I}{\sqrt{2}}\chi V(\phi)\Big|_{\rm bdry}.
\ee

Boundary supercharge $\bar\CQ_{\rm bdry}$ defines a complex of vector bundles carried by the boundary brane \cite{HHP}.

A simple and canonical example of this identification is the following.
Consider a non-singular variety $X$, and a subvariety $Y\hookrightarrow X$ of co-dimension 1 defined by an algebraic equation:
\be
V(\phi_1,\ldots,\phi_n)=0,
\ee
where $\phi_i$ are coordinates on $X$.
A boundary supercharge \eqref{stalk} defines the canonical projective resolution of the structure sheaf on $Y$ based on its Weil divisor\footnote{For a definition of Weil divisors see \cite{Watari_lectures}}:
\be\label{Weil}
\begin{tikzcd}
	0\ar[r] & \CO_X(-Y) \ar[r,"\bar\CQ_{\rm bdry}"]& \CO_X\ar[r] & \CO_Y \ar[r] & 0.
\end{tikzcd}
\ee

\subsection{Brane boundary conditions}\label{sec:brane_cond}

B-twisted brane boundary condition for vector multiplet naturally follow from constraint $\CQ_B^2=0$. They will end up in a special Lagrangian locus constraint for twisted chiral field $\Sigma$. However boundary conditions for chiral fields remain undetermined in this way.

Before proceeding we have to impose boundary conditions for other fields defining corresponding B-branes.

Let us re-consider $\CN=(2,2)$ supersymmetry in terms of superfields \cite{D-book_1,Wess:1992cp}. In addition to the real space-time coordinates $x^0$ and $x^1$ the superspace is also spanned by Grassmann coordinates:
$$
\theta^{\pm},\quad \bar\theta^{\pm}.
$$

Supercharges generate the following vector fields:
\be
\CQ_{\pm}=\frac{\p}{\p\theta^{\pm}}+\I\bar\theta^{\pm}\p_{\pm},\quad \bar\CQ_{\pm}=-\frac{\p}{\p\bar\theta^{\pm}}-\I\theta^{\pm}\p_{\pm},
\ee
where
$$
\p_{\pm}=\frac{1}{2}\left(\p_0\pm\p_1 \right),\quad x^{\pm}=x^0\pm x^1.
$$

One defines covariant derivatives:
$$
D_{\pm}=\frac{\p}{\p\theta^{\pm}}-\I\bar\theta^{\pm}\p_{\pm},\quad \bar D_{\pm}=-\frac{\p}{\p\bar\theta^{\pm}}+\I\theta^{\pm}\p_{\pm}.
$$

The chiral field is defined by a condition:
\be
\bar D_{\pm}\Phi=0.
\ee

And we can rewrite it explicitly in coordinates as:
\be
\Phi=\phi(y^{\pm})+\theta^{\alpha}\psi_{\alpha}(y^{\pm})+\theta^+\theta^-F(y^{\pm}),
\ee
where
$$
y^{\pm}=x^{\pm}-\I\theta^{\pm}\bar\theta^{\pm}.
$$

As before let us change variables as 
\be
\theta^{\pm}=\frac{\zeta^{\mp\frac{1}{2}}}{\sqrt{2}}\left(\theta_1\mp\theta_2\right),\quad \bar\theta^{\pm}=\frac{\zeta^{\pm\frac{1}{2}}}{\sqrt{2}}\left(\bar\theta_1\mp\bar\theta_2\right).
\ee
So that 
\be\nn
\begin{split}
	\theta^{\pm}\bar{\theta}^{\pm}=\frac{1}{2}\left(\theta_1\bar{\theta}_1+\theta_2\bar{\theta}_2\right)\mp \frac{1}{2}\left(\theta_1\bar{\theta}_2+\theta_2\bar{\theta}_1\right),\\
	\theta^+\theta^-=\theta_1\theta_2.
\end{split}
\ee

In these terms the supercharge fields read:
\be
\CQ_B=\sqrt{2}\left(\frac{\p}{\p\theta_1}+\I\bar\theta_1\p_0\right)+\sqrt{2}\I\bar\theta_2\p_1,\quad \bar\CQ_B=-\sqrt{2}\left(\frac{\p}{\p\bar\theta_1}+\I\theta_1\p_0\right)-\sqrt{2}\I\theta_2\p_1.
\ee

We can define new effective 1d supercharges:
$$
\CQ_0=\frac{\p}{\p\theta_1}+\I\bar\theta_1\p_0,\quad \bar\CQ_0=-\frac{\p}{\p\bar\theta_1}-\I\theta_1\p_0.
$$

As well we define 1d chiral and Fermi supermultiplets transformed by 1d supercharges:
\be
\begin{split}
	\Phi_{\rm 1d}=\phi+\theta_1\psi_1-\frac{\I}{2}\theta_1\bar\theta_1\p_0\phi,\\
	\Psi_{\rm 1d}=\psi_2-\theta_1F-\frac{\I}{2}\theta_1\bar\theta_1\p_0\psi_2.
\end{split}
\ee

So that the 2d chiral multiplet can be decomposed as:
\be
\begin{split}
	\Phi=\Phi_{\rm 1d}+\theta_2\left(\Psi_{\rm 1d}+\frac{\I}{2}\bar\theta_1\p_1\Phi_{1d}\right)+\frac{\I}{2}\theta_1\bar\theta_2\p_1\Phi_{\rm 1d}-\\
	-\frac{\I}{2}\theta_2\bar\theta_2\left(\p_0 \Phi_{\rm 1d}+\theta_1\p_1\Psi_{\rm 1d}+\frac{\I}{2}\theta_1\bar\theta_1\p_1^2 \Phi_{\rm 1d}\right).
\end{split}
\ee

The supercharges induce the following transformations on these fields:
\be
\begin{array}{ll}
	\CQ_B \Phi_{\rm 1d}=\CQ_0 \Phi_{\rm 1d}, & \bar\CQ_B\Phi_{\rm 1d}=\bar\CQ_0 \Phi_{\rm 1d},\\
	\CQ_B\Psi_{\rm 1d}=\CQ_0\Psi_{\rm 1d}, & \bar\CQ_B\Psi_{\rm 1d}=\bar\CQ_0 \Psi_{\rm 1d}+\I\p_1 \Phi_{\rm 1d}.
\end{array}
\ee

Therefore, according to \cite{Dimofte:2019zzj}, a natural choice of manifestly supersymmetric invariant boundary conditions for the chiral multiplet is the following:
\be\label{chiral_brane}
\Psi_{\rm 1d}\Big|_{\rm bdry}=\p_1 \Phi_{\rm 1d}\Big|_{\rm bdry}=0.
\ee

\section{Spectral covers and soliton counting}

\subsection{Spectral networks}\label{sec:App_SN}
Consider an $n$-dimensional K\"ahler manifold $\CM_n$ spanned by coordinates $\phi_i$, $i=1,\ldots,n$.
As well one considers a meromorphic family $\CW$ of meromorphic functions on $\CM_n$  parameterized by $z\in \IC$. 
In this section we will consider methods to count trajectories in $\CM_n$ generated by a soliton flow equation:
\be\label{gen_soliton}
\p_x \phi^i(x)=\zeta^{-1} g^{i\bar j}\overline{\p_{\phi^j}\CW(\phi(x),z)}, \quad x\in\CI\subseteq \IR.
\ee

First we describe a \emph{constant vacuum solution}. Let us consider a constant map (we will mark constant solutions by an asterisk subscript):
$$
\phi^i(x)=\phi_{*}^i={\rm const}.
$$
This map solves equation \eqref{gen_soliton} if and only if $\phi_{*}^i$ satisfies a set of algebraic equations:
$$
\p_{\phi^i}\CW(\phi_*,z_0)=0,\quad \forall i.
$$
These algebraic equations may have different roots $\phi^i_{*\alpha}(z_0)$ we label by index $\alpha\in\CV$. All of them are admissible solutions to \eqref{gen_soliton}. 
We call them \emph{vacua} and index set $\CV$ a vacuum set. We expect that the vacuum set is at least countable, and in many cases it turns out to be finite. 
In general, $\CV$ is fibered non-trivially over $z_0\in\IC$. 
This fibration is a ramified cover of $\IC$. In \emph{ramification points} two or more vacua collide. 
We can define corresponding ramification points $z_r$ algebraically from a condition that the Hessian of $\CW$ is degenerate on ramification point locus. 
We call this constraint a \emph{discriminant} constraint:
\be\label{disc}
\Delta(z_r):=\mathop{\rm Det}\lm_{i,j\in\CV}\left(\p^2_{\phi^i\phi^j}\CW(\phi_*(z_r),z_r)\right)=0.
\ee

The second type of solutions we call $(\alpha,\beta)$-\emph{soliton solutions}. 
It is defined as a boundary value problem solution for equation \eqref{gen_soliton} on an open interval $\CI=\IR$ with boundary conditions:
$$
\lim\lm_{x\to -\infty}\phi^i(x)=\phi^i_{*\alpha},\quad \lim\lm_{x\to +\infty}\phi^i(x)=\phi^i_{*\beta}.
$$
It is easy to derive a constraint that is necessary (however not sufficient) for such a solution to exist:
\be\label{BPS_constraint}
\zeta\left(\CW\left(\phi_{*\beta}(z),z\right)-\CW\left(\phi_{*\alpha}(z),z\right)\right)\in \IR_{\geq 0}.
\ee

We would like to give a geometric description to $(\alpha,\beta)$-solitons. 
The first geometrical object we would like to consider is a ring of chiral operators in the Landau-Ginzburg theory \cite{Cecotti:1991me} that is identified with a Jacobian ring of $\CW$:
\be
\CR(z):=\IC[\phi^i]/\CJ,\quad \CJ=\IC[\phi^i]\cdot \langle \p_{\phi^1}\CW,\ldots,\p_{\phi^n}\CW\rangle.
\ee
Clearly, only operators in $\CR(z)$ have a non-trivial vacuum expectation value in one of the constant vacua $\phi_*$.
Chose a basis $\CO^{\alpha}$ in $\CR(z)$. In this basis a multiplication by $\p_z\CW$ introduces a linear operator:
\be
\p_z\CW\cdot \CO^{\alpha}=C^{\alpha}_{\beta}(z)\CO^{\beta}\;{\rm mod}\; \langle \p_{\phi^1}\CW,\ldots,\p_{\phi^n}\CW\rangle.
\ee

The major geometric object of our consideration in this section is a characteristic polynomial of matrix $C(z)$:
\be
\Sigma(z,\lambda):={\rm Det}\left(C(z)-\lambda\cdot{\rm Id}\right).
\ee
A complex curve $\Sigma(z,\lambda)=0$ covering $z\in\IC$ we call a \emph{spectral cover}.

We will construct corresponding solutions to \eqref{gen_soliton} based on geometric properties of the spectral cover.

First of all notice that the discriminant constraint \eqref{disc} we used to identify ramification points coincides with the discriminant locus of the characteristic polynomial:
\be
\Delta(z)={\rm Disc}_{\lambda}\left[\Sigma(z,\lambda)\right].
\ee
In other words ramification points for branched spectral cover coincide with ramification points of the vacuum set.

Roots $\lambda^{(\alpha)}(z)$ of the spectral curve $\Sigma(z,\lambda)=0$ define corresponding vacuum expectation values:
\be
\lambda^{(\alpha)}(z_0)=\p_z\CW(\phi_{*\alpha}(z_0),z_0),\quad \alpha\in\CV.
\ee

One can choose a system of cuts. A set of roots $t^{(\alpha)}$ can be ordered on $\IC\setminus\{{\rm cuts}\}$ giving a trivialization of the vacuum bundle over $\IC\setminus\{{\rm cuts}\}$. We may assume that ramification points are all simple, so that only a pair of roots, say $\alpha$ and $\beta$ collide in a certain ramification point $z_r$:
$$\phi^i_{*\alpha}(z_r)=\phi^i_{*\beta}(z_r).$$ 
We call such a branching point of $(\alpha,\beta)$-type.

A neighborhood of a simple ramification point has a generic description in local coordinates \cite{GMW}:
$$
\CW=\frac{1}{3}(\phi^1)^3-z\phi^1+\sum\lm_{i=2}^n(\phi^i)^2.
$$
The ramification point is located at $z_r=0$. This simple model describes a behavior of two vacua $\phi_{*\pm}^1=\pm z^{\frac{1}{2}}$, $\phi_{*\pm}^{i\neq 1}=0$. 
In these local coordinates equation \eqref{gen_soliton} is integrable, therefore whenever BPS constraint \eqref{BPS_constraint} is satisfied there is a standard kink-shaped solitonic solution interpolating between $\pm$ vacua. 
We can extend these solutions to the whole plane $\IC$ spanned by $z$ using parallel transport.
A differential form of BPS constraint \eqref{BPS_constraint} reads:
\be\label{netw}
\zeta\left(\lambda^{(\beta)}(z)-\lambda^{(\alpha)}(z)\right)dz\in \IR_{\geq 0}.
\ee
From this point of view it is natural to define a meromorphic differential form:
\be
\Omega:=\zeta \;\lambda\; dz,
\ee
so that this constraint can be rewritten as:
\be
\Omega^{(\beta)}-\Omega^{(\alpha)}\in \IR_{\geq 0}.
\ee
We call this form Seiberg-Witten differential, or SW differential for short.

We will consider 1d real loci on $\IC$ spanned by $z$, defined as real lines starting from a ramification point, such that the tangent vector to this line satisfies \eqref{netw}. 
These loci define values of $z$ on $\IC$ where the corresponding solitonic solution exists. In the literature \cite{Gaiotto:2012rg} these loci are known as soliton walls, or \emph{S-walls} for brevity.
A network of all S-walls covering the spectral cover is called a \emph{spectral network}.

Let us illustrate this procedure by a quick calculation of S-walls in a neighborhood of a ramification point. In local coordinates for the spectral cover we have:
$$
\lambda^2-z=0.
$$
Vacua correspond to two roots $\lambda^{(\pm)}=\pm z^{\frac{1}{2}}$. An S-wall is defined by a differential equation \eqref{netw}:
$$
\zeta dz^{\frac{3}{2}}\in\IR_{\geq 0}.
$$
Let us choose a parameterization $s$ along the S-wall, then up to a re-parameterization we can choose a non-negative number in r.h.s. to be one. The differential equation reads:
\be\label{bp}
\zeta\frac{d}{ds}\left(z(s)\right)^{\frac{3}{2}}=1.
\ee
As a boundary condition for \eqref{bp} we choose a reflection of the fact that S-wall goes through the ramification point: $z(0)=0$. In this case S-walls are represented by three rays emanated from the ramification point (see Fig.\ref{fig:SN_CP_1}):
$$
z(s)=e^{\frac{2\pi \I}{3}n}\zeta^{-\frac{2}{3}}s^{\frac{2}{3}},\quad s\in[0,+\infty),\quad n=0,\;1,\;2\;.
$$

\subsection{Picard-Lefschetz/Berry connection, Fukaya-Seidel categories} \label{sec:App_FS_cat}
Let us consider a more generic complex space $\CP$ parameterizing the family of theories $W$.
The techniques discussed above are still applicable in this case if we choose a complex line $\ell:\,\IC\to \CP$ and pull the family of theories back to this line (see, for example, \cite[Section 2.1.2]{Galakhov:2017pod}).
One can mimic any point of $\CP$ as a point $\ell(z)$ for $z\in\IC$.

Consider a brane amplitude defined by Lagrangian cycle brane $\CL$ and operator insertion $\CO$ as a function of parameter $z$:
\be\label{brane_amp}
\Psi_{\CL}[\CO](z)=\int\lm_{\CL}\omega\; \CO(\phi)\;e^{-\zeta\beta\;\CW(\phi)},
\ee
where $\beta>0$ is a temperature circumference \cite{Cecotti:1991me} and $\omega$ is the top holomorphic form on $\CM_n$.

We choose the following bases among integration cycles and observables:
\begin{itemize}
	\item {\bf Lefschetz thimbles.} We expect $\CL_{\alpha}$ to be presented by a \emph{Lefschetz thimble} -- a union of all solutions to \eqref{gen_soliton} with a boundary condition $\lim\lm_{x\to -\infty}\phi(x)=\phi_{*\alpha}$.
	Lefschetz thimbles are Lagrangian submanifolds in $\CM_n$.
	For cycles $\CL_{\alpha}$ integrals \eqref{brane_amp} converge absolutely.
	\item {\bf Jacobian ring.} Observables $\CO^{\alpha}$ are chosen to be basis elements of $\CR(z)$.
	A difference in expectation values for different members of the same equivalence class in $\CR(z)$ is suppressed as $O(e^{-\beta \Delta})$, where
	$$
	\Delta=\min\lm_{\alpha\neq\alpha'}\left|\CW\left(\phi_{*\alpha}\right)-\CW\left(\phi_{*\alpha'}\right)\right|.
	$$
\end{itemize}

In the limit $\beta\to \infty$ the brane amplitude expectation value is given by the following expression:
\be
\Psi_{\CL_{\alpha}}[\CO^{\alpha'}]=\frac{\iota_{\p/\p\phi^1}\ldots\iota_{\p/\p\phi^n}\omega_{*\alpha}}{\sqrt{{\rm Det}_{i,j}\p^2_{ij}\CW_{*\alpha}}}\CO^{\alpha'}(\phi_{*\alpha})e^{-\beta\zeta\;\CW_{*\alpha}}\left(1+O\left(e^{-\beta\Delta}\right)\right),
\ee
where we introduce a notation $f_{*\alpha}=f(\phi_{*\alpha})$.

Bases $\{\CL_{\alpha}\}$ and $\{\CO^{\alpha'}\}$ are \emph{dual} to each other.

We could choose a basis of operators $\{\CO^{\alpha'}\}$ to derive a Berry connection (also a holomorphic part of a $tt^*$-connection \cite{Cecotti:1991me}) on brane amplitudes:
\be
\left(\nabla^{\scalebox{0.5}{Berry}}\Psi\right)^{\alpha_1}=\left(\beta^{-1}\delta_{\alpha_2}^{\alpha_1}\p_z-\zeta C_{\alpha_2}^{\alpha_1}(z)\right)\Psi_{\CL_{\alpha_3}}[\CO^{\alpha_2}](z)=O\left(\beta^{-1},e^{-\beta\Delta}\right).
\ee

However as Picard-Lefschetz theory predicts a description of parallel transport is much more transparent in the dual basis of of Lefschetz thimbles.
Let us order vacua in such a way that $\alpha<\alpha'$ if
\be
{\rm Im}\;\zeta\CW_{*\alpha}>{\rm Im}\;\zeta\CW_{*\alpha'}.
\ee
This ordering fails on spectral network lines \eqref{netw}.
When parallel transport path $\wp$ intersects such a line for a pair of $\alpha$ and $\alpha+1$ corresponding critical values are braided in $\zeta \CW$-plane clockwise, and the topology of Lefschetz thimbles in the $\zeta\CW$-plane is changed:
\be\nn
\begin{array}{c}
	\begin{tikzpicture}
		\draw[thick] (0,0) -- (0,1);
		\draw[thick, red,postaction={decorate},decoration={markings, 
			mark= at position 0.75 with {\arrow{stealth}}}] (-1,0.5) -- node[pos=0.25,above,red]{$\wp$} (1,0.5);
		\draw[thick, orange, decorate, decoration={snake, segment length=6pt, amplitude=2pt}] (0,0) -- (0,-1);
		\draw[ultra thick,orange] (-0.1,-0.1) -- (0.1,0.1) (0.1,-0.1) -- (-0.1,0.1);
		\node[below] at (0,-1) {\tiny $(\alpha,\alpha+1)$-cut};
		\draw[thick,blue, postaction={decorate},decoration={markings, 
			mark= at position 0.75 with {\arrow{stealth}}}] (0,0.5) to[out=240,in=180] (0,-0.3) to[out=0,in=300] node[pos=0.3,right,blue] {$\gamma$} (0,0.5);
		\draw[fill] (0,0.5) circle (0.04);
	\end{tikzpicture}
\end{array}\quad \begin{array}{c}
	\begin{tikzpicture}
		\node(A) at (0,0) {$\begin{array}{c}
				\begin{tikzpicture}
					\draw[thick] (0,1.5) -- (-2,1.5) (0,0.5) -- (-2,0.5) (0,-0.5) -- (-2,-0.5) (0,-1.5) -- (-2,-1.5);
					\draw[fill] (0,1.5) circle (0.06) (0,0.5) circle (0.06) (0,-0.5) circle (0.06) (0,-1.5) circle (0.06);
					\node[right] at (0,1.5) {$\zeta\CW_{*\alpha-1}$};
					\node[right] at (0,0.5) {$\zeta\CW_{*\alpha}$};
					\node[right] at (0,-0.5) {$\zeta\CW_{*\alpha+1}$};
					\node[right] at (0,-1.5) {$\zeta\CW_{*\alpha+2}$};
					\node[left] at (-2,1.5) {$\CL_{\alpha-1}$};
					\node[left] at (-2,0.5) {$\CL_{\alpha}$};
					\node[left] at (-2,-0.5) {$\CL_{\alpha+1}$};
					\node[left] at (-2,-1.5) {$\CL_{\alpha+2}$};
				\end{tikzpicture}
			\end{array}$};
		\node(B) at (6.5,0) {$\begin{array}{c}
				\begin{tikzpicture}
					\draw[thick] (0,1.5) -- (-2,1.5) (0,0.5) -- (-2,0.5) (0,-0.5) -- (-2,-0.5) (0,-1.5) -- (-2,-1.5);
					\draw[fill] (0,1.5) circle (0.06) (0,0.5) circle (0.06) (0,-0.5) circle (0.06) (0,-1.5) circle (0.06);
					\node[right] at (0,1.5) {$\zeta\CW'_{*\alpha-1}$};
					\node[right] at (0,0.5) {$\zeta\CW'_{*\alpha+1}$};
					\node[below right] at (0,-0.5) {$\zeta\CW'_{*\alpha}$};
					\node[right] at (0,-1.5) {$\zeta\CW'_{*\alpha+2}$};
					\node[left] at (-2,1.5) {$\CL'_{\alpha-1}$};
					\node[left] at (-2,0.5) {$\CL'_{\alpha+1}$};
					\node[left] at (-2,-0.5) {$\CL'_{\alpha}$};
					\node[left] at (-2,-1.5) {$\CL'_{\alpha+2}$};
					\draw[dashed] (0,-0.5) to[out=0,in=270] (1.7,0.25) to[out=90,in=0] (0,1) -- (-2,1);
					\node[right] at (1.7,0.25) {$\CL_{\alpha+1}$};
				\end{tikzpicture}
			\end{array}$};
		\path (A) edge[->] node[above,red] {$\wp$} (B);
	\end{tikzpicture}
\end{array}
\ee
All thimbles $\CL_{\epsilon}$ are transported to new ones $\CL'_{\epsilon}$ except $\epsilon=\alpha+1$ that is not a Lefschetz thimble anymore and has to be re-decomposed in a new basis.

Relation between new and old Lefshetz thimble bases is given by a Stokes matrix:
\be\label{Stokes}
\left(\begin{array}{c}
	\CL_{\alpha}'\\
	\CL_{\alpha+1}'
\end{array}\right)=\left(\begin{array}{cc}
	1 & 0\\
	e^{\pi\I \CF}e^{\zeta \tilde Z}& 1
\end{array}\right)\left(\begin{array}{c}
	\CL_{\alpha}\\
	\CL_{\alpha+1}
\end{array}\right)
\ee
where $\CF$ and $\tilde \CZ$ are a fermion number and a central charge of a BPS string solution contributing to $\fG_{\rm BPS}(\CL_{\alpha},\CL_{\alpha+1})$.
Corresponding expressions for multiple soliton contributions can be found in \cite{Gaiotto:2012rg}.
These are extensive quantities given by integrals along a cycle $\gamma$ on the spectral cover. 
Cycle $\gamma$ starts on cover sheet $\alpha$ and goes to cover sheet $\alpha+1$.
The fermion number, also identified with a Maslov index, and the central charge are defined as (see \cite{Galakhov:2016cji}):
\be
\begin{split}
	\tilde Z=\int\lm_{\gamma}\Omega,\quad
	\CF=\int\lm_{\gamma}\nu,
\end{split}
\ee
where form $\nu^{(\alpha)}$ on cover sheet $\alpha$ is defined through a basis of eigen vectors of the structure constant matrix $C(z)$:
\be\label{Maslov}
\begin{split}
	C(z)\cdot{\bf e}^{\alpha}(z)=\lambda^{(\alpha)}{\bf e}^{\alpha}(z),\\
	\nu^{(\alpha)}=\frac{1}{\pi\I}\frac{{\rm Det}\left[{\bf e}^1,\ldots,d{\bf e}^{\alpha},\ldots,{\bf e}^n\right]}{{\rm Det}\left[{\bf e}^1,\ldots,{\bf e}^{\alpha},\ldots,{\bf e}^n\right]}.
\end{split}
\ee

A categorification of formula \eqref{Stokes} is available through a Fukaya-Seidel category \cite{Seidel_book} of Lagrangian submanifolds in $\CM_n$.
As a set of exceptional objects in this category one chooses the Lefschetz thimble basis, then the parallel transport induces a permutation functor on an ordered set of Lefschetz thimbles extendable to the whole category.
Action of this functor also maps all old thimbles $\CL_{\epsilon}$ to $\CL_{\epsilon}'$ except $\epsilon=\alpha+1$. The latter is defined from the following exact triangle relations holding for arbitrary category object $X$:
\be
\begin{array}{c}
	\begin{tikzcd}
		\fG_{\rm BPS}(X,\CL_{\alpha})\otimes \fG_{\rm BPS}(\CL_{\alpha},\CL_{\alpha+1})[\CF]  \ar[rr, "{\mu^2(\cdot,X)}"] & & \fG_{\rm BPS}(X,\CL_{\alpha+1}) \ar[dl] \\
		& \fG_{\rm BPS}(X,\CL'_{\alpha+1}) \ar[ul, "{[1]}"]&
	\end{tikzcd}
\end{array},
\ee
where $\mu^2$ is a multiplication structure \cite{Auroux} defined in the following way.
Consider a moduli space $\mathscr{M}$ of Landau-Ginzburg theory disk 1-instantons \eqref{LG-instanton} with boundary conditions as depicted in the figure.
A single instanton modulus $\IR$ corresponds to disk automorphisms.
Reduced moduli space $\mathscr{M}^*$ consists of points, then $\mu^2$ is defined as:
\be
\begin{array}{c}
	\begin{tikzpicture}
		\draw[thick] (-0.2,1.5) -- (0,1.5) to[out=0,in=180]  node [above right,align=center,midway] {$\CL_{\alpha_1}$} (2,0.5) -- (2.2,0.5) (-0.2,-1.5) -- (0,-1.5) to[out=0,in=180] node [below right,align=center,midway] {$\CL_{\alpha_3}$} (2,-0.5) -- (2.2,-0.5) (-0.2,0.5) -- (0,0.5) to[out=0,in=0] node [left,align=center,midway] {$\CL_{\alpha_2}$} (0,-0.5) -- (-0.2,-0.5);
		\draw[thick, red,postaction={decorate},decoration={markings, 
			mark= at position 0.65 with {\arrow{stealth}}}] (0,1.5) -- node [left,align=center,midway] {$\xi_1$} (0,0.5);
		\draw[thick, red,postaction={decorate},decoration={markings, 
			mark= at position 0.65 with {\arrow{stealth}}}] (0,-0.5) -- node [left,align=center,midway] {$\xi_2$} (0,-1.5);
		\draw[thick, red,postaction={decorate},decoration={markings, 
			mark= at position 0.65 with {\arrow{stealth}}}] (2,0.5) -- node [left,align=center,midway] {$\xi_3$} (2,-0.5);
		\node at (0.8,0) {$\iota$};
	\end{tikzpicture}
\end{array},\quad \begin{array}{l}
	\mu^2:\;\fG_{\rm BPS}(\CL_{\alpha_1},\CL_{\alpha_2})\otimes \fG_{\rm BPS}(\CL_{\alpha_2},\CL_{\alpha_3})\longrightarrow \fG_{\rm BPS}(\CL_{\alpha_1},\CL_{\alpha_3}),\\
	\mu^2\left(|\xi_1\rangle\otimes|\xi_2\rangle\right)=\sum\lm_{|\xi_3\rangle\in \fG_{\rm BPS}(\CL_{\alpha_1},\CL_{\alpha_3})}\sum\lm_{\iota\in\mathscr{M}^*}(-1)^{\eta(\iota)}|\xi_3\rangle,
\end{array}
\ee
where we weight the instanton contributions with values of the path integral evaluated around corresponding instanton classical solution $\iota$.
It is well-known (see e.g. \cite{Witten:1982im}) that the resulting path integral localizes to a classical action contribution that can be canceled by re-scaling norms of states and a one-loop determinant that contributes with a sign.
This sign is accumulated from positive and negative fermion mode eigenvalue contributions and can be evaluated as $(-1)^{\eta(\iota)}$, where $\eta$ is the APS eta-invariant.
For alternative ways to calculate this sign by a careful analysis of the determinant bundle associated to the instanton see \cite[Section 10.5.3]{D-book_1} and \cite[Appendix F]{GMW} and references therein.

Multiplication structure $\mu^2$ together with higher multiplication structures $\mu^k$ satisfy $A_{\infty}$-relations.

\section{Brief reminder on derived categories}\label{sec:categories}
In this section we will not attempt to review this vast subject of the algebraic geometry, we will merely quote some known results. An enthusiastic reader is referred to canonical textbooks on the subject \cite{Hartshorne,D-book_1,D_book_2}.
A physicist audience oriented review of selected topics in the algebraic geometry can be found in \cite{Watari_lectures}. 

\subsection{Derived functors}

We define Abelian categories and derived functors following \cite{D_book_2}.
We should stress that for Abelian categories the set of morphisms has properties of an Abelian group.
In our case all the morphisms are associated with Hilbert spaces of BPS states, therefore they are, in fact, vector spaces over $\IC$.

An object $M$ of an Abelian category $\CA$ is called:
\begin{itemize}
	\item \emph{injective}, if for each injective  morphism $\phi:\;E\longrightarrow F$ and for any morphism $\gamma:\; E\longrightarrow M$ there is a morphism $\eta:\;F\longrightarrow M$ such that the following diagram is commutative:
	$$
	\begin{array}{c}
		\begin{tikzcd}
			E  \ar[rr, "\phi"] \ar[dr, "\gamma"']& & F \ar[dl, "\eta"] \\
			& M &
		\end{tikzcd}
	\end{array}
	$$
	\item \emph{projective}, if for each surjective morphism $\phi:\;E\longrightarrow F$ and for any morphism $\gamma:\; M \longrightarrow F$ there is a morphism $h:\;M\to E$ such that the following diagram is commutative:
	$$
	\begin{array}{c}
		\begin{tikzcd}
			&M \ar[dl,"\eta"']\ar[dr,"\gamma"]&\\
			E \ar[rr, "\phi"']& & F
		\end{tikzcd}
	\end{array}
	$$
\end{itemize}

Object $M$ of an Abelian category is said to have a projective resolution if there is an exact sequence:
\be\label{projective}
\begin{array}{c}
	\begin{tikzcd}
		\ldots \ar[r] & P_{-2} \ar[r] & P_{-1}\ar[r] &P_0\ar[r]& M\ar[r] & 0,
	\end{tikzcd}
\end{array}
\ee
where $P_i$ are projective objects.

Correspondingly, object $N$ of an Abelian category is said to have an injective resolution if there is an exact sequence:
\be\label{injective}
\begin{array}{c}
	\begin{tikzcd}
		0 \ar[r] & N \ar[r] & I_0\ar[r] &I_1\ar[r]& I_2\ar[r] & \ldots,
	\end{tikzcd}
\end{array}
\ee
where $I_i$ are injective objects.

We should stress that the situation when a category object has a resolution is \emph{not common}. 
It is said that a category has enough projectives (injectives) if all category objects have projective (injective) resolutions.

A left exact functor $\lambda$ and a right exact functor $\rho$ are defined by their action on a short exact sequence:
$$
\begin{array}{c}
	\begin{tikzpicture}
		\node(A) at (0,0) {$\begin{array}{c}
				\begin{tikzcd}
					0 \ar[r] & M_1\ar[r,"f"] & M_2\ar[r,"g"] & M_3 \ar[r] & 0
				\end{tikzcd}\\
				\mbox{is exact}
			\end{array}$};
		\node(B) at (4.5,2) {$\begin{array}{c}
				\begin{tikzcd}
					0 \ar[r] & \lambda(M_1)\ar[r,"\lambda(f)"] & \lambda(M_2)\ar[r,"\lambda(g)"] & \lambda(M_3)
				\end{tikzcd}\\
				\mbox{is exact}
			\end{array}$};
		\node(C) at (4.5,-2) {$\begin{array}{c}
				\begin{tikzcd}
					\rho(M_1)\ar[r,"\rho(f)"] & \rho(M_2)\ar[r,"\rho(g)"] & \rho(M_3) \ar[r] & 0
				\end{tikzcd}\\
				\mbox{is exact}
			\end{array}$};
		\path (A) edge[->] node[above left]{$\lambda$} (B) (A) edge[->] node[above right]{$\rho$} (C);
	\end{tikzpicture}
\end{array}
$$
A functor mapping a short exact sequence to a short exact sequence is called exact.

One defines a \emph{left derived} functor ${\bf L}\rho$ of a right exact functor $\rho$ on an object $M\in \CA$ through its projective resolution \eqref{projective} as a complex:
$$
{\bf L}\rho(M):=\left(
\begin{array}{c}
	\begin{tikzcd}
		\ldots\ar[r]&\rho(P_{-1})\ar[r]&\rho(P_0)\ar[r]& 0
	\end{tikzcd}
\end{array}
\right).
$$
Homology groups of this complex are denoted as ${\bf L}_i\rho(M)$.

One defines a \emph{right derived} functor ${\bf R}\lambda$ of a left exact functor $\lambda$ on an object $N\in \CA$ through its injective resolution \eqref{injective} as a complex:
$$
{\bf R}\lambda(N):=\left(
\begin{array}{c}
	\begin{tikzcd}
		0\ar[r]&\lambda(I_{0})\ar[r]&\lambda(I_{1})\ar[r]& \ldots
	\end{tikzcd}
\end{array}
\right).
$$
Cohomology groups of this complex are denoted as ${\bf R}^i\lambda(N)$.

Clearly, this construction incorporates chain complexes in $\CA$, so to make it functorial one needs to consider a category of complexes in $\CA$, or, more precisely, a derived category $D(\CA)$ \cite{D_book_2} whose objects are chain complexes in $\CA$.
In $D(\CA)$ both left and right derived functors are exact. 
Moreover these derived functors are independent (isomorphic) of different choices of projective and injective resolutions correspondingly.

\subsection{Derived category of coherent sheaves}\label{s:DCoh}
Following \cite{D_book_2} we call an Abelian category of coherent sheaves on $X$ a category with coherent sheaves as objects and sheaf homomorphisms as morphisms. 
A (bounded) derived category of an Abelian category of coherent sheaves on a topological space $X$ is usually denoted as:
$$
D^{(b)}{\rm Coh}(X).
$$

A practical obstacle for a direct application of the derived category machinery to coherent sheaves is a necessity to have projective and injective resolutions for an arbitrary object to calculate derived functors. 
Sometimes one could choose a ``nicer'' resolution so the calculation of a derived functor is easier. 
On the other hand there are universal algorithms to construct resolutions of generic coherent sheaves.

On a smooth projective variety $X$ a coherent sheaf $\CF$ admits \emph{a projective resolution} of length $n={\rm dim}\,X$:
$$
\begin{array}{c}
	\begin{tikzcd}
		0 \ar[r]& \CE_n \ar[r]&\ldots \ar[r]&\CE_1 \ar[r]& \CE_0 \ar[r] & \CF \ar[r] &0,
	\end{tikzcd}
\end{array}
$$
where $\CE_i$ are locally free sheaves and, therefore, projective objects. 
This statement follows from an application of the Hilbert  \emph{syzygy} theorem to coherent sheaves. 
We will not go over details of this construction referring the reader to \cite{Gunning_Rossi}. 
Notice that in our physical applications coherent sheaves will appear initially in the form of complexes of holomorphic bundles on projective varieties.

For \emph{an injective resolution} there is a variety of options \cite{Galier} usually referring to various forms of sheaf cohomology. 
We will briefly mention only two of them.

The first injective resolution is given by a $\check{\rm C}$ech complex. It is convenient form of an injective resolution usually applied in practice. 
A calculation usually reduces to a study of pole structures of meromorphic functions on $X$ (see e.g. \cite[Appendix A]{Hayashi:2008ba}).

Following \cite{Galier} for $X$ choose an ordered open covering $\CU=\{U_i\}$ of an open subset $U$. 
For the $\check{\rm C}$ech complex
$$
\begin{array}{c}
	\begin{tikzcd}
		\check C^0(U,\CF) \ar[r,"d^0"] & \check C^1(U,\CF) \ar[r,"d^1"] & \check C^2(U,\CF) \ar[r,"d^2"] &\ldots
	\end{tikzcd}
\end{array}
$$
cochains are defined as:
$$
\begin{array}{c}
	\check C^p(U,\CF)=\prod\lm_{i_0<\ldots<i_p}\CF\left(U_{i_0}\cap\ldots\cap U_{i_p}\right).
\end{array}
$$
An element of $\check C^p$ is obtained by specifying a set of elements:
$$
f_{i_0,\ldots,i_p}\in \CF\left(\bigcap\lm_{m=0}^p U_{i_m}\right),
$$
so that for a permutation $\sigma$ of indices we have:
$$
f_{\sigma(i_0),\ldots,\sigma(i_p)}={\rm sgn}(\sigma)\cdot f_{i_0,\ldots,i_p}.
$$
A complex differential is defined as:
$$
\left(d^pa\right)_{i_0,\ldots,i_{p+1}}=\sum\lm_{k=0}^{p+1}(-1)^k {\rm res}_{U_{i_0}\cap\ldots \cap U_{i_{p+1}},U_{i_0}\cap\ldots\cap\widehat{ U_{i_k}}\cap\ldots \cap U_{i_{p+1}}} a_{i_0,\ldots,\hat i_k,\ldots, i_{p+1}},
$$
where hats imply that corresponding elements are omitted.

If $\CU=\{ U_i\}$ is an open covering of $X$ define for an open subset $V\subset X$:
$$
\CU|_V=\{\mbox{all open subsets of the form }V\cap U_i\}.
$$
For a coherent sheaf $\CF$ we define an injective $\CO_X(V)$-module:
$$
\CC^p(\CU,\CF)(V):=\check C^p(\CU|_V,\CF),
$$
then a sheaf $\CF$ has the following injective resolution:
\be\nn
\begin{array}{c}
	\begin{tikzcd}
		0 \ar[r] &\CF\ar[r] &\CC^0(\CU,\CF) \ar[r] & \CC^1(\CU,\CF) \ar[r] & \CC^2(\CU,\CF) \ar[r] &\ldots
	\end{tikzcd}
\end{array}
\ee
A particularly practical example of a calculation in this framework can be found in \cite[Appendix A]{Hayashi:2008ba}.

Another injective resolution is through a de Rham complex.
This resolution is more familiar in physical applications since as we mentioned in Section~\ref{sec:Heis_loc} the majority of geometrical properties of our systems arise when we treat supercharges as modifications of de Rham differential on the target space of a quantum system.

On a smooth manifold $X$ choose an open subset $V$ and consider an injective resolution of real numbers through a de Rham complex \cite{Watari_lectures}, or, similarly, a Dolbeault complex \cite{Thomas:1999ic}:
$$
\begin{array}{c}
	\begin{tikzcd}
		0\ar[r]& \IR\ar[r] & \Omega^0(V)\ar[r,"d_{\rm dR}"] & \Omega^1(V)\ar[r,"d_{\rm dR}"] & \Omega^2(V)\ar[r,"d_{\rm dR}"]&\ldots,
	\end{tikzcd}
\end{array}
$$
where $\Omega^p(V)$ is a space of smooth $p$-forms on $V$. By tensoring with locally free sheaf $\CF$ we get an injective resolution:
$$
\begin{array}{c}
	\begin{tikzcd}
		0\ar[r]& \CF\ar[r] & \Omega^0\otimes \CF\ar[r,"\nabla_{\rm dR}"] & \Omega^1\otimes \CF\ar[r,"\nabla_{\rm dR}"] & \Omega^2\otimes \CF\ar[r,"\nabla_{\rm dR}"]&\ldots,
	\end{tikzcd}
\end{array}
$$
where $\nabla_{\rm dR}$ is a differential modified with a corresponding connection.

A cohomology theory on $X$ valued in a sheaf $\CF$ is defined in this setting in the following way \cite{D_book_2}:
$$
H^i(X,\CF):={\bf R}^i{\rm Hom}(\CO_X,\CF).
$$

\subsection{Fourier-Mukai transform}\label{sec:App_FM}

According to \cite{D_book_2} tensor product, pullback and pushforward produce derived functors.

The tensor product defines a left derived functor:
$$
\mathop{\otimes}\lm^{\bf L}\CF:\; D^b{\rm Coh}(X) \to D^b{\rm Coh}(X).
$$

For a coherent sheaf $\CF$ a continuous map of topological spaces $f:\; X\longrightarrow Y$ defines a right derived direct image functor:
$$
{\bf R}f_*:\; D^b{\rm Coh}(X) \to D^b{\rm Coh}(Y).
$$

The pullback produces a left derived functor:
$$
{\bf L}f^*:\; D^b{\rm Coh}(Y) \to D^b{\rm Coh}(X)
$$

Now consider a pair of smooth projective varieties $X$ and $Y$. Out of them one can form a product manifold and consider two projection maps:
$$
\begin{array}{c}
	\begin{tikzcd}
		& X\times Y \ar[dl,"\pi_X"']\ar[dr,"\pi_Y"]&\\
		X && Y
	\end{tikzcd}
\end{array}
$$ 
For any object $\CK\in D^b{\rm Coh}(X\times Y)$ called a \emph{Fourier-Mukai kernel} one can construct an exact functor:
$$
\bPhi_{\CK}:\; D^b{\rm Coh}(X)\longrightarrow D^b{\rm Coh}(Y),
$$
called a {\bf Fourier-Mukai transform} using an explicit expression:
\be\label{FM_transform}
\bPhi_{\CK}(\CF):={\bf R}\pi_{Y*}\left({\bf L}\pi_X^*\left(\CF\right)\mathop{\otimes}\lm^{\bf L} \CK\right).
\ee

According to \cite{Huybrechts,orlov2003derived} (Orlov's theorem) if an exact functor between two derived categories of coherent sheaves admits so called left and  right adjoints it can be represented as a Fourier-Mukai transform with some kernel.

\section{Miscellaneous calculations}
\subsection{Renormalization on the soft Higgs branch} \label{sec:Ren_soft}
First of all let us conjugate the supercharges by an exponent of the following height functional:
$$
\int d x^1\;\sigma_{\IR}\left(\sum\lm_a Q_a|\phi_a|^2-r\right).
$$
This allows one to introduce additional $\hbar$ multipliers in the underlined positions in the supercharge \eqref{supercharge}:
\be
\begin{split}
	\bar\CQ_B=\int dx^1\;\Bigg[\lambda_1\left(-\I\hbar\delta_{\sigma_{\II}}-\I\p_1\sigma_{\IR}-\hbar\delta_{A_1}+\I\hbar \theta\right)+\\
	+\lambda_2\left(\hbar\delta_{\sigma_{\IR}}-\p_1\sigma_{\II}+\underline{\hbar}\left(\sum\lm_a Q_a|\phi_a|^2-r\right)\right)
	-\I\sqrt{2}\sum\lm_{a}\bar\psi_{\dot 1,a}\left(\hbar\delta_{\bar\phi_a}+\underline{\hbar}Q_a\sigma_{\IR}\phi_a\right)-\\-\sqrt{2}\sum\lm_{a}\bar\psi_{\dot 2,a}\left(\left(\p_1\phi_a+\I Q_a A_1\phi_a\right)-Q_a\sigma_{\II}\phi_a\right)\Bigg]+\bar\CQ_{\rm bdry}.
\end{split}
\ee 
This allows one to lift constraint \eqref{Hb} at the first order contribution and deal with it at higher orders.
According to our choice of Neuman boundary conditions for the chiral fields we decompose our fields in modes in the following way:
\be
\begin{split}
	\sigma_{\IR}(x^1)&=\Sigma_{\IR}+\hbar^{\frac{1}{2}}\sum\lm_{k=1}^{\infty}s_{\IR,k}\sqrt{2}\cos \omega_k x^1;\\
	\sigma_{\II}(x^1)&=\hbar^{\frac{1}{2}}\sum\lm_{k=1}^{\infty}s_{\II,k}\sqrt{2}\sin \omega_k x^1;\\
	\lambda_1(x^1)&=\sum\lm_{k=1}^{\infty}\beta_{1,k}\sqrt{2}\sin \omega_k x^1;\\
	\lambda_2(x^1)&=\nu+\sum\lm_{k=1}^{\infty}\beta_{2,k}\sqrt{2}\cos \omega_k x^1;\\
	A_1(x^1)&=-\p_1\vartheta_0+\sum\lm_{k=1}^{\infty}a_k\sqrt{2}\sin\omega_k x^1;\\
	\phi_a(x^1)&=e^{\I Q_a\vartheta_0}\left(\Phi_a+\frac{\Phi_a}{|\Phi_a|}\hbar^{\frac{1}{2}}\sum\lm_{k=1}^{\infty}\left(X_{a,k}+\I Y_{a,k}\right)\sqrt{2}\cos\omega_k x^1\right);\\
	\psi_{1,a}(x^1)&=e^{\I Q_a\vartheta_0}\left(\eta_{a}+\frac{\Phi_a}{|\Phi_a|}\sum\lm_{k=1}^{\infty}\xi_{1,a,k}\sqrt{2}\cos\omega_k x^1\right);\\
	\psi_{2,a}(x^1)&=e^{\I Q_a\vartheta_0}\sum\lm_{k=1}^{\infty}\xi_{2,a,k}\sqrt{2}\sin\omega_k x^1.
\end{split}
\ee
where $\omega_k=\frac{\pi k}{L}$. After applying these transformations the first order contribution to the supercharge reads:
\be
\begin{split}
	\bar\CQ_B^{(0)}=\sum\lm_{k=1}^{\infty}\Bigg[&\beta_{1,k}\left(-\I\frac{\p}{\p s_{\II,k}}+\I\omega_ks_{\IR,k}-\frac{\p}{\p a_k}\right)+\beta_{2,k}\left(\frac{\p}{\p s_{\IR,k}}-\omega_ks_{\II,k}\right)-\\
	&-\frac{\I}{\sqrt{2}}\sum\lm_a\bar\xi_{\dot 1,a,k }\left(\frac{\p}{\p X_{a,k}}+\I \frac{\p}{\p Y_{a,k}}\right)+\\
	&+\sqrt{2}\sum\lm_a\bar\xi_{\dot 2,a,k}\left(\left(\omega_kX_{a,k}+Q_a|\Phi_a|s_{\II,k}\right)+\I\left(\omega_kY_{a,k}-Q_a|\Phi_a|a_k\right)\right)\Bigg].
\end{split}
\ee
And the Gauss law constraint contribution reads:
\be
\CJ_k^{(0)}=\I\left(\sum\lm_aQ_a|\Phi_a|\frac{\p}{\p Y_{a,k}}+\omega_k\frac{\p}{\p a_k}\right).
\ee
To divide variables efficiently we need to make just a few coordinate frame rotations.
Choose a vector $v^1$ in $\IR^n$ with components:
$$
v_a^1=Q_a|\Phi_a|/\Omega ,\quad a=1,\ldots,n,\quad \Omega=\sqrt{\sum\lm_{b=1}^nQ_b^2|\Phi_b|^2},
$$ 
and $n-1$ arbitrary orthonormal vectors $v^i$, $i=2,\ldots,n$ in the orthogonal complement of $v^1$ in $\IR^n$. We choose a new basis in the field space:
\be
\begin{split}
	X_{a,k}+\I Y_{a,k}=(x_k+\I y_k)v^1_a+\sum\lm_{i=2}^nz_{i,k}v^i_a,\\
	\xi_{\alpha,a,k}=\chi_{\alpha,1,k}v^1_a+\sum\lm_{i=2}^n\chi_{\alpha,i,k}v^i_a.
\end{split}
\ee
In new terms the supercharge reads:
\be
\begin{split}
	\bar\CQ_B^{(0)}=\sum\lm_{k=1}^{\infty}\Bigg[\beta_{1,k}\left(-\I\frac{\p}{\p s_{\II,k}}+\I\omega_ks_{\IR,k}-\frac{\p}{\p a_k}\right)+\beta_{2,k}\left(\frac{\p}{\p s_{\IR,k}}-\omega_ks_{\II,k}\right)-\\
	-\frac{\I}{\sqrt{2}}\bar\chi_{\dot 1,1,k }\left(\frac{\p}{\p x_k}+\I \frac{\p}{\p y_{k}}\right)+\sqrt{2}\bar\xi_{\dot 2,1,k}\left(\left(\omega_kx_{k}+\Omega s_{\II,k}\right)+\I\left(\omega_ky_k-\Omega a_k\right)\right)-\\
	-\I\sqrt{2}\sum\lm_{i=2}^n\bar\chi_{\dot 1,i,k}\frac{\p}{\p\bar z_k}+\sqrt{2}\sum\lm_{i=2}^n\bar\chi_{\dot 2,i,k}\omega_kz_k\Bigg].
\end{split}
\ee
For the Gauss law we have:
\be
\CJ_k^{(0)}=\I\left(\Omega\frac{\p}{\p y_k}+\omega_k\frac{\p}{\p a_k}\right)
\ee
If we try to solve this problem directly the resulting answer will be rather bulky and not very self-explanatory. Rather we conjugate this supercharge by another height function:
$$
\sum_k\omega_ks_{\IR,k}s_{\II,k}.
$$
This allows us to localize the theory to a locus $s_{\IR,k}=s_{\II,k}=0$. After integration over fast degrees of freedom we are left with the following supercharge:
\be
\begin{split}
	\bar\CQ_B'=\sum\lm_{k=1}^{\infty}\Bigg[
	-\frac{\I}{\sqrt{2}}\bar\chi_{\dot 1,1,k }\left(\frac{\p}{\p x_k}+\I \frac{\p}{\p y_{k}}\right)+\sqrt{2}\bar\xi_{\dot 2,1,k}\left(\omega_kx_{k}+\I\left(\omega_ky_k-\Omega a_k\right)\right)-\\
	-\I\sqrt{2}\sum\lm_{i=2}^n\bar\chi_{\dot 1,i,k}\frac{\p}{\p\bar z_k}+\sqrt{2}\sum\lm_{i=2}^n\bar\chi_{\dot 2,i,k}\omega_kz_k\Bigg].
\end{split}
\ee
The resulting wave function annihilated by this supercharge, its conjugate and by the Gauss law constraint reads:
\be
\Xi=\prod\lm_{k=1}^{\infty}\left(\prod\lm_{i=1}^n(\I\bar\chi_{\dot 1, i, k}+\bar\chi_{\dot 2, i, k})\right)e^{-\omega_k\sum\lm_{i=2}^n|z_i|^2-\omega_k\left(x_k^2+\left(y-\frac{\Omega}{\omega_k}a_k\right)^2\right)}|0\rangle.
\ee
Clearly, it is not $L^2$-integrable, there is no potential term for $2\Omega y_k+\omega_k a_k$, this direction corresponds to gauge shifts. This wave function does not contribute to the effective supercharge and Gauss charge density given by \eqref{eff_supercharge} and \eqref{chrg} respectively.

\subsection{Disorder operator on the Coulomb branch}\label{sec:disorder}

Consider a family of supercharge operators parameterized by two functions: a complex-valued function $\sigma(x)=\sigma_{\IR}(x)+\I\sigma_{\II}(x)$ and a real-valued function $\alpha(x)$:
\be
\bar{Q\!\!\!\!Q}[\sigma,\alpha]:=\int dx^1\;\left[-\I\sqrt{2}\bar\psi_{\dot 1}\left(\delta_{\bar\phi}+\sigma_{\IR}\phi\right)-\sqrt{2}\bar\psi_{\dot 2}\left(\left(\p_1\phi+\I \alpha\phi\right)-\sigma_{\II}\phi\right)\right]
\ee
A wave function annihilated by both $\bar{Q\!\!\!\!Q}$ and $\bar{Q\!\!\!\!Q}^{\dagger}$ can be represented by a functional $\Upsilon$:
$$
\Upsilon\left[\phi,\bar\phi,\bar\psi_{\dot 1},\bar\psi_{\dot 2}|\sigma,\alpha\right]|0\rangle.
$$
Let us consider a gauge operator depending on a generic real-valued phase function $\vartheta(x^1)$:
\be
\CG[\vartheta]:=\int dx^1\;(\I\vartheta)\left(\left(\bar\phi \delta_{\bar\phi}-\phi \delta_{\phi}\right)+\bar\psi_{\dot 1}\psi_{1}+\bar\psi_{\dot 2}\psi_{2}\right).
\ee
This operator shifts phases of chiral fields.
This shift can be refabricated as a shift in the gauge field $\alpha$:
\be
\begin{split}
	e^{\CG[\vartheta]}\Upsilon\left[\phi,\bar\phi,\bar\psi_{\dot 1},\bar\psi_{\dot 2}|\sigma,\alpha\right]|0\rangle&=\Upsilon\left[e^{-\I\vartheta}\phi,e^{\I\vartheta}\bar\phi,e^{\I\vartheta}\bar\psi_{\dot 1},e^{\I\vartheta}\bar\psi_{\dot 2}|\sigma,\alpha\right]|0\rangle=\\
	&=\Upsilon\left[\phi,\bar\phi,\bar\psi_{\dot 1},\bar\psi_{\dot 2}|\sigma,\alpha-\p_1\vartheta\right]|0\rangle.
\end{split}
\ee
Simply taking a variation over $\vartheta(x^1)$ on both sides we find that the wave-function is annihilated identically by the following operator:
\be\label{Gauss_Upsilon}
\left(\left(\bar\phi \delta_{\bar\phi}-\phi \delta_{\phi}\right)+\bar\psi_{\dot 1}\psi_{1}+\bar\psi_{\dot 2}\psi_{2}+\I\p_1 \delta_{\alpha}\right)\Upsilon\left[\phi,\bar\phi,\bar\psi_{\dot 1},\bar\psi_{\dot 2}|\sigma,\alpha\right]|0\rangle=0.
\ee

We define a disorder operator field in the following way:
\be\label{disord}
Y:=\langle 0|\Upsilon^{\dagger}\left(|\phi|^2-\delta_{\alpha}\right)\Upsilon|0\rangle.
\ee

To study properties of functional $\Upsilon$ we will consider a simplified case when fields $\alpha$, $\sigma$ are constant. 
Consider two bases of orthonormal modes on an interval $[0,L]$:\footnote{These modes satisfy the following relation: 
	$$
	\int\lm_0^L\tilde e_n(x)^{\dagger}(\p_x+\I\alpha-\sigma_{\II})e_m(x)\;dx=-Z_n\delta_{nm}.
	$$}
\be\label{modes}
\begin{split}
	e_0(x)=&\sqrt{\frac{2\sigma_{\II}}{e^{2\sigma_{\II}L}-1}}e^{(\sigma_{\II}-\I \alpha)x},\\
	e_n(x)=&\frac{\left(\kappa_n-\alpha-\I\sigma_{\II}\right)e^{\I\kappa_nx}+\left(\kappa_n+\alpha+\I\sigma_{\II}\right)e^{-\I\kappa_nx}}{\sqrt{2L(\kappa_n^2+\alpha^2+\sigma_{\II}^2)}},\quad n\geq 1;\\
	\tilde e_n(x)=&\sqrt{\frac{2}{L}}\sin\kappa_n x,\quad n\geq 1;
\end{split}
\ee

Decomposition of fields over normalized modes read:
\be
\phi(x^1)=\sum\lm_{n=0}^{\infty}\phi_n e_n(x^1),\quad
\psi_1(x^1)=\sum\lm_{n=0}^{\infty}\psi_{1,n} e_n(x^1),\quad \psi_2(x^1)=\sum\lm_{n=1}^{\infty}\psi_{2,n} \tilde e_n(x^1).
\ee
where $\kappa_n=\pi n/L$.

In these terms we derive:
\be
\begin{split}
	\Upsilon|0\rangle=&\left\{\begin{array}{ll}
		g(\phi_0), &\mbox{if} \; \sigma_{\IR}>0\\
		\bar g(\bar\phi_0)\bar\psi_{\dot 1,0}, &\mbox{if}\; \sigma_{\IR}<0
	\end{array}\right\}e^{-|\sigma_{\IR}||\phi_0|^2}\prod\lm_{n=1}^{\infty}\frac{(\Omega_n-\sigma_{\IR})\bar\psi_{\dot 1,n}-\I Z_n\bar\psi_{\dot 2,n}}{\sqrt{\pi(\Omega_n-\sigma_{\IR})}}e^{-\Omega_n|\phi_n|^2}|0\rangle,
\end{split}
\ee
where
\be
\begin{split}
	Z_n=\frac{\kappa_n^2+(\sigma_{\II}-\I\alpha)^2}{\sqrt{\kappa_n^2+\alpha^2+\sigma_\II^2}},\\
	\Omega_n=\sqrt{\sigma_{\IR}^2+|Z_n|^2},
\end{split}
\ee
and $g$ is an arbitrary holomorphic function.

One can use this representation to derive the expectation value of the disorder operator. When a field has only a constant mode the variation is modified as:
$$
\delta_{\alpha}=L^{-1}\p_{\alpha}.
$$
For simplicity we will perform a calculation in a point $\alpha=0$. One will need the following relations:
\be
\begin{split}
	\p_{\alpha}\phi_n\Big|_{\alpha=0}=\frac{\I \sigma_{\II}}{\kappa_n^2+\sigma_{\II}^2}\phi_n+\sum\lm_{m\neq n}\tau_{nm}\phi_m,\\
	\p_{\alpha}\psi_n\Big|_{\alpha=0}=\frac{\I \sigma_{\II}}{\kappa_n^2+\sigma_{\II}^2}\psi_n+\sum\lm_{m\neq n}\tau_{nm}\psi_m.
\end{split}
\ee
So that the result reads:
\be\label{Y_int}
Y=\frac{1}{4\pi}\int\lm_{-\Lambda/2}^{+\Lambda/2}d\kappa\left[\frac{1}{2\Omega(\kappa)}\left(1+\frac{\I\sigma_{\II}}{\Omega(\kappa)-\sigma_{\IR}}\right)+\frac{\I\sigma_{\II}}{\kappa^2+\sigma_{\II}^2}\right].
\ee
Comparing this expression with \eqref{Y_vev} we find for the disorder operator the following renormalized expression:
\be\label{Y}
Y=\langle|\phi|^2-\delta_{\alpha}\rangle=-\frac{1}{2\pi}\log\left[\frac{\sigma}{\Lambda}\right].
\ee

Effective superpotential for $\sigma$ can be constructed from duality relation $Y={\bf w}'(\sigma)$:
\be\label{small_superpotential}
{\bf w}(\sigma)=-\frac{1}{2\pi}\sigma\left(\log\left[\frac{\sigma}{\Lambda}\right]-1\right).
\ee

Despite this expression repeats the case of the theory on a cylinder there are some crucial differences. In general we allow fields to vary along the interval $[0,L]$ in an arbitrary way. There appears a problem of a smoothness of the logarithm map since it is multivalued. When field $\sigma$ winds around the zero value the disorder operator expression \eqref{Y} gets shifted by $\I$. On the other hand, the arctangent function appearing in the integration result \eqref{Y_vev} is discontinuous. We would like to argue that this discontinuity is compensated by the zeroth mode whose contribution we omitted in \eqref{Y_int}.

According to \eqref{modes} (compare to \eqref{condensate}) zero mode $e_0(x)$ is pulled to either of two ends of interval $[0,L]$ depending on the sign of $\sigma_{\II}$.
Similarly to solitons it is localized in a neighborhood of some bulk point $x_0$ if the exponent behaves as a Gaussian, in other words if $\sigma_{\II}(x_0)=0$, $\sigma'_{\II}(x_0)<0$.
Notice that roots $x_0$ defined by this constraint are positions on interval $[0,L]$ where the logarithm function \eqref{Y} defined as an integral \eqref{Y_int} acquires discontinuities.
The zero mode localized at $x_0$ creates a quantized charge contribution $\delta(x-x_0)$ to $\p_x Y$ in \eqref{Gauss_Upsilon} canceling discontinuity in \ref{Y_int}
so that 
$$
\p_x Y=-\frac{1}{2\pi}\frac{\p_x\sigma}{\sigma}
$$
is a smooth single-valued function on $[0,L]$.

\subsection{Sheaf cohomology of \texorpdfstring{$\IC\IP^n$}{CPn}} \label{s:sheaf_coho_CP_n}
In this section we review a field theoretic approach to a canonical textbook calculation of sheaf valued cohomology:
$$
H^p(\IC\IP^n,\CO(k)).
$$
Projective variety $\IC\IP^{n}$ is a moduli space for a gauged linear sigma model with $n+1$ identical chiral fields with equivalent charges $Q_a=1$ and gauge group $U(1)$. So that $\IC\IP^{n}$  will be spanned by tuples $\phi_a$, $a=1,\ldots,n+1$ modulo complexified gauge transformations $\IC^{\times}$:
\be
\left(\phi_1,\phi_2,\ldots,\phi_{n+1}\right)\sim\left(\lambda\phi_1,\lambda\phi_2,\ldots,\lambda\phi_{n+1}\right)
\ee
Let us choose a coordinate patch where $\phi_{n+1}\neq 0$ and consider projective coordinates:
$$
\xi_a=\phi_a/\phi_{n+1},\quad a=1,\ldots,n.
$$

The supercharge is given by a contribution of only constant modes (see \eqref{eff_supercharge}):
\be
\begin{split}
	\bar \CQ=\sqrt{2}\sum\lm_{a=1}^{n+1}\bar\chi_a\left(\hbar\p_{\bar\phi_a}+\hbar^{\frac{1}{2}}\sigma_{\IR}\phi_a-\hbar\tilde\mu_{\IR,a}\phi_a\right)+\lambda\left(\hbar^{\frac{1}{2}}\p_{\sigma_{\IR}}+\left(\sum\lm_a|\phi_a|^2-r\right)\right),\\
	\CQ=\sqrt{2}\sum\lm_{a=1}^{n+1}\chi_a\left(-\hbar\p_{\phi_a}+\hbar^{\frac{1}{2}}\sigma_{\IR}\bar\phi_a-\hbar\tilde\mu_{\IR,a}\bar\phi_a\right)+\lambda\left(-\hbar^{\frac{1}{2}}\p_{\sigma_{\IR}}+\left(\sum\lm_a|\phi_a|^2-r\right)\right);
\end{split}
\ee
where we have saved the contribution of real masses $\mu_{\IR,a}$. In our physical realization $\IC\IP^n$ appears as an object of differential geometry rather than algebraic geometry. The action of $\IC^{\times}$ is split in action of $U(1)$ and a sphere constraint:
$$
\sum\lm_{a=1}^{n+1} |\phi_a|^2=r.
$$
We can split these actions explicitly using projective coordinates:
\be
\phi_a=\frac{\xi_a \rho e^{\I\vartheta}}{\sqrt{1+\sum\lm_b|\xi_b|^2}},\;a=1,\ldots,n,\quad \phi_{n+1}=\frac{\rho e^{\I\vartheta}}{\sqrt{1+\sum\lm_b|\xi_b|^2}},
\ee
where $\xi_a\in\IC$, $\rho,\;\vartheta\in\IR$.
For derivatives we have:
\be
\begin{split}
	\p_{\phi_a}=\rho^{-1}e^{-\I\vartheta}\sqrt{1+\sum\lm_b|\xi_b|^2}\p_{\xi_a}+\frac{\bar\xi_a}{2\sqrt{1+\sum\lm_b|\xi_b|^2}}e^{-\I\vartheta}\p_{\rho},\;a=1,\ldots,n;\\
	\p_{\phi_{n+1}}=-\rho^{-1}e^{-\I\vartheta}\sqrt{1+\sum\lm_b|\xi_b|^2}\sum\lm_a\xi_a\p_{\xi_a}+\frac{e^{-\I\vartheta}}{2\sqrt{1+\sum\lm_b|\xi_b|^2}}\p_{\rho}-\frac{\I \rho^{-1}e^{-\I\vartheta}}{2}\sqrt{1+\sum\lm_b|\xi_b|^2}\p_{\vartheta}.
\end{split}
\ee
The integration measure includes the Jacobian contribution:
\be
\prod\lm_{a=1}^{n+1}d\phi_ad\bar\phi_a\sim\rho^{2n+1}d\rho\times \frac{\prod\lm_{a=1}^nd\xi_ad\bar\xi_a}{\left(1+\sum\lm_b|\xi_b|^2\right)^{n+1}}
\ee
Let us redefine fermion fields in the following way:\footnote{Inverse transform: $$\chi_{n+1}=\frac{\psi}{\sqrt{1+\sum\lm_b|\xi_b|^2}}-\frac{\sum\lm_a\bar \xi_a\nu_a}{1+\sum\lm_b|\xi_b|^2},\quad \chi_a=\nu_a+\xi_a\chi_{n+1}.$$}
\be
\begin{split}
	\psi:=\frac{\sum\lm_a\bar\xi_a \chi_a+\chi_{n+1}}{\sqrt{1+\sum\lm_b|\xi_b|^2}};\quad
	\nu_a:=\chi_a-\xi_a\chi_{n+1},\;a=1,\ldots,n;
\end{split}
\ee
The commutation relations for the new fermion field read:\footnote{Initial fermion anti-commutation relations are:
	$$
	\left\{\lambda,\bar\lambda\right\}=1,\quad \left\{\chi_a,\bar\chi_b\right\}=\delta_{ab}.
	$$}\footnote{$\mathop{\rm Det}\lm_{a,b}\left\{\nu_a,\bar\nu_b\right\}=1+\sum\lm_c|\xi_c|^2.$}
\be
\left\{\psi,\bar\psi\right\}=1,\quad \left\{\nu_a,\bar\nu_b\right\}=\delta_{ab}+\xi_a\bar\xi_b=:\frac{g^{a\bar b}}{1+\sum\lm_c|\xi_c|^2},
\ee
where $g^{a\bar b}$ is the inverse of the canonical Fubini-Study metric on $\IC\IP^n$:
\be
ds^2=\sum\lm_{a,b=1}^n g_{a\bar b}d\xi_ad\bar\xi_b=\frac{\left(1+\sum\lm_b|\xi_b|^2\right)\sum\lm_a|d\xi_a|-\left|\sum\lm_a\bar\xi_ad\xi_a\right|^2}{\left(1+\sum\lm_b|\xi_b|^2\right)^2}.
\ee

The resulting supercharge reads:
\be
\begin{split}
	\bar \CQ=\lambda\left(\hbar^{\frac{1}{2}}\p_{\sigma_{\IR}}+\rho^2-r\right)+\\
	+\sqrt{2}\hbar\sum\lm_{a}\bar\nu_ae^{\I\vartheta}\rho^{-1}\sqrt{1+\sum\lm_b|\xi_b|^2}\left(\p_{\bar\xi_a}-\frac{\I\xi_a\p_{\vartheta}}{2\left(1+\sum\lm_b|\xi_b|^2\right)}-\frac{\xi_a\rho^2\left(\tilde\mu_{\IR,a}-\tilde\mu_{\IR,n+1}\right)}{\left(1+\sum\lm_b|\xi_b|^2\right)^2}\right)+\\
	+\sqrt{2}\bar\psi e^{\I\vartheta}\left(\frac{\hbar}{2}\p_{\rho}+\hbar^{\frac{1}{2}}\sigma_{\IR}\rho+\frac{\I \hbar}{2\rho}\p_{\vartheta}-\hbar\rho\frac{\tilde\mu_{\IR,n+1}+\sum\lm_b|\xi_b|^2\tilde\mu_{\IR,b}}{1+\sum\lm_b|\xi_b|^2}\right).
\end{split}
\ee
We expand field $\rho$ around its vacuum value:
$$
\rho\to\sqrt{r}+\hbar^{\frac{1}{2}}\rho
$$
The zeroth order supercharge and corresponding ground state wave function read:
\be
\begin{split}
	\bar\CQ^{(0)}=\lambda\left(\p_{\sigma_{\IR}}+2\sqrt{r}\rho\right)+\sqrt{2}\bar\psi e^{\I\vartheta}\left(\frac{1}{2}\p_{\rho}+\sqrt{r}\sigma_{\IR}\right),\\
	\Psi^{(0)}=e^{-\sqrt{\frac{r}{2}}\sigma_{\IR}^2-\sqrt{2r}\rho^2}\left(1+\bar\lambda\bar\psi e^{\I\vartheta}\right)|0\rangle.
\end{split}
\ee

After renormalization we find the following supercharges:\footnote{After renormalization the fermionic fields satisfy effectively vielbein equations (where we just annihilated perpendicular $\psi$-component):
	$$\p_{\xi_c}\nu_a=\delta_{ac}\frac{\sum\lm_b\bar\xi_b\nu_b}{1+\sum\lm_b|\xi_b|^2},\quad \p_{\bar\xi_c}\nu_a=0.$$}
\be
\begin{split}
	\bar\CQ_{\rm eff}=e^{-\fH}\sqrt{2}\sqrt{1+\sum\lm_b|\xi_b|^2}e^{\I\vartheta}\sum\lm_a\bar\nu_a\left(\p_{\bar\xi_a}-\frac{\I \xi_a}{2\left(1+\sum\lm_b|\xi_b|^2\right)}\p_{\vartheta}\right)e^{\fH},\\
	\CQ_{\rm eff}=e^{\fH}\sqrt{2}\sqrt{1+\sum\lm_b|\xi_b|^2}e^{-\I\vartheta}\sum\lm_a\nu_a\left(-\p_{\xi_a}-\frac{\I \bar\xi_a}{2\left(1+\sum\lm_b|\xi_b|^2\right)}\p_{\vartheta}\right)e^{-\fH},
\end{split}
\ee
where the height function:
$$
\fH=-r\frac{\sum\lm_{a=1}^n\mu_{\IR,a}|\xi_a|^2+\mu_{\IR,n+1}}{1+\sum\lm_{a=1}^n|\xi_a|^2}
$$
is a smooth function on the whole $\IC\IP^n$, therefore it does not affect the cohomologies and can be set to 0.

Electric charge operator is also constructed form constant modes:
\be
{\bf j}=\sum\lm_a\left(\bar\phi_a\p_{\bar\phi_a}-\phi_a\p_{\phi_a}+\bar\chi_a\chi_a\right)=\I\p_{\vartheta}+\bar\psi\psi+\left(1+\sum\lm_c|\xi_c|^2\right)\sum\lm_{a,b}g_{\bar ab}\bar\nu_a\nu_b.
\ee
For $\fH=0$ we find the following BPS states:
\be
\begin{split}
	\left|\underline{\ell}\right\rangle=\frac{e^{\I \ell_{n+1}\vartheta}\prod\lm_{b=1}^{n}\left(e^{\I\vartheta}\xi_b\right)^{\ell_b}}{\left(1+\sum\lm_{c=1}^n|\xi_c|^2\right)^{\frac{1}{2}\sum\lm_{a=1}^{n+1}\ell_a}}|0\rangle,\;{\bf j}\left|\underline{\ell}\right\rangle=\left(-\sum\lm_{a=1}^{n+1}\ell_a\right)\left|\underline{\ell}\right\rangle,\\ \ell_a\geq 0,\;a=1,\ldots,n+1;\\
	\left|\underline{\tilde\ell}\right\rangle=\frac{e^{-\I \left(\tilde\ell_{n+1}+1\right)\vartheta}\prod\lm_{b=1}^{n}\left(e^{-\I\vartheta}\bar\xi_a\right)^{\tilde\ell_a}\bar\nu_a}{\left(1+\sum\lm_{c=1}^n|\xi_c|^2\right)^{\frac{1}{2}+\frac{1}{2}\sum\lm_{b=1}^{n+1}\tilde\ell_b}}|0\rangle,\;{\bf j}\left|\underline{\tilde\ell}\right\rangle=\left(n+1+\sum\lm_{a=1}^{n+1}\tilde\ell_a\right)\left|\underline{\tilde\ell}\right\rangle,\\ \tilde\ell_a\geq 0,\;a=1,\ldots,n+1.
\end{split}
\ee
Comparing with the sheaf cohomologies calculated using $\check{\rm C}$ech cohomology machinery (see e.g. \cite[Theorem 4.66]{D_book_2}) we find a complete agreement:
\be
\fG_{\rm BPS}^{(f,k)}\cong H^f\left(\IC\IP^n,\CO(-k)\right),
\ee
where $f$ is a fermion number, and $k$ is $\bf j$-eigenvalue.

\subsection{Details of Bondal-Kapranov-Schechtman calculation}\label{s:BKS}

Here we give another proof of Proposition 1.14 and Corollary 1.21 in \cite{Bondal_Kapranov_Schecntman} combining Cartier divisor \cite{Watari_lectures} and $\check{\rm C}$ech cohomology (see Appendix \ref{s:DCoh}) tools.\footnote{
	The author would like to thank Taizan Watari for a suggestion to apply this technique in the calculation.}

\subsubsection{Varieties}

Consider three Calabi-Yau 3-folds: 
\begin{itemize}
	\item $X_+:$ $A\in{\rm Mat}_{2\times 2}(\IC)$, $\ell \in\IP^1_+$, $A\cdot\ell=0$ -- 3-fold.
	\item $X_-:$ $A'\in{\rm Mat}_{2\times 2}(\IC)$, $\ell' \in\IP^1_+$, $A'\cdot\ell'=0$ -- 3-fold.
	\item $X_0:=X_+\times_{(A=A'^T)}X_-$ -- 3-fold.\footnote{In this short-hand notation it is implied that $X_0$ is a fiber product along the subset of points $((A,\ell),(A^{T},\ell'))\in X_+\times X_-$}.
\end{itemize}

These varieties can be realized as the following complex bundles:
\begin{itemize}
	\item $X_+\cong\left(\CO(-1)^{\oplus 2}\to \IC\IP_+^1\right).$
	\item $X_-\cong\left(\CO(-1)^{\oplus 2}\to \IC\IP_-^1\right).$
	\item $X_0\cong\left(\CO(-1,-1)\to \IC\IP_+^1\times \IC\IP_-^1\right).$
\end{itemize}

\subsubsection{Projection maps and line bundles}

Consider the following map diagram:
$$
\begin{array}{c}
	\begin{tikzcd}
		{} & X_0 \ar[d,"\pi_0"] \ar[dl,"p_+"'] \ar[dr,"p_-"]& {}\\
		X_+\ar[d,"\pi_+"'] & \IC\IP^1_+\times \IC\IP^1_- \ar[dl]\ar[dr] & X_- \ar[d,"\pi_-"]\\
		\IC\IP^1_+ & {}& \IC\IP^1_-
	\end{tikzcd}
\end{array},
$$
where $\pi_i$ are just projections to the base, and $p_\pm$ is a forgetful projection omitting the first or the second factor in $X_0$.

We define the following line bundles:
$$
\CY_+(k):=\pi_+^*\left(\CO_{\IP_+^1}(k)\right),\quad \CY_0(k,m):=\pi_0^*\left(\CO_{\IP_+^1\times \IP_-^1}(k,m)\right),\quad \CY_-(m):=\pi_-^*\left(\CO_{\IP_-^1}(m)\right).
$$

Proposition 1.14 and Corollary 1.21 in \cite{Bondal_Kapranov_Schecntman} can be summarized in the following relations:
\begin{subequations}
	\begin{align}
		{\bf L}p_-^*\CY_-(k)&\cong\CY_0(0,k), \label{BKS1}\\
		{\bf R}p_{+*}\CY_0(k,0)&\cong\CY_+(k), \label{BKS2} \\
		{\bf R}p_{+*}\CY_0(k,1)&\cong\CJ\otimes \CY_+(k-1), \label{BKS3}
	\end{align}
\end{subequations}
where all complexes are concentrated in elements of the zeroth degree, and $\CJ$ is an ideal sheaf of zero sections of $\CO(-1)^{\oplus 2}$-bundle.

\subsubsection{Open covers}

Define an open cover $X_+=U_x\cup U_y$.

Local coordinates on $U_x$ are:
$$
\ell=\left(\begin{array}{c}
	x \\ 1
\end{array}\right),\quad A=\left(\begin{array}{cc}
	\alpha_x & -\alpha_x x\\
	\beta_x & -\beta_x x
\end{array}\right),\quad \CO_{X_+}(U_x)\cong \IC[x,\alpha_x,\beta_x].
$$

Local coordinates on $U_y$ are:
$$
\ell=\left(\begin{array}{c}
	1 \\ y
\end{array}\right),\quad A=\left(\begin{array}{cc}
	y\alpha_y & -\alpha_y\\
	y\beta_y & -\beta_y
\end{array}\right),\quad \CO_{X_+}(U_y)\cong \IC[y,\alpha_y,\beta_y].
$$

Over $U_x\cap U_y$ the following relations hold:
$$
xy=1,\quad \alpha_x=\alpha_y y,\quad \beta_x=\beta_y y.
$$

Define an open cover $X_-=U_z\cup U_w$.

Local coordinates on $U_z$ are:
$$
\ell'=\left(\begin{array}{cc}
	z & 1
\end{array}\right),\quad A'^T=\left(\begin{array}{cc}
	\gamma_z & \delta_z\\
	-z\gamma_z & -z \delta_z
\end{array}\right),\quad \CO_{X_-}(U_z)\cong \IC[z,\gamma_z,\delta_z].
$$

Local coordinates on $U_w$ are:
$$
\ell'=\left(\begin{array}{cc}
	1 & w
\end{array}\right),\quad A'^T=\left(\begin{array}{cc}
	w\gamma_w & w\delta_w\\
	-\gamma_w & -\delta_w
\end{array}\right),\quad \CO_{X_-}(U_w)\cong \IC[w,\gamma_w,\delta_w].
$$

Over $U_z\cap U_w$ the following relations hold:
$$
zw=1,\quad \gamma_z=w\gamma_w,\quad \delta_z=w\delta_w.
$$

Define an open cover $X_0=V_{xz}\cup V_{xw}\cup V_{yz}\cup V_{yw}$.

Local coordinates on $V_{xz}$ are:
$$
\ell=\left(\begin{array}{c}
	x \\ 1
\end{array}\right),\quad \ell'=\left(\begin{array}{cc}
	z & 1
\end{array}\right),\quad A=\zeta_{xz}\left(\begin{array}{cc}
	1 & -x\\
	-z & xz
\end{array}\right),\quad \CO_{X_0}(V_{xz})\cong \IC[x,z,\zeta_{xz}].
$$

Local coordinates on $V_{xw}$ are:
$$
\ell=\left(\begin{array}{c}
	x \\ 1
\end{array}\right),\quad \ell'=\left(\begin{array}{cc}
	1 & w
\end{array}\right),\quad A=\zeta_{xw}\left(\begin{array}{cc}
	w & -xw\\
	-1 & x
\end{array}\right),\quad \CO_{X_0}(V_{xw})\cong \IC[x,w,\zeta_{xw}].
$$

Local coordinates on $V_{yz}$ are:
$$
\ell=\left(\begin{array}{c}
	1 \\ y
\end{array}\right),\quad \ell'=\left(\begin{array}{cc}
	z & 1
\end{array}\right),\quad A=\zeta_{yz}\left(\begin{array}{cc}
	y & -1\\
	-yz & z
\end{array}\right),\quad \CO_{X_0}(V_{yz})\cong \IC[y,z,\zeta_{yz}].
$$

Local coordinates on $V_{yw}$ are:
$$
\ell=\left(\begin{array}{c}
	1 \\ y
\end{array}\right),\quad \ell'=\left(\begin{array}{cc}
	1 & w
\end{array}\right),\quad A=\zeta_{yw}\left(\begin{array}{cc}
	yw & -w\\
	-y & 1
\end{array}\right),\quad \CO_{X_0}(V_{yw})\cong \IC[y,w,\zeta_{yw}].
$$

On $V_{xz}\cap V_{xw}\cap V_{yz}\cap V_{yw}$ the following relations hold:
$$
xy=1,\quad zw=1,\quad \zeta_{xz}=w\,\zeta_{xw}=y\,\zeta_{yz}=yw\,\zeta_{yw}.
$$

\subsubsection{Line bundles}

\begin{itemize}
	\item Line bundle $\CY_+(k)$ on $X_+$ is given by a Cartier divisor:
	$$
	D_k^{(+)}=\left\{\left(U_x,\frac{1}{x^k}\right),\left(U_y,1\right) \right\}.
	$$
	\item Line bundle $\CY_-(k)$ on $X_-$ is given by a Cartier divisor:
	$$
	D_k^{(-)}=\left\{\left(U_z,\frac{1}{z^k}\right),\left(U_w,1\right) \right\}.
	$$
	\item Line bundle $\CY_0(k,m)$ on $X_0$ is given by a Cartier divisor:
	$$
	D_{k,m}^{(0)}=\left\{\left(V_{xz},\frac{1}{x^kz^m}\right),\left(V_{xw},\frac{1}{x^k}\right),\left(V_{yz},\frac{1}{z^m}\right),\left(V_{yw},1\right) \right\}.
	$$
\end{itemize}

$\CY_-(k)$ is a line bundle therefore it is locally free, so its projective resolution contains itself as a single $0^{\rm th}$ element.
Thus for \eqref{BKS1} we have:
$$
{\bf L}_ip_-^*\CY_-(k)={\bf L}_0p_-^*\CY_-(k)=p_-^*\CY_-(k)=\CY_0(0,k).
$$

\subsubsection{Direct images}

Since ${\rm dim}_{\IC}\; X_0={\rm dim}_{\IC}\; X_+=3$ the fiber of $p_+:\;X_0\to X_+$ is zero dimensional. 
Therefore only ${\bf R}^0p_{+*}$ has a chance to be non-trivial. 
Thus we have for a sheaf $\CF$ the following relations hold:
$$
{\bf R}^0p_{+*}(\CF)=p_{+*}(\CF),
$$
$$
\left[p_{+*}(\CF)\right](U)=\CF\left(p_+^{-1}\left(U\right) \right).
$$

\paragraph{Direct image \eqref{BKS2}.}

We have:
\be\nn
\begin{split}
	\CY_0(k,0)(V_{xz})& \cong\frac{\IC[x,z,\zeta_{xz}]}{x^k},\\
	\CY_0(k,0)(V_{xw})& \cong\frac{\IC[x,w,\zeta_{xw}]}{x^k},\\
	\CY_0(k,0)(V_{yz})& \cong\IC[y,z,\zeta_{yz}],\\
	\CY_0(k,0)(V_{yw})& \cong\IC[y,w,\zeta_{yw}].
\end{split}
\ee

Then we calculate: 
$$
\left[{\bf R}^0p_{+*}\CY_0(k,0)\right](U_x)=\CY_0(k,0)\left(V_{xz}\cup V_{xw}\right)=\frac{\IC[x,z,\zeta_{xz}]}{x^k}\cap \frac{\IC[x,w,\zeta_{xw}]}{x^k}
$$

Calculate corresponding sections:
$$
\psi=\frac{1}{x^k}\sum\lm_{i,j,k\geq 0}c_{ijk}x^iz^j\zeta_{xz}^k=\frac{1}{x^k}\sum\lm_{i',j',k'\geq 0}c_{i'j'k'}x^{i'}\frac{1}{z^{j'}}\left(z\zeta_{xz}\right)^{k'}.
$$

Then we have:
\be\nn
\begin{split}
	i=i',\quad
	k=k',\quad
	j'=k-j.
\end{split}
\ee

We have to constrain a region for admissible index values $j,j',k\geq 0$:
$$
\begin{array}{c}
	\begin{tikzpicture}
		\draw[->] (-2,0) -- (2,0);
		\draw[->] (0,-2) -- (0,2);
		\node[right] at (2,0) {$j$};
		\node[above] at (0,2) {$k$};
		\draw[dashed, fill=red, opacity=0.5] (0,0) -- (2,2) -- (0,2) -- cycle;
	\end{tikzpicture}
\end{array}
$$

For sections we have after substitution $k\to k+j$:
$$
\psi=\frac{1}{x^k}\sum\lm_{i,j,k\geq 0}c_{ijk}x^i z^j\zeta_{xz}^{k+j}=\frac{1}{x^k}\sum\lm_{i,j,k\geq 0}c_{ijk}x^i \left(z\zeta_{xz}\right)^j\zeta_{xz}^k=\frac{1}{x^k}\sum\lm_{i,j,k\geq 0}c_{ijk}x^i\alpha_x^k(-\beta_x)^j
$$

We conclude:
$$
\left[{\bf R}^0p_{+*}\CY_0(k,0)\right](U_x)\cong\frac{1}{x^k}\IC[x,\alpha_x,\beta_x]
$$

Similarly we calculate:
$$
\left[{\bf R}^0p_{+*}\CY_0(k,0)\right](U_y)\cong\IC[y,\alpha_y,\beta_y]
$$

This bundle is described by a divisor:
$$
D=\left\{\left(U_x,\frac{1}{x^k}\right),\left(U_y,1\right)\right\}
$$

So we arrive to the following conclusion:
$$
{\bf R}^0p_{+*}\CY_0(k,0)=\CO_{X_+}(D)=\CY_+(k)
$$

\paragraph{Direct image \eqref{BKS3}.} 

We have:
\be\nn
\begin{split}
	\CY_0(k,1)(V_{xz})& \cong\frac{1}{x^kz}\IC[x,z,\zeta_{xz}]\\
	\CY_0(k,1)(V_{xw})& \cong\frac{1}{x^k}\IC[x,w,\zeta_{xw}]\\
	\CY_0(k,1)(V_{yz})& \cong\frac{1}{z}\IC[y,z,\zeta_{yz}]\\
	\CY_0(k,1)(V_{yw})& \cong\IC[y,w,\zeta_{yw}]
\end{split}
\ee

Then we calculate: 
$$
\left[{\bf R}^0p_{+*}\CY_0(k,1)\right](U_x)=\CY_0(k,1)\left(V_{xz}\cup V_{xw}\right)=\frac{\IC[x,z,\zeta_{xz}]}{x^kz}\cap \frac{\IC[x,w,\zeta_{xw}]}{x^k}.
$$

We calculate corresponding sections:
$$
\psi=\frac{1}{x^kz}\sum\lm_{i,j,k\geq 0}c_{ijk}x^iz^j\zeta_{xz}^k=\frac{1}{x^k}\sum\lm_{i',j',k'\geq 0}c_{i'j'k'}x^{i'}\frac{1}{z^{j'}}\left(z\zeta_{xz}\right)^{k'}.
$$

Then we have:
\be\nn
\begin{split}
	i=i',\quad
	k=k',\quad
	j'=k+1-j.
\end{split}
\ee

We have to constrain a region for admissible index values $j,j',k\geq 0$:
$$
\begin{array}{c}
	\begin{tikzpicture}
		\draw[->] (-2,0) -- (2,0);
		\draw[->] (0,-2) -- (0,2);
		\node[right] at (2,0) {$j$};
		\node[above] at (0,2) {$k$};
		\draw[dashed] (0,-0.5) -- (2,1.5);
		\draw[dashed, fill=red, opacity=0.5] (0,0) -- (0.5,0) -- (2,1.5) -- (2,2) -- (0,2) -- cycle;
		\node[below right] at (0.5,0) {$1$};
		\node[left] at (0,-0.5) {$-1$};
	\end{tikzpicture}
\end{array}
$$

We organize these series in the following way:
$$
\psi=\frac{1}{x^k z}\left(\sum\lm_{i,j,k\geq 0}c_{ijk}x^i z^j\zeta_{xz}^{k+j}+z \sum\lm_{i,j,k\geq 0}\tilde c_{ijk}x^i z^j\zeta_{xz}^{k+j}\right).
$$

We conclude:
\be\label{integral_1}
\begin{split}
	\left[{\bf R}^0p_{+*}\CY_0(k,1)\right](U_x)&\cong\frac{\alpha_x\IC[x,\alpha_x,\beta_x]\oplus\beta_x\IC[x,\alpha_x,\beta_x]}{x^k\beta_x}=\\
	&=\left(\alpha_x\IC[x,\alpha_x,\beta_x]\oplus\beta_x\IC[x,\alpha_x,\beta_x]\right)\otimes \frac{\IC[x,\alpha_x,\beta_x]}{x^k\beta_x}.
\end{split}
\ee

Similarly we derive:
\be\label{integral_2}
\begin{split}
	\left[{\bf R}^0p_{+*}\CY_0(k,1)\right](U_y)&\cong\frac{\alpha_y\IC[y,\alpha_y,\beta_y]\oplus\beta_y\IC[y,\alpha_y,\beta_y]}{\beta_y}=\\
	&=\left(\alpha_y\IC[y,\alpha_y,\beta_y]\oplus\beta_y\IC[y,\alpha_y,\beta_y]\right)\otimes \frac{\IC[y,\alpha_y,\beta_y]}{\beta_y}.
\end{split}
\ee

We think of $X_+$ as a $\CO(-1)^{\oplus 2}$-bundle over $\IC\IP^1$ with $\alpha_x$, $\beta_x$ being fiber coordinates. 
An ideal sheaf corresponding to zero sections of  $\CO(-1)^{\oplus 2}$ we denote as $\CJ$:
$$
\CJ(U_x)\cong \alpha_x\IC[x,\alpha_x,\beta_x]\oplus\beta_x\IC[x,\alpha_x,\beta_x],\quad
\CJ(U_y)\cong \alpha_y\IC[y,\alpha_y,\beta_y]\oplus\beta_y\IC[y,\alpha_y,\beta_y]\,.
$$

For  the direct image functor we have:
$$
{\bf R}^0p_{+*}\CY_0(k,1)\cong \CJ\otimes \CO_{X_+}(D),
$$
where 
$$
D=\left\{\left(U_x,\frac{1}{x^k\beta_x}\right),\left(U_y,\frac{1}{\beta_y}\right)\right\}.
$$

Notice that divisor $D$ is linear equivalent to divisor $\tilde D$:
$$
\tilde D=\left\{\left(U_x,\frac{1}{x^{k-1}}\right),\left(U_y,1\right)\right\},
$$
therefore
$$
\CO_{X_+}(D)\cong \CO_{X_+}(\tilde D)\cong \CY_+(k-1).
$$

Thus we conclude:
$$
{\bf R}^0p_{+*}\CY_0(k,1)=\CJ\otimes \CY_{+}(k-1)\,.
$$

\subsubsection{Grothendieck groups}\label{Groth_groups}
Suppose $C$ is an image of the zero section of the bundle $\CO(-1)^{\oplus 2}$. 
Then we have an exact sequence:
$$
\begin{array}{c}
	\begin{tikzcd}
		0\ar[r]& \CO_{X_+}(U)\ar[rr,"\CJ(U)\otimes"]& &\CO_{X_+}(U)\ar[r]& \CO_C(U)\ar[r]& 0
	\end{tikzcd}
\end{array}\,.
$$
Thus we conclude that in $K(X_+)$:
$$
\left[\CO_C\right]=0\,,
$$
and
$$
\left[\CJ\otimes \CY_+(k)\right]=\left[\CY_+(k)\right]\,.
$$

As well we have a conclusion from the Kozsul complex:
\be\nn
\left[\CY(k)\right]-2\left[\CY(k+1)\right]+\left[\CY(k+2)\right]=0\,.
\ee

Thus we derive the following expression for the Fourier-Mukai transform on K-theory classes:
$$
\left(\left[\CY_-(0)\right],\left[\CY_-(1)\right]\right)\mapsto\left({\rm dim}\,{\rm Hom}(\CO,\CO(1))\cdot \left[\CY_+(0)\right]+(-1)\cdot\left[\CY_+(1)\right],\left[\CY_+(0)\right]\right)\,.
$$

\def\mygreen{black!40!green}
\subsection{Braiding fixed points on a quiver variety}\label{sec:app_quiver}
Here we will consider braiding on quiver varieties discussed in Section \ref{sec:quiver} in details for an example of $\CS_{3,4}$:
\be
\begin{array}{c}
	\begin{tikzpicture}[scale=0.4]
		\draw[thick] (0,0) -- (15,0) (9,0) -- (9,-1.3) (6,0) -- (6,-1.3);
		\draw[fill=white] (0,0) circle (0.3) (3,0) circle (0.3) (6,0) circle (0.3) (9,0) circle (0.3) (12,0) circle (0.3) (15,0) circle (0.3);
		\node[above] at (0,0.3) {$1$};
		\node[above] at (3,0.3) {$2$};
		\node[above] at (6,0.3) {$3$};
		\node[above] at (9,0.3) {$3$};
		\node[above] at (12,0.3) {$2$};
		\node[above] at (15,0.3) {$1$};
		\begin{scope}[shift={(9,-1.3)}]
			\draw[fill=white] (-0.3,-0.3) -- (-0.3,0.3) -- (0.3,0.3) -- (0.3,-0.3) -- cycle;
		\end{scope}
		\begin{scope}[shift={(6,-1.3)}]
			\draw[fill=white] (-0.3,-0.3) -- (-0.3,0.3) -- (0.3,0.3) -- (0.3,-0.3) -- cycle;
		\end{scope}
		\node[below] at (9,-1.6) {$1$};
		\node[below] at (6,-1.6) {$1$};
		\node[left] at (-0.5,0) {$\CS_{3,4}=$};
	\end{tikzpicture}
\end{array}
\ee

In this framework we will consider braiding action along $\wp_{3,4}$ on the following subspace:
\be
\begin{array}{c}
	\begin{tikzpicture}
		\node(0) at (0,0) {$-$};
		\node[left] at (0) {$($};
		\node[below right] at (0) {$,$};
		\node(1) at (0.5,0) {$+$};
		\node[below right] at (1) {$,$};
		\node(2) at (1,0) {$-$};
		\node[below right] at (2) {$,$};
		\node(3) at (1.5,0) {$-$};
		\node[below right] at (3) {$,$};
		\node(4) at (2,0) {$+$};
		\node[below right] at (4) {$,$};
		\node(5) at (2.5,0) {$+$};
		\node[below right] at (5) {$,$};
		\node(6) at (3,0) {$-$};
		\node[right] at (6) {$)$};
		\begin{scope}[shift={(0,-1.5)}]
			\node(0) at (0,0) {$-$};
			\node[left] at (0) {$($};
			\node[below right] at (0) {$,$};
			\node(1) at (0.5,0) {$+$};
			\node[below right] at (1) {$,$};
			\node(2) at (1,0) {$-$};
			\node[below right] at (2) {$,$};
			\node(3) at (1.5,0) {$+$};
			\node[below right] at (3) {$,$};
			\node(4) at (2,0) {$-$};
			\node[below right] at (4) {$,$};
			\node(5) at (2.5,0) {$+$};
			\node[below right] at (5) {$,$};
			\node(6) at (3,0) {$-$};
			\node[right] at (6) {$)$};
		\end{scope}
		\begin{scope}[shift={(0,-2)}]
			\node(0) at (0,0) {\tiny 0};
			\node(1) at (0.5,0) {\tiny 1};
			\node(2) at (1,0) {\tiny 2};
			\node(3) at (1.5,0) {\tiny 3};
			\node(4) at (2,0) {\tiny\color{orange} 4};
			\node(5) at (2.5,0) {\tiny 5};
			\node(6) at (3,0) {\tiny 6};
		\end{scope}
		\begin{scope}[shift={(-1.5,0)}]
			\draw[thick] (3,-0.3) -- (3,-0.5) -- (3.5,-0.5) -- (3.5,-0.3) (3,-1.2) -- (3,-1) -- (3.5,-1) -- (3.5,-1.2) (3.25,-0.5) -- (3.25,-1);
		\end{scope}
	\end{tikzpicture}
\end{array}
\ee

Corresponding diagram with migrating cell marked by the orange marker is depicted in Fig.\ref{fig:App_G}(a).
\begin{figure}[h!]
	\begin{center}
		\begin{tikzpicture}
			\node at (0,0) {$\begin{array}{c}
					\begin{tikzpicture}[scale=0.7]
						\draw[thick,fill=gray] (0,0) -- (0,3) -- (1,3) -- (1,2) -- (2,2) -- (2,1) -- (3,1) -- (3,0) -- cycle;
						\draw[thick, fill=white!40!blue] (4,3) -- (1,3) -- (1,2) -- (3,2) -- (3,0) -- (4,0) -- cycle;
						\draw[thick, fill=orange] (2,1) -- (3,1) -- (3,2) -- (2,2) -- cycle;
						\foreach \i in {1,...,3}
						{
							\draw (\i,0) -- (\i,3);
						}
						\foreach \i in {1,...,2}
						{
							\draw (0,\i) -- (4,\i);
						}
						\node at (0.5,0.5) {3};
						\node at (1.5,0.5) {4};
						\node at (2.5,0.5) {5};
						\node at (3.5,0.5) {6};
						\node at (0.5,1.5) {2};
						\node at (1.5,1.5) {3};
						\node at (2.5,1.5) {4};
						\node at (3.5,1.5) {5};
						\node at (0.5,2.5) {1};
						\node at (1.5,2.5) {2};
						\node at (2.5,2.5) {3};
						\node at (3.5,2.5) {4};
					\end{tikzpicture}
				\end{array}$};
			\node at (5,0) {$\begin{array}{c}
					\begin{tikzpicture}[scale=0.7]
						\draw[thick,fill=gray] (0,0) -- (0,3) -- (1,3) -- (1,2) -- (2,2) -- (2,1) -- (3,1) -- (3,0) -- cycle;
						\draw[thick, fill=white!40!blue] (4,3) -- (1,3) -- (1,2) -- (3,2) -- (3,0) -- (4,0) -- cycle;
						\draw[thick, fill=orange] (2,1) -- (3,1) -- (3,2) -- (2,2) -- cycle;
						\foreach \i in {1,...,3}
						{
							\draw (\i,0) -- (\i,3);
						}
						\foreach \i in {1,...,2}
						{
							\draw (0,\i) -- (4,\i);
						}
						\node at (0.5,0.5) {$e^{(3)}_{1}$};
						\node at (1.5,0.5) {$e^{(4)}_{2}$};
						\node at (2.5,0.5) {$e^{(5)}_{2}$};
						\node at (3.5,0.5) {$e^{(6)}_{1}$};
						\node at (0.5,1.5) {$e^{(2)}_{1}$};
						\node at (1.5,1.5) {$e^{(3)}_{3}$};
						\node at (2.5,1.5) {$e^{(4)}_{3}$};
						\node at (3.5,1.5) {$e^{(5)}_{1}$};
						\node at (0.5,2.5) {$e^{(1)}_{1}$};
						\node at (1.5,2.5) {$e^{(2)}_{2}$};
						\node at (2.5,2.5) {$e^{(3)}_{2}$};
						\node at (3.5,2.5) {$e^{(4)}_{1}$};
					\end{tikzpicture}
				\end{array}$};
			\node at (0,-1.5) {(a)};
			\node at (5,-1.5) {(b)};
		\end{tikzpicture}
	\end{center}
	\caption{Migrating cell and vector assignment.}\label{fig:App_G}
\end{figure}

Cells of the diagrams define diagonal expectation values of the scalar fields in the gauge multiplets according to the rule \eqref{eigen}. 
We also identify vectors of vector spaces $V_a$ associated with gauge nodes as eigen vectors of matrices $\langle\sigma^{(c)}\rangle$ so that a single eigen value and single vector corresponds to each cell of the diagram:
\be
\begin{split}
	&\langle\sigma^{(c)}\rangle\; e^{(c)}_{\alpha}=\sigma_{\alpha}^{(c)}\; e^{(c)}_{\alpha},\\
	&V_a={\rm Span}\left\{ e_i^{(a)}\right\}_{i=1}^{v_a},\quad a=1,\ldots,m-1.
\end{split}
\ee
Vector assignment is depicted in Fig.\ref{fig:App_G}(b).

Let us express $\IC\IP^1$-volumes for the first fixed point in terms of field expectation values:
\be
\begin{split}
	&{\vec U}=\big(1,\; \bar B_{1,1,1} \bar  B_{2,1,1} \bar \Gamma _{3,1,1},\; \bar B_{2,2,2} \bar B_{3,2,1} \bar \Gamma _{4,1,1},\; \bar B_{3,2,1} \bar \Gamma _{4,1,1},\\
	& \bar A_{2,3,1}\bar  A_{3,3,3} \bar B_{2,1,1} \bar \Gamma _{3,1,1},\; \bar A_{3,2,1}\bar  A_{4,2,2} \bar \Gamma _{3,1,1}, \; \bar A_{4,1,1} \bar A_{5,1,1} \bar \Gamma _{4,1,1}\big),
\end{split}
\ee
where $\bar X_{a,i,j}$ is an expectation value of $(i,j)$-matrix element of field $X_a$. 
To define these elements one has to solve algebraic equations \eqref{Nakajima} explicitly, it is hard to do analytically even with the knowledge that only a single solution exists, therefore we solve them numerically in the limit $r_4=s^2$, $s\to 0$ and for some generic assignment of other FI parameters:
\be
\begin{split}
	& {\vec r}=\left(1.45,1.17,1.07,{\color{red} s^2},1.70,1.03\right),\\
	& {\vec U}({\vec r})=\left(1.00,5.36,3.61,3.34,2.60\;{\color{red}s},5.05,3.76\right).
\end{split}
\ee
Form these expressions it is clear that the flag variety becomes singular as $s$ approaches 0.

As it is clear form Fig.\ref{fig:App_G}(b) the migrating scalar field is eigenvalue  ${\color{red} \sigma_{3}^{(4)}}$.  

Some chiral fields produce vacuum condensates that do not disappear in the limit $s\to 0$. We denote these condensates in the following way:
\be
{\color{blue} x_1},\;{\color{blue} x_2},\;{\color{blue} x_3},\;{\color{blue} x_4},\;{\color{blue} x_5},\;{\color{blue} x_6},\;{\color{blue} x_7},\;{\color{blue} x_8},\;{\color{blue} x_9},\;{\color{blue} x_{10}},\;{\color{blue} x_{11}}.
\ee
These degrees of freedom are not dynamical and not independent moduli. 
They are constrained further by some remnant $\bf D$-term relations, however it is hard to resolve these relations analytically, so we leave these fields in expressions assuming they are some generic variables of order $O(1)$.

There are field condensates that scale as $s$. 
It is not hard to define what are these fields form diagram in Fig.\ref{fig:App_G}(b).
These fields correspond to matrix elements mapping vectors associated with horizontal and vertical nearest neighbors of the migrating cell to the very vector associated with the migrating cell:
\be
\begin{split}
	&A_{3,3,3}\sim\nu_1s:\;e_3^{(3)}\mapsto e_3^{(4)},\quad  B_{5,2,3}\sim\nu_2s:\;e_2^{(5)}\mapsto e_3^{(4)},\\
	&A_{3,2,3}\sim\nu_3s:\;e_2^{(3)}\mapsto e_3^{(4)},\quad B_{5,1,3}\sim\nu_4s:\;e_1^{(5)}\mapsto e_3^{(4)}.
\end{split}
\ee
Due to $\bf F$-term constraint these fields are not independent.
There are two independent combinations we denote as $\color{orange} y_1$, $\color{orange} y_2$.

Quantum fields are tangent field to the quiver variety locus \eqref{Nakajima}, where we can substitute simultaneously the $\bf D$-term constraint by stability conditions (all our representations are small perturbations of stable representations and are automatically stable) and the action of the gauge group by the complexified gauge group.
To characterize  these degrees of freedom we use another small parameter $t$ for ``tangent''.
We will work with expansions in both $s$ and $t$.
The complexified gauge group acts on quantum degrees of freedom by corresponding algebraic shift actions.
So that a quantum field $\delta \phi_{a\to b}$ associated with arrow $a\to b$ in the quiver is shifted by an algebraic element $g$ as
\be
\left\{g_a\right\}_{a=1}^{m-1}\in\fg:\quad t\;\delta \phi_{a\to b}\mapsto t\;\delta \phi_{a\to b}+t\; g_b\cdot\langle \phi_{a\to b}\rangle- t\; \langle \phi_{a\to b}\rangle\cdot g_a,
\ee
where $\langle \phi_{a\to b}\rangle$ is a classical vacuum average for the field associated with arrow $a\to b$.
First we find independent deformation fields describing the tangent vector space to $\bf F$-term constraint then use the gauge transformations to eliminate gauge-dependent degrees of freedom.
Remaining gauge-invariant degrees of freedom are:
\be\nn
\begin{array}{|c||c|c|c|c|}
	\hline
	\mbox{Fields} & {\color{black!40!green} u_1} &{\color{black!40!green} u_2} & {\color{black!40!green} u_3} & {\color{black!40!green} u_4} \\
	\hline
	\mbox{Masses}&-\mu _-+\mu _+-\epsilon _1& -\mu _-+\mu _+-\epsilon _1& \mu _--\mu _+-\epsilon _2 & \mu _+-{\color{red} \sigma_3^{(4)}} \\
	\hline
	\hline
	\mbox{Fields} & {\color{\mygreen} u_5}& {\color{black!40!green} u_6} & {\color{black!40!green} u_7} & {\color{black!40!green} u_8}\\
	\hline
	\mbox{Masses}& \mu _--{\color{red} \sigma_3^{(4)}}+\epsilon _1& -\mu _++{\color{red} \sigma_3^{(4)}}-\epsilon _1-\epsilon _2& -\mu _-+{\color{red} \sigma_3^{(4)}}-2 \epsilon _1-\epsilon _2 & \mu _--\mu _+-\epsilon _2\\
	\hline
\end{array}
\ee

In these terms the quiver maps read:
\be\nn
A_1=\left(\begin{array}{c}   0\\  \frac{{\color{blue} x_7} {\color{blue} x_8} {\color{\mygreen} u_1} t}{{\color{blue} x_5} {\color{blue} x_6}}\\  \end{array} \right),\;
B_1=\left(\begin{array}{cc}   {\color{blue} x_5} & {\color{\mygreen} u_3} t\\  \end{array} \right),\;
A_5=\left(\begin{array}{cc}   {\color{blue} x_4} & 0\\  \end{array} \right),\;
B_5=\left(\begin{array}{c}   0\\  {\color{\mygreen} u_8} t\\  \end{array} \right),
\ee
\be\nn
A_2=\left(\begin{array}{cc}   -\frac{{\color{blue} x_3} {\color{\mygreen} u_5} {\color{orange} y_1} s t}{{\color{blue} x_6} {\color{blue} x_9}} & 0\\  \frac{{\color{blue} x_8} {\color{\mygreen} u_1} t}{{\color{blue} x_6}} & 0\\  \frac{{\color{blue} x_1} {\color{blue} x_9}}{{\color{blue} x_6}} & 0\\  \end{array} \right),\;
B_2=\left(\begin{array}{ccc}   {\color{blue} x_6} & \frac{{\color{blue} x_2} {\color{blue} x_6} {\color{\mygreen} u_5} {\color{orange} y_2} s t}{{\color{blue} x_1} {\color{blue} x_8} {\color{blue} x_9}} & \frac{{\color{blue} x_3} {\color{blue} x_6} {\color{\mygreen} u_5} {\color{orange} y_1} s t}{{\color{blue} x_1} {\color{blue} x_9}^2}\\  0 & {\color{blue} x_7} & 0\\  \end{array} \right),\;
B_3=\left(\begin{array}{ccc}   \frac{{\color{blue} x_2} {\color{blue} x_4} {\color{\mygreen} u_8} t}{{\color{blue} x_1} {\color{blue} x_3}} & 0 & 0\\  {\color{blue} x_8} & 0 & {\color{\mygreen} u_4} t\\  0 & {\color{blue} x_9} & {\color{\mygreen} u_5} t\\  \end{array} \right),
\ee
\be\nn
A_3=\left(\begin{array}{ccc}   {\color{\mygreen} u_1} t & -\frac{{\color{blue} x_2} {\color{\mygreen} u_4} {\color{orange} y_2} s t}{	{\color{blue} x_8}^2} & -\frac{{\color{blue} x_3} {\color{\mygreen} u_4} {\color{orange} y_1} s t}{{\color{blue} x_8} {\color{blue} x_9}}\\  {\color{blue} x_1} & 0 & 0\\  0 & \frac{{\color{blue} x_2} {\color{gray} g_{4,3,3}} {\color{orange} y_2} s t}{{\color{blue} x_8}}+\frac{{\color{blue} x_2} {\color{\mygreen} u_6} t}{{\color{blue} x_8}}+\frac{{\color{blue} x_2} {\color{orange} y_2} s}{{\color{blue} x_8}} & \frac{{\color{blue} x_3} {\color{gray} g_{4,3,3}} {\color{orange} y_1} s t}{{\color{blue} x_9}}+ \frac{({\color{blue} x_3} {\color{\mygreen} u_7} + {\color{\mygreen} u_2} {\color{orange} y_2} s) t}{{\color{blue} x_9}^2}+\frac{{\color{blue} x_3} {\color{orange} y_1} s}{{\color{blue} x_9}}\\  \end{array} \right),
\ee
\be\nn
A_4=\left(\begin{array}{ccc}   {\color{blue} x_2} & {\color{\mygreen} u_2} t & 0\\  0 & {\color{blue} x_3} & 0\\  \end{array} \right),\;
B_4=\left(\begin{array}{cc}   0 & 0\\  \frac{{\color{blue} x_4} {\color{\mygreen} u_8} t}{{\color{blue} x_3}} & 0\\  {\color{orange} y_2} s+{\color{\mygreen} u_6} t+{\color{gray} g_{4,3,3}} {\color{orange} y_2} s t & {\color{orange} y_1} s+{\color{\mygreen} u_7} t+{\color{gray} g_{4,3,3}} {\color{orange} y_1} s t\\  \end{array} \right),
\ee
\be\nn
\Gamma_3=\left(\begin{array}{c}   {\color{blue} x_{10}}\\  0\\  0\\  \end{array} \right),\;
\Delta_3=\left(\begin{array}{ccc}   \frac{{\color{blue} x_3} {\color{\mygreen} u_5} {\color{orange} y_1} s t}{{\color{blue} x_{10}} {\color{blue} x_9}} & 0 & 0\\  \end{array} \right),\;
\Gamma_4=\left(\begin{array}{c}   {\color{blue} x_{11}}\\  0\\  0\\  \end{array} \right),\;
\Delta_4=\left(\begin{array}{ccc}   \frac{{\color{blue} x_2} {\color{\mygreen} u_4} {\color{orange} y_2} s t}{{\color{blue} x_{11}} {\color{blue} x_8}} & \frac{{\color{blue} x_3} {\color{\mygreen} u_4} {\color{orange} y_1} s t}{{\color{blue} x_{11}} {\color{blue} x_8}} & 0\\  \end{array} \right).
\ee

We have eliminated all the gauge algebra variables except $\color{gray} g_{4,3,3}$ that has the same quantum numbers as migrating scalar field $\sigma_3^{(4)}$, in particular, this algebraic element maps vector $e_3^{(4)}$ to itself.
This happens since it is coupled to fields $s{\color{orange} y_{1,2}}$. 
If $s$ is not small and one of fields $y_{1,2}$, say, $y_1$ is non-zero we could use this expectation value to cancel $\color{gray} g_{4,3,3}$ with one of tangent degrees of freedom:
$$
g_{4,3,3}=-\frac{u_7}{s y_1},
$$
however in the limit $s\to 0$ this gauge transform is singular.
Similarly a Higgs mass for the corresponding gauge degree of freedom is generated:
$$
m_{\rm Higgs}\sim s\sqrt{|{\color{orange} y_1}|^2+|{\color{orange} y_2}|^2}
$$
Clearly in the limit $s\to 0$ this gauge degree of freedom remains massless and contributes to the unbroken $U(1)$ gauge symmetry of effective $\CS_{1,1}$ theory.

Similarly, fields ${\color{\mygreen} u_6}$ and ${\color{\mygreen} u_7}$ correspond to quantum fluctuations of classical condensates $s{\color{orange} y_{1,2}}$, in the limit $s\to 0$ they get mixed.
Since we assume that $s{\color{orange} y_{1,2}}$ are also of quantum order in what follows we change variables as:
\be
s {\color{orange} y_1}+t {\color{\mygreen} u_7}+s t {\color{orange} y_1}{\color{gray} g_{4,3,3}}\to s {\color{orange} y_1},\quad s {\color{orange} y_2}+t {\color{\mygreen} u_6}+s t {\color{orange} y_2}{\color{gray} g_{4,3,3}}\to s {\color{orange} y_2}.
\ee
Mentioned fields do not appear in combinations other than the presented one.

We derive the following $\bf D$-term and $\bf F$-term equations for these fields by expanding initial $\bf D$-term and $\bf F$-term to corresponding orders in $s$ and $t$:
\be
\begin{split}
	&{\bf D}\mbox{-term:}\; s^2\left(1+\frac{|{\color{blue} x_3}|^2}{|{\color{blue} x_9}|^2}\right)|{\color{orange} y_1}|^2+s^2\left(1+\frac{|{\color{blue} x_2}|^2}{|{\color{blue} x_8}|^2}\right)|{\color{orange} y_2}|^2-\\
	&\qquad\qquad\qquad-t^2|{\color{\mygreen} u_4}|^2-t^2|{\color{\mygreen} u_4}|^2-r_4+O(s^2t,t^2s)=0,\\
	&{\bf F}\mbox{-term:}\;-st\left(\frac{{\color{blue} x_3}}{{\color{blue} x_9}}{\color{orange} y_1}{\color{\mygreen} u_5}+\frac{{\color{blue} x_2}}{{\color{blue} x_8}}{\color{orange} y_2}{\color{\mygreen} u_4}\right)+O(s^2,t^2)=0.
\end{split}
\ee
We conclude that in the IR the theory flows to the following effective theory:
\be
\begin{array}{c}
	\begin{tikzpicture}
		\node (A) {$\begin{array}{c}
				\begin{array}{c}
					\begin{tikzpicture}
						\node at (-2,0) {$1$};
						\node at (0,0) {$1$};
						\node at (2,0) {$1$};
						\draw (0,0) circle (0.3);
						\begin{scope}[shift={(2,0)}]
							\draw (-0.3,-0.3) -- (-0.3,0.3) -- (0.3,0.3) -- (0.3,-0.3) -- cycle;
						\end{scope}
						\begin{scope}[shift={(-2,0)}]
							\draw (-0.3,-0.3) -- (-0.3,0.3) -- (0.3,0.3) -- (0.3,-0.3) -- cycle;
						\end{scope}
						\draw[<-] (1.7,0.1) -- (0.3,0.1);
						\draw[->] (1.7,-0.1) -- (0.3,-0.1);
						\draw[->] (-1.7,0.1) -- (-0.3,0.1);
						\draw[<-] (-1.7,-0.1) -- (-0.3,-0.1);
						\node[above] at (-1,0.1) {$\color{orange} y_1$};
						\node[below] at (-1,-0.1) {$\color{\mygreen} u_5$};
						\node[above] at (1,0.1) {$\color{\mygreen} u_4$};
						\node[below] at (1,-0.1) {$\color{orange} y_2$};
					\end{tikzpicture}
				\end{array}\oplus\mbox{Free } \left({\color{\mygreen}u_1},{\color{\mygreen}u_2},{\color{\mygreen}u_3},{\color{\mygreen}u_8}\right)
			\end{array}$};
		\node at ([shift={(-1.75,0)}]A.north) {${\color{red} \sigma_{3}^{(4)}}$};
	\end{tikzpicture}
\end{array}
\ee

General prescription for Maffei's map construction is rather involved, however for fixed points there is a simpler solution \cite[Lemma 3.2]{2019arXiv191003010I}.
To find the cotangent bundle we have to use perturbation theory and \cite[Lemma 18]{maffei2005quiver} to restrict perturbative degrees of freedom. 
\cite[Lemma 18]{maffei2005quiver} guarantees that we will find a unique up to gauge transformations solution.
We will not list here the results of our calculations since manipulations are rather simple however expressions are rather bulky.
The result for orthogonal decomposition reads up to $O(s^2,t^2)$:
\be
\begin{split}
	&\ell_0=\left(\begin{array}{ccccccc}   1 & 0 & 0 & 0 & 0 & 0 & 0\\  \end{array} \right),\\
	&\ell_1=\left(\begin{array}{ccccccc}   \frac{{\color{blue} x_{11}} {\color{blue} x_2} {\color{blue} x_4} {\color{\mygreen} u_8} t}{{\color{blue} x_1} {\color{blue} x_{10}} {\color{blue} x_3}}+\frac{{\color{blue} x_{11}} {\color{blue} x_2} {\color{\mygreen} u_5} {\color{orange} y_2} s t}{{\color{blue} x_1} {\color{blue} x_{10}} {\color{blue} x_9}}+\frac{{\color{blue} x_{11}} {\color{blue} x_7} {\color{blue} x_8} {\color{\mygreen} u_3} t}{{\color{blue} x_{10}} {\color{blue} x_5} {\color{blue} x_6}} & 0 & 0 & 0 & 1 & 0 & 0\\  \end{array} \right),\\
	& \ell_2=\left(\begin{array}{ccccccc}   \frac{{\color{blue} x_2} {\color{\mygreen} u_4} {\color{orange} y_2} s t}{{\color{blue} x_8}} & 1 & 0 & 0 & \frac{{\color{blue} x_{10}} {\color{\mygreen} u_1} t}{{\color{blue} x_{11}}} & 0 & 0\\  \end{array}\right), \\
	&E_{3,4}=\left(\begin{array}{ccccccc}   0 & \frac{{\color{blue} x_2} {\color{\mygreen} u_4} {\color{orange} y_2} s t}{{\color{blue} x_8}} & 1 & 0 & \frac{{\color{blue} x_1} {\color{blue} x_{10}} {\color{blue} x_3} {\color{\mygreen} u_4} {\color{orange} y_1} s t}{{\color{blue} x_{11}} {\color{blue} x_8}} & 0 & 0\\  0 & \frac{{\color{blue} x_{11}} {\color{blue} x_2} {\color{blue} x_4} {\color{\mygreen} u_8} t}{{\color{blue} x_1} {\color{blue} x_{10}} {\color{blue} x_3}}+\frac{{\color{blue} x_{11}} {\color{blue} x_2} {\color{\mygreen} u_5} {\color{orange} y_2} s t}{{\color{blue} x_1} {\color{blue} x_{10}} {\color{blue} x_9}} & 0 & 0 & \frac{{\color{blue} x_3} {\color{\mygreen} u_5} {\color{orange} y_1} s t}{{\color{blue} x_9}} & 1 & 0\\  \end{array} \right),\\
	&\ell_5=\left(\begin{array}{ccccccc}   0 & 0 & \frac{{\color{blue} x_{11}} {\color{blue} x_2} {\color{blue} x_4} {\color{\mygreen} u_8} t}{{\color{blue} x_1} {\color{blue} x_{10}} {\color{blue} x_3}} & 0 & 0 & 0 & 1\\  \end{array} \right),\\
	&\ell_6=\left(\begin{array}{ccccccc}   0 & 0 & 0 & 1 & 0 & 0 & \frac{{\color{blue} x_{10}} {\color{blue} x_2} {\color{blue} x_4} {\color{\mygreen} u_1} t + {\color{blue} x_1} {\color{blue} x_{10}} {\color{blue} x_4} {\color{\mygreen} u_2} t}{ {\color{blue} x_{11}} {\color{blue} x_2} {\color{blue} x_4}}\\  \end{array} \right).
\end{split}
\ee
Nilpotent element $z$ reads:
\be
z=\left(\begin{array}{ccccccc}   \frac{{\color{blue} x_2} {\color{\mygreen} u_4} {\color{orange} y_2} s t}{{\color{blue} x_8}} & 0 & 0 & 0 & 0 & 0 & 0\\  1 & 0 & 0 & 0 & 0 & 0 & 0\\  0 & 1 & 0 & 0 & 0 & 0 & 0\\  0 & 0 & 1 & 0 & 0 & 0 & 0\\  0 & \frac{{\color{blue} x_1} {\color{blue} x_{10}} {\color{blue} x_3} {\color{\mygreen} u_4} {\color{orange} y_1} s t}{{\color{blue} x_{11}} {\color{blue} x_8}} & 0 & 0 & \frac{{\color{blue} x_3} {\color{\mygreen} u_5} {\color{orange} y_1} s t}{{\color{blue} x_9}} & 0 & 0\\  0 & 0 & 0 & 0 & 1 & 0 & 0\\  0 & 0 & 0 & 0 & 0 & 1 & 0\\  \end{array} \right).
\ee
This orthogonal decomposition is completely parameterized by neutral mesons:
\be
{\color{orange} y_1}{\color{\mygreen} u_4},\;{\color{orange} y_2}{\color{\mygreen} u_4},\;{\color{orange} y_1}{\color{\mygreen} u_5},\; {\color{orange} y_2}{\color{\mygreen} u_5},\;{\color{\mygreen} u_1},\; {\color{\mygreen} u_2},\;{\color{\mygreen} u_3},\;{\color{\mygreen} u_8}.
\ee
Charged chiral fields define embedding of a subspace
\be
\ell_4=\left(\begin{array}{ccccccc}   0 & {\color{blue} x_{11}} {\color{blue} x_2} {\color{blue} x_4} {\color{\mygreen} u_8} {\color{orange} y_1} s t & {\color{blue} x_{11}} {\color{blue} x_2} {\color{orange} y_2} s & 0 & 0 & {\color{blue} x_1} {\color{blue} x_{10}} {\color{blue} x_3} {\color{orange} y_1} s+{\color{blue} x_{10}} {\color{blue} x_2} {\color{\mygreen} u_1} {\color{orange} y_2} s t+{\color{blue} x_1} {\color{blue} x_{10}} {\color{\mygreen} u_2} {\color{orange} y_2} s t & 0\\  \end{array} \right)
\ee
into $E_{3,4}$:
\be
\begin{array}{c}
	\begin{tikzcd}
		\ell_4 \ar[rr,hookrightarrow,"\mbox{\scalebox{0.8}{$\left(\begin{array}{c}
					{\color{orange} y_2}s\\
					{\color{orange} y_1}s
				\end{array}\right)$}}"]& & E_{3,4}.
	\end{tikzcd}
\end{array}
\ee

\newpage

\bibliographystyle{utphys}
\bibliography{biblio}

\providecommand{\href}[2]{#2}\begingroup\raggedright\begin{thebibliography}{100}

\bibitem{D-book_1}
K.~Hori, S.~Katz, A.~Klemm, R.~Pandharipande, R.~Thomas, C.~Vafa, R.~Vakil, and
  E.~Zaslow, {\em {Mirror symmetry}}, vol.~1 of {\em Clay mathematics
  monographs}.
\newblock AMS, Providence, USA, 2003.

\bibitem{D_book_2}
P.~S. Aspinwall, T.~Bridgeland, A.~Craw, M.~R. Douglas, A.~Kapustin, G.~W.
  Moore, M.~Gross, G.~Segal, B.~Szendr\"oi, and P.~M.~H. Wilson, {\em
  {Dirichlet branes and mirror symmetry}}, vol.~4 of {\em Clay Mathematics
  Monographs}.
\newblock AMS, Providence, RI, 2009.

\bibitem{Strominger:1996it}
A.~Strominger, S.-T. Yau, and E.~Zaslow, ``{Mirror symmetry is T duality},''
  \href{http://dx.doi.org/10.1016/0550-3213(96)00434-8}{{\em Nucl. Phys. B}
  {\bfseries 479} (1996) 243--259},
  \href{http://arxiv.org/abs/hep-th/9606040}{{\ttfamily arXiv:hep-th/9606040}}.

\bibitem{Bullimore:2016nji}
M.~Bullimore, T.~Dimofte, D.~Gaiotto, and J.~Hilburn, ``{Boundaries, Mirror
  Symmetry, and Symplectic Duality in 3d $\mathcal{N}=4$ Gauge Theory},''
  \href{http://dx.doi.org/10.1007/JHEP10(2016)108}{{\em JHEP} {\bfseries 10}
  (2016) 108}, \href{http://arxiv.org/abs/1603.08382}{{\ttfamily
  arXiv:1603.08382 [hep-th]}}.

\bibitem{Dimofte:2019zzj}
T.~Dimofte, N.~Garner, M.~Geracie, and J.~Hilburn, ``{Mirror symmetry and line
  operators},'' \href{http://dx.doi.org/10.1007/JHEP02(2020)075}{{\em JHEP}
  {\bfseries 02} (2020) 075}, \href{http://arxiv.org/abs/1908.00013}{{\ttfamily
  arXiv:1908.00013 [hep-th]}}.

\bibitem{Rimanyi:2019zyi}
R.~Rim\'anyi, A.~Smirnov, A.~Varchenko, and Z.~Zhou, ``{3d Mirror Symmetry and
  Elliptic Stable Envelopes},''
  \href{http://arxiv.org/abs/1902.03677}{{\ttfamily arXiv:1902.03677
  [math.AG]}}.

\bibitem{Aganagic:2016jmx}
M.~Aganagic and A.~Okounkov, ``{Elliptic stable envelopes},''
  \href{http://dx.doi.org/10.1090/jams/954}{{\em J. Am. Math. Soc.} {\bfseries
  34} no.~1, (2021) 79--133}, \href{http://arxiv.org/abs/1604.00423}{{\ttfamily
  arXiv:1604.00423 [math.AG]}}.

\bibitem{Braverman:2016pwk}
A.~Braverman, M.~Finkelberg, and H.~Nakajima, ``{Coulomb branches of $3d$
  $\mathcal{N}=4$ quiver gauge theories and slices in the affine
  Grassmannian},'' \href{http://dx.doi.org/10.4310/ATMP.2019.v23.n1.a3}{{\em
  Adv. Theor. Math. Phys.} {\bfseries 23} (2019) 75--166},
  \href{http://arxiv.org/abs/1604.03625}{{\ttfamily arXiv:1604.03625
  [math.RT]}}.

\bibitem{Gu:2018fpm}
W.~Gu and E.~Sharpe, ``{A proposal for nonabelian mirrors},''
  \href{http://arxiv.org/abs/1806.04678}{{\ttfamily arXiv:1806.04678
  [hep-th]}}.

\bibitem{Fan:2007ba}
H.~Fan, T.~J. Jarvis, and Y.~Ruan, ``{The Witten equation, mirror symmetry and
  quantum singularity theory},''
  \href{http://arxiv.org/abs/0712.4021}{{\ttfamily arXiv:0712.4021 [math.AG]}}.

\bibitem{Kontsevich:1994dn}
M.~Kontsevich, ``{Homological Algebra of Mirror Symmetry},''
  \href{http://arxiv.org/abs/alg-geom/9411018}{{\ttfamily
  arXiv:alg-geom/9411018}}.

\bibitem{GMW}
D.~Gaiotto, G.~W. Moore, and E.~Witten, ``{Algebra of the Infrared: String
  Field Theoretic Structures in Massive ${\cal N}=(2,2)$ Field Theory In Two
  Dimensions},'' \href{http://arxiv.org/abs/1506.04087}{{\ttfamily
  arXiv:1506.04087 [hep-th]}}.

\bibitem{Kapranov:2014uwa}
M.~Kapranov, M.~Kontsevich, and Y.~Soibelman, ``{Algebra of the infrared and
  secondary polytopes},''
  \href{http://dx.doi.org/10.1016/j.aim.2016.03.028}{{\em Adv. Math.}
  {\bfseries 300} (2016) 616--671},
  \href{http://arxiv.org/abs/1408.2673}{{\ttfamily arXiv:1408.2673 [math.SG]}}.

\bibitem{Kapranov:2020zoa}
M.~Kapranov, Y.~Soibelman, and L.~Soukhanov, ``{Perverse schobers and the
  Algebra of the Infrared},'' \href{http://arxiv.org/abs/2011.00845}{{\ttfamily
  arXiv:2011.00845 [math.AG]}}.

\bibitem{Khan:2020hir}
A.~Z. Khan and G.~W. Moore, ``{Categorical Wall-Crossing in Landau-Ginzburg
  Models},'' \href{http://arxiv.org/abs/2010.11837}{{\ttfamily arXiv:2010.11837
  [hep-th]}}.

\bibitem{2018arXiv181008776S}
L.~{Soukhanov}, ``{2-Morse Theory and the Algebra of the Infrared},''
  \href{http://arxiv.org/abs/1810.08776}{{\ttfamily arXiv:1810.08776
  [math.AG]}}.

\bibitem{Witten:1982im}
E.~Witten, ``{Supersymmetry and Morse theory},'' {\em J. Diff. Geom.}
  {\bfseries 17} no.~4, (1982) 661--692.

\bibitem{Carqueville:2016kdq}
N.~Carqueville, C.~Meusburger, and G.~Schaumann, ``{3-dimensional defect TQFTs
  and their tricategories},''
  \href{http://dx.doi.org/10.1016/j.aim.2020.107024}{{\em Adv. Math.}
  {\bfseries 364} (2020) 107024},
  \href{http://arxiv.org/abs/1603.01171}{{\ttfamily arXiv:1603.01171
  [math.QA]}}.

\bibitem{Leinster}
T.~{Leinster}, ``{A Survey of Definitions of n-Category},''
  \href{http://arxiv.org/abs/math/0107188}{{\ttfamily arXiv:math/0107188
  [math.CT]}}.

\bibitem{Gukov:2017kmk}
S.~Gukov, D.~Pei, P.~Putrov, and C.~Vafa, ``{BPS spectra and 3-manifold
  invariants},'' \href{http://dx.doi.org/10.1142/S0218216520400039}{{\em J.
  Knot Theor. Ramifications} {\bfseries 29} no.~02, (2020) 2040003},
  \href{http://arxiv.org/abs/1701.06567}{{\ttfamily arXiv:1701.06567
  [hep-th]}}.

\bibitem{Moore:2017byz}
G.~W. Moore, ``{A Comment On Berry Connections},''
  \href{http://arxiv.org/abs/1706.01149}{{\ttfamily arXiv:1706.01149
  [hep-th]}}.

\bibitem{Cecotti:1992rm}
S.~Cecotti and C.~Vafa, ``{On classification of N=2 supersymmetric theories},''
  \href{http://dx.doi.org/10.1007/BF02096804}{{\em Commun. Math. Phys.}
  {\bfseries 158} (1993) 569--644},
  \href{http://arxiv.org/abs/hep-th/9211097}{{\ttfamily arXiv:hep-th/9211097}}.

\bibitem{tarasov2002duality}
V.~Tarasov and A.~Varchenko, ``{Duality for Knizhnik--Zamolodchikov and
  dynamical equations},'' {\em Acta Applicandae Mathematica} {\bfseries 73}
  no.~1, (2002) 141--154, \href{http://arxiv.org/abs/math/0112005}{{\ttfamily
  arXiv:math/0112005 [math.QA]}}.

\bibitem{MaulikOkounkov}
D.~{Maulik} and A.~{Okounkov}, ``{Quantum Groups and Quantum Cohomology},''
  \href{http://arxiv.org/abs/1211.1287}{{\ttfamily arXiv:1211.1287 [math.AG]}}.

\bibitem{Okounkov:2016sya}
A.~Okounkov and A.~Smirnov, ``{Quantum difference equation for Nakajima
  varieties},'' \href{http://arxiv.org/abs/1602.09007}{{\ttfamily
  arXiv:1602.09007 [math-ph]}}.

\bibitem{Aganagic:2020olg}
M.~Aganagic, ``{Knot Categorification from Mirror Symmetry, Part I: Coherent
  Sheaves},'' \href{http://arxiv.org/abs/2004.14518}{{\ttfamily
  arXiv:2004.14518 [hep-th]}}.

\bibitem{aganagic2021knot}
M.~Aganagic, ``{Knot Categorification from Mirror Symmetry, Part II:
  Lagrangians},'' \href{http://arxiv.org/abs/2105.06039}{{\ttfamily
  arXiv:2105.06039 [hep-th]}}.

\bibitem{Cordes:1994fc}
S.~Cordes, G.~W. Moore, and S.~Ramgoolam, ``{Lectures on 2-d Yang-Mills theory,
  equivariant cohomology and topological field theories},''
  \href{http://dx.doi.org/10.1016/0920-5632(95)00434-B}{{\em Nucl. Phys. B
  Proc. Suppl.} {\bfseries 41} (1995) 184--244},
  \href{http://arxiv.org/abs/hep-th/9411210}{{\ttfamily arXiv:hep-th/9411210}}.

\bibitem{Pestun:2016qko}
V.~Pestun, ``{Review of localization in geometry},''
  \href{http://dx.doi.org/10.1088/1751-8121/aa6161}{{\em J. Phys. A} {\bfseries
  50} no.~44, (2017) 443002}, \href{http://arxiv.org/abs/1608.02954}{{\ttfamily
  arXiv:1608.02954 [hep-th]}}.

\bibitem{Douglas:1996sw}
M.~R. Douglas and G.~W. Moore, ``{D-branes, quivers, and ALE instantons},''
  \href{http://arxiv.org/abs/hep-th/9603167}{{\ttfamily arXiv:hep-th/9603167}}.

\bibitem{nakajima1996varieties}
H.~Nakajima, ``Varieties associated with quivers,'' {\em Representation theory
  of algebras and related topics (Mexico City, 1994)} {\bfseries 19} (1996)
  139--157.

\bibitem{donaldson1983new}
S.~K. Donaldson, ``{A new proof of a theorem of Narasimhan and Seshadri},''
  {\em Journal of Differential Geometry} {\bfseries 18} no.~2, (1983) 269--277.

\bibitem{deBoer:1996mp}
J.~de~Boer, K.~Hori, H.~Ooguri, and Y.~Oz, ``{Mirror symmetry in
  three-dimensional gauge theories, quivers and D-branes},''
  \href{http://dx.doi.org/10.1016/S0550-3213(97)00125-9}{{\em Nucl. Phys. B}
  {\bfseries 493} (1997) 101--147},
  \href{http://arxiv.org/abs/hep-th/9611063}{{\ttfamily arXiv:hep-th/9611063}}.

\bibitem{Hori:2013ika}
K.~Hori and M.~Romo, ``{Exact Results In Two-Dimensional (2,2) Supersymmetric
  Gauge Theories With Boundary},''
  \href{http://arxiv.org/abs/1308.2438}{{\ttfamily arXiv:1308.2438 [hep-th]}}.

\bibitem{Witten:1993yc}
E.~Witten, ``{Phases of N=2 theories in two-dimensions},''
  \href{http://dx.doi.org/10.1016/0550-3213(93)90033-L}{{\em Nucl. Phys. B}
  {\bfseries 403} (1993) 159--222},
  \href{http://arxiv.org/abs/hep-th/9301042}{{\ttfamily arXiv:hep-th/9301042}}.

\bibitem{Manschot:2013sya}
J.~Manschot, B.~Pioline, and A.~Sen, ``{On the Coulomb and Higgs branch
  formulae for multi-centered black holes and quiver invariants},''
  \href{http://dx.doi.org/10.1007/JHEP05(2013)166}{{\em JHEP} {\bfseries 05}
  (2013) 166}, \href{http://arxiv.org/abs/1302.5498}{{\ttfamily arXiv:1302.5498
  [hep-th]}}.

\bibitem{HHP}
M.~Herbst, K.~Hori, and D.~Page, ``{Phases Of N=2 Theories In 1+1 Dimensions
  With Boundary},'' \href{http://arxiv.org/abs/0803.2045}{{\ttfamily
  arXiv:0803.2045 [hep-th]}}.

\bibitem{Hori:2000kt}
K.~Hori and C.~Vafa, ``{Mirror symmetry},''
  \href{http://arxiv.org/abs/hep-th/0002222}{{\ttfamily arXiv:hep-th/0002222}}.

\bibitem{Alim:2011kw}
M.~Alim, S.~Cecotti, C.~Cordova, S.~Espahbodi, A.~Rastogi, and C.~Vafa,
  ``{$\mathcal{N} = 2$ quantum field theories and their BPS quivers},''
  \href{http://dx.doi.org/10.4310/ATMP.2014.v18.n1.a2}{{\em Adv. Theor. Math.
  Phys.} {\bfseries 18} no.~1, (2014) 27--127},
  \href{http://arxiv.org/abs/1112.3984}{{\ttfamily arXiv:1112.3984 [hep-th]}}.

\bibitem{Polyakov:1987ez}
A.~M. Polyakov, {\em {Gauge Fields and Strings}}, vol.~3 of {\em Contemporary
  concepts in physics}.
\newblock Hardwood academic publishers, New York, NY, 1987.

\bibitem{Ooguri:1996ck}
H.~Ooguri, Y.~Oz, and Z.~Yin, ``{D-branes on Calabi-Yau spaces and their
  mirrors},'' \href{http://dx.doi.org/10.1016/0550-3213(96)00379-3}{{\em Nucl.
  Phys. B} {\bfseries 477} (1996) 407--430},
  \href{http://arxiv.org/abs/hep-th/9606112}{{\ttfamily arXiv:hep-th/9606112}}.

\bibitem{Kontsevich:2008fj}
M.~Kontsevich and Y.~Soibelman, ``{Stability structures, motivic
  Donaldson-Thomas invariants and cluster transformations},''
  \href{http://arxiv.org/abs/0811.2435}{{\ttfamily arXiv:0811.2435 [math.AG]}}.

\bibitem{Gukov:2010sw}
S.~Gukov, ``{Quantization via Mirror Symmetry},''
  \href{http://arxiv.org/abs/1011.2218}{{\ttfamily arXiv:1011.2218 [hep-th]}}.

\bibitem{2008arXiv0811.1228F}
B.~{Fang}, C.-C.~M. {Liu}, D.~{Treumann}, and E.~{Zaslow}, ``{T-Duality and
  Homological Mirror Symmetry of Toric Varieties},''
  \href{http://arxiv.org/abs/0811.1228}{{\ttfamily arXiv:0811.1228 [math.AG]}}.

\bibitem{Galakhov:2016cji}
D.~Galakhov and G.~W. Moore, ``{Comments On The Two-Dimensional Landau-Ginzburg
  Approach To Link Homology},''
  \href{http://arxiv.org/abs/1607.04222}{{\ttfamily arXiv:1607.04222
  [hep-th]}}.

\bibitem{Chun:2015gda}
S.~Chun, S.~Gukov, and D.~Roggenkamp, ``{Junctions of surface operators and
  categorification of quantum groups},''
  \href{http://arxiv.org/abs/1507.06318}{{\ttfamily arXiv:1507.06318
  [hep-th]}}.

\bibitem{Honda:2013uca}
D.~Honda and T.~Okuda, ``{Exact results for boundaries and domain walls in 2d
  supersymmetric theories},''
  \href{http://dx.doi.org/10.1007/JHEP09(2015)140}{{\em JHEP} {\bfseries 09}
  (2015) 140}, \href{http://arxiv.org/abs/1308.2217}{{\ttfamily arXiv:1308.2217
  [hep-th]}}.

\bibitem{Seidel_book}
P.~Seidel, \href{http://dx.doi.org/10.4171/063}{{\em Fukaya categories and
  {P}icard-{L}efschetz theory}}.
\newblock Zurich Lectures in Advanced Mathematics. European Mathematical
  Society (EMS), Z\"{u}rich, 2008.
\newblock \url{https://doi.org/10.4171/063}.

\bibitem{gorodentsev2004helix}
A.~L. Gorodentsev and S.~A. Kuleshov, ``Helix theory,'' {\em Moscow
  Mathematical Journal} {\bfseries 4} no.~2, (2004) 377--440.

\bibitem{Aganagic:2009kf}
M.~Aganagic, H.~Ooguri, C.~Vafa, and M.~Yamazaki, ``{Wall Crossing and
  M-theory},'' {\em Publ. Res. Inst. Math. Sci. Kyoto} {\bfseries 47} (2011)
  569, \href{http://arxiv.org/abs/0908.1194}{{\ttfamily arXiv:0908.1194
  [hep-th]}}.

\bibitem{Gaiotto:2008cd}
D.~Gaiotto, G.~W. Moore, and A.~Neitzke, ``{Four-dimensional wall-crossing via
  three-dimensional field theory},''
  \href{http://dx.doi.org/10.1007/s00220-010-1071-2}{{\em Commun. Math. Phys.}
  {\bfseries 299} (2010) 163--224},
  \href{http://arxiv.org/abs/0807.4723}{{\ttfamily arXiv:0807.4723 [hep-th]}}.

\bibitem{Galakhov:2017pod}
D.~Galakhov, ``{Why Is Landau-Ginzburg Link Cohomology Equivalent To Khovanov
  Homology?},'' \href{http://dx.doi.org/10.1007/JHEP05(2019)085}{{\em JHEP}
  {\bfseries 05} (2019) 085}, \href{http://arxiv.org/abs/1702.07086}{{\ttfamily
  arXiv:1702.07086 [hep-th]}}.

\bibitem{Kerr:2017usa}
G.~Kerr and Y.~Soibelman, ``{On 2d-4d motivic wall-crossing formulas},''
  \href{http://arxiv.org/abs/1711.03695}{{\ttfamily arXiv:1711.03695
  [math.AG]}}.

\bibitem{Clingempeel:2018iub}
J.~Clingempeel, B.~Le~Floch, and M.~Romo, ``{Brane transport in anomalous (2,2)
  models and localization},'' \href{http://arxiv.org/abs/1811.12385}{{\ttfamily
  arXiv:1811.12385 [hep-th]}}.

\bibitem{Brunner:2020miu}
I.~Brunner, I.~Mayer, and C.~Schmidt-Colinet, ``{Topological defects and SUSY
  RG flow},'' \href{http://dx.doi.org/10.1007/JHEP03(2021)098}{{\em JHEP}
  {\bfseries 03} (2021) 098}, \href{http://arxiv.org/abs/2007.02353}{{\ttfamily
  arXiv:2007.02353 [hep-th]}}.

\bibitem{Brunner:2021cga}
I.~Brunner, F.~Klos, and D.~Roggenkamp, ``{Phase transitions in GLSMs and
  defects},'' \href{http://arxiv.org/abs/2101.12315}{{\ttfamily
  arXiv:2101.12315 [hep-th]}}.

\bibitem{Chen:2020iyo}
Z.~Chen, J.~Guo, and M.~Romo, ``{A GLSM view on Homological Projective
  Duality},'' \href{http://arxiv.org/abs/2012.14109}{{\ttfamily
  arXiv:2012.14109 [hep-th]}}.

\bibitem{Brunner:2009zt}
I.~Brunner, D.~Roggenkamp, and S.~Rossi, ``{Defect Perturbations in
  Landau-Ginzburg Models},''
  \href{http://dx.doi.org/10.1007/JHEP03(2010)015}{{\em JHEP} {\bfseries 03}
  (2010) 015}, \href{http://arxiv.org/abs/0909.0696}{{\ttfamily arXiv:0909.0696
  [hep-th]}}.

\bibitem{Brunner:2008fa}
I.~Brunner, H.~Jockers, and D.~Roggenkamp, ``{Defects and D-Brane
  Monodromies},'' \href{http://dx.doi.org/10.4310/ATMP.2009.v13.n4.a4}{{\em
  Adv. Theor. Math. Phys.} {\bfseries 13} no.~4, (2009) 1077--1135},
  \href{http://arxiv.org/abs/0806.4734}{{\ttfamily arXiv:0806.4734 [hep-th]}}.

\bibitem{Aspinwall:2004jr}
P.~S. Aspinwall,
  \href{http://dx.doi.org/10.1142/9789812775108_0001}{``{D-branes on Calabi-Yau
  manifolds},''} in {\em {Theoretical Advanced Study Institute in Elementary
  Particle Physics (TASI 2003): Recent Trends in String Theory}}.
\newblock 3, 2004.
\newblock \href{http://arxiv.org/abs/hep-th/0403166}{{\ttfamily
  arXiv:hep-th/0403166}}.

\bibitem{Knapp:2016rec}
J.~Knapp, M.~Romo, and E.~Scheidegger, ``{Hemisphere Partition Function and
  Analytic Continuation to the Conifold Point},''
  \href{http://dx.doi.org/10.4310/CNTP.2017.v11.n1.a3}{{\em Commun. Num. Theor.
  Phys.} {\bfseries 11} (2017) 73--164},
  \href{http://arxiv.org/abs/1602.01382}{{\ttfamily arXiv:1602.01382
  [hep-th]}}.

\bibitem{Erkinger:2017aaa}
D.~Erkinger and J.~Knapp, ``{Hemisphere Partition Function and Monodromy},''
  \href{http://dx.doi.org/10.1007/JHEP05(2017)150}{{\em JHEP} {\bfseries 05}
  (2017) 150}, \href{http://arxiv.org/abs/1704.00901}{{\ttfamily
  arXiv:1704.00901 [hep-th]}}.

\bibitem{Knapp:2020oba}
J.~Knapp, M.~Romo, and E.~Scheidegger, ``{D-Brane Central Charge and
  Landau\textendash{}Ginzburg Orbifolds},''
  \href{http://dx.doi.org/10.1007/s00220-021-04042-w}{{\em Commun. Math. Phys.}
  {\bfseries 384} no.~1, (2021) 609--697},
  \href{http://arxiv.org/abs/2003.00182}{{\ttfamily arXiv:2003.00182
  [hep-th]}}.

\bibitem{Hartshorne}
R.~Hartshorne, {\em Algebraic geometry}.
\newblock Springer-Verlag, New York-Heidelberg, 1977.
\newblock Graduate Texts in Mathematics, No. 52.

\bibitem{Flop}
Y.-W. {Fan}, H.~{Hong}, S.-C. {Lau}, and S.-T. {Yau}, ``{Mirror of Atiyah flop
  in symplectic geometry and stability conditions},''
  \href{http://arxiv.org/abs/1706.02942}{{\ttfamily arXiv:1706.02942
  [math.SG]}}.

\bibitem{Segal:2010cz}
E.~Segal, ``{Equivalences between GIT quotients of Landau-Ginzburg B-models},''
  \href{http://dx.doi.org/10.1007/s00220-011-1232-y}{{\em Commun. Math. Phys.}
  {\bfseries 304} (2011) 411--432},
  \href{http://arxiv.org/abs/0910.5534}{{\ttfamily arXiv:0910.5534 [math.AG]}}.

\bibitem{2012arXiv1206.0219D}
W.~{Donovan} and E.~{Segal}, ``{Window shifts, flop equivalences and
  Grassmannian twists},'' {\em Compositio Mathematica} {\bfseries 150} no.~6,
  (2014) 942--978, \href{http://arxiv.org/abs/1206.0219}{{\ttfamily
  arXiv:1206.0219 [math.AG]}}.

\bibitem{DuninBarkowski:2012na}
P.~Dunin-Barkowski, A.~Sleptsov, and A.~Smirnov, ``{Explicit computation of
  Drinfeld associator in the case of the fundamental representation of
  gl(N)},'' \href{http://dx.doi.org/10.1088/1751-8113/45/38/385204}{{\em J.
  Phys. A} {\bfseries 45} (2012) 385204},
  \href{http://arxiv.org/abs/1201.0025}{{\ttfamily arXiv:1201.0025 [hep-th]}}.

\bibitem{Belavin:1984vu}
A.~A. Belavin, A.~M. Polyakov, and A.~B. Zamolodchikov, ``{Infinite Conformal
  Symmetry in Two-Dimensional Quantum Field Theory},''
  \href{http://dx.doi.org/10.1016/0550-3213(84)90052-X}{{\em Nucl. Phys. B}
  {\bfseries 241} (1984) 333--380}.

\bibitem{DiFrancesco:639405}
P.~Di~Francesco, P.~Mathieu, and D.~Sénéchal,
  \href{http://dx.doi.org/10.1007/978-1-4612-2256-9}{{\em {Conformal field
  theory}}}.
\newblock Graduate texts in contemporary physics. Springer, New York, NY, 1997.
\newblock \url{https://cds.cern.ch/record/639405}.

\bibitem{Witten:1988hf}
E.~Witten, ``{Quantum Field Theory and the Jones Polynomial},''
  \href{http://dx.doi.org/10.1007/BF01217730}{{\em Commun. Math. Phys.}
  {\bfseries 121} (1989) 351--399}.

\bibitem{Kirillov:191317}
A.~N. Kirillov and N.~Y. Reshetikhin, ``{Representations of the algebra
  U$_{q}$(sl(2)), q orthogonal polynomials and invariants of links},''.
  \url{https://cds.cern.ch/record/191317}.

\bibitem{drinfeld1989quasi}
V.~Drinfeld, ``{Quasi-Hopf Algebras and Knizhnik-Zamolodchikov Equations},''
  {\em Problems of Modern Quantum Field Theory. Series: Research Reports in
  Physics} (1989) 1--13.

\bibitem{Khovanov:1999qla}
M.~Khovanov, ``{A categorification of the Jones polynomial},''
  \href{http://dx.doi.org/10.1215/S0012-7094-00-10131-7}{{\em Duke Math. J.}
  {\bfseries 101} (2000) 359},
  \href{http://arxiv.org/abs/math/9908171}{{\ttfamily arXiv:math/9908171}}.

\bibitem{KR}
M.~{Khovanov} and L.~{Rozansky}, ``{Matrix factorizations and link homology},''
  \href{http://arxiv.org/abs/math/0401268}{{\ttfamily arXiv:math/0401268
  [math.QA]}}.

\bibitem{Ooguri:1999bv}
H.~Ooguri and C.~Vafa, ``{Knot invariants and topological strings},''
  \href{http://dx.doi.org/10.1016/S0550-3213(00)00118-8}{{\em Nucl. Phys. B}
  {\bfseries 577} (2000) 419--438},
  \href{http://arxiv.org/abs/hep-th/9912123}{{\ttfamily arXiv:hep-th/9912123}}.

\bibitem{Gukov:2004hz}
S.~Gukov, A.~S. Schwarz, and C.~Vafa, ``{Khovanov-Rozansky homology and
  topological strings},''
  \href{http://dx.doi.org/10.1007/s11005-005-0008-8}{{\em Lett. Math. Phys.}
  {\bfseries 74} (2005) 53--74},
  \href{http://arxiv.org/abs/hep-th/0412243}{{\ttfamily arXiv:hep-th/0412243}}.

\bibitem{Gukov:2007tf}
S.~Gukov, A.~Iqbal, C.~Kozcaz, and C.~Vafa, ``{Link Homologies and the Refined
  Topological Vertex},''
  \href{http://dx.doi.org/10.1007/s00220-010-1045-4}{{\em Commun. Math. Phys.}
  {\bfseries 298} (2010) 757--785},
  \href{http://arxiv.org/abs/0705.1368}{{\ttfamily arXiv:0705.1368 [hep-th]}}.

\bibitem{Haydys:2010dv}
A.~Haydys, ``{Fukaya-Seidel category and gauge theory},''
  \href{http://dx.doi.org/10.4310/JSG.2015.v13.n1.a5}{{\em J. Sympl. Geom.}
  {\bfseries 13} (2015) 151--207},
  \href{http://arxiv.org/abs/1010.2353}{{\ttfamily arXiv:1010.2353 [math.SG]}}.

\bibitem{WItten:2011pz}
E.~Witten, ``{Khovanov Homology And Gauge Theory},''
  \href{http://arxiv.org/abs/1108.3103}{{\ttfamily arXiv:1108.3103 [math.GT]}}.

\bibitem{Witten:2011zz}
E.~Witten, ``{Fivebranes and Knots},''
  \href{http://arxiv.org/abs/1101.3216}{{\ttfamily arXiv:1101.3216 [hep-th]}}.

\bibitem{Anokhina:2017iui}
A.~Anokhina, ``{Towards formalization of the soliton counting technique for the
  Khovanov-Rozansky invariants in the deformed $\mathcal{R}$-matrix
  approach},'' \href{http://dx.doi.org/10.1142/S0217751X18502214}{{\em Int. J.
  Mod. Phys. A} {\bfseries 33} no.~36, (2019) 1850221},
  \href{http://arxiv.org/abs/1710.07306}{{\ttfamily arXiv:1710.07306
  [hep-th]}}.

\bibitem{Anokhina:2021lbh}
A.~Anokhina, A.~Morozov, and A.~Popolitov, ``{Khovanov polynomials for
  satellites and asymptotic adjoint polynomials},''
  \href{http://arxiv.org/abs/2104.14491}{{\ttfamily arXiv:2104.14491
  [hep-th]}}.

\bibitem{Dolotin:2013osa}
V.~Dolotin and A.~Morozov, ``{Introduction to Khovanov Homologies. III. A new
  and simple tensor-algebra construction of Khovanov-Rozansky invariants},''
  \href{http://dx.doi.org/10.1016/j.nuclphysb.2013.11.007}{{\em Nucl. Phys. B}
  {\bfseries 878} (2014) 12--81},
  \href{http://arxiv.org/abs/1308.5759}{{\ttfamily arXiv:1308.5759 [hep-th]}}.

\bibitem{Arthamonov:2017oxw}
S.~Arthamonov and S.~Shakirov, ``{Genus Two Generalization of $A_1$ spherical
  DAHA},'' \href{http://arxiv.org/abs/1704.02947}{{\ttfamily arXiv:1704.02947
  [math.QA]}}.

\bibitem{2007arXiv0708.2228M}
M.~{Mackaay}, M.~{Stosic}, and P.~{Vaz}, ``{Sl(N) link homology using foams and
  the Kapustin-Li formula},'' \href{http://arxiv.org/abs/0708.2228}{{\ttfamily
  arXiv:0708.2228 [math.GT]}}.

\bibitem{2005math......8510R}
J.~{Rasmussen}, ``{Khovanov-Rozansky homology of two-bridge knots and links},''
  \href{http://arxiv.org/abs/math/0508510}{{\ttfamily arXiv:math/0508510
  [math.GT]}}.

\bibitem{2016arXiv160807308G}
E.~{Gorsky}, A.~{Negu{\c{t}}}, and J.~{Rasmussen}, ``{Flag Hilbert schemes,
  colored projectors and Khovanov-Rozansky homology},''
  \href{http://arxiv.org/abs/1608.07308}{{\ttfamily arXiv:1608.07308
  [math.GT]}}.

\bibitem{2019arXiv190208281G}
E.~{Gorsky}, M.~{Hogancamp}, A.~{Mellit}, and K.~{Nakagane}, ``{Serre duality
  for Khovanov-Rozansky homology},''
  \href{http://arxiv.org/abs/1902.08281}{{\ttfamily arXiv:1902.08281
  [math.RT]}}.

\bibitem{2019arXiv190506511O}
A.~{Oblomkov} and L.~{Rozansky}, ``{Dualizable link homology},''
  \href{http://arxiv.org/abs/1905.06511}{{\ttfamily arXiv:1905.06511
  [math.GN]}}.

\bibitem{2007arXiv0710.4300O}
P.~{Ozsvath}, J.~{Rasmussen}, and Z.~{Szabo}, ``{Odd Khovanov homology},''
  \href{http://arxiv.org/abs/0710.4300}{{\ttfamily arXiv:0710.4300 [math.QA]}}.

\bibitem{Abouzaid:2017clg}
M.~Abouzaid and C.~Manolescu, ``{A sheaf-theoretic model for SL(2,C) Floer
  homology},'' \href{http://arxiv.org/abs/1708.00289}{{\ttfamily
  arXiv:1708.00289 [math.GT]}}.

\bibitem{2008arXiv0803.4121K}
M.~{Khovanov} and A.~D. {Lauda}, ``{A diagrammatic approach to categorification
  of quantum groups I},'' \href{http://arxiv.org/abs/0803.4121}{{\ttfamily
  arXiv:0803.4121 [math.QA]}}.

\bibitem{2008arXiv0812.5023R}
R.~{Rouquier}, ``{2-Kac-Moody algebras},''
  \href{http://arxiv.org/abs/0812.5023}{{\ttfamily arXiv:0812.5023 [math.RT]}}.

\bibitem{seidel2001braid}
P.~Seidel and R.~Thomas, ``Braid group actions on derived categories of
  coherent sheaves,'' {\em Duke Mathematical Journal} {\bfseries 108} no.~1,
  (1, 2001) 37--108, \href{http://arxiv.org/abs/math/0001043}{{\ttfamily
  arXiv:math/0001043}}.

\bibitem{2011arXiv1104.0352C}
S.~{Cautis}, J.~{Kamnitzer}, and A.~{Licata}, ``{Coherent Sheaves on Quiver
  Varieties and Categorification},''
  \href{http://arxiv.org/abs/1104.0352}{{\ttfamily arXiv:1104.0352 [math.AG]}}.

\bibitem{2012arXiv1212.6076L}
A.~D. {Lauda}, H.~{Queffelec}, and D.~E.~V. {Rose}, ``{Khovanov homology is a
  skew Howe 2-representation of categorified quantum sl(m)},''
  \href{http://arxiv.org/abs/1212.6076}{{\ttfamily arXiv:1212.6076 [math.QA]}}.

\bibitem{CautisKamnitzerI}
S.~{Cautis} and J.~{Kamnitzer}, ``{Knot homology via derived categories of
  coherent sheaves I, sl(2) case},''
  \href{http://arxiv.org/abs/math/0701194}{{\ttfamily arXiv:math/0701194
  [math.AG]}}.

\bibitem{2013arXiv1306.6242M}
M.~{Mackaay} and Y.~{Yonezawa}, ``{sl(N)-Web categories},''
  \href{http://arxiv.org/abs/1306.6242}{{\ttfamily arXiv:1306.6242 [math.QA]}}.

\bibitem{2015arXiv150206011M}
M.~{Mackaay} and B.~{Webster}, ``{Categorified skew Howe duality and comparison
  of knot homologies},'' \href{http://arxiv.org/abs/1502.06011}{{\ttfamily
  arXiv:1502.06011 [math.GT]}}.

\bibitem{2014arXiv1409.4461W}
B.~{Webster}, ``{On generalized category $\mathcal{O}$ for a quiver variety},''
  \href{http://arxiv.org/abs/1409.4461}{{\ttfamily arXiv:1409.4461 [math.AG]}}.

\bibitem{Gaiotto:2015zna}
D.~Gaiotto, G.~W. Moore, and E.~Witten, ``{An Introduction To The Web-Based
  Formalism},'' \href{http://arxiv.org/abs/1506.04086}{{\ttfamily
  arXiv:1506.04086 [hep-th]}}.

\bibitem{nakajima1994instantons}
H.~Nakajima {\em et~al.}, ``{Instantons on ALE spaces, quiver varieties, and
  Kac-Moody algebras},'' {\em Duke Mathematical Journal} {\bfseries 76} no.~2,
  (1994) 365--416.

\bibitem{nakajima1998quiver}
H.~Nakajima {\em et~al.}, ``Quiver varieties and kac-moody algebras,'' {\em
  Duke Mathematical Journal} {\bfseries 91} no.~3, (1998) 515--560.

\bibitem{nakajima2001quiver}
H.~Nakajima, ``Quiver varieties and finite dimensional representations of
  quantum affine algebras,'' {\em Journal of the American Mathematical Society}
  (2001) 145--238.

\bibitem{2009arXiv0905.0686G}
V.~{Ginzburg}, ``{Lectures on Nakajima's Quiver Varieties},''
  \href{http://arxiv.org/abs/0905.0686}{{\ttfamily arXiv:0905.0686 [math.RT]}}.

\bibitem{Ooguri:2008yb}
H.~Ooguri and M.~Yamazaki, ``{Crystal Melting and Toric Calabi-Yau
  Manifolds},'' \href{http://dx.doi.org/10.1007/s00220-009-0836-y}{{\em Commun.
  Math. Phys.} {\bfseries 292} (2009) 179--199},
\href{http://arxiv.org/abs/0811.2801}{{\ttfamily arXiv:0811.2801 [hep-th]}}.

\bibitem{Okounkov:2003sp}
A.~Okounkov, N.~Reshetikhin, and C.~Vafa, ``{Quantum Calabi-Yau and classical
  crystals},'' \href{http://dx.doi.org/10.1007/0-8176-4467-9_16}{{\em Prog.
  Math.} {\bfseries 244} (2006) 597},
  \href{http://arxiv.org/abs/hep-th/0309208}{{\ttfamily arXiv:hep-th/0309208}}.

\bibitem{Yamazaki:2010fz}
M.~Yamazaki, ``{Crystal Melting and Wall Crossing Phenomena},''
  \href{http://dx.doi.org/10.1142/S0217751X11051482}{{\em Int. J. Mod. Phys. A}
  {\bfseries 26} (2011) 1097--1228},
  \href{http://arxiv.org/abs/1002.1709}{{\ttfamily arXiv:1002.1709 [hep-th]}}.

\bibitem{Li:2020rij}
W.~Li and M.~Yamazaki, ``{Quiver Yangian from Crystal Melting},''
  \href{http://dx.doi.org/10.1007/JHEP11(2020)035}{{\em JHEP} {\bfseries 11}
  (2020) 035}, \href{http://arxiv.org/abs/2003.08909}{{\ttfamily
  arXiv:2003.08909 [hep-th]}}.

\bibitem{Aganagic:2010qr}
M.~Aganagic and K.~Schaeffer, ``{Wall Crossing, Quivers and Crystals},''
  \href{http://dx.doi.org/10.1007/JHEP10(2012)153}{{\em JHEP} {\bfseries 10}
  (2012) 153}, \href{http://arxiv.org/abs/1006.2113}{{\ttfamily arXiv:1006.2113
  [hep-th]}}.

\bibitem{Borokhov:2002ib}
V.~Borokhov, A.~Kapustin, and X.-k. Wu, ``{Topological disorder operators in
  three-dimensional conformal field theory},''
  \href{http://dx.doi.org/10.1088/1126-6708/2002/11/049}{{\em JHEP} {\bfseries
  11} (2002) 049}, \href{http://arxiv.org/abs/hep-th/0206054}{{\ttfamily
  arXiv:hep-th/0206054}}.

\bibitem{Behtash:2017rqj}
A.~Behtash, ``{More on Homological Supersymmetric Quantum Mechanics},''
  \href{http://dx.doi.org/10.1103/PhysRevD.97.065002}{{\em Phys. Rev. D}
  {\bfseries 97} no.~6, (2018) 065002},
  \href{http://arxiv.org/abs/1703.00511}{{\ttfamily arXiv:1703.00511
  [hep-th]}}.

\bibitem{landau1958quantum}
L.~D. Landau and E.~M. Lifshitz, {\em Quantum Mechanics: Non-relativistic
  Theory. V. 3 of Course of Theoretical Physics}.
\newblock Pergamon Press, 1958.

\bibitem{ZinnJustin:2002ru}
J.~Zinn-Justin, ``{Quantum field theory and critical phenomena},'' {\em Int.
  Ser. Monogr. Phys.} {\bfseries 113} (2002) 1--1054.

\bibitem{Auroux}
D.~{Auroux}, ``{A beginner's introduction to Fukaya categories},''
  \href{http://arxiv.org/abs/1301.7056}{{\ttfamily arXiv:1301.7056 [math.SG]}}.

\bibitem{Equivariant_sheves}
J.~Bernstein and V.~Lunts, \href{http://dx.doi.org/10.1007/BFb0073549}{{\em
  Equivariant sheaves and functors}}, vol.~1578 of {\em Lecture Notes in
  Mathematics}.
\newblock Springer-Verlag, Berlin, 1994.
\newblock \url{https://doi.org/10.1007/BFb0073549}.

\bibitem{Denef:2007vg}
F.~Denef and G.~W. Moore, ``{Split states, entropy enigmas, holes and halos},''
  \href{http://dx.doi.org/10.1007/JHEP11(2011)129}{{\em JHEP} {\bfseries 11}
  (2011) 129}, \href{http://arxiv.org/abs/hep-th/0702146}{{\ttfamily
  arXiv:hep-th/0702146}}.

\bibitem{Eto:2006pg}
M.~Eto, Y.~Isozumi, M.~Nitta, K.~Ohashi, and N.~Sakai, ``{Solitons in the Higgs
  phase: The Moduli matrix approach},''
  \href{http://dx.doi.org/10.1088/0305-4470/39/26/R01}{{\em J. Phys. A}
  {\bfseries 39} (2006) R315--R392},
  \href{http://arxiv.org/abs/hep-th/0602170}{{\ttfamily arXiv:hep-th/0602170}}.

\bibitem{Eto:2007aw}
M.~Eto, T.~Fujimori, M.~Nitta, K.~Ohashi, K.~Ohta, and N.~Sakai, ``{Statistical
  mechanics of vortices from D-branes and T-duality},''
  \href{http://dx.doi.org/10.1016/j.nuclphysb.2007.06.020}{{\em Nucl. Phys. B}
  {\bfseries 788} (2008) 120--136},
  \href{http://arxiv.org/abs/hep-th/0703197}{{\ttfamily arXiv:hep-th/0703197}}.

\bibitem{Eto:2007uc}
M.~Eto, T.~Fujimori, T.~Nagashima, M.~Nitta, K.~Ohashi, and N.~Sakai,
  ``{Dynamics of Domain Wall Networks},''
  \href{http://dx.doi.org/10.1103/PhysRevD.76.125025}{{\em Phys. Rev. D}
  {\bfseries 76} (2007) 125025},
  \href{http://arxiv.org/abs/0707.3267}{{\ttfamily arXiv:0707.3267 [hep-th]}}.

\bibitem{Shifman:2010id}
M.~Shifman and A.~Yung, ``{Non-Abelian Confinement in N=2 Supersymmetric QCD:
  Duality and Kinks on Confining Strings},''
  \href{http://dx.doi.org/10.1103/PhysRevD.81.085009}{{\em Phys. Rev. D}
  {\bfseries 81} (2010) 085009},
  \href{http://arxiv.org/abs/1002.0322}{{\ttfamily arXiv:1002.0322 [hep-th]}}.

\bibitem{Fujimori:2016ljw}
T.~Fujimori, S.~Kamata, T.~Misumi, M.~Nitta, and N.~Sakai, ``{Nonperturbative
  contributions from complexified solutions in $\mathbb{C}P^{N-1}$models},''
  \href{http://dx.doi.org/10.1103/PhysRevD.94.105002}{{\em Phys. Rev. D}
  {\bfseries 94} no.~10, (2016) 105002},
  \href{http://arxiv.org/abs/1607.04205}{{\ttfamily arXiv:1607.04205
  [hep-th]}}.

\bibitem{Harland:2009mf}
D.~Harland, ``{Kinks, chains, and loop groups in the CP**n sigma models},''
  \href{http://dx.doi.org/10.1063/1.3266172}{{\em J. Math. Phys.} {\bfseries
  50} (2009) 122902}, \href{http://arxiv.org/abs/0902.2303}{{\ttfamily
  arXiv:0902.2303 [hep-th]}}.

\bibitem{Manschot:2010qz}
J.~Manschot, B.~Pioline, and A.~Sen, ``{Wall Crossing from Boltzmann Black Hole
  Halos},'' \href{http://dx.doi.org/10.1007/JHEP07(2011)059}{{\em JHEP}
  {\bfseries 07} (2011) 059}, \href{http://arxiv.org/abs/1011.1258}{{\ttfamily
  arXiv:1011.1258 [hep-th]}}.

\bibitem{Gaiotto:2012rg}
D.~Gaiotto, G.~W. Moore, and A.~Neitzke, ``{Spectral networks},''
  \href{http://dx.doi.org/10.1007/s00023-013-0239-7}{{\em Annales Henri
  Poincare} {\bfseries 14} (2013) 1643--1731},
  \href{http://arxiv.org/abs/1204.4824}{{\ttfamily arXiv:1204.4824 [hep-th]}}.

\bibitem{Galakhov:2014xba}
D.~Galakhov, P.~Longhi, and G.~W. Moore, ``{Spectral Networks with Spin},''
  \href{http://dx.doi.org/10.1007/s00220-015-2455-0}{{\em Commun. Math. Phys.}
  {\bfseries 340} no.~1, (2015) 171--232},
  \href{http://arxiv.org/abs/1408.0207}{{\ttfamily arXiv:1408.0207 [hep-th]}}.

\bibitem{Gaiotto:2009hg}
D.~Gaiotto, G.~W. Moore, and A.~Neitzke, ``{Wall-crossing, Hitchin Systems, and
  the WKB Approximation},'' \href{http://arxiv.org/abs/0907.3987}{{\ttfamily
  arXiv:0907.3987 [hep-th]}}.

\bibitem{Klemm:1996bj}
A.~Klemm, W.~Lerche, P.~Mayr, C.~Vafa, and N.~P. Warner, ``{Selfdual strings
  and N=2 supersymmetric field theory},''
  \href{http://dx.doi.org/10.1016/0550-3213(96)00353-7}{{\em Nucl. Phys. B}
  {\bfseries 477} (1996) 746--766},
  \href{http://arxiv.org/abs/hep-th/9604034}{{\ttfamily arXiv:hep-th/9604034}}.

\bibitem{Gaiotto:2011nm}
D.~Gaiotto and E.~Witten, ``{Knot Invariants from Four-Dimensional Gauge
  Theory},'' \href{http://dx.doi.org/10.4310/ATMP.2012.v16.n3.a5}{{\em Adv.
  Theor. Math. Phys.} {\bfseries 16} no.~3, (2012) 935--1086},
  \href{http://arxiv.org/abs/1106.4789}{{\ttfamily arXiv:1106.4789 [hep-th]}}.

\bibitem{Zenkevich:2017ylb}
A.~Nedelin, S.~Pasquetti, and Y.~Zenkevich, ``{T[SU(N)] duality webs: mirror
  symmetry, spectral duality and gauge/CFT correspondences},''
  \href{http://dx.doi.org/10.1007/JHEP02(2019)176}{{\em JHEP} {\bfseries 02}
  (2019) 176}, \href{http://arxiv.org/abs/1712.08140}{{\ttfamily
  arXiv:1712.08140 [hep-th]}}.

\bibitem{Bondal_Kapranov_Schecntman}
A.~{Bondal}, M.~{Kapranov}, and V.~{Schechtman}, ``{Perverse schobers and
  birational geometry},'' \href{http://arxiv.org/abs/1801.08286}{{\ttfamily
  arXiv:1801.08286 [math.AG]}}.

\bibitem{Huybrechts}
D.~Huybrechts,
  \href{http://dx.doi.org/10.1093/acprof:oso/9780199296866.001.0001}{{\em
  Fourier-{M}ukai transforms in algebraic geometry}}.
\newblock Oxford Mathematical Monographs. The Clarendon Press, Oxford
  University Press, Oxford, 2006.
\newblock \url{https://doi.org/10.1093/acprof:oso/9780199296866.001.0001}.

\bibitem{orlov2003derived}
D.~O. Orlov, ``Derived categories of coherent sheaves and equivalences between
  them,'' {\em Russian Mathematical Surveys} {\bfseries 58} no.~3, (2003) 511.

\bibitem{Knizhnik:1984nr}
V.~G. Knizhnik and A.~B. Zamolodchikov, ``{Current Algebra and Wess-Zumino
  Model in Two-Dimensions},''
  \href{http://dx.doi.org/10.1016/0550-3213(84)90374-2}{{\em Nucl. Phys. B}
  {\bfseries 247} (1984) 83--103}.

\bibitem{Dotsenko:1984nm}
V.~S. Dotsenko and V.~A. Fateev, ``{Conformal Algebra and Multipoint
  Correlation Functions in Two-Dimensional Statistical Models},''
  \href{http://dx.doi.org/10.1016/0550-3213(84)90269-4}{{\em Nucl. Phys. B}
  {\bfseries 240} (1984) 312}.

\bibitem{1990IJMPA...5.2495G}
A.~{Gerasimov}, A.~{Morozov}, M.~{Olshanetsky}, A.~{Marshakov}, and
  S.~{Shatashvili}, ``{Wess-Zumino Model as a Theory of Free Fields},''
  \href{http://dx.doi.org/10.1142/S0217751X9000115X}{{\em International Journal
  of Modern Physics A} {\bfseries 5} no.~13, (Jan., 1990) 2495--2589}.

\bibitem{schechtman1991arrangements}
V.~V. Schechtman and A.~N. Varchenko, ``Arrangements of hyperplanes and lie
  algebra homology,'' {\em Inventiones mathematicae} {\bfseries 106} no.~1,
  (1991) 139--194.

\bibitem{etingof1998lectures}
P.~I. Etingof, I.~Frenkel, and A.~A. Kirillov, {\em Lectures on representation
  theory and Knizhnik-Zamolodchikov equations}.
\newblock No.~58. American Mathematical Soc., 1998.

\bibitem{Reshetikhin:1994qw}
N.~Reshetikhin and A.~Varchenko, ``{Quasiclassical asymptotics of solutions to
  the KZ equations},'' \href{http://arxiv.org/abs/hep-th/9402126}{{\ttfamily
  arXiv:hep-th/9402126}}.

\bibitem{Galakhov:2014aha}
D.~Galakhov, A.~Mironov, and A.~Morozov, ``{Wall Crossing Invariants: from
  quantum mechanics to knots},''
  \href{http://dx.doi.org/10.1134/S1063776115030206}{{\em J. Exp. Theor. Phys.}
  {\bfseries 120} no.~3, (2015) 549--577},
  \href{http://arxiv.org/abs/1410.8482}{{\ttfamily arXiv:1410.8482 [hep-th]}}.

\bibitem{Nekrasov:2002qd}
N.~A. Nekrasov, ``{Seiberg-Witten prepotential from instanton counting},''
  \href{http://dx.doi.org/10.4310/ATMP.2003.v7.n5.a4}{{\em Adv. Theor. Math.
  Phys.} {\bfseries 7} no.~5, (2003) 831--864},
  \href{http://arxiv.org/abs/hep-th/0206161}{{\ttfamily arXiv:hep-th/0206161}}.

\bibitem{Nekrasov:2003rj}
N.~Nekrasov and A.~Okounkov, ``{Seiberg-Witten theory and random partitions},''
  \href{http://dx.doi.org/10.1007/0-8176-4467-9_15}{{\em Prog. Math.}
  {\bfseries 244} (2006) 525--596},
  \href{http://arxiv.org/abs/hep-th/0306238}{{\ttfamily arXiv:hep-th/0306238}}.

\bibitem{Moore:1998et}
G.~W. Moore, N.~Nekrasov, and S.~Shatashvili, ``{D particle bound states and
  generalized instantons},''
  \href{http://dx.doi.org/10.1007/s002200050016}{{\em Commun. Math. Phys.}
  {\bfseries 209} (2000) 77--95},
  \href{http://arxiv.org/abs/hep-th/9803265}{{\ttfamily arXiv:hep-th/9803265}}.

\bibitem{Galakhov:2020vyb}
D.~Galakhov and M.~Yamazaki, ``{Quiver Yangian and Supersymmetric Quantum
  Mechanics},'' \href{http://arxiv.org/abs/2008.07006}{{\ttfamily
  arXiv:2008.07006 [hep-th]}}.

\bibitem{Mironov:2012uh}
A.~Mironov, A.~Morozov, Y.~Zenkevich, and A.~Zotov, ``{Spectral Duality in
  Integrable Systems from AGT Conjecture},''
  \href{http://dx.doi.org/10.1134/S0021364013010062}{{\em JETP Lett.}
  {\bfseries 97} (2013) 45--51},
  \href{http://arxiv.org/abs/1204.0913}{{\ttfamily arXiv:1204.0913 [hep-th]}}.

\bibitem{Gaiotto:2013bwa}
D.~Gaiotto and P.~Koroteev, ``{On Three Dimensional Quiver Gauge Theories and
  Integrability},'' \href{http://dx.doi.org/10.1007/JHEP05(2013)126}{{\em JHEP}
  {\bfseries 05} (2013) 126}, \href{http://arxiv.org/abs/1304.0779}{{\ttfamily
  arXiv:1304.0779 [hep-th]}}.

\bibitem{Bullimore:2014awa}
M.~Bullimore, H.-C. Kim, and P.~Koroteev, ``{Defects and Quantum Seiberg-Witten
  Geometry},'' \href{http://dx.doi.org/10.1007/JHEP05(2015)095}{{\em JHEP}
  {\bfseries 05} (2015) 095}, \href{http://arxiv.org/abs/1412.6081}{{\ttfamily
  arXiv:1412.6081 [hep-th]}}.

\bibitem{Intriligator:1996ex}
K.~A. Intriligator and N.~Seiberg, ``{Mirror symmetry in three-dimensional
  gauge theories},'' \href{http://dx.doi.org/10.1016/0370-2693(96)01088-X}{{\em
  Phys. Lett. B} {\bfseries 387} (1996) 513--519},
  \href{http://arxiv.org/abs/hep-th/9607207}{{\ttfamily arXiv:hep-th/9607207}}.

\bibitem{deBoer:1996ck}
J.~de~Boer, K.~Hori, H.~Ooguri, Y.~Oz, and Z.~Yin, ``{Mirror symmetry in
  three-dimensional theories, SL(2,Z) and D-brane moduli spaces},''
  \href{http://dx.doi.org/10.1016/S0550-3213(97)00115-6}{{\em Nucl. Phys. B}
  {\bfseries 493} (1997) 148--176},
  \href{http://arxiv.org/abs/hep-th/9612131}{{\ttfamily arXiv:hep-th/9612131}}.

\bibitem{Kapustin:2006pk}
A.~Kapustin and E.~Witten, ``{Electric-Magnetic Duality And The Geometric
  Langlands Program},''
  \href{http://dx.doi.org/10.4310/CNTP.2007.v1.n1.a1}{{\em Commun. Num. Theor.
  Phys.} {\bfseries 1} (2007) 1--236},
  \href{http://arxiv.org/abs/hep-th/0604151}{{\ttfamily arXiv:hep-th/0604151}}.

\bibitem{Aganagic:2017smx}
M.~Aganagic, E.~Frenkel, and A.~Okounkov, ``{Quantum $q$-Langlands
  Correspondence},'' \href{http://dx.doi.org/10.1090/mosc/278}{{\em Trans.
  Moscow Math. Soc.} {\bfseries 79} (2018) 1--83},
  \href{http://arxiv.org/abs/1701.03146}{{\ttfamily arXiv:1701.03146
  [hep-th]}}.

\bibitem{2020arXiv201108603D}
H.~{Dinkins}, ``{3d mirror symmetry of the cotangent bundle of the full flag
  variety},'' \href{http://arxiv.org/abs/2011.08603}{{\ttfamily
  arXiv:2011.08603 [math.AG]}}.

\bibitem{Aprile:2018oau}
F.~Aprile, S.~Pasquetti, and Y.~Zenkevich, ``{Flipping the head of $T[SU(N)]$:
  mirror symmetry, spectral duality and monopoles},''
  \href{http://dx.doi.org/10.1007/JHEP04(2019)138}{{\em JHEP} {\bfseries 04}
  (2019) 138}, \href{http://arxiv.org/abs/1812.08142}{{\ttfamily
  arXiv:1812.08142 [hep-th]}}.

\bibitem{Aganagic:2001uw}
M.~Aganagic, K.~Hori, A.~Karch, and D.~Tong, ``{Mirror symmetry in
  (2+1)-dimensions and (1+1)-dimensions},''
  \href{http://dx.doi.org/10.1088/1126-6708/2001/07/022}{{\em JHEP} {\bfseries
  07} (2001) 022}, \href{http://arxiv.org/abs/hep-th/0105075}{{\ttfamily
  arXiv:hep-th/0105075}}.

\bibitem{Koroteev:2021lvp}
P.~Koroteev and A.~M. Zeitlin, ``{3d Mirror Symmetry for Instanton Moduli
  Spaces},'' \href{http://arxiv.org/abs/2105.00588}{{\ttfamily arXiv:2105.00588
  [math.AG]}}.

\bibitem{Affleck:2021ypq}
I.~Affleck, D.~Bykov, and K.~Wamer, ``{Flag manifold sigma models: spin chains
  and integrable theories},'' \href{http://arxiv.org/abs/2101.11638}{{\ttfamily
  arXiv:2101.11638 [hep-th]}}.

\bibitem{maffei2005quiver}
A.~{Maffei}, ``{Quiver varieties of type A},''
  \href{http://arxiv.org/abs/math/9812142}{{\ttfamily arXiv:math/9812142
  [math.AG]}}.

\bibitem{CautisKamnitzerII}
S.~{Cautis} and J.~{Kamnitzer}, ``{Knot homology via derived categories of
  coherent sheaves II, $\mathfrak{sl}_m$ case},''
  \href{http://dx.doi.org/10.1007/s00222-008-0138-6}{{\em Inventiones
  Mathematicae} {\bfseries 174} no.~1, (July, 2008) 165--232},
  \href{http://arxiv.org/abs/0710.3216}{{\ttfamily arXiv:0710.3216 [math.AG]}}.

\bibitem{Mirkovic_Vybornov}
I.~{Mirkovi{\'c}} and M.~{Vybornov}, ``{On quiver varieties and affine
  Grassmannians of type A},''
  \href{http://arxiv.org/abs/math/0206084}{{\ttfamily arXiv:math/0206084
  [math.AG]}}.

\bibitem{Khovanov:2004bc}
M.~Khovanov and L.~Rozansky, ``{Topological Landau-Ginzburg models on a
  world-sheet foam},''
  \href{http://dx.doi.org/10.4310/ATMP.2007.v11.n2.a2}{{\em Adv. Theor. Math.
  Phys.} {\bfseries 11} no.~2, (2007) 233--259},
  \href{http://arxiv.org/abs/hep-th/0404189}{{\ttfamily arXiv:hep-th/0404189}}.

\bibitem{2015arXiv150803010S}
E.~{Smirnov}, ``{Grassmannians, flag varieties, and Gelfand-Zetlin
  polytopes},'' \href{http://arxiv.org/abs/1508.03010}{{\ttfamily
  arXiv:1508.03010 [math.AG]}}.

\bibitem{Dijkgraaf:2009pc}
R.~Dijkgraaf and C.~Vafa, ``{Toda Theories, Matrix Models, Topological Strings,
  and N=2 Gauge Systems},'' \href{http://arxiv.org/abs/0909.2453}{{\ttfamily
  arXiv:0909.2453 [hep-th]}}.

\bibitem{Itoyama:2009sc}
H.~Itoyama, K.~Maruyoshi, and T.~Oota, ``{The Quiver Matrix Model and 2d-4d
  Conformal Connection},'' \href{http://dx.doi.org/10.1143/PTP.123.957}{{\em
  Prog. Theor. Phys.} {\bfseries 123} (2010) 957--987},
  \href{http://arxiv.org/abs/0911.4244}{{\ttfamily arXiv:0911.4244 [hep-th]}}.

\bibitem{Eguchi:2009gf}
T.~Eguchi and K.~Maruyoshi, ``{Penner Type Matrix Model and Seiberg-Witten
  Theory},'' \href{http://dx.doi.org/10.1007/JHEP02(2010)022}{{\em JHEP}
  {\bfseries 02} (2010) 022}, \href{http://arxiv.org/abs/0911.4797}{{\ttfamily
  arXiv:0911.4797 [hep-th]}}.

\bibitem{Schiappa:2009cc}
R.~Schiappa and N.~Wyllard, ``{An A(r) threesome: Matrix models, 2d CFTs and 4d
  N=2 gauge theories},'' \href{http://dx.doi.org/10.1063/1.3449328}{{\em J.
  Math. Phys.} {\bfseries 51} (2010) 082304},
  \href{http://arxiv.org/abs/0911.5337}{{\ttfamily arXiv:0911.5337 [hep-th]}}.

\bibitem{Mironov:2010ym}
A.~Mironov, A.~Morozov, and A.~Morozov, ``{Conformal blocks and generalized
  Selberg integrals},''
  \href{http://dx.doi.org/10.1016/j.nuclphysb.2010.10.016}{{\em Nucl. Phys. B}
  {\bfseries 843} (2011) 534--557},
  \href{http://arxiv.org/abs/1003.5752}{{\ttfamily arXiv:1003.5752 [hep-th]}}.

\bibitem{Mironov:2011jn}
A.~Mironov, A.~Morozov, A.~Popolitov, and S.~Shakirov, ``{Resolvents and
  Seiberg-Witten representation for Gaussian beta-ensemble},''
  \href{http://dx.doi.org/10.1007/s11232-012-0049-y}{{\em Theor. Math. Phys.}
  {\bfseries 171} (2012) 505--522},
  \href{http://arxiv.org/abs/1103.5470}{{\ttfamily arXiv:1103.5470 [hep-th]}}.

\bibitem{CautisKamnitzerIV}
S.~{Cautis} and J.~{Kamnitzer}, ``{Knot homology via derived categories of
  coherent sheaves IV, coloured links},'' {\em arXiv e-prints} (2014) ,
  \href{http://arxiv.org/abs/1410.7156}{{\ttfamily arXiv:1410.7156 [math.AG]}}.

\bibitem{MR1492989}
A.~Klimyk and K.~Schm\"{u}dgen,
  \href{http://dx.doi.org/10.1007/978-3-642-60896-4}{{\em Quantum groups and
  their representations}}.
\newblock Texts and Monographs in Physics. Springer-Verlag, Berlin, 1997.
\newblock \url{https://doi.org/10.1007/978-3-642-60896-4}.

\bibitem{Mironov:2010su}
A.~Mironov, A.~Morozov, and S.~Shakirov, ``{On 'Dotsenko-Fateev' representation
  of the toric conformal blocks},''
  \href{http://dx.doi.org/10.1088/1751-8113/44/8/085401}{{\em J. Phys. A}
  {\bfseries 44} (2011) 085401},
  \href{http://arxiv.org/abs/1010.1734}{{\ttfamily arXiv:1010.1734 [hep-th]}}.

\bibitem{Galakhov:2015fna}
D.~Galakhov, D.~Melnikov, A.~Mironov, and A.~Morozov, ``{Knot invariants from
  Virasoro related representation and pretzel knots},''
  \href{http://dx.doi.org/10.1016/j.nuclphysb.2015.07.035}{{\em Nucl. Phys. B}
  {\bfseries 899} (2015) 194--228},
  \href{http://arxiv.org/abs/1502.02621}{{\ttfamily arXiv:1502.02621
  [hep-th]}}.

\bibitem{Costello:2020ndc}
K.~Costello, T.~Dimofte, and D.~Gaiotto, ``{Boundary Chiral Algebras and
  Holomorphic Twists},'' \href{http://arxiv.org/abs/2005.00083}{{\ttfamily
  arXiv:2005.00083 [hep-th]}}.

\bibitem{Carqueville:2011zea}
N.~Carqueville and D.~Murfet, ``{Computing Khovanov\textendash{}Rozansky
  homology and defect fusion},''
  \href{http://dx.doi.org/10.2140/agt.2014.14.489}{{\em Algebr. Geom. Topol.}
  {\bfseries 14} no.~1, (2014) 489--537},
  \href{http://arxiv.org/abs/1108.1081}{{\ttfamily arXiv:1108.1081 [math.QA]}}.

\bibitem{Dedushenko:2018icp}
M.~Dedushenko, Y.~Fan, S.~S. Pufu, and R.~Yacoby, ``{Coulomb Branch
  Quantization and Abelianized Monopole Bubbling},''
  \href{http://dx.doi.org/10.1007/JHEP10(2019)179}{{\em JHEP} {\bfseries 10}
  (2019) 179}, \href{http://arxiv.org/abs/1812.08788}{{\ttfamily
  arXiv:1812.08788 [hep-th]}}.

\bibitem{reshetikhin1991invariants}
N.~Reshetikhin and V.~G. Turaev, ``Invariants of 3-manifolds via link
  polynomials and quantum groups,'' {\em Inventiones mathematicae} {\bfseries
  103} no.~1, (1991) 547--597.

\bibitem{Wang:2020lqp}
D.~Wang, ``{Monopoles and Landau-Ginzburg Models I},''
  \href{http://arxiv.org/abs/2004.06227}{{\ttfamily arXiv:2004.06227
  [math.GT]}}.

\bibitem{Rapcak:2020ueh}
M.~Rapcak, Y.~Soibelman, Y.~Yang, and G.~Zhao, ``{Cohomological Hall algebras
  and perverse coherent sheaves on toric Calabi-Yau 3-folds},''
  \href{http://arxiv.org/abs/2007.13365}{{\ttfamily arXiv:2007.13365
  [math.QA]}}.

\bibitem{Nekrasov:2018xsb}
N.~Nekrasov and N.~Piazzalunga, ``{Magnificent Four with Colors},''
  \href{http://dx.doi.org/10.1007/s00220-019-03426-3}{{\em Commun. Math. Phys.}
  {\bfseries 372} no.~2, (2019) 573--597},
  \href{http://arxiv.org/abs/1808.05206}{{\ttfamily arXiv:1808.05206
  [hep-th]}}.

\bibitem{Kanno:2020ybd}
H.~Kanno, ``{Quiver matrix model of ADHM type and BPS state counting in diverse
  dimensions},'' \href{http://dx.doi.org/10.1093/ptep/ptaa079}{{\em PTEP}
  {\bfseries 2020} no.~11, (2020) 11B104},
  \href{http://arxiv.org/abs/2004.05760}{{\ttfamily arXiv:2004.05760
  [hep-th]}}.

\bibitem{Cao:2019tvv}
Y.~Cao, M.~Kool, and S.~Monavari, ``{$K$-theoretic DT/PT correspondence for
  toric Calabi-Yau 4-folds},''
  \href{http://arxiv.org/abs/1906.07856}{{\ttfamily arXiv:1906.07856
  [math.AG]}}.

\bibitem{Oh:2020rnj}
J.~Oh and R.~P. Thomas, ``{Counting sheaves on Calabi-Yau 4-folds, I},''
  \href{http://arxiv.org/abs/2009.05542}{{\ttfamily arXiv:2009.05542
  [math.AG]}}.

\bibitem{Gradshteyn:1702455}
I.~S. Gradshteyn, I.~M. Ryzhik, D.~Zwillinger, and V.~Moll,
  \href{http://dx.doi.org/0123849330}{{\em {Table of integrals, series, and
  products; 8th ed.}}}
\newblock Academic Press, Amsterdam, Sep, 2014.
\newblock \url{https://cds.cern.ch/record/1702455}.

\bibitem{Wess:1992cp}
J.~Wess and J.~Bagger, {\em {Supersymmetry and supergravity}}.
\newblock Princeton University Press, Princeton, NJ, USA, 1992.

\bibitem{Hori:2000ic}
K.~Hori, ``{Linear models of supersymmetric D-branes},'' in {\em {KIAS Annual
  International Conference on Symplectic Geometry and Mirror Symmetry}}.
\newblock 12, 2000.
\newblock \href{http://arxiv.org/abs/hep-th/0012179}{{\ttfamily
  arXiv:hep-th/0012179}}.

\bibitem{Watari_lectures}
T.~Watari, ``An algebraic geometry primer (for physicists).'' Lecture course
  ``{I}ntroductory {A}lgebraic {G}eometry for {Y}oung {S}tring {T}heorists{"}
  available at \url{
  http://member.ipmu.jp/taizan.watari/Lectures/16a-KIPMU-ag4string/16a-KIPMU-ag4string.html}
  (2021/04/22).

\bibitem{Cecotti:1991me}
S.~Cecotti and C.~Vafa, ``{Topological antitopological fusion},''
  \href{http://dx.doi.org/10.1016/0550-3213(91)90021-O}{{\em Nucl. Phys. B}
  {\bfseries 367} (1991) 359--461}.

\bibitem{Gunning_Rossi}
R.~C. Gunning and H.~Rossi, {\em Analytic functions of several complex
  variables}.
\newblock Prentice-Hall, Inc., Englewood Cliffs, N.J., 1965.

\bibitem{Galier}
J.~Gallier and J.~Quaintance, ``{H}omology, {C}ohomology, and {S}heaf
  cohomology.'' Available at
  \url{https://www.cis.upenn.edu/~jean/gbooks/sheaf-coho.html} (2021/04/22).

\bibitem{Hayashi:2008ba}
H.~Hayashi, R.~Tatar, Y.~Toda, T.~Watari, and M.~Yamazaki, ``{New Aspects of
  Heterotic--F Theory Duality},''
  \href{http://dx.doi.org/10.1016/j.nuclphysb.2008.07.031}{{\em Nucl. Phys. B}
  {\bfseries 806} (2009) 224--299},
  \href{http://arxiv.org/abs/0805.1057}{{\ttfamily arXiv:0805.1057 [hep-th]}}.

\bibitem{Thomas:1999ic}
R.~P. Thomas, ``{Derived categories for the working mathematician},'' {\em
  AMS/IP Stud. Adv. Math.} {\bfseries 23} (2001) 349--361,
  \href{http://arxiv.org/abs/math/0001045}{{\ttfamily arXiv:math/0001045}}.

\bibitem{2019arXiv191003010I}
M.~S. {Im}, C.-J. {Lai}, and A.~{Wilbert}, ``{Irreducible components of two-row
  Springer fibers and Nakajima quiver varieties},''
  \href{http://arxiv.org/abs/1910.03010}{{\ttfamily arXiv:1910.03010
  [math.RT]}}.

\end{thebibliography}\endgroup

\end{document}